\documentclass[a4paper,12pt]{article}
\pdfoutput=1
\usepackage{epsfig,amsmath,amsfonts,amsthm}
\usepackage[figuresright]{rotating}
\usepackage[left=28mm,right=20mm,top=30mm,bottom=20mm]{geometry}
\usepackage{subfig}
\usepackage{graphicx}
\usepackage{rotating}
\usepackage{booktabs}
\usepackage{moreverb}

\DeclareGraphicsExtensions{.pdf,.png,.jpg}
\usepackage{natbib}
\numberwithin{equation}{section}

\def\e{\hbox{E}}

\def\var{\hbox{Var}}

\def\max{\hbox{max}}

\def\Bin{\hbox{Bin}}
\def\logit{\hbox{logit}}
\def\expit{\hbox{expit}}

\providecommand{\keywords}[1]{\textbf{{Keywords}} #1}
\usepackage[nokeyprefix]{refstyle}
\usepackage{varioref}
\usepackage{xr}
\usepackage{hyperref}

\begin{document}

\title{Simulations for estimation of heterogeneity variance $\tau^2$ in constant and inverse variance weights meta-analysis of log-odds-ratios}

\author{Elena Kulinskaya and  David C. Hoaglin }

\date{\today}

\maketitle

\begin{abstract}
A number of  popular estimators of the between-study variance, $\tau^2$, are based on the  Cochran's $Q$ statistic for testing heterogeneity in meta analysis.
We introduce new point and interval estimators of $\tau^2$ for log-odds-ratio. These include new DerSimonian-Kacker-type moment estimators based on the first moment of  $Q_F$, the $Q$ statistic with effective-sample-size weights, and  novel median-unbiased estimators. We study, by simulation, bias and coverage of these new  estimators of $\tau^2$ and, for comparative purposes, bias and coverage of a number of well-known estimators based on the $Q$ statistic with inverse-variance weights, $Q_{IV}$, such as the Mandel-Paule, DerSimonian-Laird, and restricted-maximum-likelihood estimators, and an estimator based on the Kulinskaya-Dollinger (2015) improved approximation to $Q_{IV}$.

% new DerSimonian-Kacker-type moment estimators based on the first moment of  $Q_F$, and a novel median-unbiased estimator.

%The results show that: an approximation based on an algorithm of Farebrother follows both the null and the alternative  distributions of $Q_F$ reasonably well, whereas the usual chi-square approximation for the null distribution of $Q_{IV}$  and the Biggerstaff-Jackson approximation to its alternative distribution are poor;
%he test for heterogeneity based on $Q_F$ and F SW has error rates that somewhat exceed the nominal 5\%, and the test based on $Q_{IV}$ and the chi-square approximation has error rates that are noticeably too low;
%in estimating $\tau^2$ our moment estimator  based on $Q_F$ is almost unbiased, the Mandel-Paule estimator has some negative bias in some situations, and the DerSimonian-Laird and restricted-maximum-likelihood estimators have considerable negative bias; and all 95\% interval estimators have coverage that is too high when $\tau^2 =0$, but otherwise the Q-profile interval performs very well.
\end{abstract}


\section{Introduction}

When the individual studies in a meta-analysis report binary outcomes in the treatment and control arms, the most common measure of effect is the odds ratio (OR) or its log (LOR).  % The LOR is popular in medical research, but some substantive arguments favor the relative risk or risk ratio (RR) . Popular measures of effect also include the risk difference (RD).

In studying estimation of the overall effect in random-effects meta-analyses of the mean difference (MD),  the standardized mean difference (SMD), and LOR, we found that SSW, a weighted mean whose weights involve only the studies' arm-level sample sizes, performed well, avoiding shortcomings associated with estimators that use inverse-variance weights based on estimated variances (\cite{BHK2018SMD, BHK2020LOR}).

We also previously studied $Q_F$, a version of Cochran's $Q$ statistic \citep{cochran1954combination} for assessment of heterogeneity  that uses those constant weights. That work  draws on favorable results for $Q$ with effective-sample-size-based weights when the measure of effect is the mean difference (MD) (\cite{Qfixed}) or the standardized mean difference (SMD) (\cite{Qfixed}). Here we introduce and investigate several new estimators of $\tau^2$ for LOR, based on $Q_F$. These include new DerSimonian-Kacker-type moment estimators based on the first moment of  $Q_F$, the $Q$ statistic with effective-sample-size weights, and novel median-unbiased estimators. % RR, and RD.% For meta-analysis of LOR, from $Q_F$ we also derive new point and interval estimators of the between-study variance, $\tau^2$.

%Simulation of the actual distribution of $Q$ for LOR, RR, and RD enables us to study the accuracy of approximations for its null distribution ($\tau^2 = 0$) and  the empirical level when $\tau^2 = 0$. % and when $\tau^2 > 0$.
We  study, by simulation, the bias of point estimators of $\tau^2$ and the coverage of confidence intervals for $\tau^2$. For comparison we include the usual version of $Q$ ($Q_{IV}$), which uses inverse-variance weights, and familiar point and interval estimators of $\tau^2$.

%Section~\ref{sec:EffectLOR} briefly reviews study-level estimation of LOR, RD, and RR. Section~\ref{sec:REM} provides a detailed introduction to random-effects models for LOR. Section~\ref{sec:REMandQ} reviews the general random-effects model and describes the $Q$ statistic.  Section~\ref{sec:Approx} discusses approximations to the distributions of $Q_F$ and $Q_{IV}$. Section~\ref{sec:PtandIntvl} introduces new point and interval estimators of $\tau^2$ for LOR. Section~\ref{simul} describes the simulation design and summarizes the results. Section~\ref{sec:Example} examines an example of meta-analysis using LOR and RR. Section~\ref{sec:Discussion} offers a summary and discussion. %An appendix gives the derivation of conditional and unconditional moments of Hedges's estimator of study-level SMD.
\section{Study-level estimation of log-odds-ratio} \label{sec:EffectLOR}

Consider $K$ studies that used a particular individual-level binary outcome.
Study $i$ ($i = 1, \ldots, K$) reports $X_{iT}$ and $X_{iC}$, the numbers of events in the $n_{iT}$ subjects in the Treatment arm and the $n_{iC}$ subjects in the Control arm. It is customary to treat $X_{iT}$ and $X_{iC}$ as independent binomial variables:
\begin{equation} \label{eq:binomialXs}
X_{iT}\sim {\Bin}(n_{iT},p_{iT})\qquad \text{and}\qquad X_{iC}\sim {\Bin}(n_{iC},p_{iC}).
\end{equation}
The log-odds-ratio for Study $i$ is
\begin{equation} \label{eq:psi}
\theta_{i} = \log_{e} \left(\frac{p_{iT}(1 - p_{iC})} {p_{iC}(1 - p_{iT})}\right) \qquad \text{estimated by} \qquad
\hat\theta_{i} = \log_{e} \left(\frac{\hat p_{iT}(1 - \hat p_{iC})} {\hat p_{iC}(1 - \hat p_{iT})}\right).
\end{equation}

As inputs, a two-stage meta-analysis uses estimates of the $\theta_i$ ($\hat{\theta}_i$) and estimates of their variances ($\hat{v}_i^2$).  The maximum likelihood estimator of $p$ is $\tilde p=X/n$. It is helpful to have an unbiased estimator of $\theta$.
%For a binomial random variable $X \sim {\Bin}(n,p)$, the maximum-likelihood (ML) estimator of $p$ is  $\tilde p = X/n$ .  \cite{Bohning2005}  studied estimators of $p$ of the form $(X+a)/(n+2a)$.
The use of  $\hat p = (X + 0.5)/(n + 1)$ eliminates $O(1/n)$ bias and provides the least biased estimator of log-odds (\cite{G-P-T-1985}).
    	% and also Staudte (2012), who also suggests using $\tilde p=(X+1)/(n+2)$ when estimating  variances both for LOR and for LRR.
We use $\hat p$ when estimating LOR in the $Q_F$.

The (conditional, given $p_{ij}$ and $n_{ij}$) asymptotic variance of $\hat{\theta}_i$, derived by the delta method, is
\begin{equation} \label{eq:sigma}
v_{i}^2 = {\var}(\hat{\theta}_{i}) = \frac{1} {n_{iT} {p}_{iT} (1 - {p}_{iT})} + \frac{1}{n_{iC} {p}_{iC} (1 - {p}_{iC})},
\end{equation}
estimated by substituting $\hat{p}_{ij}$ for $p_{ij}$. This estimator of the variance is unbiased in large samples, but \cite{G-P-T-1985} note that it overestimates the variance for small sample sizes.
%, and recommend the use of $\\bar p=(X+1)/(n+2)$ in this case.%
%(In analyses, we follow the particular method's procedure for calculating $\hat{p}_{ij}$.)
%\cite{G-P-T-1985} also give approximate conditional higher moments of LOR. % \cite{kulinskaya2015accurate} implemented exact calculation of conditional central moments of $\logit(\hat p)$ for LOR. We follow this approach in our simulations. % through the routine given below.

%\section{Study-level estimation of risk difference and relative risk} \label{sec:RDandRR}

%We consider the  binomial data generation model ({\ref{eq:binomialXs}). % For simplicity of notation, we delete the subscript $i$ denoting the individual studies.

%All three binary effect measures (RD, LRR, and LOR) have the form $\eta = h(p_T) - h(p_C)$, where the function $h(p) = p$ for RD, $h(p) = \log(p)$ for LRR, and $h(p) = \logit(p)$ for LOR. This relation facilitates calculation of conditional moments of $\hat\eta$ from the moments of $h(p)$. The Appendix gives the details.  Even when they are not available in closed form,
Our simulations yield an exact calculation of conditional central moments of LOR, following  the implementation of \cite{kulinskaya2015accurate}.

\section{Random-effects model and the $Q$ statistic} \label{sec:REMandQ}

We consider a generic random-effects model (REM): For Study $i$ ($i = 1,\ldots,K$)  the estimate of the effect is $\hat\theta_i \sim G(\theta_i, v_i^2)$, where the effect-measure-specific distribution $G$ has mean $\theta_i$ and variance $v_i^2$, and $\theta_i \sim N(\theta, \tau^2)$.
Thus, the $\hat{\theta}_i$ are unbiased estimates of the true conditional effects $\theta_i$, and the $v_i^2 = \var(\hat{\theta}_i | \theta_i)$ are the true conditional variances.

Cochran's $Q$ statistic is a weighted sum of the squared deviations of the estimated effects $\hat\theta_i$ from their weighted mean $\bar\theta_w = \sum w_i\hat\theta_i / \sum w_i$:
\begin{equation} \label{Q}
Q=\sum w_i (\hat{\theta}_i - \bar{\theta}_w)^2.
\end{equation}
In \cite{cochran1954combination} $w_i$ is the reciprocal of the \textit{estimated} variance of $\hat{\theta}_i$, hence the notation $Q_{IV}$. In meta-analysis those $w_i$ come from the fixed-effect model.
In what follows, we derive estimators of $\tau^2$ based on  $Q_F$, discussed by \cite{dersimonian2007random} and further studied by \cite{Qfixed}, in which the $w_i$ are arbitrary positive constants. A particular choice of the fixed weights which we usually specify in $Q_F$  is $w_i = \tilde{n}_i$, where  $\tilde{n}_i = n_{iC} n_{iT} / n_i$, the effective sample size in Study $i$ ($n_i = n_{iC} + n_{iT}$).

Define $W = \sum w_i$,  $q_i = w_i / W$, and $\Theta_i = \hat\theta_i - \theta$.  In this notation, and expanding $\bar\theta_w$, Equation (\ref{Q}) can be written  as
\begin{equation} \label{Q1}
Q = W \left[ \sum q_i (1 - q_i) \Theta_i^2 - \sum_{i \not = j} q_i q_j \Theta_i \Theta_j \right].
\end{equation}
We distinguish between the conditional distribution of $Q$ (given the $\theta_i$) and the unconditional distribution, and the respective moments of $\Theta_i$. For instance, the conditional second moment of $\Theta_i$ is $M_{2i}^c = v_i^2$, and the unconditional second moment  is  $M_{2i} = \e(\Theta_i^2) = \var(\hat{\theta}_i) = \e(v_i^2)$, given in Equation~(\ref{M_2i_F}).

Under the above REM, it is straightforward to obtain the first moment of $Q_F$ as
\begin{equation} \label{M1Q}
\e(Q_F) = W \left[ \sum q_i (1 - q_i) \var(\Theta_i) \right] = W \left[ \sum q_i (1 - q_i) (\e(v_i^2)) \right].
\end{equation}
This expression is similar to Equation (4) in \cite{dersimonian2007random};  they use the conditional variance  $v_i^2+ \tau^2$ instead of the unconditional mean $\e(v_i^2)$.

%\cite{Qfixed} also provide expressions for the second and third moments of $Q_F$, but these moments require higher moments of $\Theta$, up to the fourth and the sixth moments, respectively.

%In the fixed-intercept models (i.e., when the $p_{iC}$ are fixed), assuming also homogeneity of effects ($\tau^2 = 0$), unconditional and conditional moments of each binary effect measure coincide. Therefore, the conditional moments of $Q$  are sufficient to obtain a moment-based approximation to the distribution of $Q_F$ under homogeneity.

The full (unconditional) variance of $\hat{\theta}_i$
depends on the generation mechanism and was derived in \cite{BHK2021SIM} %
  under  a bivariate normal model  for $(\logit(p_{iC}), \logit(p_{iT}))$.%  with variances $\sigma^2, \tau^2$ and correlation $\rho$, where $c$ defines so called type 1 or 2 model with $c=0$ or $c=1/2$.

In the conventionally assumed fixed-intercept model (i.e., when the $p_{iC}$ are fixed),% with $c = 0$,% $\sigma^2 = 0$ and  $\rho = 0$, resulting in
\begin{equation}\label{M_2i_F}
\e(v_i^2) = \frac{1}{n_{iT} \breve{p}_{iT} (1 - \breve{p}_{iT})} + \frac{1}{n_{iC} {p}_{iC} (1 - {p}_{iC})} + \tau^2 (1+
  \frac{1}{2 n_{iT}} \left( [ \breve{p}_{iT} (1 - \breve{p}_{iT})]^{-1} -2\right)))
\end{equation}
  for   $\breve{p}_{iT}=\expit (\alpha_i+\theta)$. This unconditional variance can be estimated by substituting $\hat p_{iC}$ for $p_{iC}$ and $\expit(\hat\alpha_i+\hat\theta)$ for $\breve{p}_{iT}$, where  $\hat\alpha_i=\logit (\hat p_{iC})$ and   $\hat\theta$ is the  estimated LOR. We refer to this  estimate of $p_{iT}$ as {\bf model-based}. Alternatively, a {\bf na\"{i}ve} estimate would use $\hat p_{iT}$. An advantage of such a na\"{i}ve estimate is that it maintains the variance inflation of $\e(v_i^2)$ in comparison with $v_i^2+\tau^2$.

\section{Approximations to the distributions of $Q_F$ and $Q_{IV}$} \label{sec:Approx}

Standard approximation to the distribution of $Q_{IV}$ is the chi-square distribution with $K-1$ degrees of freedom. A few estimators of $\tau^2$, such as DerSimonisn-Laird and Mandel-Paule, are based on the chi-square moments. However, it is well known that this approximation is not satisfactory for small to moderate sample sizes.
 \cite{kulinskaya2015accurate} provided an improved approximation to the null distribution of $Q_{IV}$ based on a two-moment gamma approximation; we  study  point and interval estimators of $\tau^2$ based on this approximation, denoted by KD.

%For meta-analysis of mean differences (MD), \cite{Qfixed} considered the distribution of $Q_F$, a quadratic form in normal variables, which has the form $Q = \Theta^{T}A\Theta$ for a symmetric matrix $A$ of rank $K - 1$. Because the vector $\Theta$ has a multivariate normal distribution, $N(\mu,\Sigma)$, the distribution of $Q_F$ can be evaluated by the algorithm of \cite{Farebrother1984} (after determining the eigenvalues of $A \Sigma$ and some other inputs). %If the variances in $\Sigma$ are the true variances, Farebrother's algorithm evaluates the exact distribution of $Q$.
% In practice, it is necessary to plug in estimated variances. The resulting approximation is quite accurate for MD.  %\cite{Qfixed} also considered a two-moment approximation and a three-moment approximation. %The three-moment approximation regularly encountered numerical problems, so we do not include it here.

For LOR, $Q_F$ is a quadratic form in asymptotically normal variables.  The Farebrother algorithm \citep{Farebrother1984}, applicable for quadratic forms in normal variables,  may provide a satisfactory approximation for larger sample sizes, though it may not behave well for small $n$. To apply it, we  plug in  estimated variances.  We study point and interval estimators of $\tau^2$ based on this approximation to the distribution of $Q_F$.  %  We investigate the quality of that approximation, which we denote by F SSW, and the two-moment  approximation (2M SSW), which is based on the gamma distribution.
%We investigate two approaches to estimation of $p_{iT}$ to plug into the calculation of the second and the fourth central moments of LOR: a \lq na\"{i}ve'  approach, where $p_{iT}$ is estimated from $X_{iT}$ and $n_{iT}$, and a \lq model-based' approach, which uses the relation $\hat h(p_{iT})=\hat h(p_{iC})+\bar\theta$ for a fixed-weights mean effect $\bar\theta$. % Thus, we study four new approximations to the null distribution of $Q_F$: F SSW Na\"{i}ve, F SSW Model, 2M SSW Na\"{i}ve, and  2M SSW Model.
  %\cite{biggerstaff2008exact} used the Farebrother approximation to the distribution of a quadratic form in normal variables as the ``exact'' distribution of $Q_{IV}$.  \cite{Jacksen2014metaregr} extended this approach to a $Q$ with  arbitrary weights in a meta-regression setting. % We denote this approximation by BJ.
%When $\tau^2 = 0$, the \cite{biggerstaff2008exact} approximation to the distribution of $Q_{IV}$ is the $\chi^2_{K - 1}$  distribution. % For comparison, our simulations include these three approximations.

\section{Point and interval estimators of $\tau^2$ for LOR} \label{sec:PtandIntvl}

\subsection{Point estimators}

The unconditional variance of $\hat\theta$ in Equation~(\ref{M_2i_F})  can be written as a sum of two terms,
\begin{equation}\label{eq:M2iLOR} \e(v_i^2)=v_i^2+\tau^2(1+
  \frac{1}{2 n_{iT}} \left( [ {p}_{iT} (1 - {p}_{iT})]^{-1} -2\right)).\end{equation}
%  Therefore, we  substitute $C_i= (1+
 % \frac{1}{2 n_{iT}} \left( [ {p}_{T} (1 - {p}_{T})]^{-1} -2\right))$ when calculating the SSU estimator of $\tau^2$ in (1.4) for the case of LOR.
Rearranging the terms in Equation~(\ref{Q1}) gives the moment-based estimator of $\tau^2$:

  \begin{equation} \label{tau_SSU}
\hat\tau^2_{U} = \frac{Q / W - \sum q_i (1 - q_i) \hat v_i^2} {\sum q_i (1 - q_i)C_i},
\end{equation}
for $C_i= 1+
  \frac{1}{2 n_{iT}} \left( [ \hat{p}_{iT} (1 - \hat{p}_{iT})]^{-1} -2\right)$.

\cite{dersimonian2007random} obtain a similar result; they use the conditional estimate, $\hat v_i^2+\hat \tau^2$, instead of the unconditional estimate, $\widehat \e(v_i^2)$, obtaining
\begin{equation} \label{tau_DSK}
\hat\tau^2_{M} = \frac{Q / W - \sum q_i (1 - q_i) \hat v_i^2} {\sum q_i (1 - q_i)}.
\end{equation}
%For MD the two estimators are the same, because then $\e(v_i^2) = v_i^2$.
We study both estimators with effective-sample-size weights. With the conditional estimated variances in Equation~(\ref{tau_DSK}), we denote the estimator by SSC for \lq\lq Sample Sizes Conditional"; with the unconditional estimated variances, as in Equation~(\ref{tau_SSU}), it is SSU for \lq\lq Sample Sizes Unconditional".  These estimators differ by a term of order $O(1/n_i)$ and will be very similar for large sample sizes.

The estimators $\hat{\tau}^2_{U}$ and $\hat\tau^2_{M}$
arose from setting the observed value of $Q$ equal to its expected value and solving for $\tau^2$. Instead of the expected value, one could use the median of the distribution of $Q$ given $\tau^2$ (\cite{QfixedSMD, Brown_1947_AMSX_582, ViechtMedian}).  If the true (or approximate) cumulative distribution function is $F(\cdot | \tau^2)$, a point estimator of $\tau^2$ can be found as
$$\hat{\tau}^2_{med} = \max(0,\; \{\tau^2: F(Q | \tau^2) = 0.5\} ).$$
In the Farebrother approximation to the distribution of $Q$ (Section~\ref{sec:Approx}), one can use either the conditional estimated variances or the unconditional estimated variances. We denote the resulting estimators by SMC and SMU (\lq\lq Sample sizes Median (Un)Conditional"), respectively. Depending on the chosen estimate of $p_{iT}$ in $\e(v_i^2)$ (\ref{M_2i_F}), we have \lq\lq model-based" or \lq\lq na\"{i}ve" versions of SSU and SMU: SSU model or SSU na\"{i}ve and SMU model or SMU na\"{i}ve. The SSC and the SMC estimators can be obtained from the procedure \textit{rma} in  \textit{metafor} by choosing the method as GENQ or GENQM, respectively, and specifying $nbar$ weights (\cite{metafor}).

For comparison, our simulations (Section~\ref{simul}) include four estimators that use inverse-variance weights: \cite{dersimonian1986meta} (DL), REML, \cite{mandel1970interlaboratory} (MP), and an estimator (KD) based on the work of \cite{kulinskaya2015accurate} and discussed by \cite{BHK2020LOR}. KD uses an improved non-null first moment of $Q$ and has better performance than most other estimators.

A perennial question involves whether analysts should add $1/2$ to each of $X_{iT},\;X_{iC},\; n_{iT}-X_{iT},\; n_{iC}-X_{iC}$ only when one of them is zero, or in all studies. To obtain evidence on this issue, we included the corresponding two versions, \lq\lq only" and \lq\lq always", of DL, REML, MP, SSC, and SMC. (We follow the prevalent practice of omitting \lq\lq double-zero" studies, in which two of those cell counts are zero.)

\subsection{Interval estimators}

Straightforward use of the cumulative distribution function $F(\cdot | \tau^2)$ also yields a $ 100(1 - \alpha)\% $ confidence interval for $\tau^2$: $$\{\tau^2\geq 0: F(Q | \tau^2) \in [\alpha/2, 1 - \alpha/2] \}.$$
We use both the conditional estimated variances and the unconditional estimated variances in the Farebrother approximation to $F$; we refer to the resulting profile estimators as FPC and FPU (\lq\lq Farebrother Profile (Un)Conditional" intervals). \cite{Jackson2013CI} introduced a similar approach using conditional variances. The FPC interval can be obtained from \textit{confint} procedure in \textit{metafor} for GENQ or GENQM objects which used $nbar$ weights (\cite{metafor}).   For FPU intervals, we further distinguish between  \lq\lq model-based" or \lq\lq na\"{i}ve" versions, FPU model or FMU na\"{i}ve, respectively, depending on the chosen estimate of $p_{iT}$ in $M_{2i}$ (\ref{M_2i_F}).

Our simulations (Section~\ref{simul}) also include the Q-profile (QP) interval (\cite{viechtbauer2007confidence}), the profile-likelihood (PL) interval (\cite{Hardy_1996_StatMed_619}), and the KD interval, which is based on the chi-square distribution with the corrected first moment developed by \cite{kulinskaya2015accurate}.

For evidence on whether to add  $1/2$ to all four observed numbers $X_{iT},\;X_{iC},\; n_{iT}-X_{iT},\; n_{iC}-X_{iC}$ only when one of these is zero,  or always, we included two versions of the PL, QP, and FPC methods.

\section{Simulation design  and results} \label{simul}

\subsection{Simulation design}

Our simulation design follows that described in \cite{BHK2020LOR}. Briefly,
we varied five parameters: the overall true effect ($\theta$), the between-studies variance ($\tau^2$), the number of studies ($K$), the studies' total sample size ($n$ or $\bar{n}$), and the probability in the control arm ($p_{iC}$). We kept the proportion of observations in the control arm ($f$) at $1/2$.  %For RD and LRR, we considered only approximations to the null distribution of $Q_F$ ($\tau^2 = 0$). % Table~\ref{tab:altdataSMD} lists the values of each parameter.

The values of LOR $\theta$ (0, 0.1, 0.5, 1, 1.5, and 2) aim to represent the range containing most values encountered in practice. LOR is a symmetric effect measure, so the sign of $\theta$ is not relevant.
The choices of sample sizes corresponding to $\bar{n}$ follow a suggestion of \cite{sanches-2000}, who constructed the studies' sample sizes to have skewness 1.464, which they regarded as typical in behavioral and health sciences. For $K = 5$, Table~\ref{tab:design} lists the sets of five sample sizes. The simulations for $K = 10$ and $K = 30$ used each set of unequal sample sizes twice and six times, respectively. % The meta-analyses studied by Rubio-Aparicio et al. had median study-level sample sizes from 16 to 87.5; within those meta-analyses, the sample sizes varied substantially.

%Many studies allocate subjects equally to the two groups ($f = 1/2$), and rough equality holds more widely (as in the studies analyzed by Rubio-Aparicio et al.). Unequal allocations, either planned or observed, are not uncommon. To investigate potential impacts of such situations, we also used $f = 3/4$, a substantial departure from equality.

%The values of $p_{iC}$ are 0.1, 0.2, and 0.5. %\textbf{Explain why we chose them.}

We generated the true effect sizes $\theta_{i}$ from a normal distribution:  $\theta_{i} \sim N(\theta, \tau^2)$. The  values of $p_{iC}$ and $\theta_i$ defined the probabilities $p_{iT}$, and the counts $X_{iC}$ and $X_{iT}$ were generated from the respective binomial distributions.  We used a total of $10,000$ repetitions for each combination of parameters. We discarded \lq double-zero' or \lq double-$n$' studies and reduced the observed  value of $K$ accordingly. We discarded repetitions with $K<3$ and used the observed number of repetitions for analysis.

We used R statistical software \citep{rrr} for the simulations. We used {\it metafor}  for all methods of interest that it implemented. The user-friendly R programs implementing our methods are available in \cite{Rprogs_tau2_Kulinskaya_Hoaglin_2022}.

%{\bf Instead of this Section (mostly copied from MD-SMD paper), we can just say that our simulations are described in MD-SMD paper. LET US DO THIS! But we probably still want to include the table and write a para.}
%, and obtain their estimated within-study variances from Equation~(\ref{eq:g_var}).

%We study eight point estimators of $\tau^2$ (DL, REML, MP,  KDB, SC, SU, SMEDC and SMEDU) and five interval estimators of $\tau^2$ (QP, PL, KDB, FPC and FPU).

\begin{table}[ht]
	\caption{ \label{tab:design} \emph{Values of parameters in the simulations}}
	\begin{footnotesize}
		\begin{center}
			\begin{tabular}
				{|l|l|l|}
				\hline
				Parameter & Equal study sizes & Unequal study sizes \\
                                	& & \\
				\hline
				$K$ (number of studies) & 5, 10, 30 & 5, 10, 30 \\
				\hline
				$n$ or $\bar{n}$ (average (individual) study size ---  & 20, 40, 100, 250 & 30 (12,16,18,20,84), \\
				total of the two arms) &  & 60 (24,32,36,40,168), \\
				For  $K = 10$ and $K = 30$, the same set of unequal & & 100 (64,72,76,80,208), \\
				 study sizes is used twice or six times, respectively. & &160 (124,132,136,140,268) \\
                			\hline
				$f$ (proportion of observations in the control arm) & 1/2 &1/2  \\
				\hline
                			$p_{iC}$ (probability in the control arm) & .1, .2, .5 & .1, .2, .5 \\
                			\hline
				$\theta$ (true value of LOR) & 0, 0.5, 1, 1.5, 2 &  0, 0.5, 1, 1.5, 2 \\
                			$\tau^{2}$ (variance of random effects) & 0(0.1)1& 0(0.1)1 \\
                			\hline                			
			\end{tabular}
		\end{center}
	\end{footnotesize}
\end{table}

\subsection{Summary of simulation results}

\subsubsection{ Bias of point estimators of $\tau^2$ for LOR (Appendix A)}

For small sample sizes, all estimators of $\tau^2$ are considerably biased, positively at $\tau^2=0$ and linearly decreasing to negative values for larger $\tau^2$. The lines for different estimators are  parallel.
When $p_{iC} = .1$ and  $n = 100$,  the bias of all estimators is almost constant in $\tau^2$; for $p_{iC} = .5$  this happens by  $n=40$.  For larger sample sizes, different estimators acquire positive or negative trends in bias.

The best estimators, almost unbiased by $n=250$ when $p_{iC} = .1$, are MP \lq\lq only", KD, SSU model and SSC \lq\lq always".  The same estimators are recommended for larger $P_{iC}$ values, where they are, in general, less biased earlier.

\subsubsection{ Median bias of point estimators of $\tau^2$ for LOR (Appendix B)}
We define median bias as $P(\hat\tau^2 \geq \tau^2) - P(\hat\tau^2 \leq \tau^2)$. For a median, the median bias is zero.

For  $0.1 \leq \tau^2 \leq 1$, all estimators of $\tau^2$ are  negatively biased for small to medium sample sizes  $n \leq 40$.  Interestingly, the median bias is almost constant across the range of $\tau^2$ values.  The negative bias is also present for the majority of the standard estimators of $\tau^2$ for larger values of $n$. This means that the standard estimators of $\tau^2$ usually provide values of $\hat\tau^2$ that are too low. The KD estimator  is the least median-biased for $n=20$. For larger values of $n$, from $n= 40$ at $p_{iC}\geq .2$ and $n \geq 100$ for $P_{iC} = .1$, the new median-unbiased estimators perform well. We recommend the SMC estimator with added $1/2$ and the SMU model estimator. Both consistently result in almost median-unbiased estimation across the range of $p_{iC}$ values.

\subsubsection{ Coverage of interval estimators of $\tau^2$ for LOR (Appendix C)}

When $P_{iC} = .1$, all confidence intervals but PL provide good coverage when $K = 5$ and $K = 10$.  PL is too conservative, with constant coverage of 1. When $K=30$, PL and, to a lesser degree, QP have much too low coverage for $n$ up to 40, or even 100. This is exacerbated when $1/2$ is added. Then the coverage of PL deteriorates dramatically.  For $K=30$, only KD provides good coverage for $n=20$. The coverage of FPC \lq\lq only" is satisfactory from $n=40$, and  from $n=20$ for $p_{iC} \geq .2$. FPC \lq\lq only" has somewhat higher coverage than FPC \lq\lq always" for $n \leq 40$, but this is reversed from $n=100$.  QP works well from $n=100$, and from $n=20$ for $p_{iC} = .5$. We do not recommend FPU.

To summarize, KD and FPC intervals are recommended for use in practice.

\subsubsection{ Left coverage error (Appendix D)}
A left coverage error (or \lq\lq miss left") occurs when the value of the parameter is to the left of the lower confidence limit of the confidence interval.  For central confidence intervals the probability of this should be near the nominal 2.5\% level.

When $p_{iC} = .1$, the observed miss-left probability is very close to zero for all estimators,  for  sample sizes up to 40, unless $\theta \geq 1.5$. KD is the only interval with non-zero, though also rather low, probability. For larger $p_{iC}$ values, this probability is somewhat higher but still very low for $n \leq 40$.  KD, FPU model and FPC \lq\lq always" levels are closer to 2.5\% from $n=100$, and QP levels are typically lower than  nominal when $p_{iC} = .1$ but improve for larger $p_{iC}$ values. PL levels are especially low for all sample sizes.

\subsubsection{ Right coverage error (Appendix E)}
A right coverage error (or \lq\lq miss right") occurs when the value of the parameter is to the right of the upper confidence limit of the confidence interval. For central confidence intervals, the miss-right probability should be near 2.5\%.

For $p_{iC} = .1$,  KD, QP \lq\lq always", and FPC \lq\lq always" are the closest when $n=20$ and $K=5$, but  they acquire some positive bias for larger $n$. FPC \lq\lq only" and QP \lq\lq only" are reasonably close to 2.5\% from $n=40$ when $K=5$ or $K = 10$. KD is consistently close to 2.5\% for $K=10$. For $K=30$, the results are more varied. KD, QP \lq\lq only" and FPC \lq\lq always" are close to 2.5\% from $n=100$. Miss-right probabilities for all other estimators are typically higher. PL has erratic levels, from 0 to above .1. For larger values of $p_{iC}$, the behavior of all estimators,  with the exception of PL,  improves earlier, so that KD, QP \lq\lq only" and FPC \lq\lq always" are all close to 2.5\% when $n=40$ and $p_{iC} = .5$, even when $K=30$.

\section*{Acknowledgements}
The work by E. Kulinskaya was supported by the Economic and Social Research Council
[grant number ES/L011859/1].

\clearpage
\bibliography{Qfixed_LOR.bib}

\begin{thebibliography}{21}
\providecommand{\natexlab}[1]{#1}
\providecommand{\url}[1]{\texttt{#1}}
\expandafter\ifx\csname urlstyle\endcsname\relax
  \providecommand{\doi}[1]{doi: #1}\else
  \providecommand{\doi}{doi: \begingroup \urlstyle{rm}\Url}\fi

\bibitem[Bakbergenuly et~al.(2020{\natexlab{a}})Bakbergenuly, Hoaglin, and
  Kulinskaya]{BHK2018SMD}
Ilyas Bakbergenuly, David~C. Hoaglin, and Elena Kulinskaya.
\newblock Estimation in meta-analyses of mean difference and standardized mean
  difference.
\newblock \emph{Statistics in Medicine}, 39\penalty0 (2):\penalty0 171--191,
  2020{\natexlab{a}}.

\bibitem[Bakbergenuly et~al.(2020{\natexlab{b}})Bakbergenuly, Hoaglin, and
  Kulinskaya]{BHK2020LOR}
Ilyas Bakbergenuly, David~C. Hoaglin, and Elena Kulinskaya.
\newblock Methods for estimating between-study variance and overall effect in
  meta-analysis of odds ratios.
\newblock \emph{Research Synthesis Methods}, 11\penalty0 (3):\penalty0
  426--442, 2020{\natexlab{b}}.
\newblock \doi{https://doi.org/10.1002/jrsm.1404}.
\newblock URL \url{https://onlinelibrary.wiley.com/doi/abs/10.1002/jrsm.1404}.

\bibitem[Bakbergenuly et~al.(2021)Bakbergenuly, Hoaglin, and
  Kulinskaya]{QfixedSMD}
Ilyas Bakbergenuly, David~C. Hoaglin, and Elena Kulinskaya.
\newblock On the {$Q$} statistic with constant weights for standardized mean
  difference.
\newblock \emph{British Journal of Mathematical and Statistical Psychology},
  2021.
\newblock \doi{https://doi.org/10.1111/bmsp.12263}.

\bibitem[Brown(1947)]{Brown_1947_AMSX_582}
George~W. Brown.
\newblock On small-sample estimation.
\newblock \emph{Annals of Mathematical Statistics}, 18:\penalty0 582--585,
  1947.

\bibitem[Cochran(1954)]{cochran1954combination}
William~G. Cochran.
\newblock The combination of estimates from different experiments.
\newblock \emph{Biometrics}, 10\penalty0 (1):\penalty0 101--129, 1954.

\bibitem[DerSimonian and Kacker(2007)]{dersimonian2007random}
Rebecca DerSimonian and Raghu Kacker.
\newblock Random-effects model for meta-analysis of clinical trials: an update.
\newblock \emph{Contemporary Clinical Trials}, 28\penalty0 (2):\penalty0
  105--114, 2007.

\bibitem[DerSimonian and Laird(1986)]{dersimonian1986meta}
Rebecca DerSimonian and Nan Laird.
\newblock Meta-analysis in clinical trials.
\newblock \emph{Controlled Clinical Trials}, 7\penalty0 (3):\penalty0 177--188,
  1986.

\bibitem[Farebrother(1984)]{Farebrother1984}
R.~W. Farebrother.
\newblock Algorithm {AS} 204: The distribution of a positive linear combination
  of $\chi^2$ random variables.
\newblock \emph{Journal of the Royal Statistical Society, {S}eries {C}},
  33\penalty0 (3):\penalty0 332--339, 1984.

\bibitem[Gart et~al.(1985)Gart, Pettigrew, and Thomas]{G-P-T-1985}
John~J. Gart, Hugh~M. Pettigrew, and Donald~G. Thomas.
\newblock The effect of bias, variance estimation, skewness and kurtosis of the
  empirical logit on weighted least squares analyses.
\newblock \emph{Biometrika}, 72\penalty0 (1):\penalty0 179--190, 1985.

\bibitem[Hardy and Thompson(1996)]{Hardy_1996_StatMed_619}
Rebecca~J. Hardy and Simon~G. Thompson.
\newblock A likelihood approach to meta-analysis with random effects.
\newblock \emph{Statistics in Medicine}, 15:\penalty0 619--629, 1996.

\bibitem[Jackson(2013)]{Jackson2013CI}
Dan Jackson.
\newblock Confidence intervals for the between-study variance in random effects
  meta-analysis using generalised {C}ochran heterogeneity statistics.
\newblock \emph{Research Synthesis Methods}, 4\penalty0 (3):\penalty0 220--229,
  2013.
\newblock \doi{https://doi.org/10.1002/jrsm.1081}.

\bibitem[Kulinskaya and Dollinger(2015)]{kulinskaya2015accurate}
Elena Kulinskaya and Michael~B. Dollinger.
\newblock An accurate test for homogeneity of odds ratios based on {C}ochran's
  {$Q$}-statistic.
\newblock \emph{BMC Medical Research Methodology}, 15\penalty0 (1):\penalty0
  49, 2015.

\bibitem[Kulinskaya and Hoaglin(2022)]{Rprogs_tau2_Kulinskaya_Hoaglin_2022}
Elena Kulinskaya and David~C. Hoaglin.
\newblock R programs for estimation of heterogeneity variance $\tau^2$ for
  log-odds-ratio using the generalised {Q} statistic with constant and inverse
  variance weights.
\newblock OSF, https://osf.io/5n3vd, July 27 2022.
\newblock URL \url{osf.io/yqgsk}.

\bibitem[Kulinskaya et~al.(2021{\natexlab{a}})Kulinskaya, Hoaglin, and
  Bakbergenuly]{BHK2021SIM}
Elena Kulinskaya, David~C. Hoaglin, and Ilyas Bakbergenuly.
\newblock Exploring consequences of simulation design for apparent performance
  of methods of meta-analysis.
\newblock \emph{Statistical Methods in Medical Research}, 30\penalty0
  (7):\penalty0 1667--1690, 2021{\natexlab{a}}.
\newblock \doi{10.1177/09622802211013065}.
\newblock URL \url{https://doi.org/10.1177/09622802211013065}.
\newblock PMID: 34110941.

\bibitem[Kulinskaya et~al.(2021{\natexlab{b}})Kulinskaya, Hoaglin,
  Bakbergenuly, and Newman]{Qfixed}
Elena Kulinskaya, David~C. Hoaglin, Ilyas Bakbergenuly, and Joseph Newman.
\newblock A {$Q$} statistic with constant weights for assessing heterogeneity
  in meta-analysis.
\newblock \emph{Research Synthesis Methods}, 12:\penalty0 711--730,
  2021{\natexlab{b}}.
\newblock \doi{https://doi.org/10.1002/jrsm.1491}.

\bibitem[Mandel and Paule(1970)]{mandel1970interlaboratory}
John Mandel and Robert~C. Paule.
\newblock Interlaboratory evaluation of a material with unequal numbers of
  replicates.
\newblock \emph{Analytical Chemistry}, 42\penalty0 (11):\penalty0 1194--1197,
  1970.

\bibitem[{R Core Team}(2016)]{rrr}
{R Core Team}.
\newblock \emph{R: A Language and Environment for Statistical Computing}.
\newblock R Foundation for Statistical Computing, Vienna, Austria, 2016.
\newblock URL \url{https://www.R-project.org/}.

\bibitem[S{\'a}nchez-Meca and Mar{\'\i}n-Mart{\'\i}nez(2000)]{sanches-2000}
Julio S{\'a}nchez-Meca and Fulgencio Mar{\'\i}n-Mart{\'\i}nez.
\newblock Testing the significance of a common risk difference in
  meta-analysis.
\newblock \emph{Computational Statistics \& Data Analysis}, 33\penalty0
  (3):\penalty0 299--313, 2000.

\bibitem[Viechtbauer(2007)]{viechtbauer2007confidence}
Wolfgang Viechtbauer.
\newblock Confidence intervals for the amount of heterogeneity in
  meta-analysis.
\newblock \emph{Statistics in Medicine}, 26\penalty0 (1):\penalty0 37--52,
  2007.

\bibitem[Viechtbauer(2010)]{metafor}
Wolfgang Viechtbauer.
\newblock Conducting meta-analyses in {R} with the {metafor} package.
\newblock \emph{Journal of Statistical Software}, 36:\penalty0 3, 2010.
\newblock URL \url{https://doi.org/10.18637/jss.v036.i03}.
\newblock Website: https://www.metafor-project.org.

\bibitem[Viechtbauer(2021)]{ViechtMedian}
Wolfgang Viechtbauer.
\newblock Median-unbiased estimators for the amount of heterogeneity in
  meta-analysis.
\newblock 9th European Congress of Methodology. European Association of
  Methodology, 2021.
\newblock URL
  \url{https://www.wvbauer.com/lib/exe/fetch.php/talks:2021_viechtbauer_eam_median_tau2.pdf}.

\end{thebibliography}
\clearpage

\section*{Appendices}
\begin{itemize}
\item Appendix A: Bias in point estimators of between-study variance
\item Appendix B: Median bias in point estimators of between-study variance
\item Appendix C: Coverage of 95\% confidence intervals for between-study variance
\item Appendix D: Sample miss-left probability for 95\% confidence intervals for between-study variance
\item Appendix E: Sample miss-right probability for 95\% confidence intervals for between-study variance
\end{itemize}

\setcounter{figure}{0}
\setcounter{section}{0}
\clearpage

%\begin{figure}[t]
%	\centering
%	\includegraphics[scale=0.33]{Figure5.pdf}
%	\caption{Coverage at 95\% nominal level of confidence of five interval estimators of between-study variance $\tau^2$:  QP, PL, KDB, FPC, and FPU. First two rows: equal sample sizes, ${n} = 20,\; f = .5,\; \delta = 0$ and $\delta = 2$;  second two rows: unequal sample sizes, $\bar n = 30,\; f = .75, \delta = 0$ and $\delta = 2$.}
%	\label{F5}
%\end{figure}
\setcounter{figure}{0}
\setcounter{section}{0}

\section*{Appendix A: Bias in point estimators of between-study variance}

Each figure corresponds to a value of the probability of an event in the Control arm $p_{iC}$  (= .1, .2, .5) . \\
The fraction of each study's sample size in the Control arm ($f$) is held constant at 0.5. For each combination of a value of $n$ (= 20, 40, 100, 250) or  $\bar{n}$ (= 30, 60, 100, 160) and a value of $K$ (= 5, 10, 30), a panel plots bias versus $\tau^2$ (= 0.0(0.1)1).\\
The point estimators of $\tau^2$ are
\begin{itemize}
\item DL (DerSimonian-Laird method, inverse-variance weights)
\item REML method, inverse-variance weights)
\item MP (Mandel-Paule method, inverse-variance weights)
\item KD (Kulinskaya-Dollinger (2015) approximation, inverse-variance weights)
\item SSC method, effective-sample-size weights, conditional variance of LOR
\item SMC method, median-unbiased, effective-sample-size weights, conditional variance of LOR
\item SSU method, effective-sample-size weights, unconditional variance of LOR
\item SMU method, median-unbiased, effective-sample-size weights, unconditional variance of LOR
\end{itemize}
The plots include two versions of DL, REML, MP, SSC, and SMC: adding $1/2$ to all four of $X_{iT},\;X_{iC},\; n_{iT}-X_{iT},\; n_{iC}-X_{iC}$ only when one of these is zero (solid lines) or always (dashed lines).\\
Plots also include two versions of SSU and SMU: model-based estimation of $p_{iT}$ (solid lines) or na\"{i}ve estimation (dashed lines).

\clearpage
\renewcommand{\thefigure}{A.\arabic{figure}}

\begin{figure}[ht]
	\centering
	\includegraphics[scale=0.33]{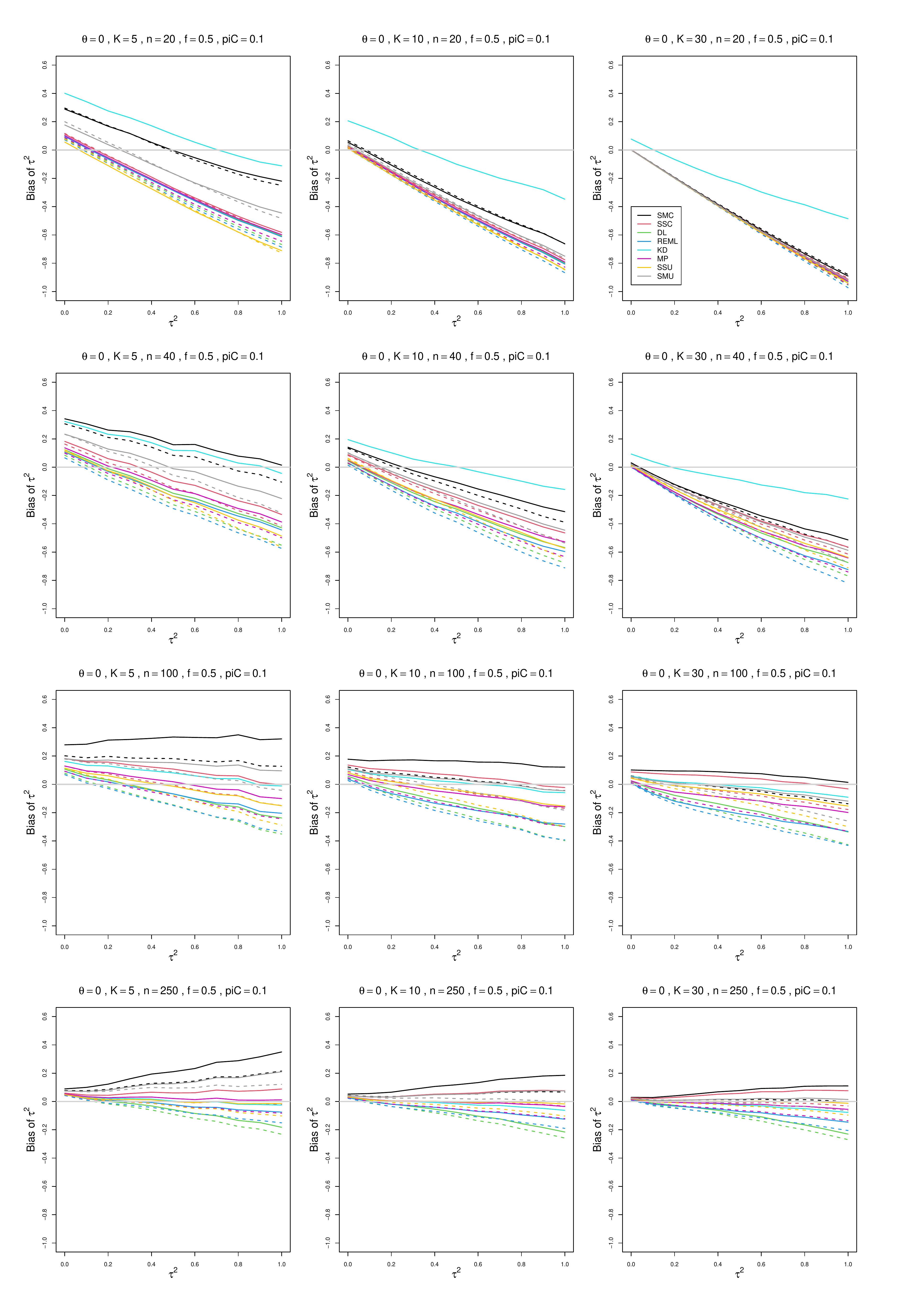}
	\caption{Bias  of estimators of between-study variance of LOR (DL, REML, KD, MP, SMC, SSC, SMU, and SSU) vs $\tau^2$, for equal sample sizes $n=20,\;40,\;100$ and $250$, $p_{iC} = .1$, $\theta=0$ and  $f = 0.5$.   Solid lines: DL, REML, MP, SSC, SMC \lq\lq only"; KD; SSU and SMU model-based. Dashed lines: DL, REML, MP, SSC, SMC  \lq\lq always"; SSU and SMU na\"ive. }
	\label{PlotBiasOfTau2_piC_01theta=0_LOR_equal_sample_sizes}
\end{figure}

\begin{figure}[ht]
	\centering
	\includegraphics[scale=0.33]{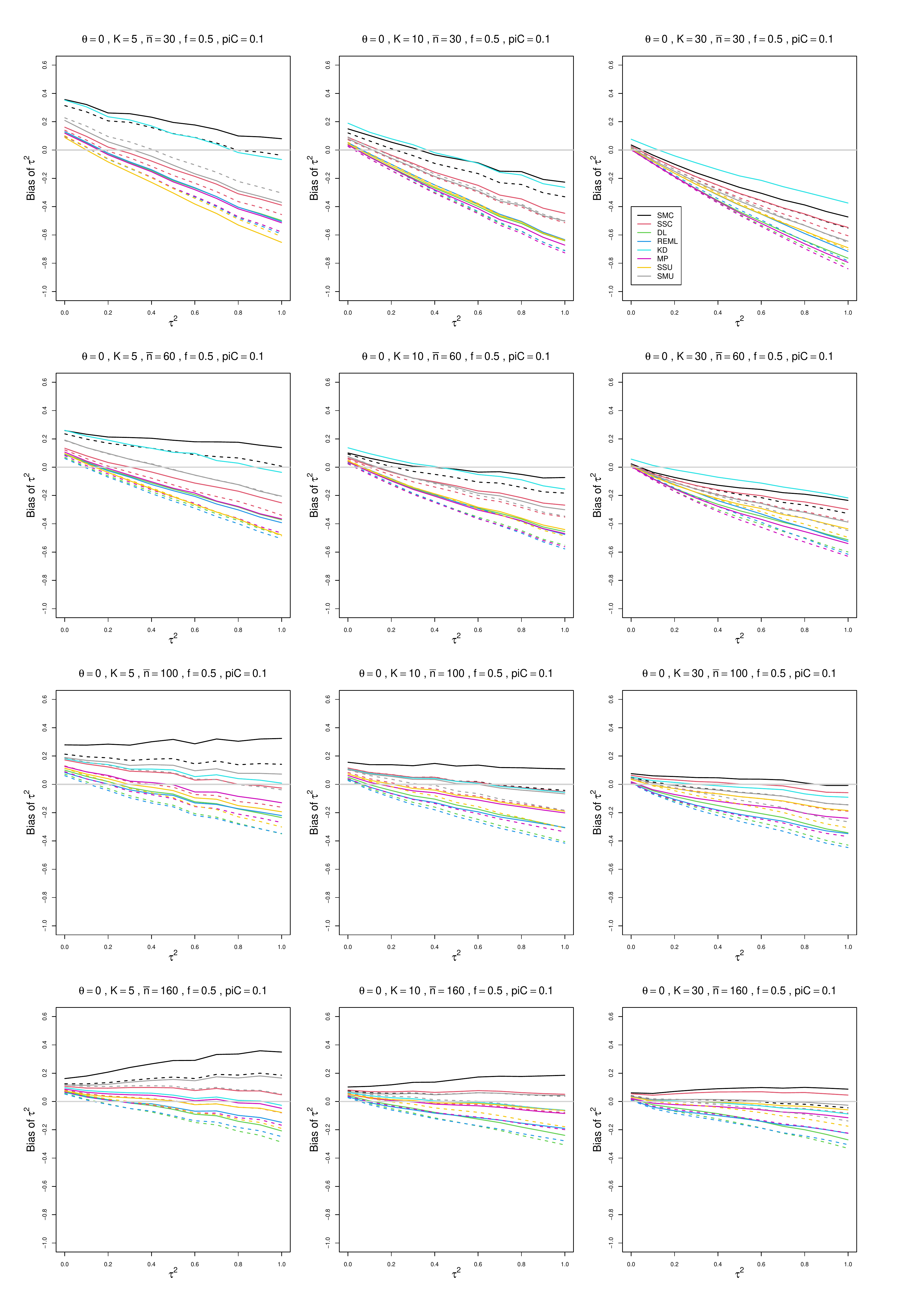}
	\caption{Bias  of estimators of between-study variance of LOR (DL, REML, KD, MP, SMC, SSC, SMU, and SSU) vs $\tau^2$, for unequal sample sizes $\bar{n}=30,\;60,\;100$ and $160$, $p_{iC} = .1$, $\theta=0$ and  $f=0.5$.   Solid lines: DL, REML, MP, SSC, SMC \lq\lq only"; KD; SSU and SMU model-based. Dashed lines: DL, REML, MP, SSC, SMC  \lq\lq always"; SSU and SMU na\"ive.  }
	\label{PlotBiasOfTau2_piC_01theta=0_LOR_unequal_sample_sizes}
\end{figure}

\begin{figure}[ht]
	\centering
	\includegraphics[scale=0.33]{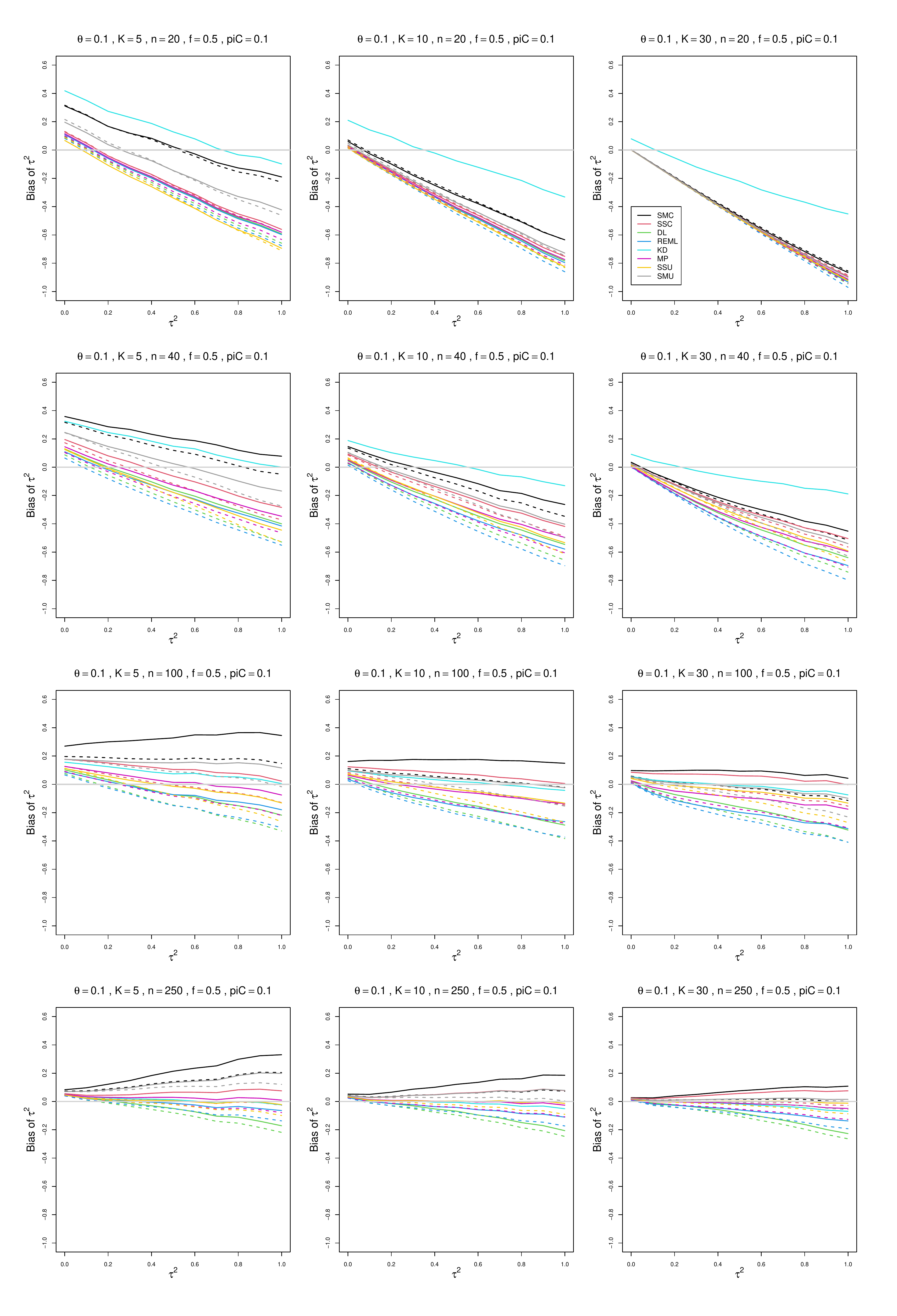}
	\caption{Bias  of estimators of between-study variance of LOR (DL, REML, KD, MP, SMC, SSC, SMU, and SSU) vs $\tau^2$, for equal sample sizes $n=20,\;40,\;100$ and $250$, $p_{iC} = .1$, $\theta=0.1$ and  $f=0.5$.   Solid lines: DL, REML, MP, SSC, SMC \lq\lq only"; KD; SSU and SMU model-based. Dashed lines: DL, REML, MP, SSC, SMC  \lq\lq always"; SSU and SMU na\"ive.  }
	\label{PlotBiasOfTau2_piC_01theta=0.1_LOR_equal_sample_sizes}
\end{figure}

\begin{figure}[ht]
	\centering
	\includegraphics[scale=0.33]{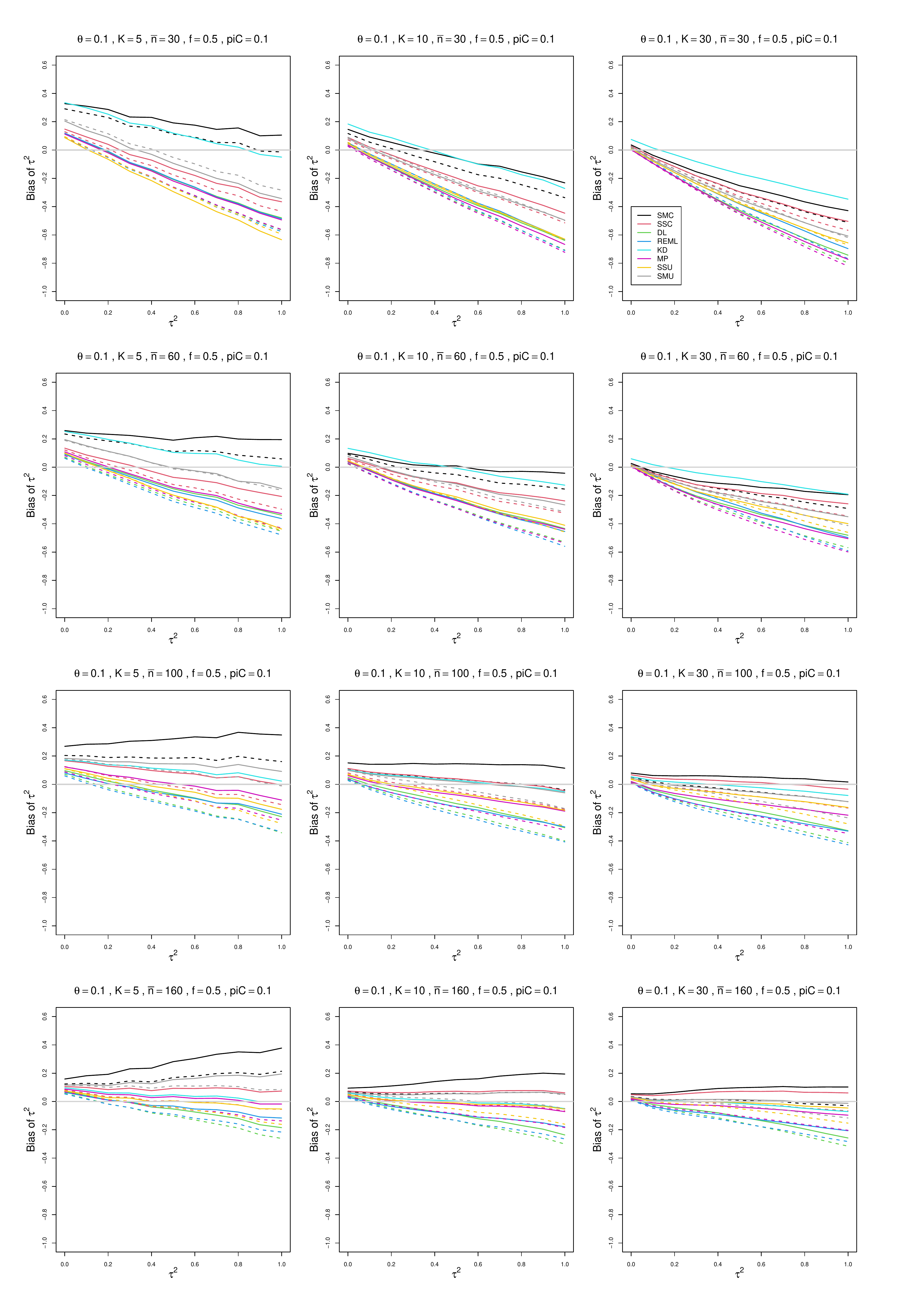}
	\caption{Bias  of estimators of between-study variance of LOR (DL, REML, KD, MP, SMC, SSC, SMU, and SSU) vs $\tau^2$, for unequal sample sizes $\bar{n}=30,\;60,\;100$ and $160$, $p_{iC} = .1$, $\theta=0.1$ and  $f=0.5$.   Solid lines: DL, REML, MP, SSC, SMC \lq\lq only"; KD; SSU and SMU model-based. Dashed lines: DL, REML, MP, SSC, SMC  \lq\lq always"; SSU and SMU na\"ive. }
	\label{PlotBiasOfTau2_piC_01theta=0.1_LOR_unequal_sample_sizes}
\end{figure}

\begin{figure}[ht]
	\centering
	\includegraphics[scale=0.33]{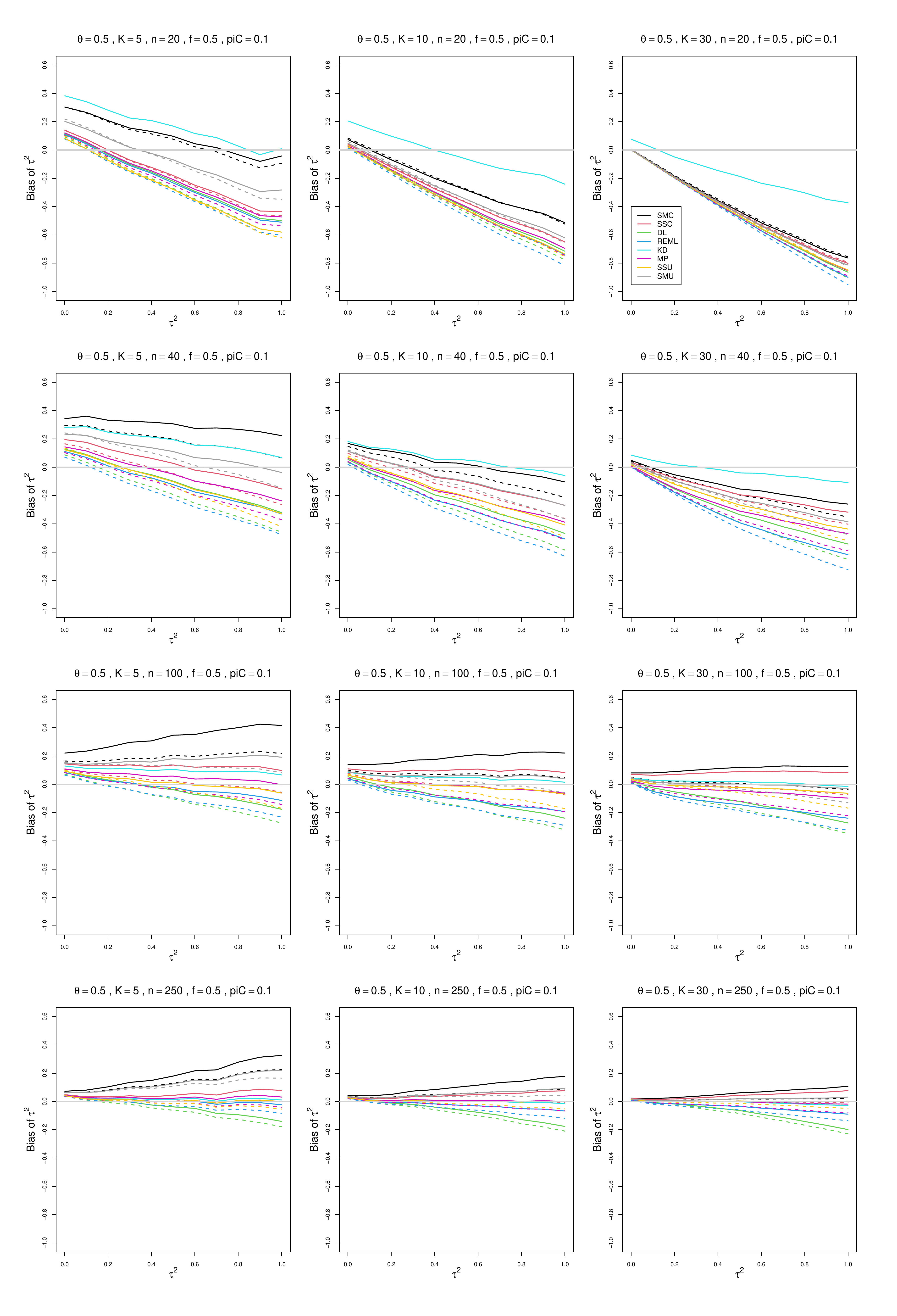}
	\caption{Bias  of estimators of between-study variance of LOR (DL, REML, KD, MP, SMC, SSC, SMU, and SSU) vs $\tau^2$, for equal sample sizes $n=20,\;40,\;100$ and $250$, $p_{iC} = .1$, $\theta=0.5$ and  $f=0.5$.   Solid lines: DL, REML, MP, SSC, SMC \lq\lq only"; KD; SSU and SMU model-based. Dashed lines: DL, REML, MP, SSC, SMC  \lq\lq always"; SSU and SMU na\"ive.  }
	\label{PlotBiasOfTau2_piC_01theta=0.5_LOR_equal_sample_sizes}
\end{figure}

\begin{figure}[ht]
	\centering
	\includegraphics[scale=0.33]{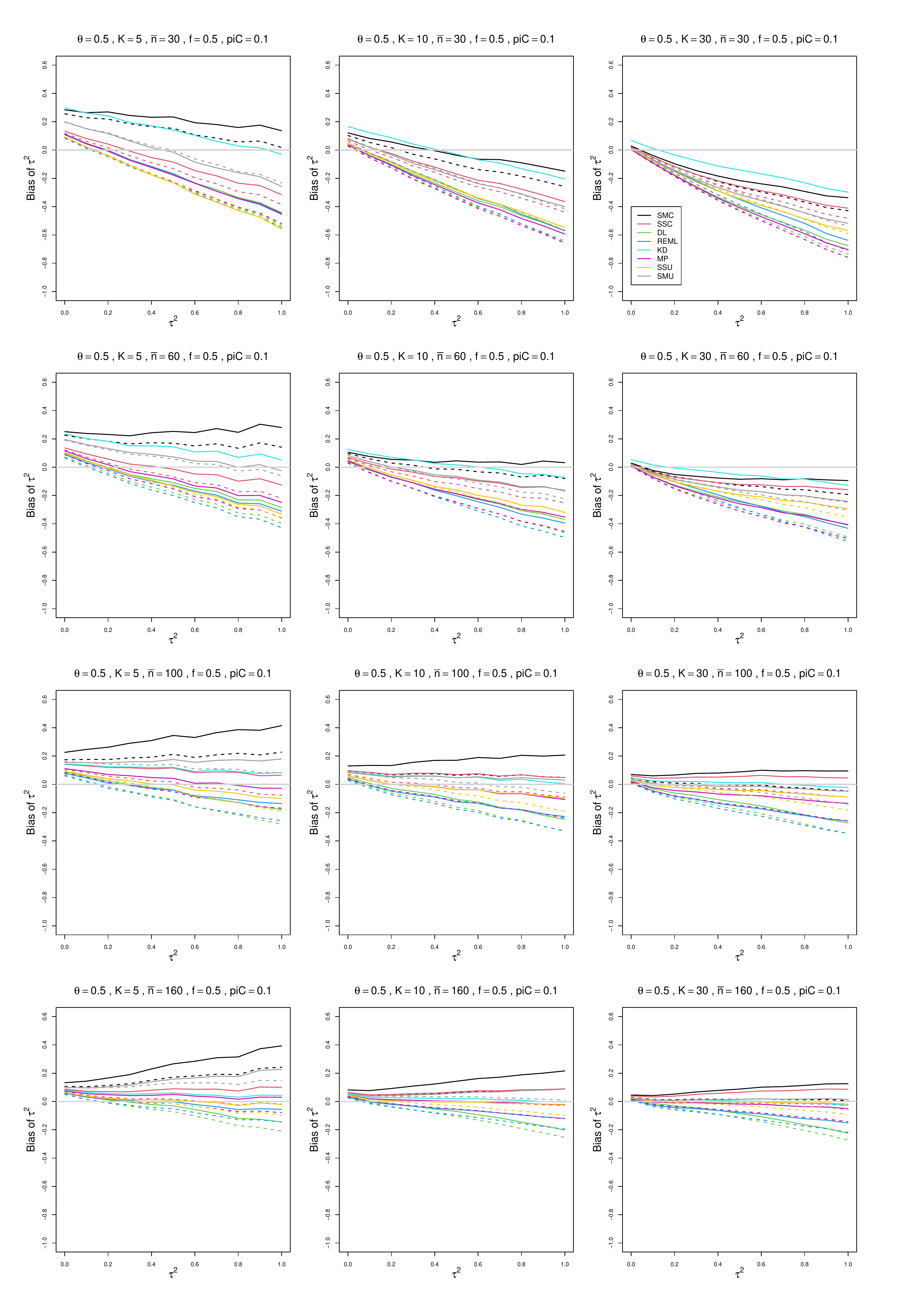}
	\caption{Bias  of estimators of between-study variance of LOR (DL, REML, KD, MP, SMC, SSC, SMU, and SSU) vs $\tau^2$, for unequal sample sizes $\bar{n}=30,\;60,\;100$ and $160$, $p_{iC} = .1$, $\theta=0.5$ and  $f=0.5$.   Solid lines: DL, REML, MP, SSC, SMC \lq\lq only"; KD; SSU and SMU model-based. Dashed lines: DL, REML, MP, SSC, SMC  \lq\lq always"; SSU and SMU na\"ive. }
	\label{PlotBiasOfTau2_piC_01theta=0.5_LOR_unequal_sample_sizes}
\end{figure}

\begin{figure}[ht]
	\centering
	\includegraphics[scale=0.33]{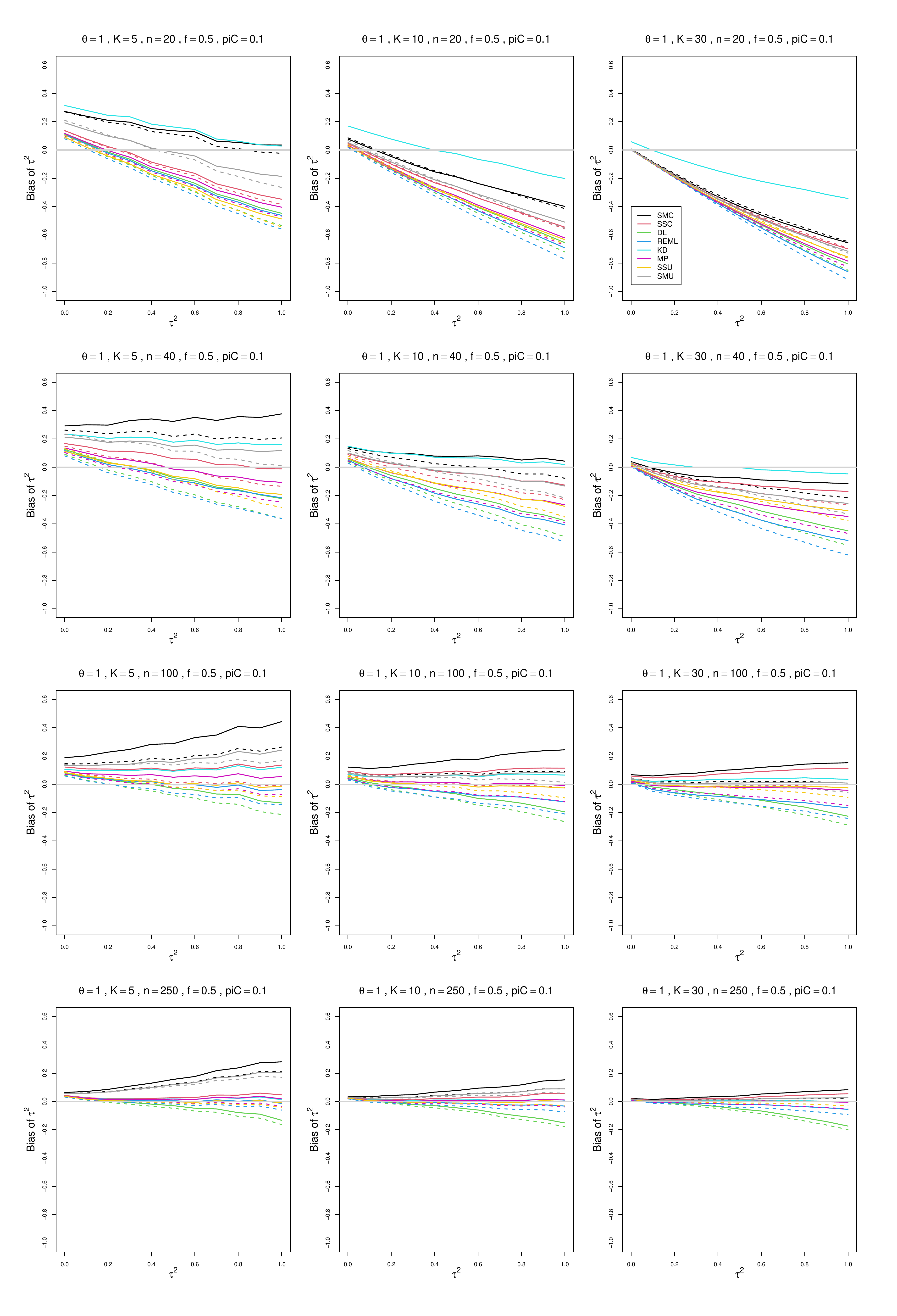}
	\caption{Bias  of estimators of between-study variance of LOR (DL, REML, KD, MP, SMC, SSC, SMU, and SSU) vs $\tau^2$, for equal sample sizes $n=20,\;40,\;100$ and $250$, $p_{iC} = .1$, $\theta=1$ and  $f=0.5$.   Solid lines: DL, REML, MP, SSC, SMC \lq\lq only"; KD; SSU and SMU model-based. Dashed lines: DL, REML, MP, SSC, SMC  \lq\lq always"; SSU and SMU na\"ive. }
	\label{PlotBiasOfTau2_piC_01theta=1_LOR_equal_sample_sizes}
\end{figure}

\begin{figure}[ht]
	\centering
	\includegraphics[scale=0.33]{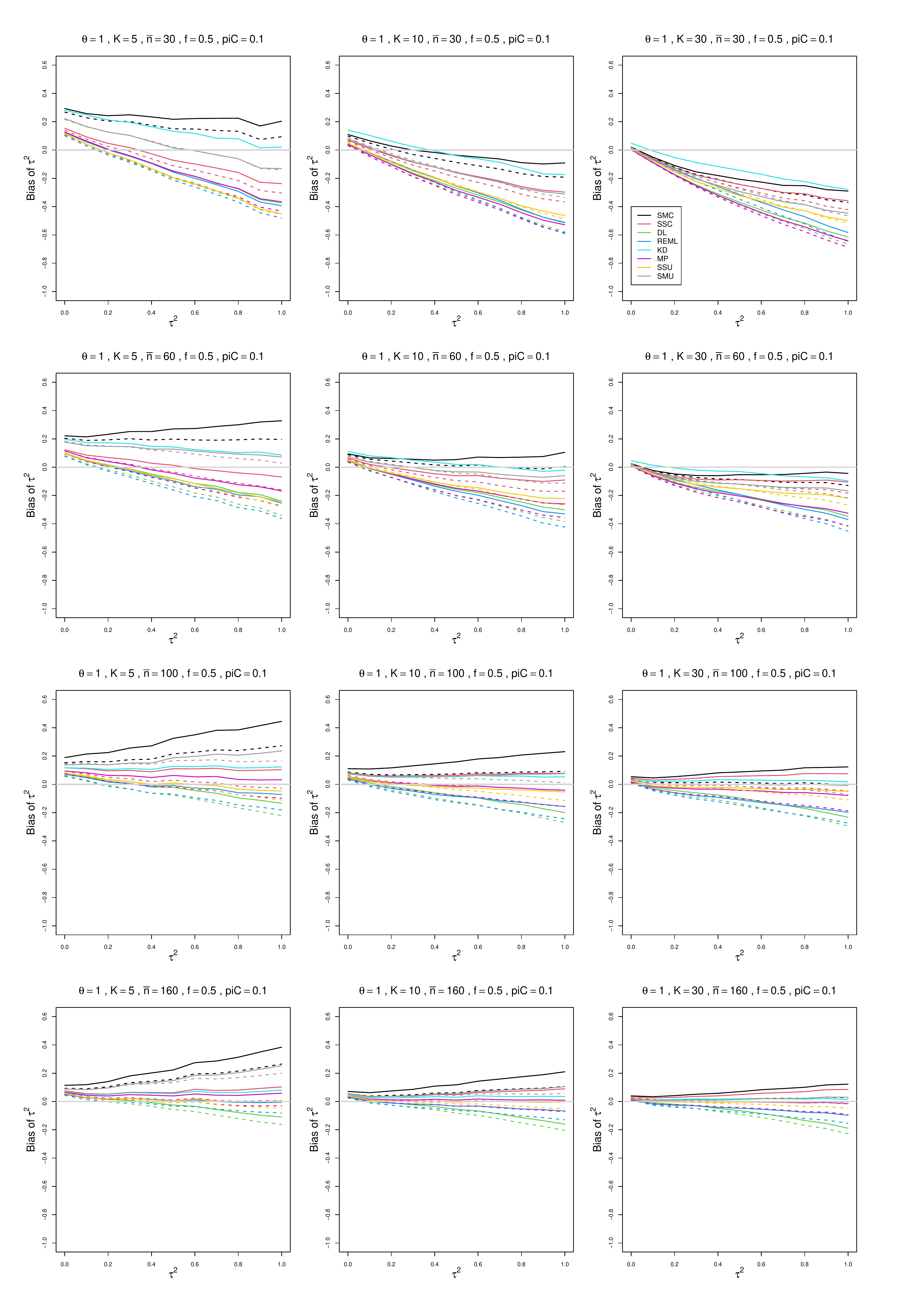}
	\caption{Bias  of estimators of between-study variance of LOR (DL, REML, KD, MP, SMC, SSC, SMU, and SSU) vs $\tau^2$, for unequal sample sizes $\bar{n}=30,\;60,\;100$ and $160$, $p_{iC} = .1$, $\theta=1$ and  $f=0.5$.   Solid lines: DL, REML, MP, SSC, SMC \lq\lq only"; KD; SSU and SMU model-based. Dashed lines: DL, REML, MP, SSC, SMC  \lq\lq always"; SSU and SMU na\"ive.  }
	\label{PlotBiasOfTau2_piC_01theta=1_LOR_unequal_sample_sizes}
\end{figure}

\begin{figure}[ht]
	\centering
	\includegraphics[scale=0.33]{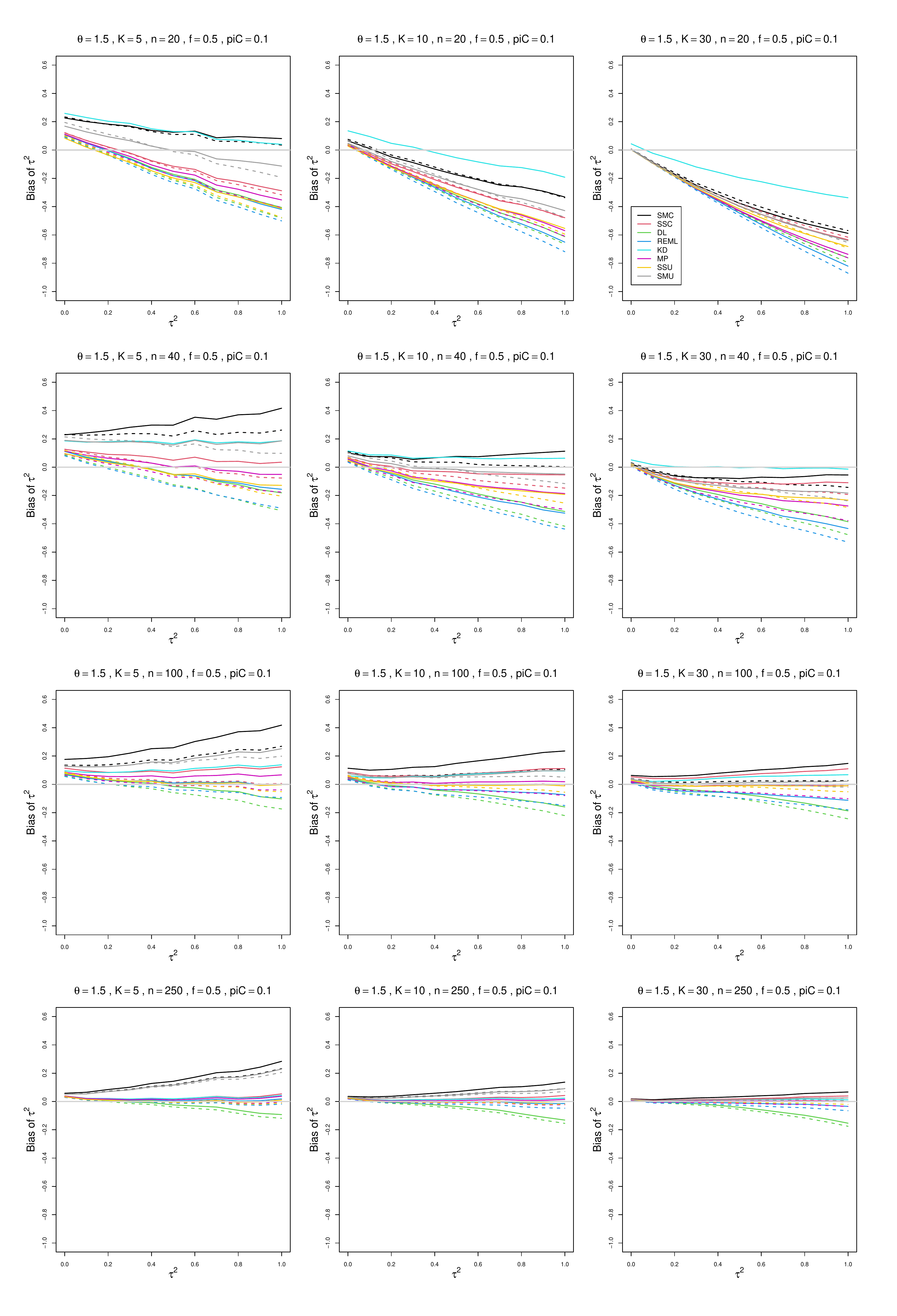}
	\caption{Bias  of estimators of between-study variance of LOR (DL, REML, KD, MP, SMC, SSC, SMU, and SSU) vs $\tau^2$, for equal sample sizes $n=20,\;40,\;100$ and $250$, $p_{iC} = .1$, $\theta=1.5$ and  $f=0.5$.   Solid lines: DL, REML, MP, SSC, SMC \lq\lq only"; KD; SSU and SMU model-based. Dashed lines: DL, REML, MP, SSC, SMC  \lq\lq always"; SSU and SMU na\"ive. }
	\label{PlotBiasOfTau2_piC_01theta=1.5_LOR_equal_sample_sizes}
\end{figure}

\begin{figure}[ht]
	\centering
	\includegraphics[scale=0.33]{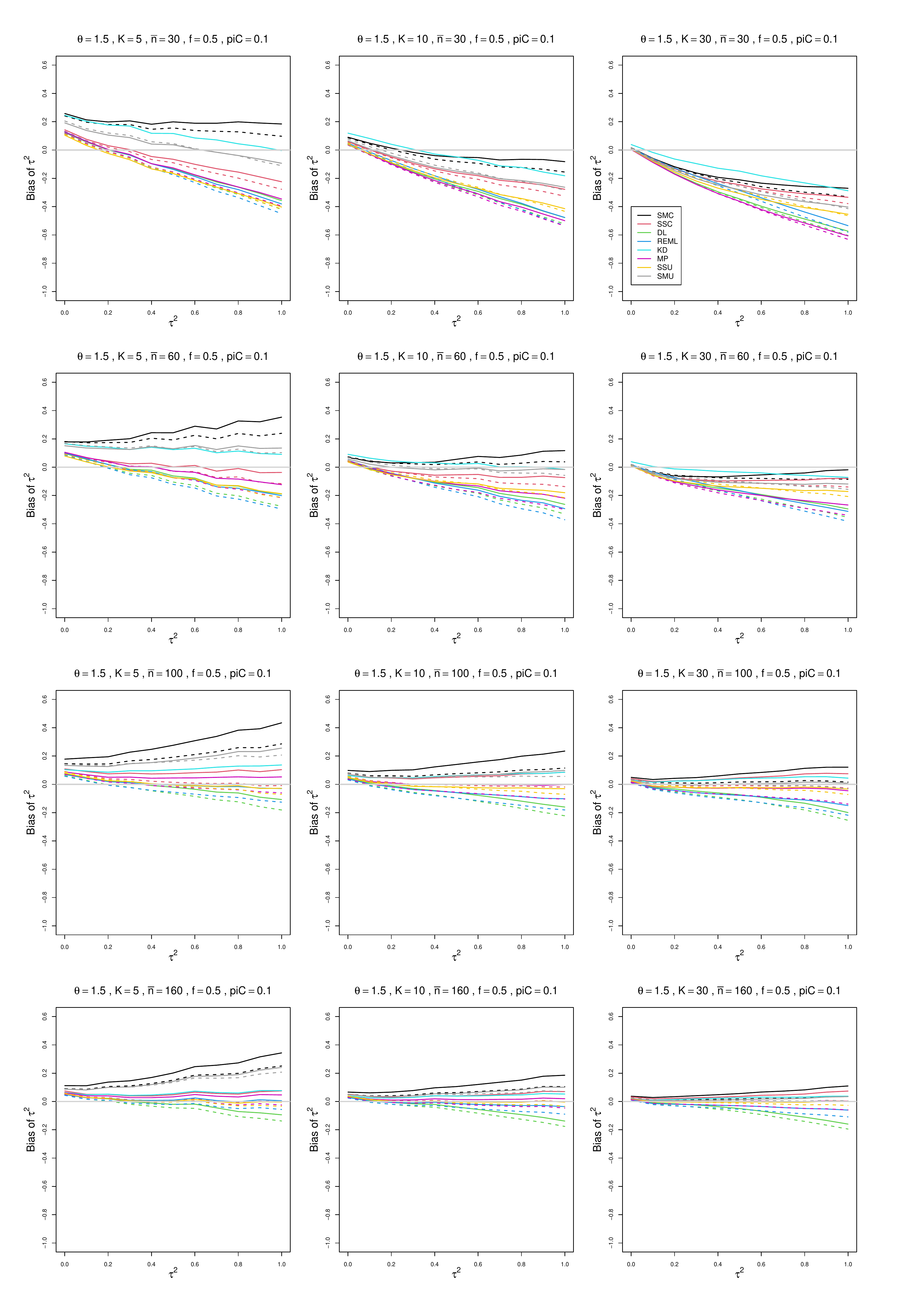}
	\caption{Bias  of estimators of between-study variance of LOR (DL, REML, KD, MP, SMC, SSC, SMU, and SSU) vs $\tau^2$, for unequal sample sizes $\bar{n}=30,\;60,\;100$ and $160$, $p_{iC} = .1$, $\theta=1.5$ and  $f=0.5$.   Solid lines: DL, REML, MP, SSC, SMC \lq\lq only"; KD; SSU and SMU model-based. Dashed lines: DL, REML, MP, SSC, SMC  \lq\lq always"; SSU and SMU na\"ive. }
	\label{PlotBiasOfTau2_piC_01theta=1.5_LOR_unequal_sample_sizes}
\end{figure}

\begin{figure}[ht]
	\centering
	\includegraphics[scale=0.33]{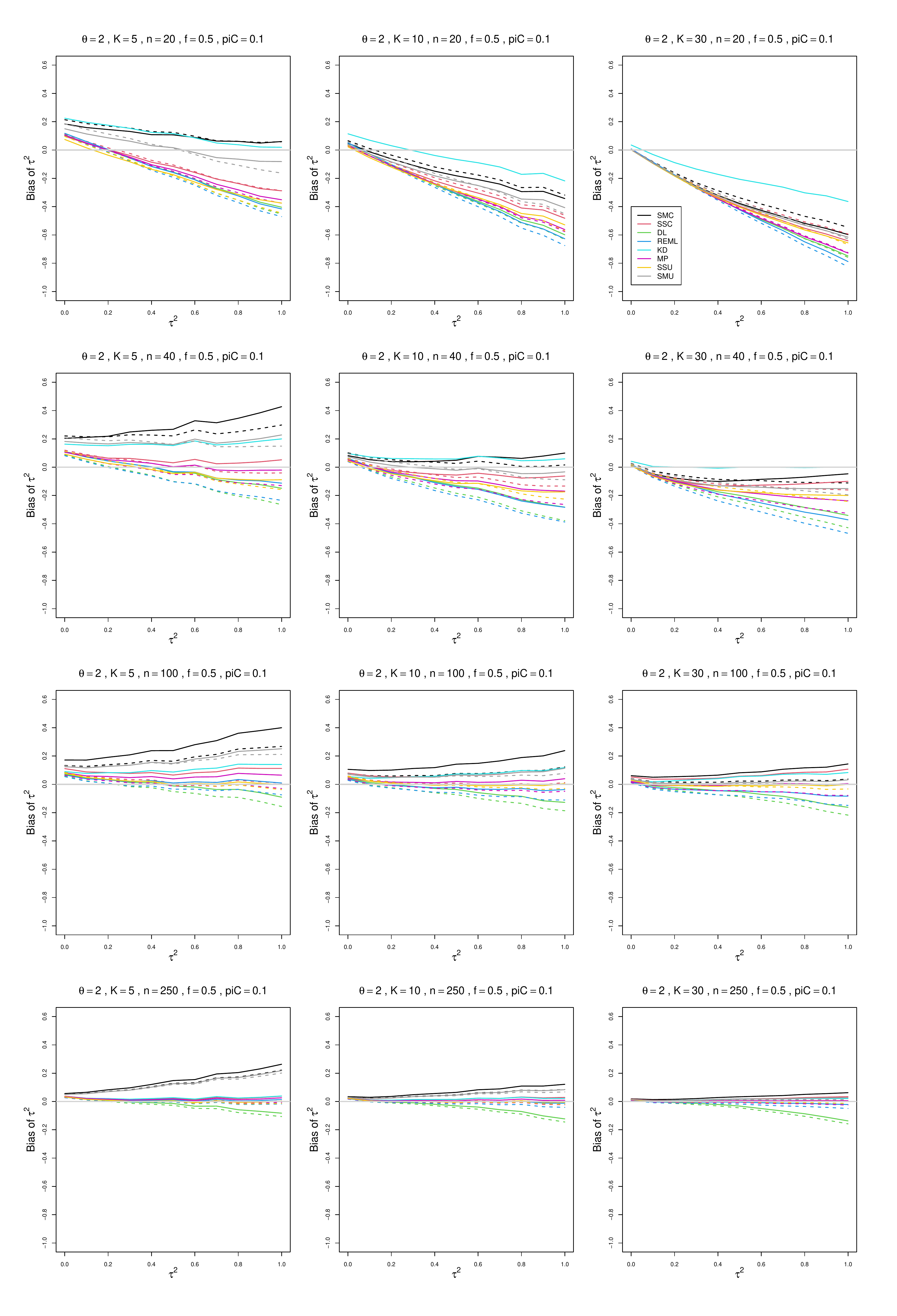}
	\caption{Bias  of estimators of between-study variance of LOR (DL, REML, KD, MP, SMC, SSC, SMU, and SSU) vs $\tau^2$, for equal sample sizes $n=20,\;40,\;100$ and $250$, $p_{iC} = .1$, $\theta=2$ and  $f=0.5$.   Solid lines: DL, REML, MP, SSC, SMC \lq\lq only"; KD; SSU and SMU model-based. Dashed lines: DL, REML, MP, SSC, SMC  \lq\lq always"; SSU and SMU na\"ive.  }
	\label{PlotBiasOfTau2_piC_01theta=2_LOR_equal_sample_sizes}
\end{figure}

\begin{figure}[ht]
	\centering
	\includegraphics[scale=0.33]{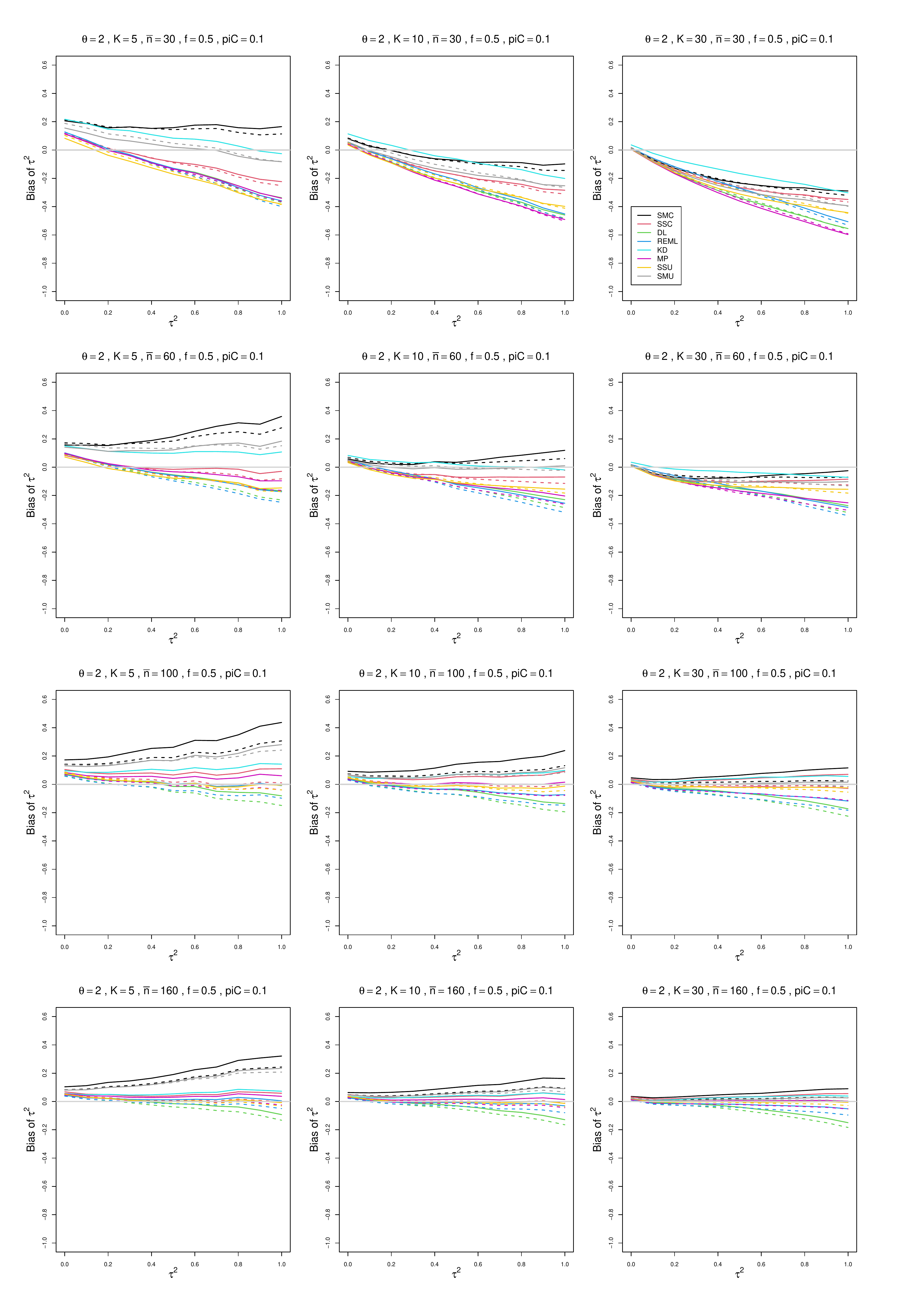}
	\caption{Bias  of estimators of between-study variance of LOR (DL, REML, KD, MP, SMC, SSC, SMU, and SSU) vs $\tau^2$, for unequal sample sizes $\bar{n}=30,\;60,\;100$ and $160$, $p_{iC} = .1$, $\theta=2$ and  $f=0.5$.   Solid lines: DL, REML, MP, SSC, SMC \lq\lq only"; KD; SSU and SMU model-based. Dashed lines: DL, REML, MP, SSC, SMC  \lq\lq always"; SSU and SMU na\"ive.  }
	\label{PlotBiasOfTau2_piC_01theta=2_LOR_unequal_sample_sizes}
\end{figure}

\begin{figure}[ht]
	\centering
	\includegraphics[scale=0.33]{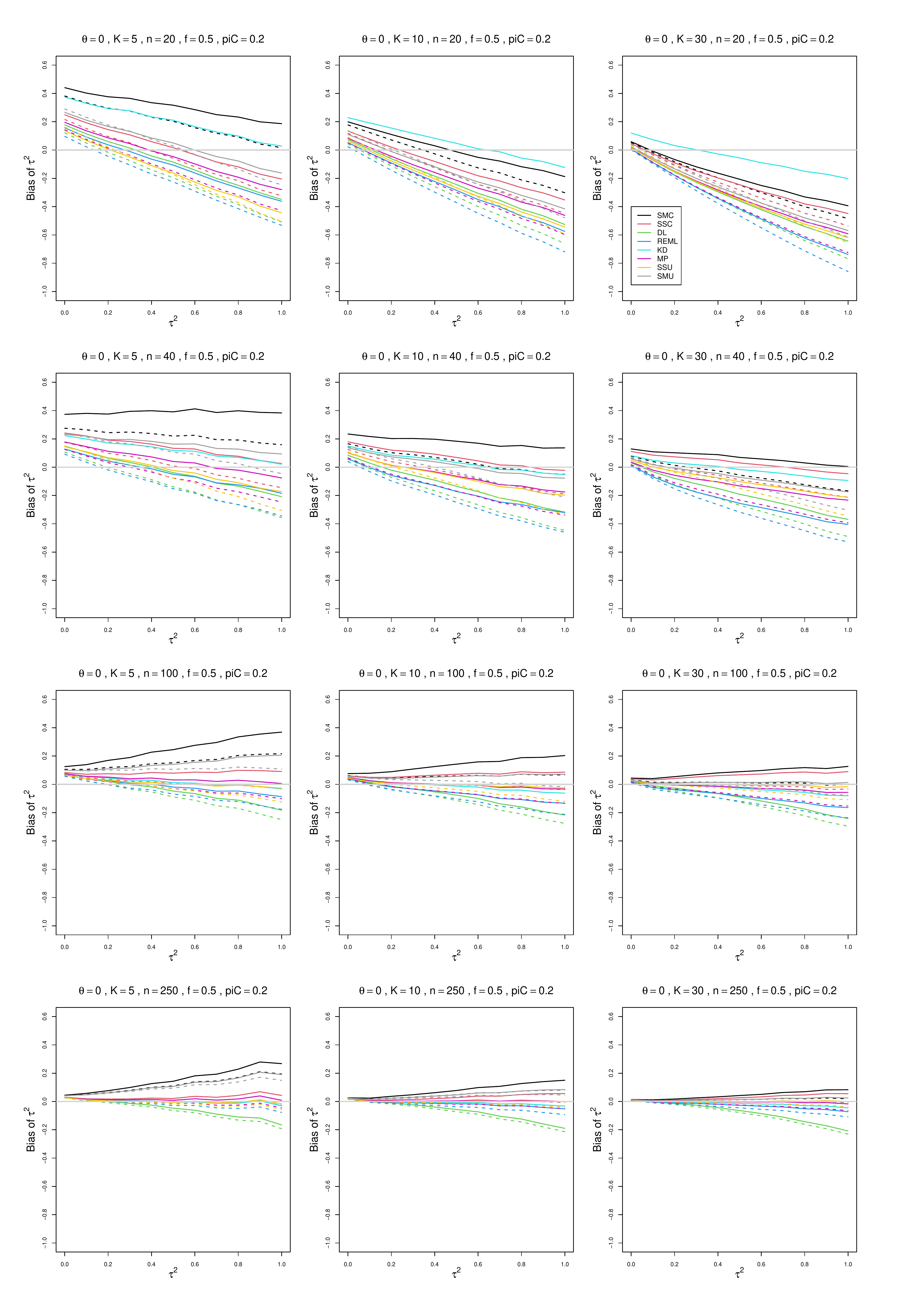}
	\caption{Bias  of estimators of between-study variance of LOR (DL, REML, KD, MP, SMC, SSC, SMU, and SSU) vs $\tau^2$, for equal sample sizes $n=20,\;40,\;100$ and $250$, $p_{iC} = .2$, $\theta=0$ and  $f=0.5$.   Solid lines: DL, REML, MP, SSC, SMC \lq\lq only"; KD; SSU and SMU model-based. Dashed lines: DL, REML, MP, SSC, SMC  \lq\lq always"; SSU and SMU na\"ive.  }
	\label{PlotBiasOfTau2_piC_02theta=0_LOR_equal_sample_sizes}
\end{figure}

\begin{figure}[ht]
	\centering
	\includegraphics[scale=0.33]{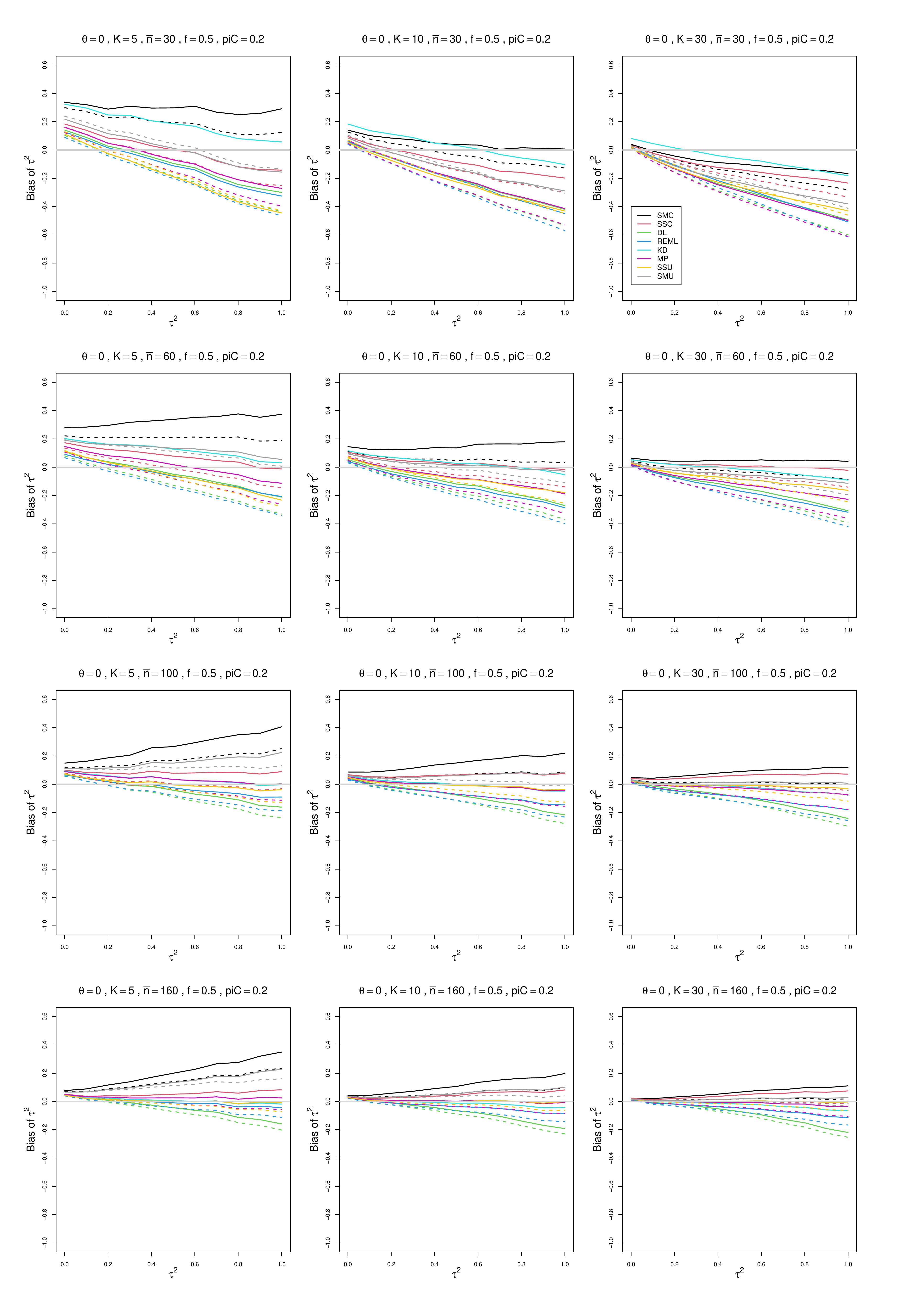}
	\caption{Bias  of estimators of between-study variance of LOR (DL, REML, KD, MP, SMC, SSC, SMU, and SSU) vs $\tau^2$, for unequal sample sizes $\bar{n}=30,\;60,\;100$ and $160$, $p_{iC} = .2$, $\theta=0$ and  $f=0.5$.   Solid lines: DL, REML, MP, SSC, SMC \lq\lq only"; KD; SSU and SMU model-based. Dashed lines: DL, REML, MP, SSC, SMC  \lq\lq always"; SSU and SMU na\"ive.  }
	\label{PlotBiasOfTau2_piC_02theta=0_LOR_unequal_sample_sizes}
\end{figure}

\begin{figure}[ht]
	\centering
	\includegraphics[scale=0.33]{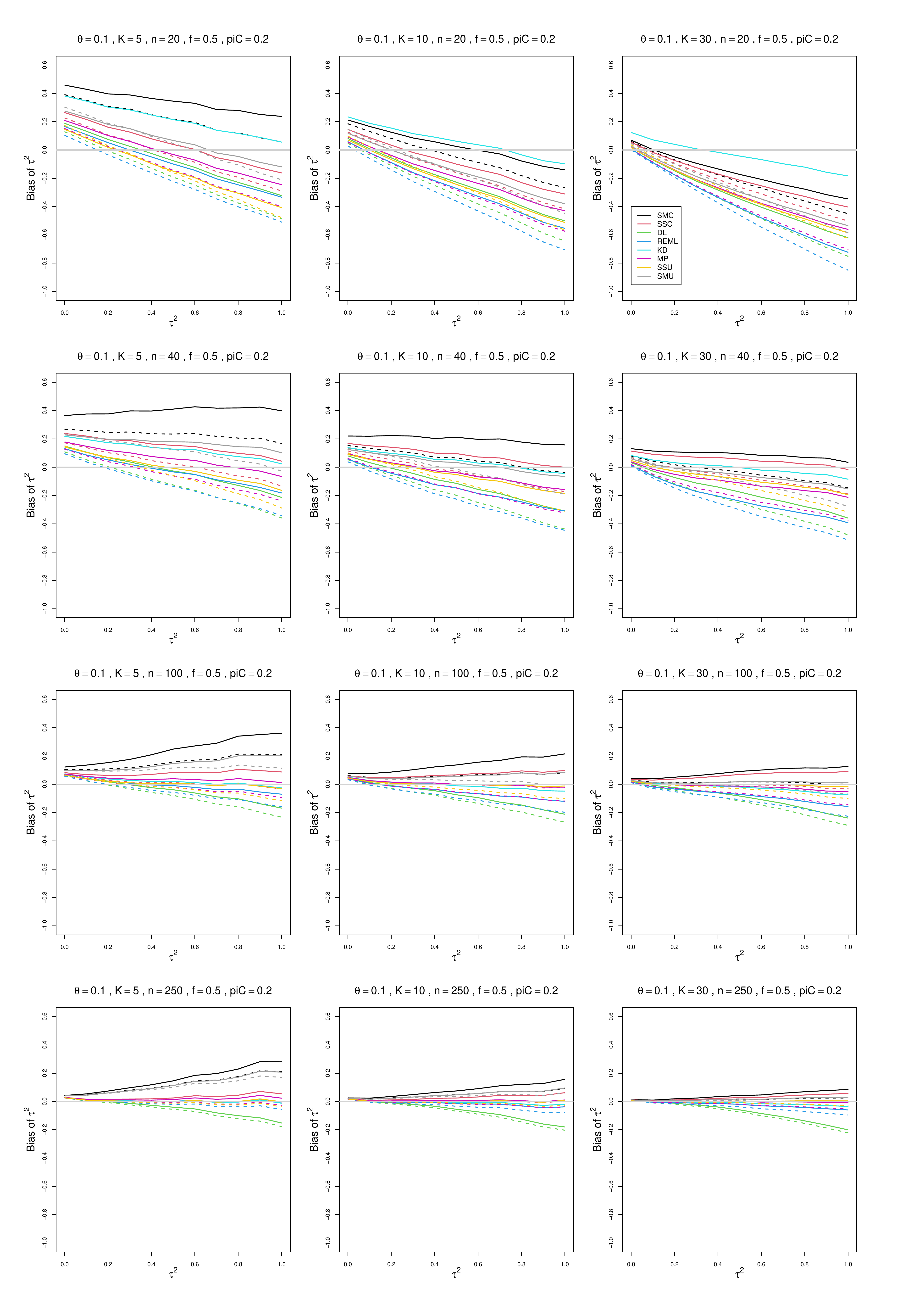}
	\caption{Bias  of estimators of between-study variance of LOR (DL, REML, KD, MP, SMC, SSC, SMU, and SSU) vs $\tau^2$, for equal sample sizes $n=20,\;40,\;100$ and $250$, $p_{iC} = .2$, $\theta=0.1$ and  $f=0.5$.   Solid lines: DL, REML, MP, SSC, SMC \lq\lq only"; KD; SSU and SMU model-based. Dashed lines: DL, REML, MP, SSC, SMC  \lq\lq always"; SSU and SMU na\"ive.  }
	\label{PlotBiasOfTau2_piC_02theta=0.1_LOR_equal_sample_sizes}
\end{figure}

\begin{figure}[ht]
	\centering
	\includegraphics[scale=0.33]{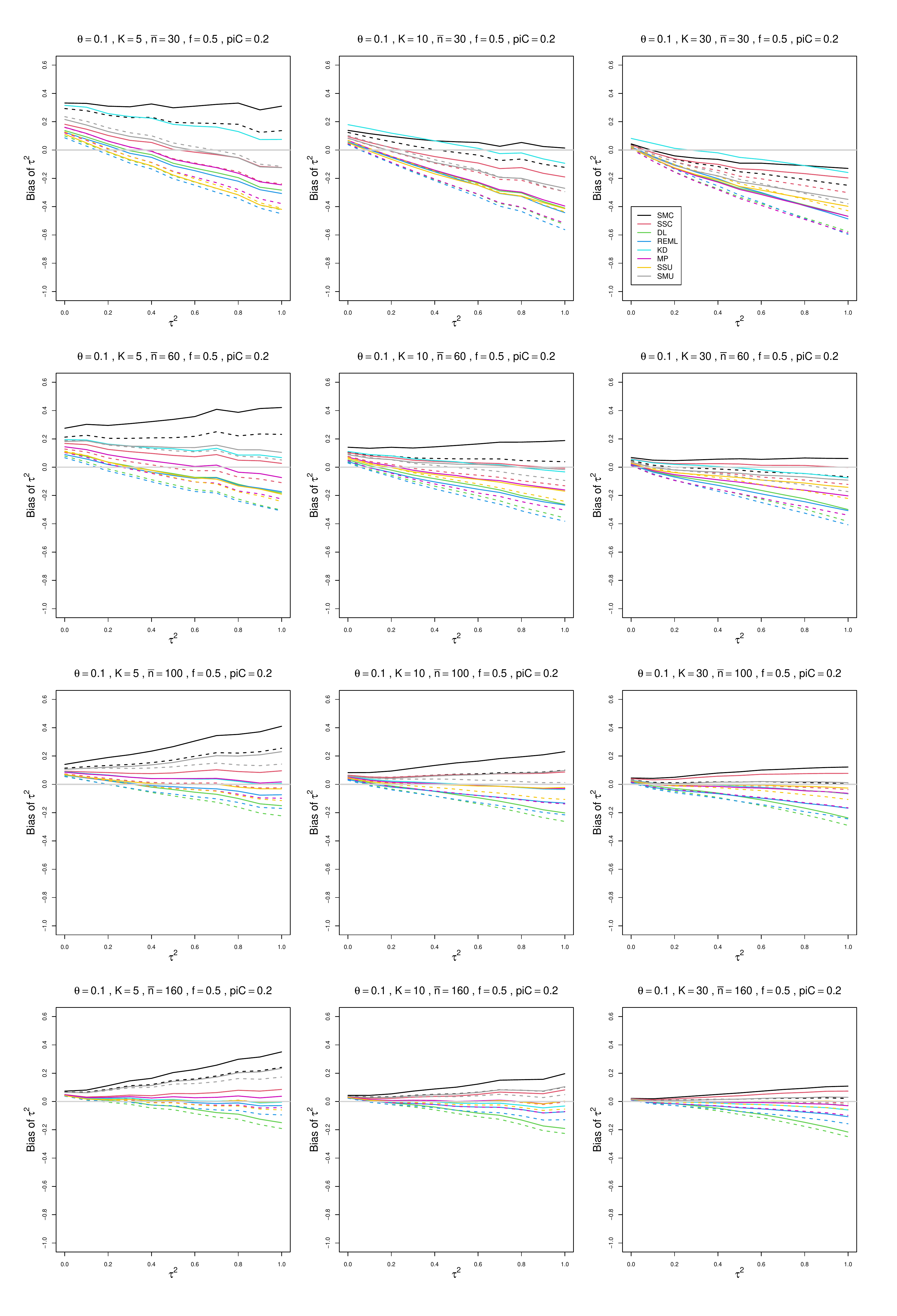}
	\caption{Bias  of estimators of between-study variance of LOR (DL, REML, KD, MP, SMC, SSC, SMU, and SSU) vs $\tau^2$, for unequal sample sizes $\bar{n}=30,\;60,\;100$ and $160$, $p_{iC} = .2$, $\theta=0.1$ and  $f=0.5$.   Solid lines: DL, REML, MP, SSC, SMC \lq\lq only"; KD; SSU and SMU model-based. Dashed lines: DL, REML, MP, SSC, SMC  \lq\lq always"; SSU and SMU na\"ive.  }
	\label{PlotBiasOfTau2_piC_02theta=0.1_LOR_unequal_sample_sizes}
\end{figure}

\begin{figure}[ht]
	\centering
	\includegraphics[scale=0.33]{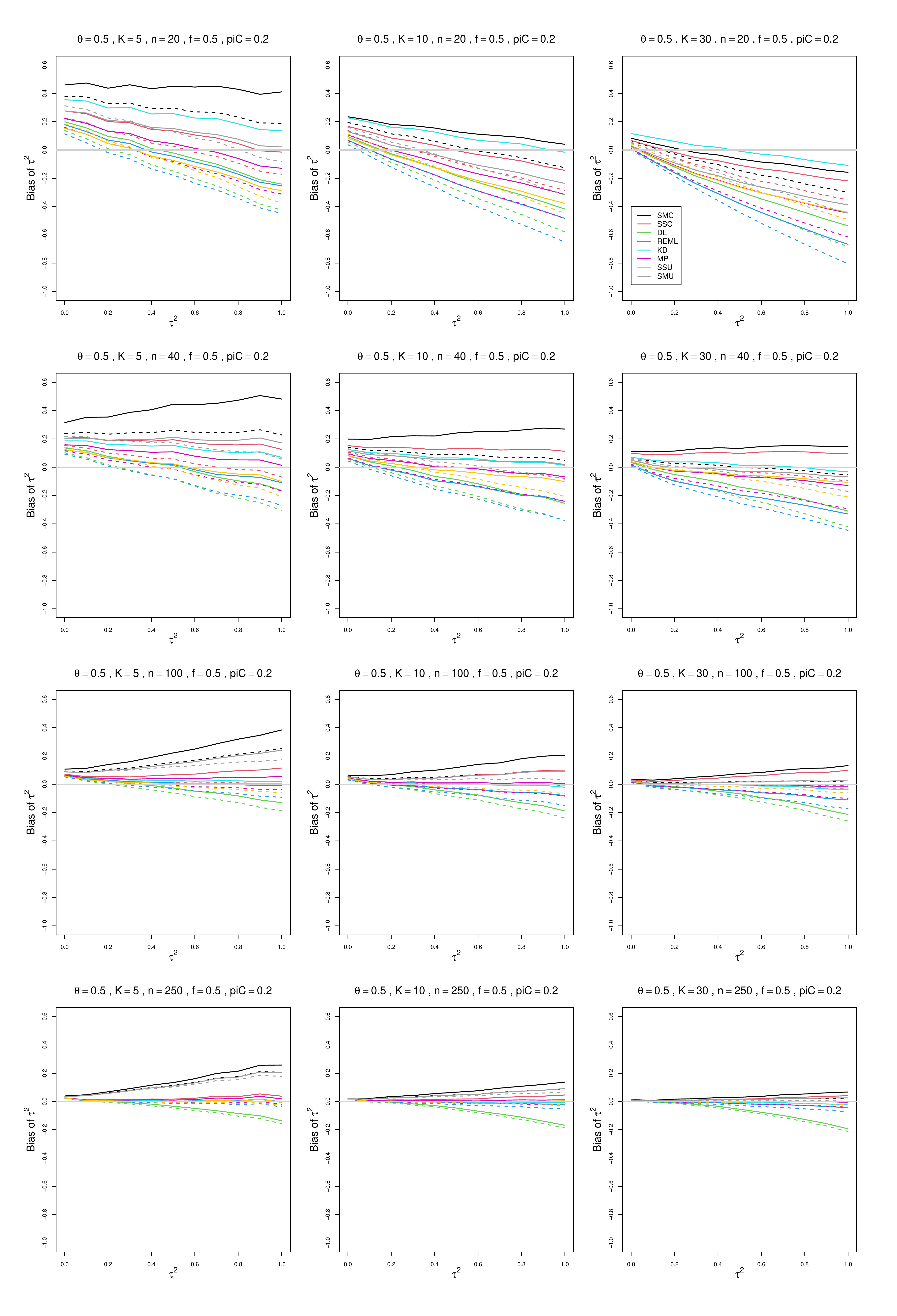}
	\caption{Bias  of estimators of between-study variance of LOR (DL, REML, KD, MP, SMC, SSC, SMU, and SSU) vs $\tau^2$, for equal sample sizes $n=20,\;40,\;100$ and $250$, $p_{iC} = .2$, $\theta=0.5$ and  $f=0.5$.   Solid lines: DL, REML, MP, SSC, SMC \lq\lq only"; KD; SSU and SMU model-based. Dashed lines: DL, REML, MP, SSC, SMC  \lq\lq always"; SSU and SMU na\"ive.  }
	\label{PlotBiasOfTau2_piC_02theta=0.5_LOR_equal_sample_sizes}
\end{figure}

\begin{figure}[ht]
	\centering
	\includegraphics[scale=0.33]{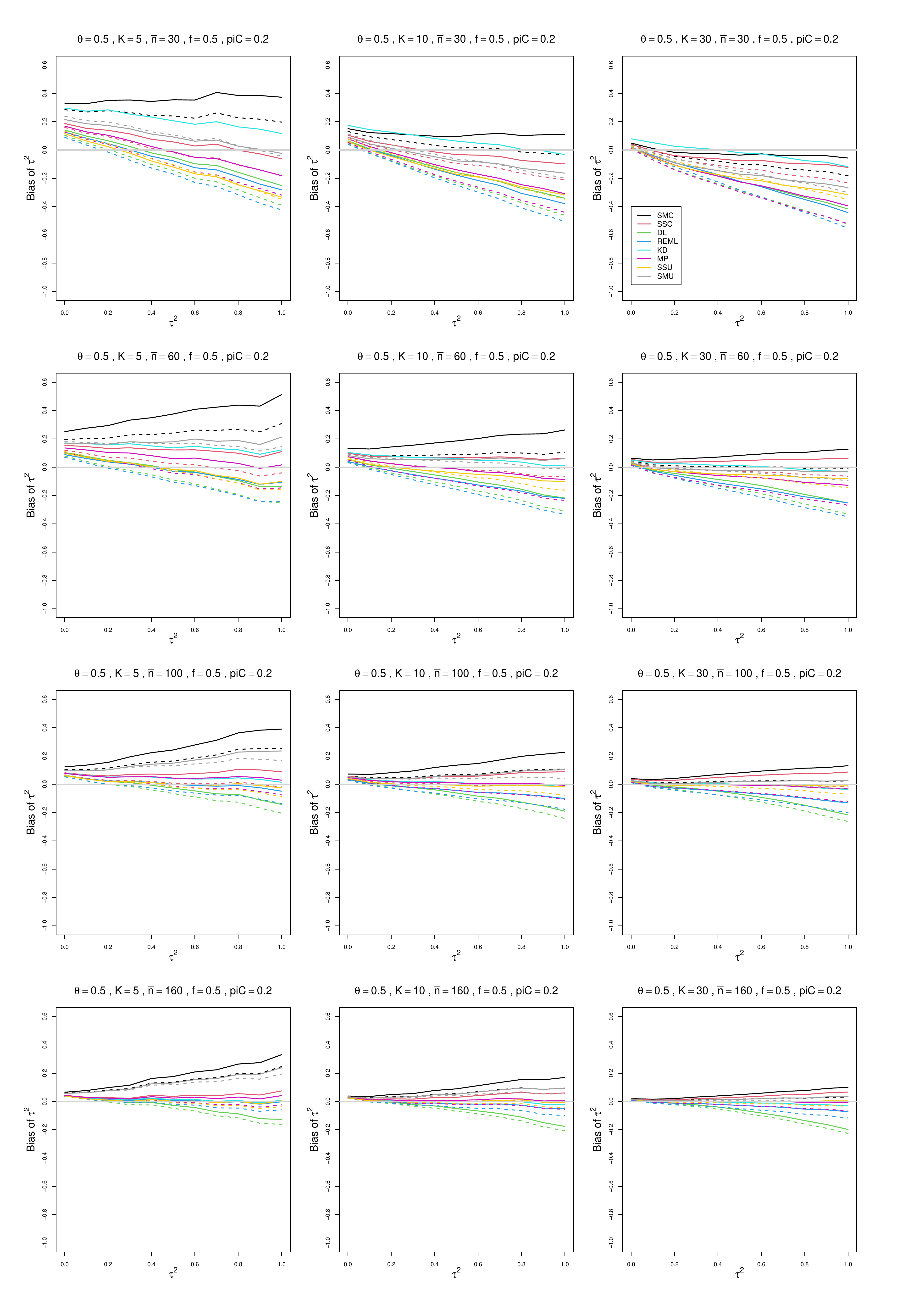}
	\caption{Bias  of estimators of between-study variance of LOR (DL, REML, KD, MP, SMC, SSC, SMU, and SSU) vs $\tau^2$, for unequal sample sizes $\bar{n}=30,\;60,\;100$ and $160$, $p_{iC} = .2$, $\theta=0.5$ and  $f=0.5$.   Solid lines: DL, REML, MP, SSC, SMC \lq\lq only"; KD; SSU and SMU model-based. Dashed lines: DL, REML, MP, SSC, SMC  \lq\lq always"; SSU and SMU na\"ive. }
	\label{PlotBiasOfTau2_piC_02theta=0.5_LOR_unequal_sample_sizes}
\end{figure}

\begin{figure}[ht]
	\centering
	\includegraphics[scale=0.33]{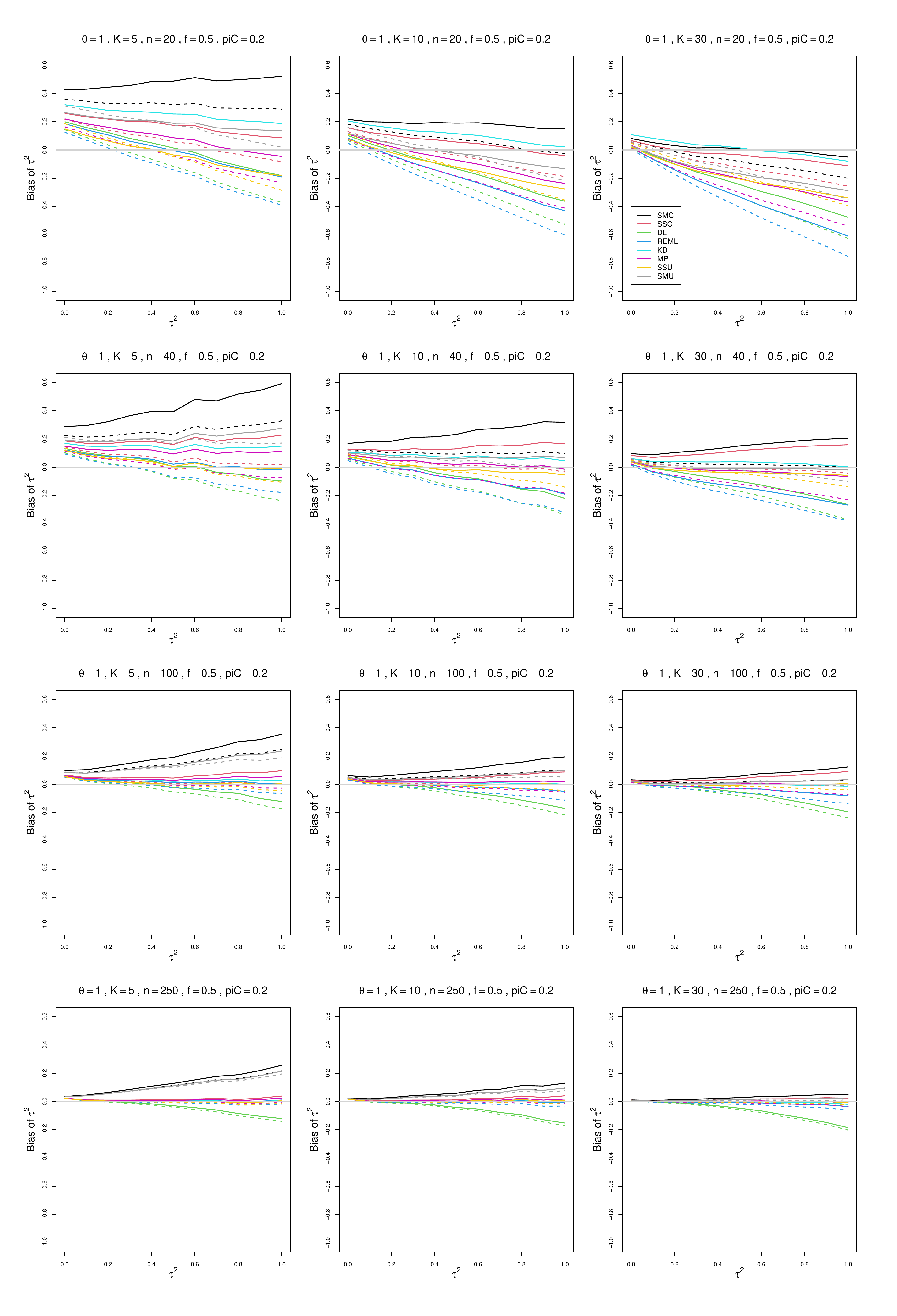}
	\caption{Bias  of estimators of between-study variance of LOR (DL, REML, KD, MP, SMC, SSC, SMU, and SSU) vs $\tau^2$, for equal sample sizes $n=20,\;40,\;100$ and $250$, $p_{iC} = .2$, $\theta=1$ and  $f=0.5$.   Solid lines: DL, REML, MP, SSC, SMC \lq\lq only"; KD; SSU and SMU model-based. Dashed lines: DL, REML, MP, SSC, SMC  \lq\lq always"; SSU and SMU na\"ive.   }
	\label{PlotBiasOfTau2_piC_02theta=1_LOR_equal_sample_sizes}
\end{figure}

\begin{figure}[ht]
	\centering
	\includegraphics[scale=0.33]{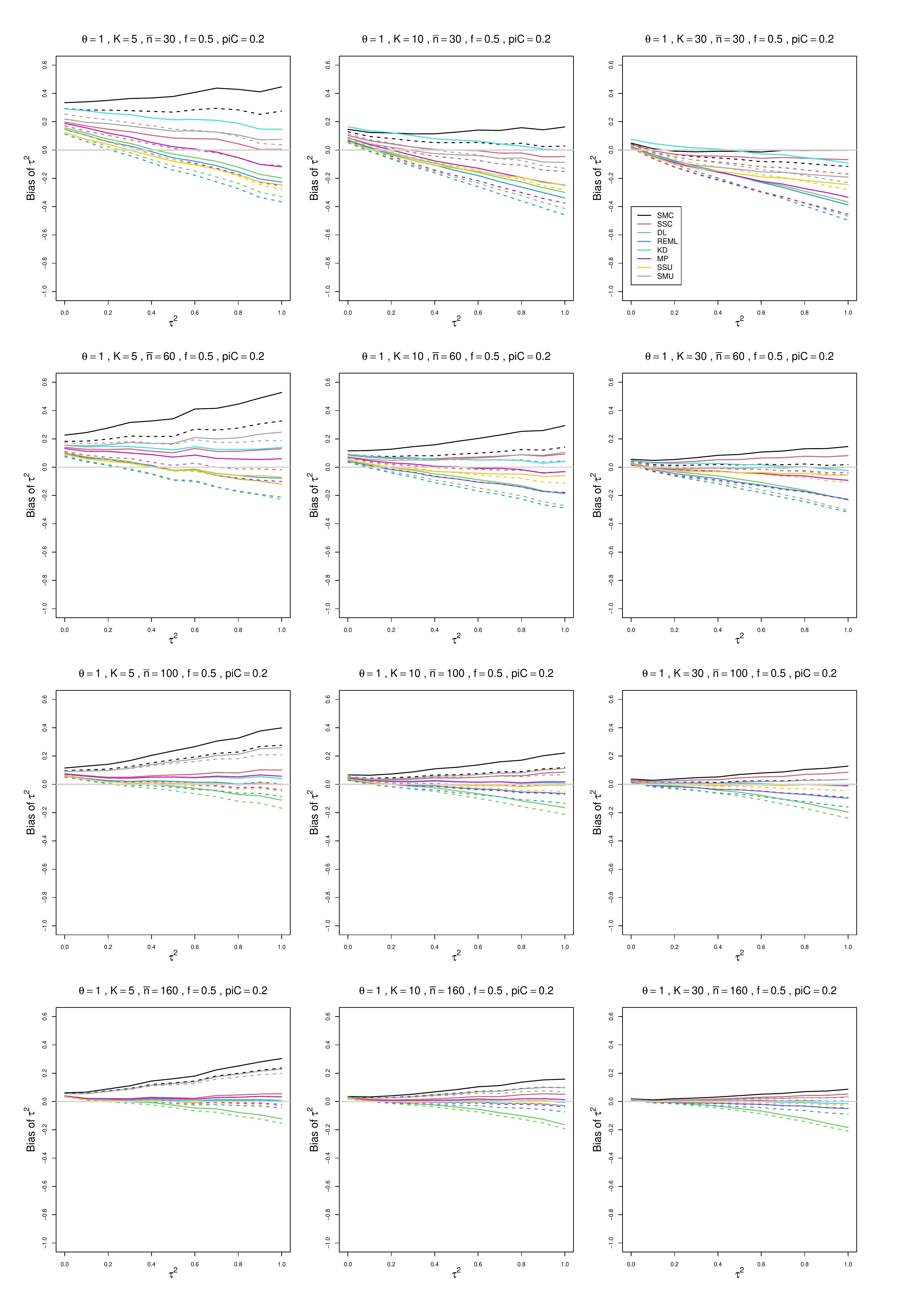}
	\caption{Bias  of estimators of between-study variance of LOR (DL, REML, KD, MP, SMC, SSC, SMU, and SSU) vs $\tau^2$, for unequal sample sizes $\bar{n}=30,\;60,\;100$ and $160$, $p_{iC} = .2$, $\theta=1$ and  $f=0.5$.   Solid lines: DL, REML, MP, SSC, SMC \lq\lq only"; KD; SSU and SMU model-based. Dashed lines: DL, REML, MP, SSC, SMC  \lq\lq always"; SSU and SMU na\"ive.  }
	\label{PlotBiasOfTau2_piC_02theta=1_LOR_unequal_sample_sizes}
\end{figure}

\begin{figure}[ht]
	\centering
	\includegraphics[scale=0.33]{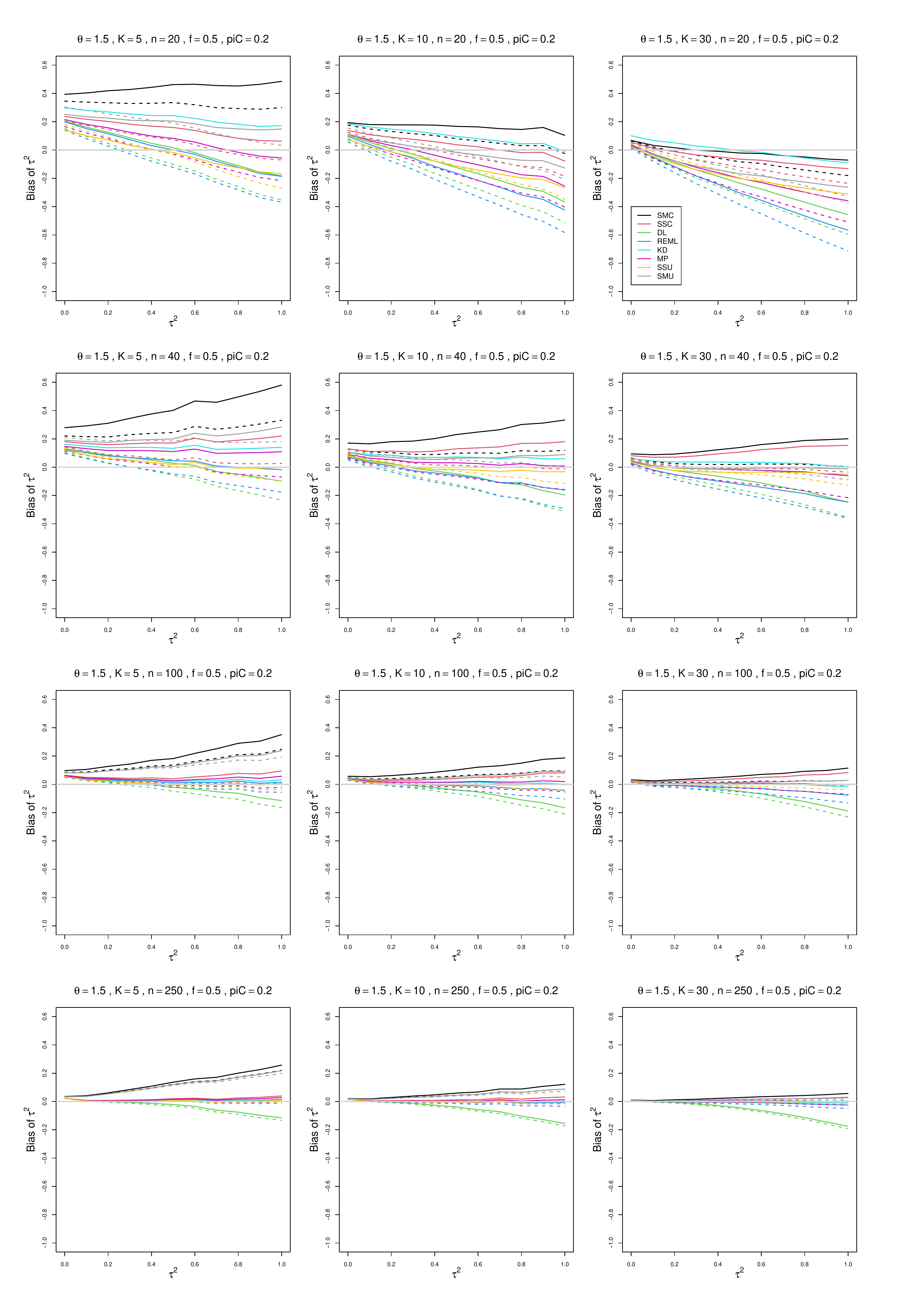}
	\caption{Bias  of estimators of between-study variance of LOR (DL, REML, KD, MP, SMC, SSC, SMU, and SSU) vs $\tau^2$, for equal sample sizes $n=20,\;40,\;100$ and $250$, $p_{iC} = .2$, $\theta=1.5$ and  $f=0.5$.   Solid lines: DL, REML, MP, SSC, SMC \lq\lq only"; KD; SSU and SMU model-based. Dashed lines: DL, REML, MP, SSC, SMC  \lq\lq always"; SSU and SMU na\"ive.  }
	\label{PlotBiasOfTau2_piC_02theta=1.5_LOR_equal_sample_sizes}
\end{figure}

\begin{figure}[ht]
	\centering
	\includegraphics[scale=0.33]{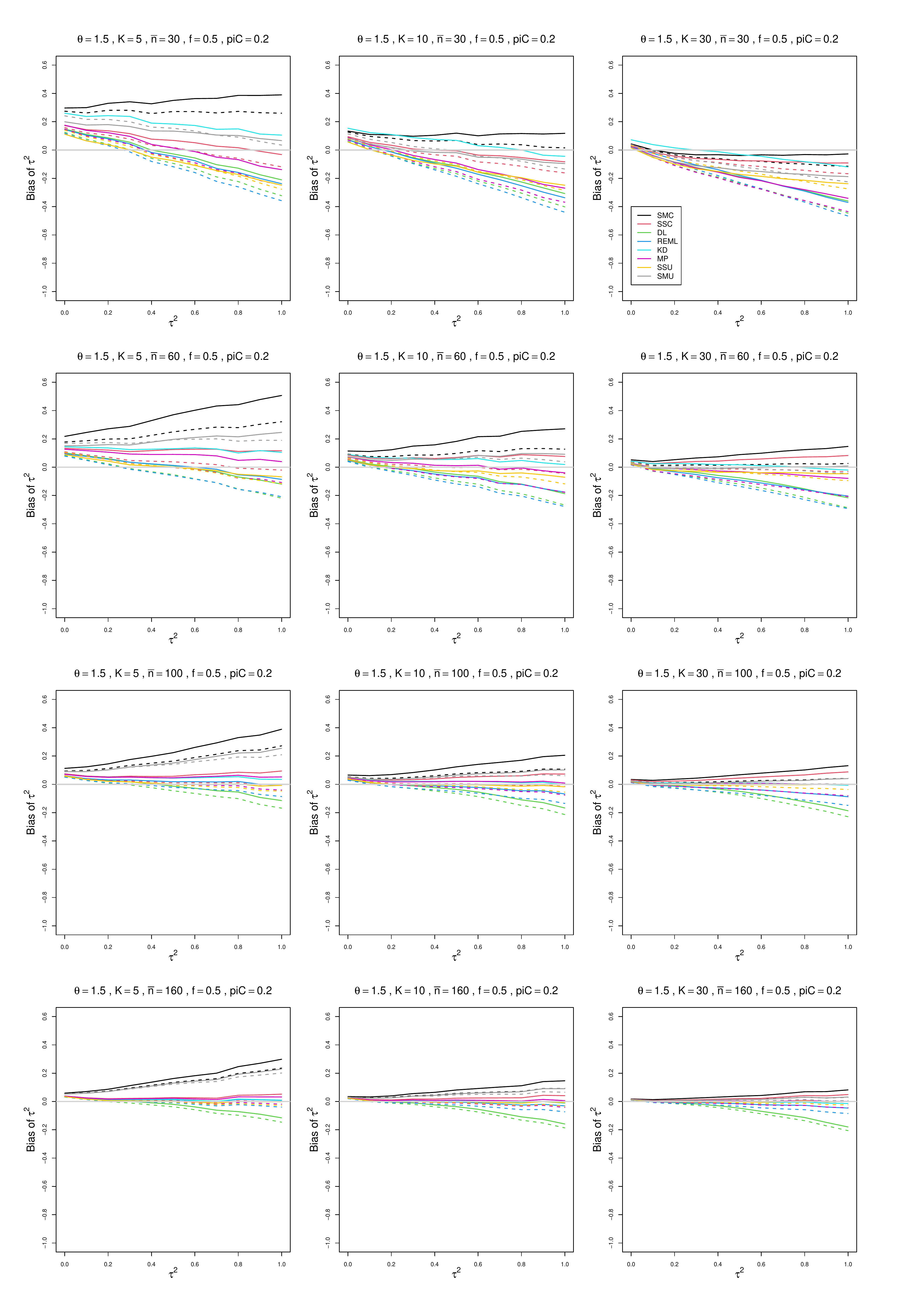}
	\caption{Bias  of estimators of between-study variance of LOR (DL, REML, KD, MP, SMC, SSC, SMU, and SSU) vs $\tau^2$, for unequal sample sizes $\bar{n}=30,\;60,\;100$ and $160$, $p_{iC} = .2$, $\theta=1.5$ and  $f=0.5$.   Solid lines: DL, REML, MP, SSC, SMC \lq\lq only"; KD; SSU and SMU model-based. Dashed lines: DL, REML, MP, SSC, SMC  \lq\lq always"; SSU and SMU na\"ive. }
	\label{PlotBiasOfTau2_piC_02theta=1.5_LOR_unequal_sample_sizes}
\end{figure}

\begin{figure}[ht]
	\centering
	\includegraphics[scale=0.33]{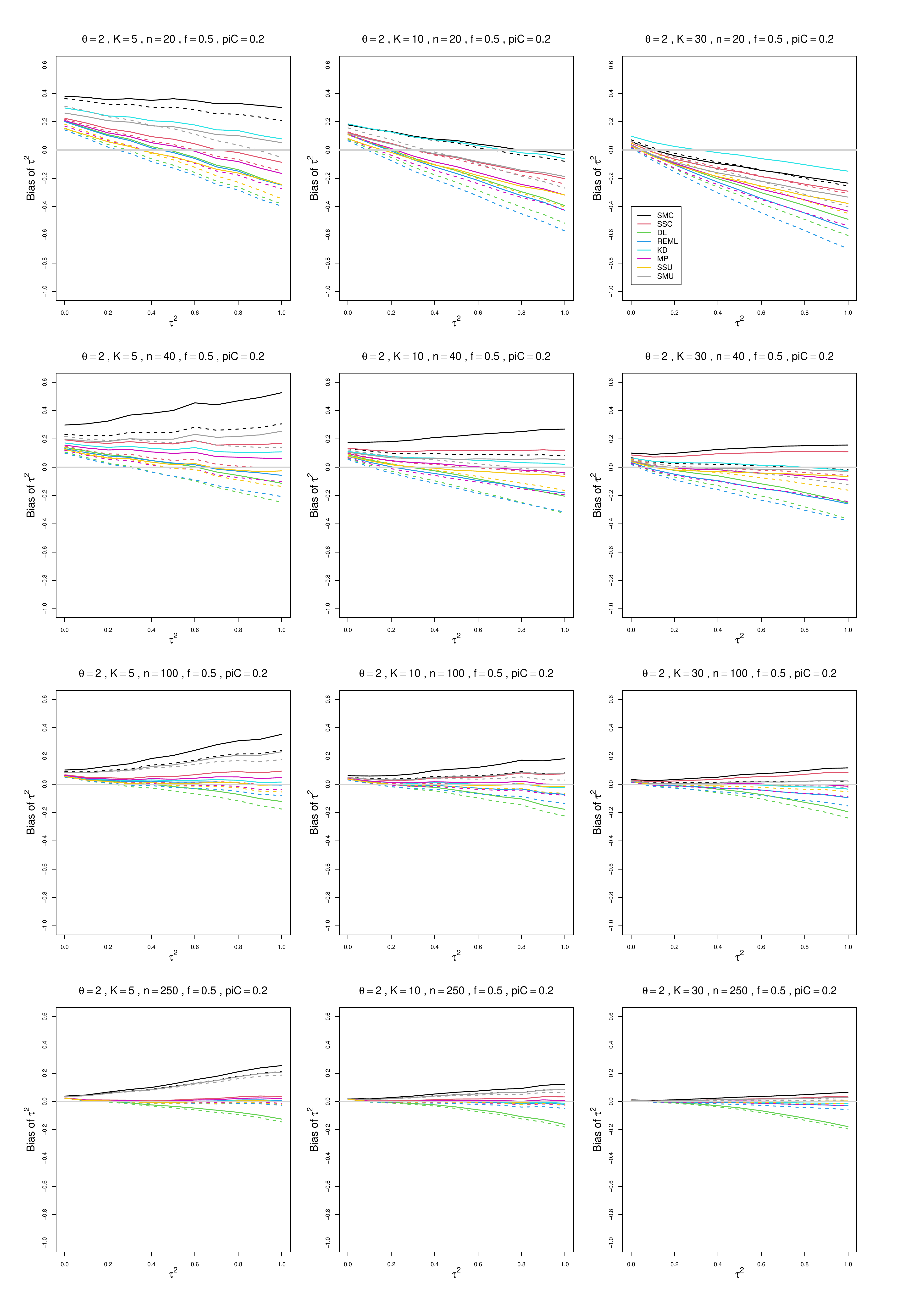}
	\caption{Bias  of estimators of between-study variance of LOR (DL, REML, KD, MP, SMC, SSC, SMU, and SSU) vs $\tau^2$, for equal sample sizes $n=20,\;40,\;100$ and $250$, $p_{iC} = .2$, $\theta=2$ and  $f=0.5$.   Solid lines: DL, REML, MP, SSC, SMC \lq\lq only"; KD; SSU and SMU model-based. Dashed lines: DL, REML, MP, SSC, SMC  \lq\lq always"; SSU and SMU na\"ive.  }
	\label{PlotBiasOfTau2_piC_02theta=2_LOR_equal_sample_sizes}
\end{figure}

\begin{figure}[ht]
	\centering
	\includegraphics[scale=0.33]{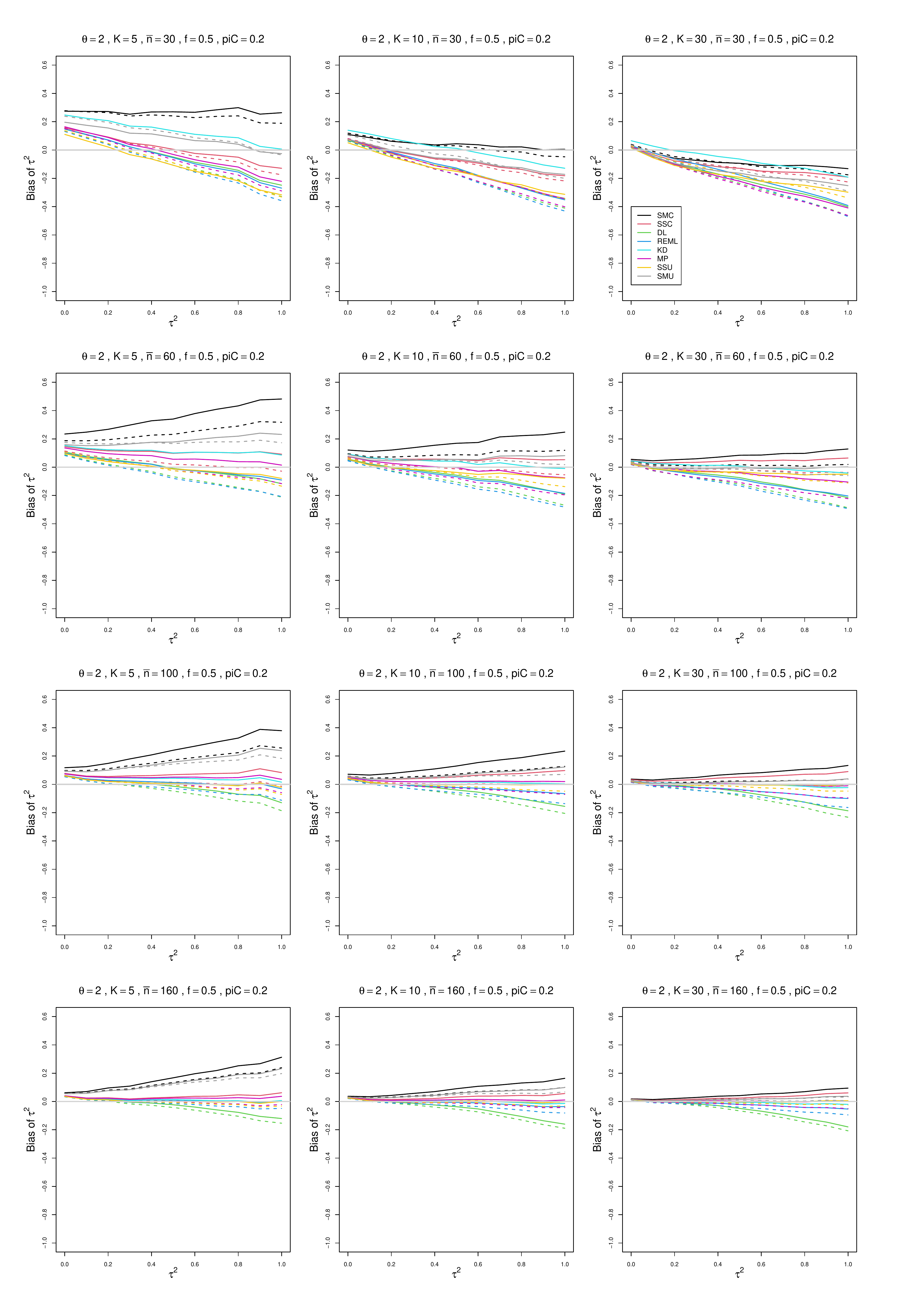}
	\caption{Bias  of estimators of between-study variance of LOR (DL, REML, KD, MP, SMC, SSC, SMU, and SSU) vs $\tau^2$, for unequal sample sizes $\bar{n}=30,\;60,\;100$ and $160$, $p_{iC} = .2$, $\theta=2$ and  $f=0.5$.   Solid lines: DL, REML, MP, SSC, SMC \lq\lq only"; KD; SSU and SMU model-based. Dashed lines: DL, REML, MP, SSC, SMC  \lq\lq always"; SSU and SMU na\"ive.  }
	\label{PlotBiasOfTau2_piC_02theta=2_LOR_unequal_sample_sizes}
\end{figure}

\begin{figure}[ht]
	\centering
	\includegraphics[scale=0.33]{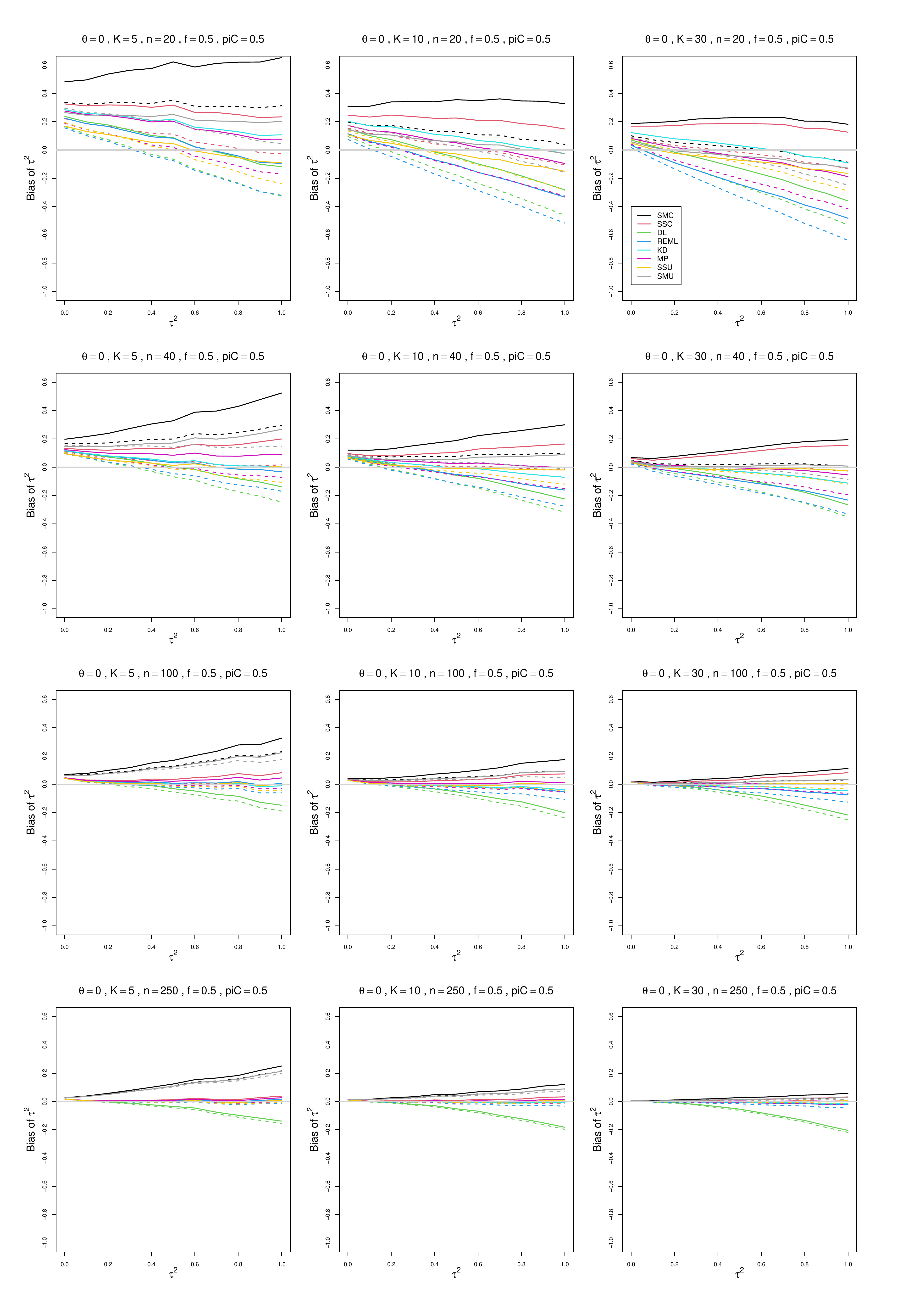}
	\caption{Bias  of estimators of between-study variance of LOR (DL, REML, KD, MP, SMC, SSC, SMU, and SSU) vs $\tau^2$, for equal sample sizes $n=20,\;40,\;100$ and $250$, $p_{iC} = .5$, $\theta=0$ and  $f=0.5$.   Solid lines: DL, REML, MP, SSC, SMC \lq\lq only"; KD; SSU and SMU model-based. Dashed lines: DL, REML, MP, SSC, SMC  \lq\lq always"; SSU and SMU na\"ive.  }
	\label{PlotBiasOfTau2_piC_05theta=0_LOR_equal_sample_sizes}
\end{figure}

\begin{figure}[ht]
	\centering
	\includegraphics[scale=0.33]{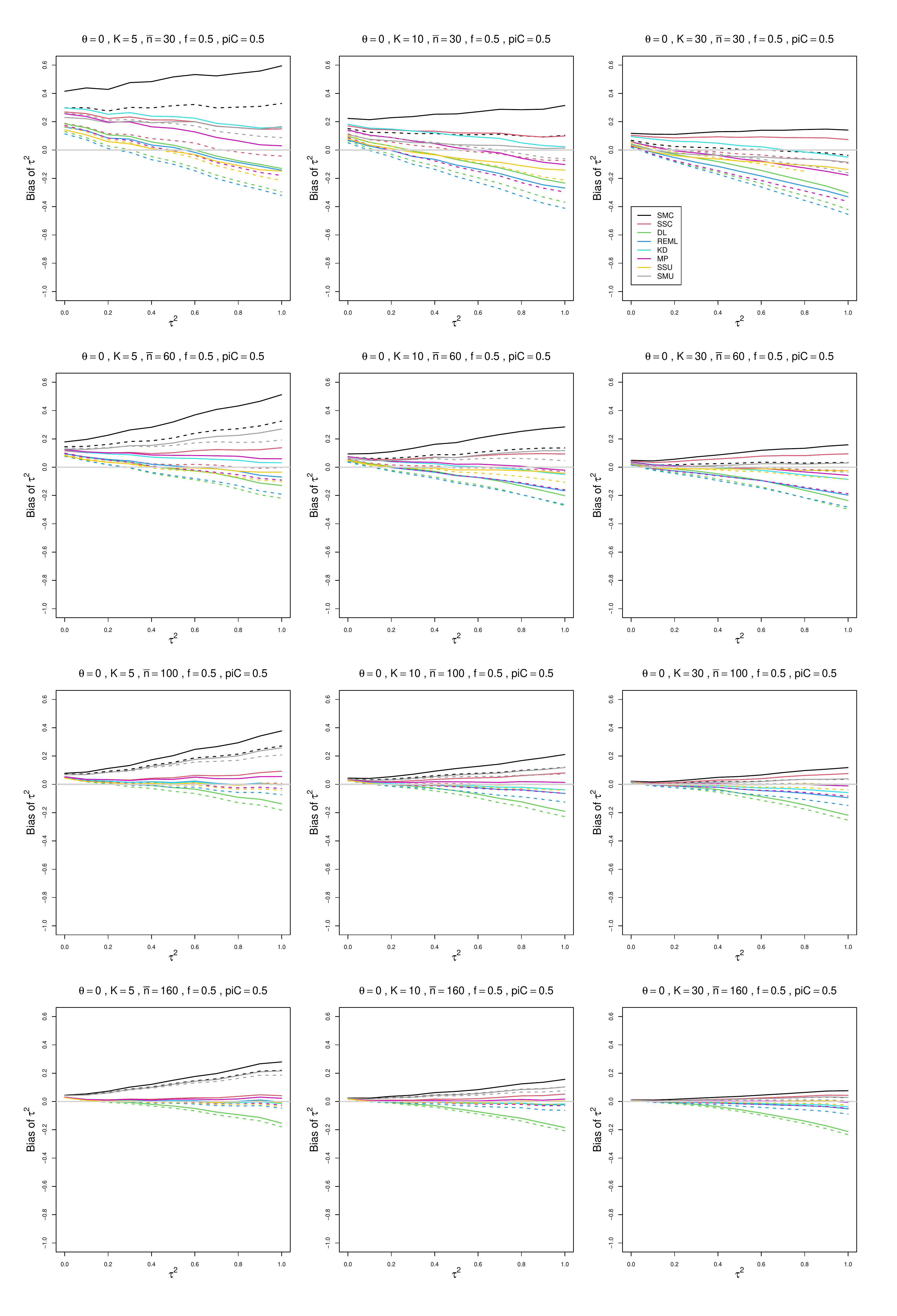}
	\caption{Bias  of estimators of between-study variance of LOR (DL, REML, KD, MP, SMC, SSC, SMU, and SSU) vs $\tau^2$, for unequal sample sizes $\bar{n}=30,\;60,\;100$ and $160$, $p_{iC} = .5$, $\theta=0$ and  $f=0.5$.   Solid lines: DL, REML, MP, SSC, SMC \lq\lq only"; KD; SSU and SMU model-based. Dashed lines: DL, REML, MP, SSC, SMC  \lq\lq always"; SSU and SMU na\"ive.  }
	\label{PlotBiasOfTau2_piC_05theta=0_LOR_unequal_sample_sizes}
\end{figure}

\begin{figure}[ht]
	\centering
	\includegraphics[scale=0.33]{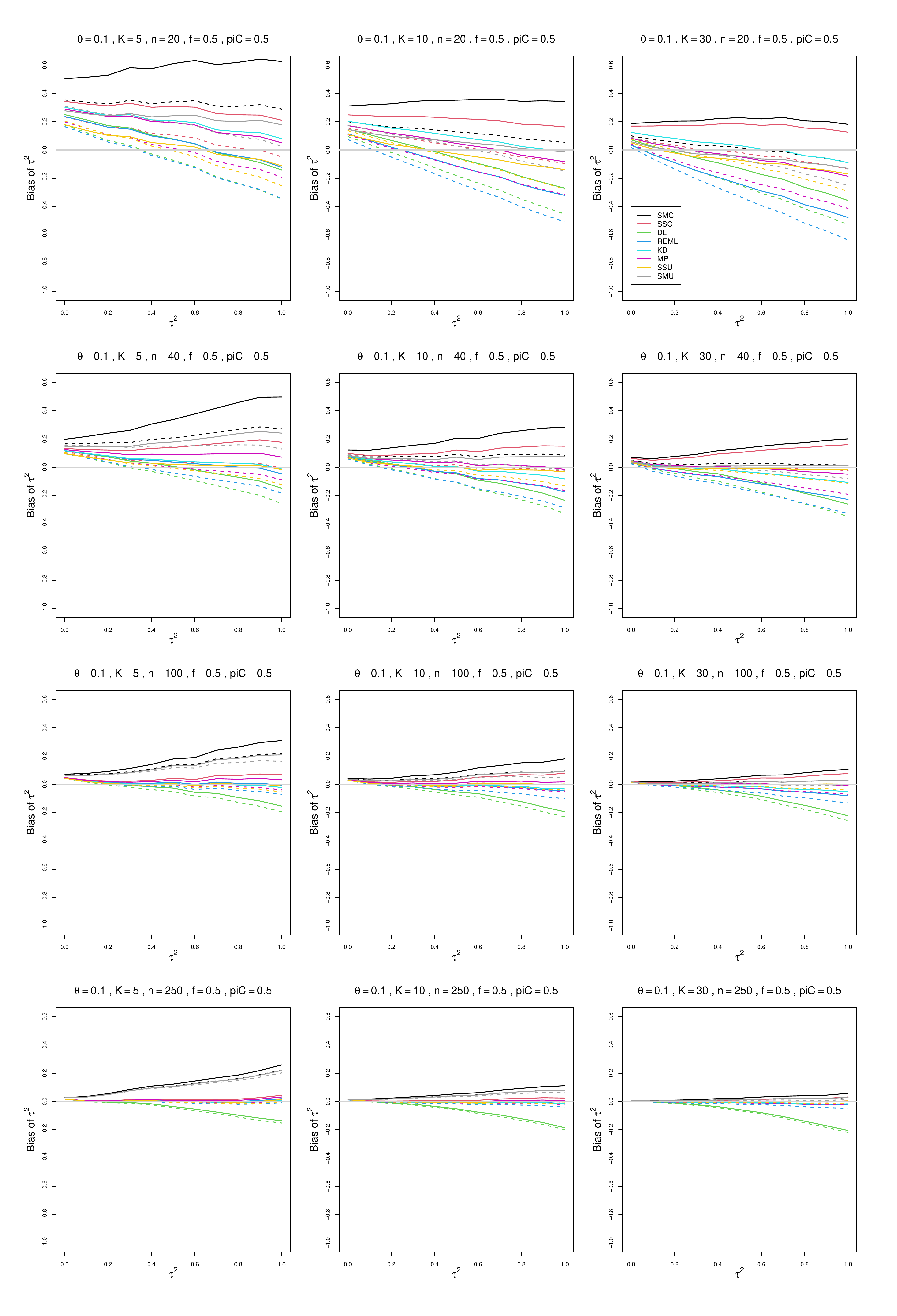}
	\caption{Bias  of estimators of between-study variance of LOR (DL, REML, KD, MP, SMC, SSC, SMU, and SSU) vs $\tau^2$, for equal sample sizes $n=20,\;40,\;100$ and $250$, $p_{iC} = .5$, $\theta=0.1$ and  $f=0.5$.   Solid lines: DL, REML, MP, SSC, SMC \lq\lq only"; KD; SSU and SMU model-based. Dashed lines: DL, REML, MP, SSC, SMC  \lq\lq always"; SSU and SMU na\"ive.  }
	\label{PlotBiasOfTau2_piC_05theta=0.1_LOR_equal_sample_sizes}
\end{figure}

\begin{figure}[ht]
	\centering
	\includegraphics[scale=0.33]{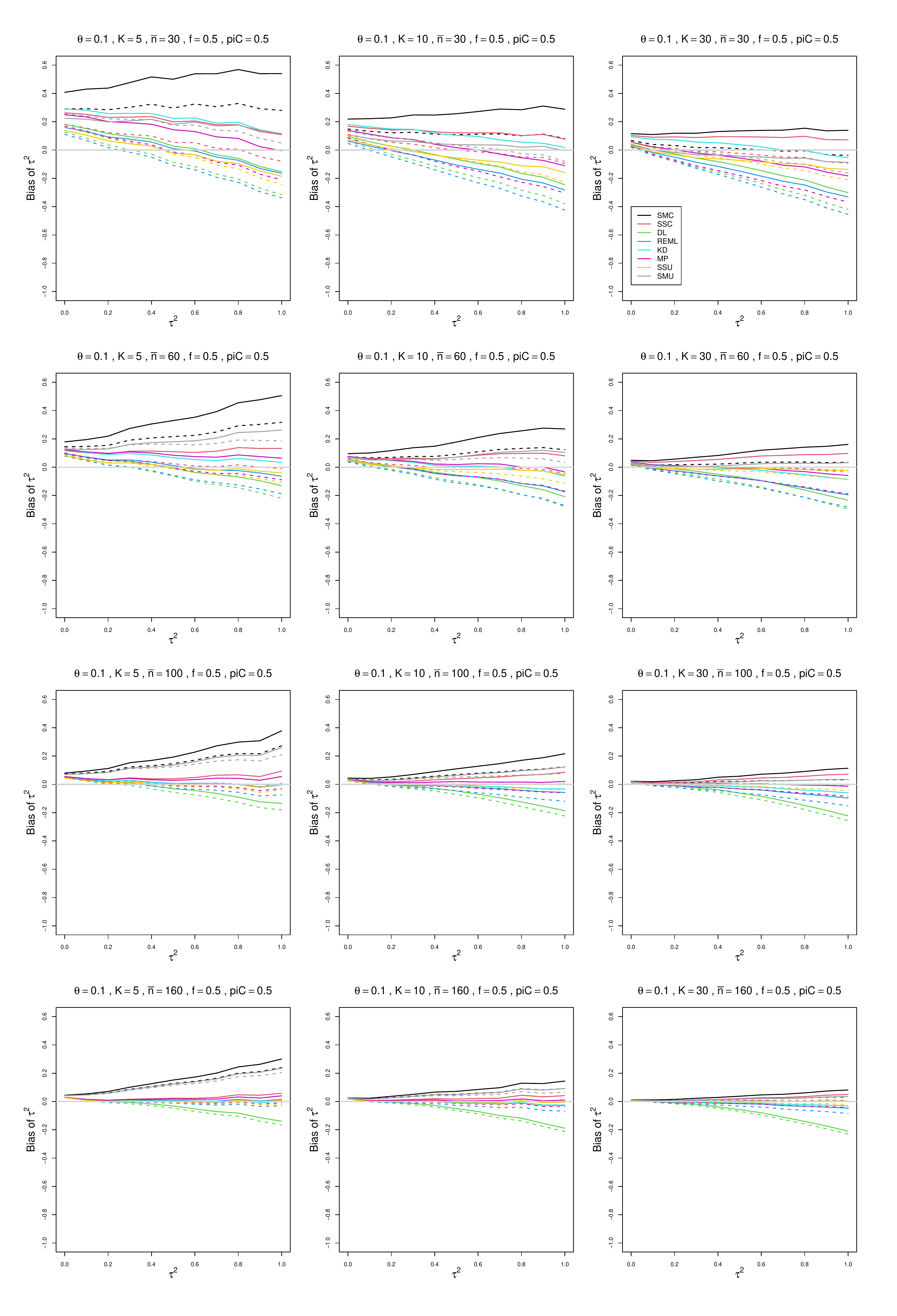}
	\caption{Bias  of estimators of between-study variance of LOR (DL, REML, KD, MP, SMC, SSC, SMU, and SSU) vs $\tau^2$, for unequal sample sizes $\bar{n}=30,\;60,\;100$ and $160$, $p_{iC} = .5$, $\theta=0.1$ and  $f=0.5$.   Solid lines: DL, REML, MP, SSC, SMC \lq\lq only"; KD; SSU and SMU model-based. Dashed lines: DL, REML, MP, SSC, SMC  \lq\lq always"; SSU and SMU na\"ive.   }
	\label{PlotBiasOfTau2_piC_05theta=0.1_LOR_unequal_sample_sizes}
\end{figure}

\begin{figure}[ht]
	\centering
	\includegraphics[scale=0.33]{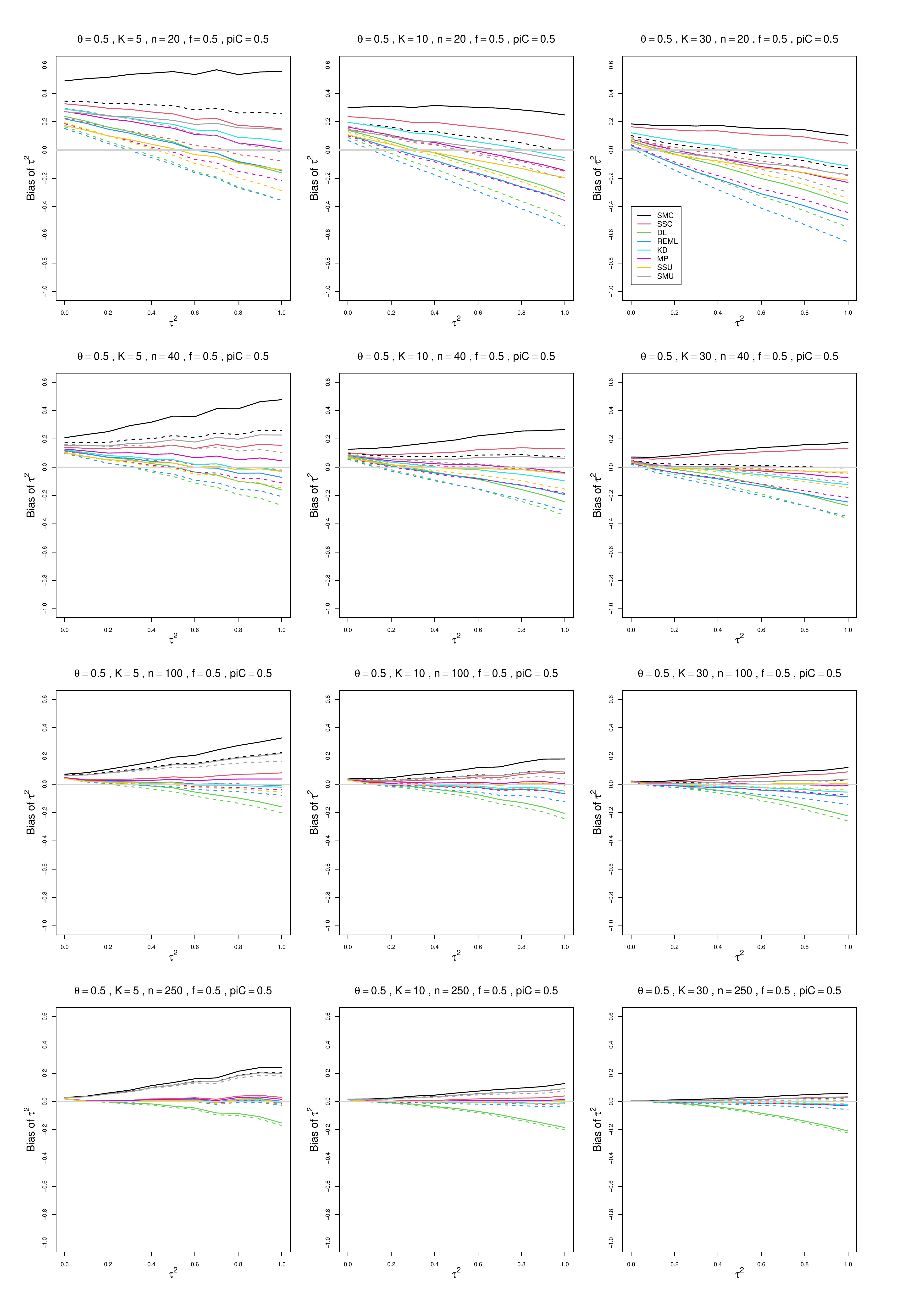}
	\caption{Bias of estimators of between-study variance of LOR (DL, REML, KD, MP, SMC, SSC, SMU, and SSU) vs $\tau^2$, for equal sample sizes $n=20,\;40,\;100$ and $250$, $p_{iC} = .5$, $\theta=0.5$ and  $f=0.5$.   Solid lines: DL, REML, MP, SSC, SMC \lq\lq only"; KD; SSU and SMU model-based. Dashed lines: DL, REML, MP, SSC, SMC  \lq\lq always"; SSU and SMU na\"ive.  }
	\label{PlotBiasOfTau2_piC_05theta=0.5_LOR_equal_sample_sizes}
\end{figure}

\begin{figure}[ht]
	\centering
	\includegraphics[scale=0.33]{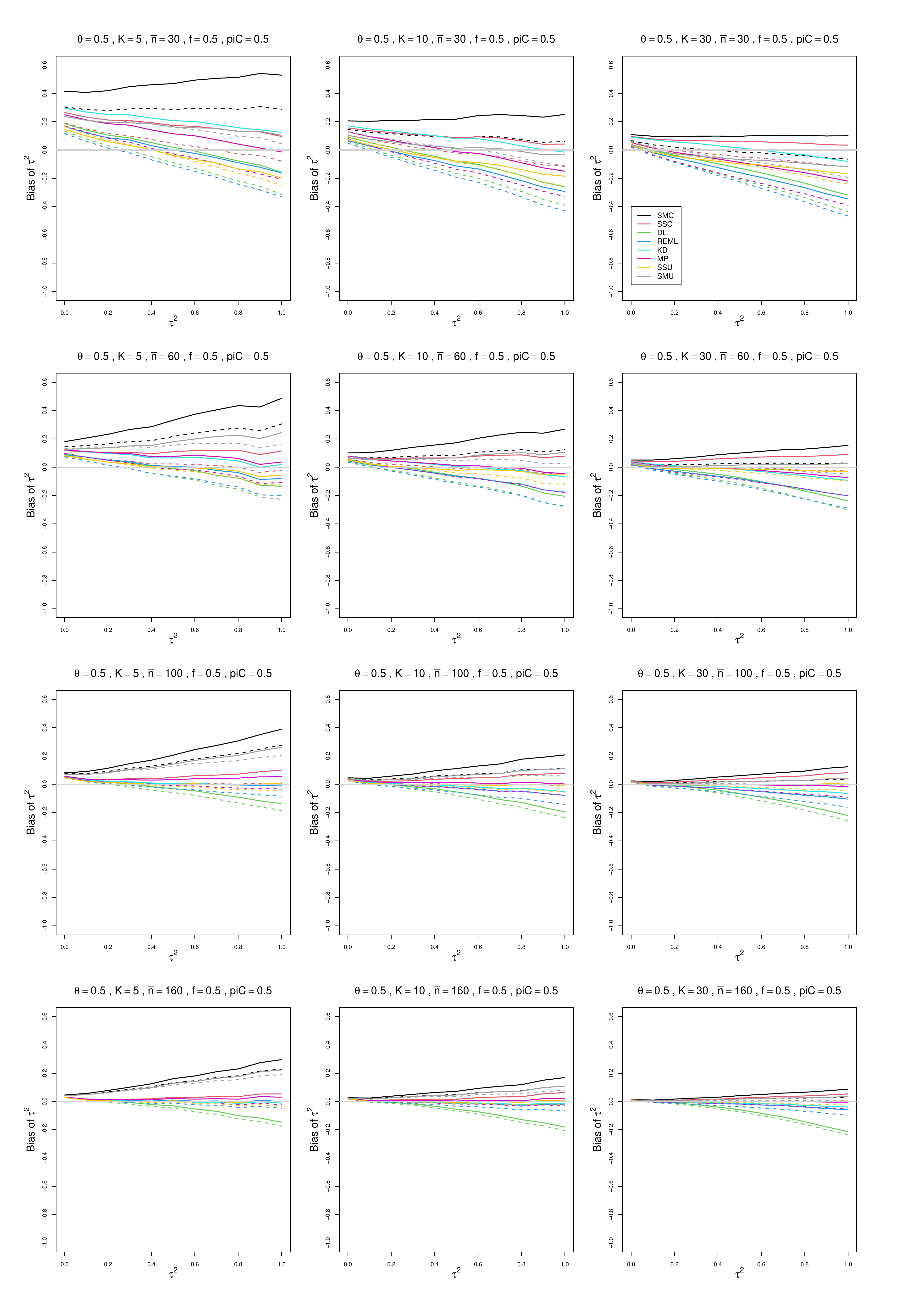}
	\caption{Bias  of estimators of between-study variance of LOR (DL, REML, KD, MP, SMC, SSC, SMU, and SSU) vs $\tau^2$, for unequal sample sizes $\bar{n}=30,\;60,\;100$ and $160$, $p_{iC} = .5$, $\theta=0.5$ and  $f=0.5$.   Solid lines: DL, REML, MP, SSC, SMC \lq\lq only"; KD; SSU and SMU model-based. Dashed lines: DL, REML, MP, SSC, SMC  \lq\lq always"; SSU and SMU na\"ive.  }
	\label{PlotBiasOfTau2_piC_05theta=0.5_LOR_unequal_sample_sizes}
\end{figure}

\begin{figure}[ht]
	\centering
	\includegraphics[scale=0.33]{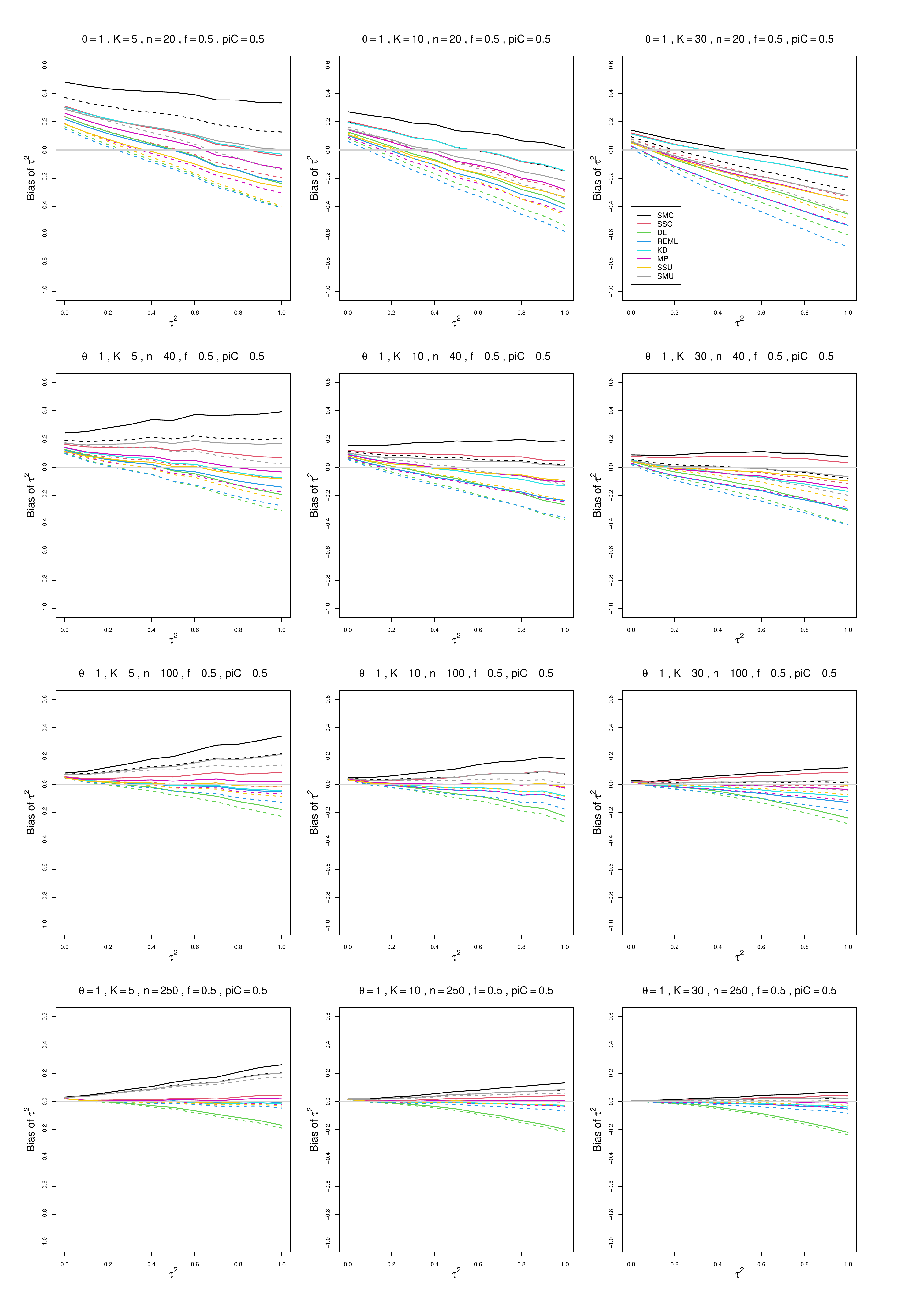}
	\caption{Bias  of estimators of between-study variance of LOR (DL, REML, KD, MP, SMC, SSC, SMU, and SSU) vs $\tau^2$, for equal sample sizes $n=20,\;40,\;100$ and $250$, $p_{iC} = .5$, $\theta=1$ and  $f=0.5$.   Solid lines: DL, REML, MP, SSC, SMC \lq\lq only"; KD; SSU and SMU model-based. Dashed lines: DL, REML, MP, SSC, SMC  \lq\lq always"; SSU and SMU na\"ive.  }
	\label{PlotBiasOfTau2_piC_05theta=1_LOR_equal_sample_sizes}
\end{figure}

\begin{figure}[ht]
	\centering
	\includegraphics[scale=0.33]{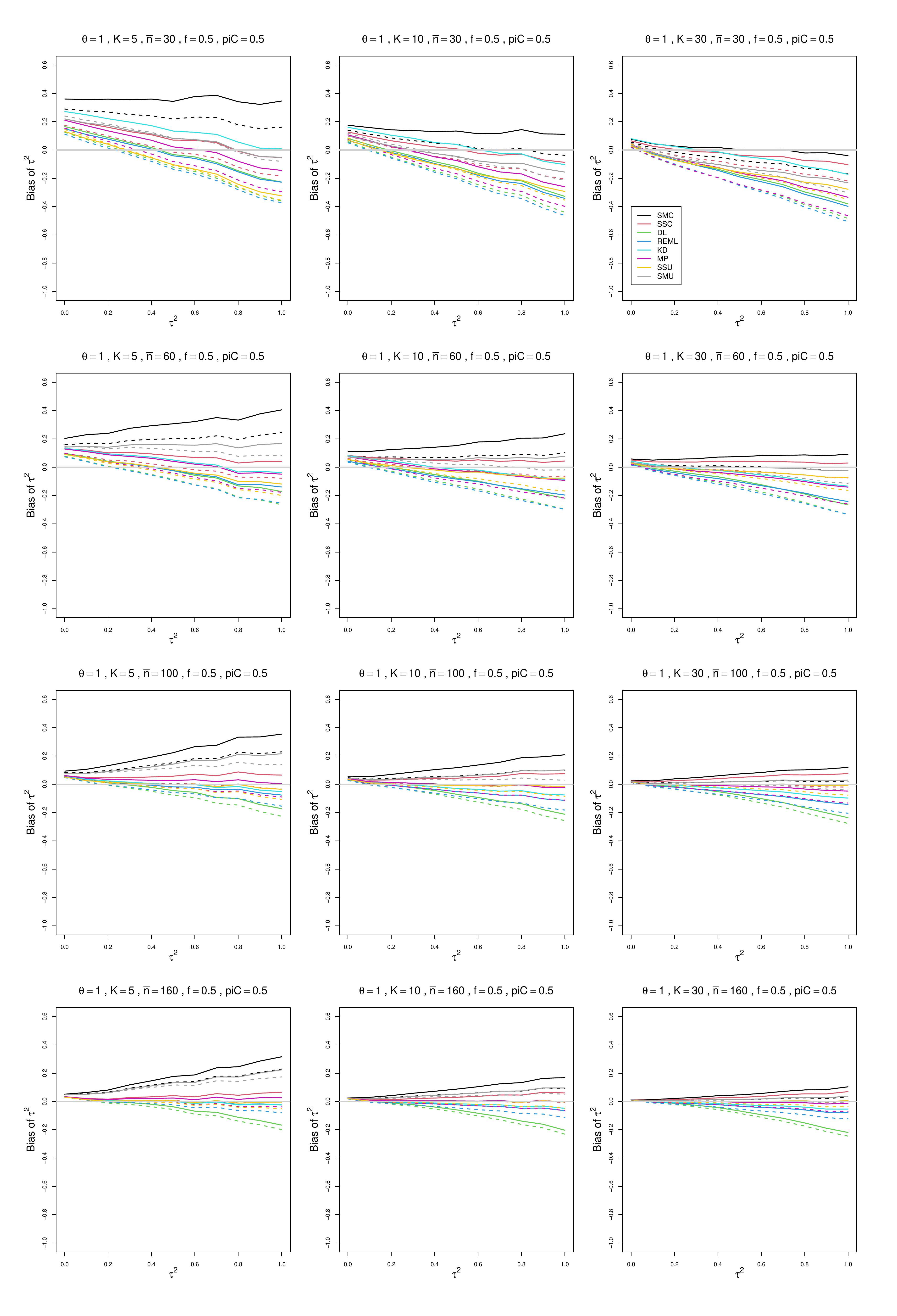}
	\caption{Bias  of estimators of between-study variance of LOR (DL, REML, KD, MP, SMC, SSC, SMU, and SSU) vs $\tau^2$, for unequal sample sizes $\bar{n}=30,\;60,\;100$ and $160$, $p_{iC} = .5$, $\theta=1$ and  $f=0.5$.   Solid lines: DL, REML, MP, SSC, SMC \lq\lq only"; KD; SSU and SMU model-based. Dashed lines: DL, REML, MP, SSC, SMC  \lq\lq always"; SSU and SMU na\"ive.  }
	\label{PlotBiasOfTau2_piC_05theta=1_LOR_unequal_sample_sizes}
\end{figure}

\begin{figure}[ht]
	\centering
	\includegraphics[scale=0.33]{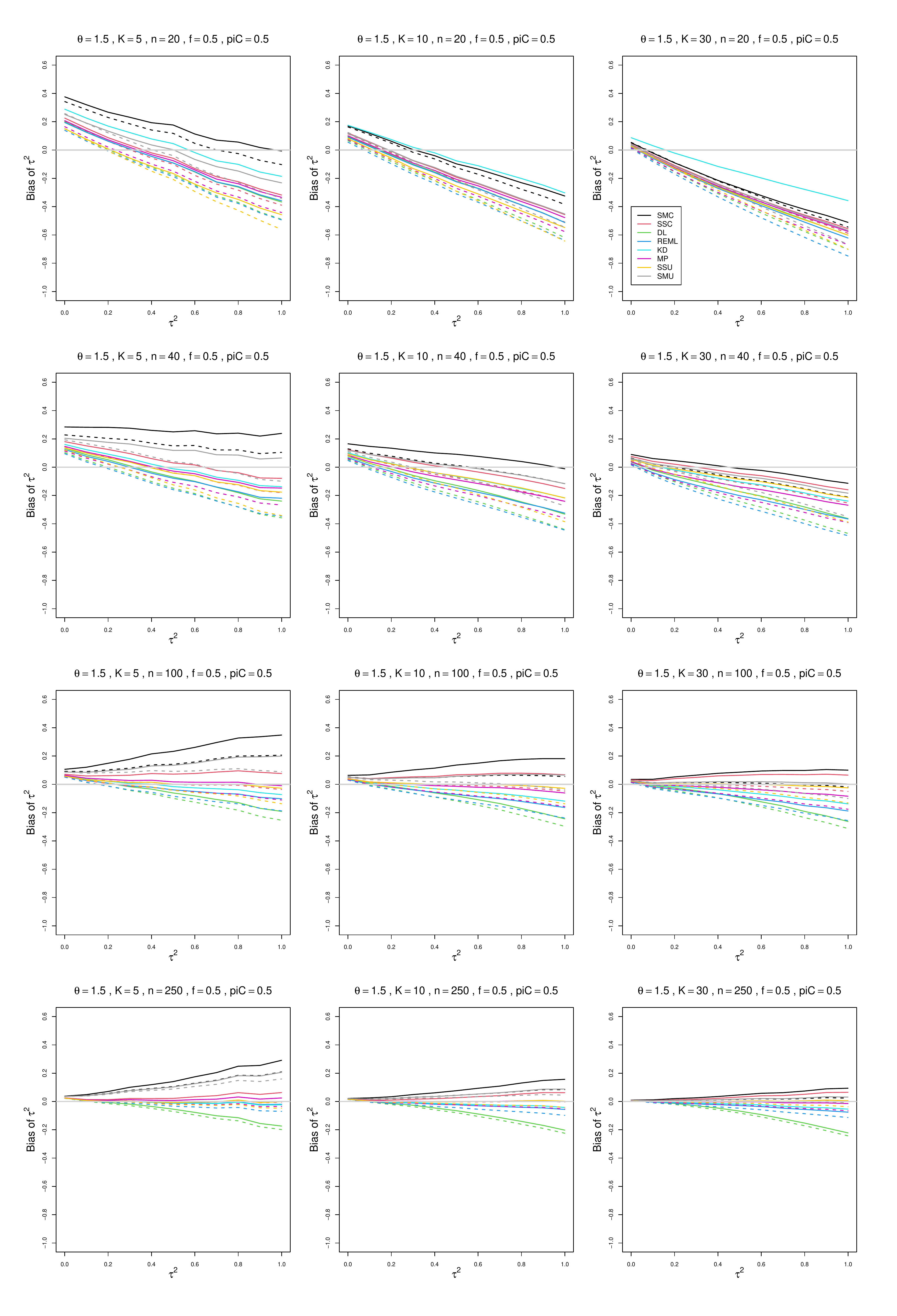}
	\caption{Bias  of estimators of between-study variance of LOR (DL, REML, KD, MP, SMC, SSC, SMU, and SSU) vs $\tau^2$, for equal sample sizes $n=20,\;40,\;100$ and $250$, $p_{iC} = .5$, $\theta=1.5$ and  $f=0.5$.   Solid lines: DL, REML, MP, SSC, SMC \lq\lq only"; KD; SSU and SMU model-based. Dashed lines: DL, REML, MP, SSC, SMC  \lq\lq always"; SSU and SMU na\"ive.  }
	\label{PlotBiasOfTau2_piC_05theta=1.5_LOR_equal_sample_sizes}
\end{figure}

\begin{figure}[ht]
	\centering
	\includegraphics[scale=0.33]{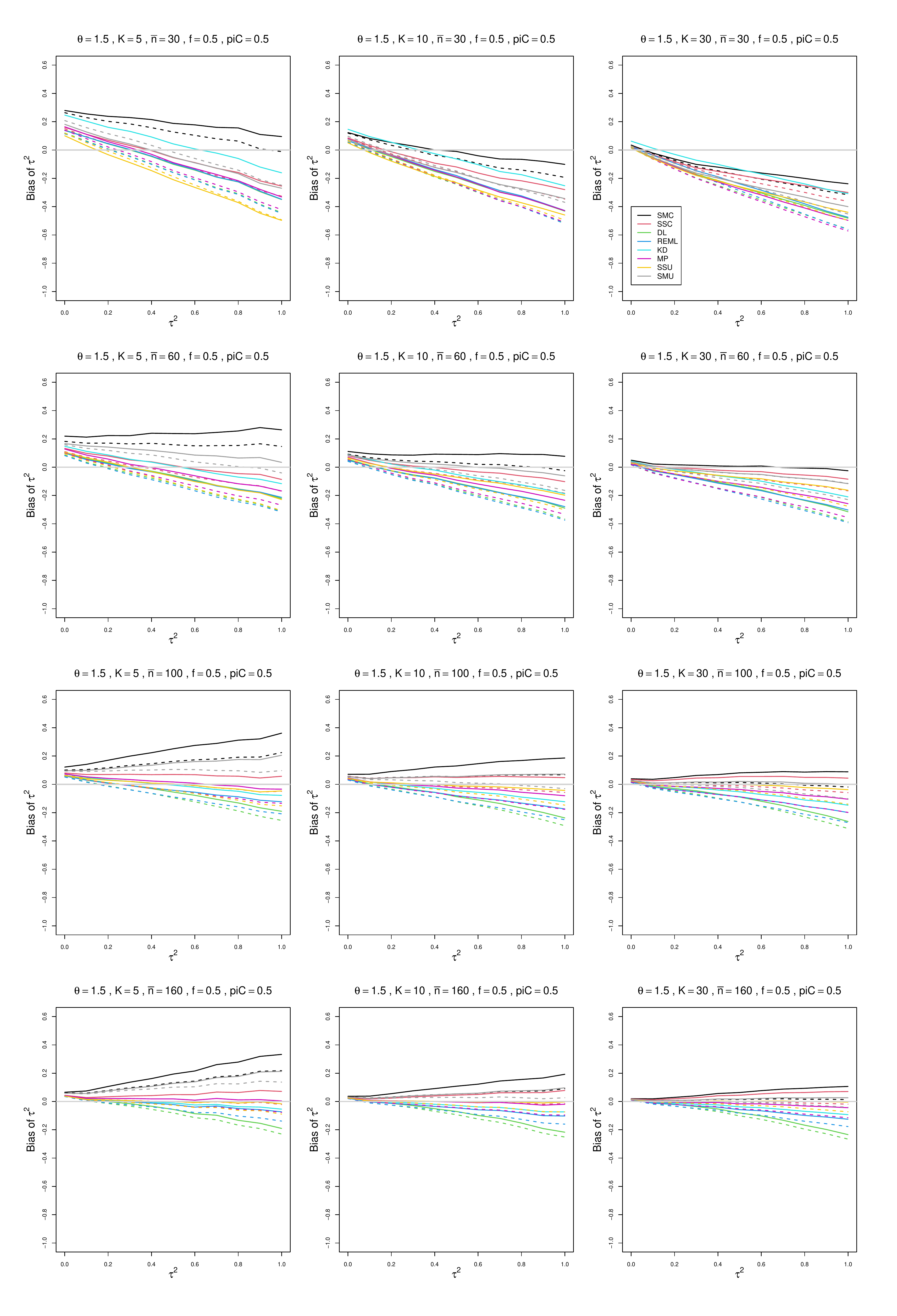}
	\caption{Bias  of estimators of between-study variance of LOR (DL, REML, KD, MP, SMC, SSC, SMU, and SSU) vs $\tau^2$, for unequal sample sizes $\bar{n}=30,\;60,\;100$ and $160$, $p_{iC} = .5$, $\theta=1.5$ and  $f=0.5$.   Solid lines: DL, REML, MP, SSC, SMC \lq\lq only"; KD; SSU and SMU model-based. Dashed lines: DL, REML, MP, SSC, SMC  \lq\lq always"; SSU and SMU na\"ive.  }
	\label{PlotBiasOfTau2_piC_05theta=1.5_LOR_unequal_sample_sizes}
\end{figure}

\begin{figure}[ht]
	\centering
	\includegraphics[scale=0.33]{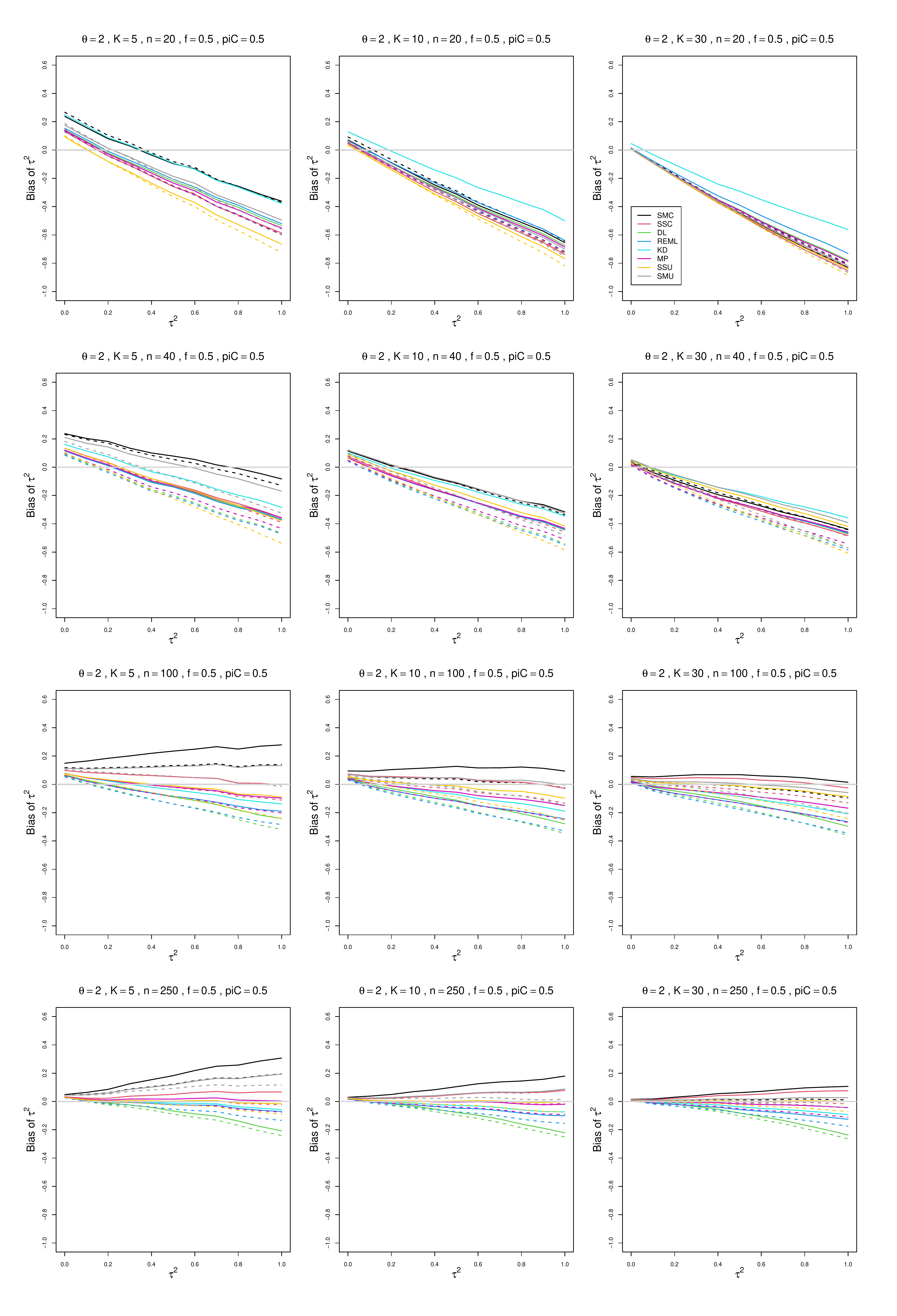}
	\caption{Bias  of estimators of between-study variance of LOR (DL, REML, KD, MP, SMC, SSC, SMU, and SSU) vs $\tau^2$, for equal sample sizes $n=20,\;40,\;100$ and $250$, $p_{iC} = .5$, $\theta=2$ and  $f=0.5$.   Solid lines: DL, REML, MP, SSC, SMC \lq\lq only"; KD; SSU and SMU model-based. Dashed lines: DL, REML, MP, SSC, SMC  \lq\lq always"; SSU and SMU na\"ive.  }
	\label{PlotBiasOfTau2_piC_05theta=2_LOR_equal_sample_sizes}
\end{figure}

\begin{figure}[ht]
	\centering
	\includegraphics[scale=0.33]{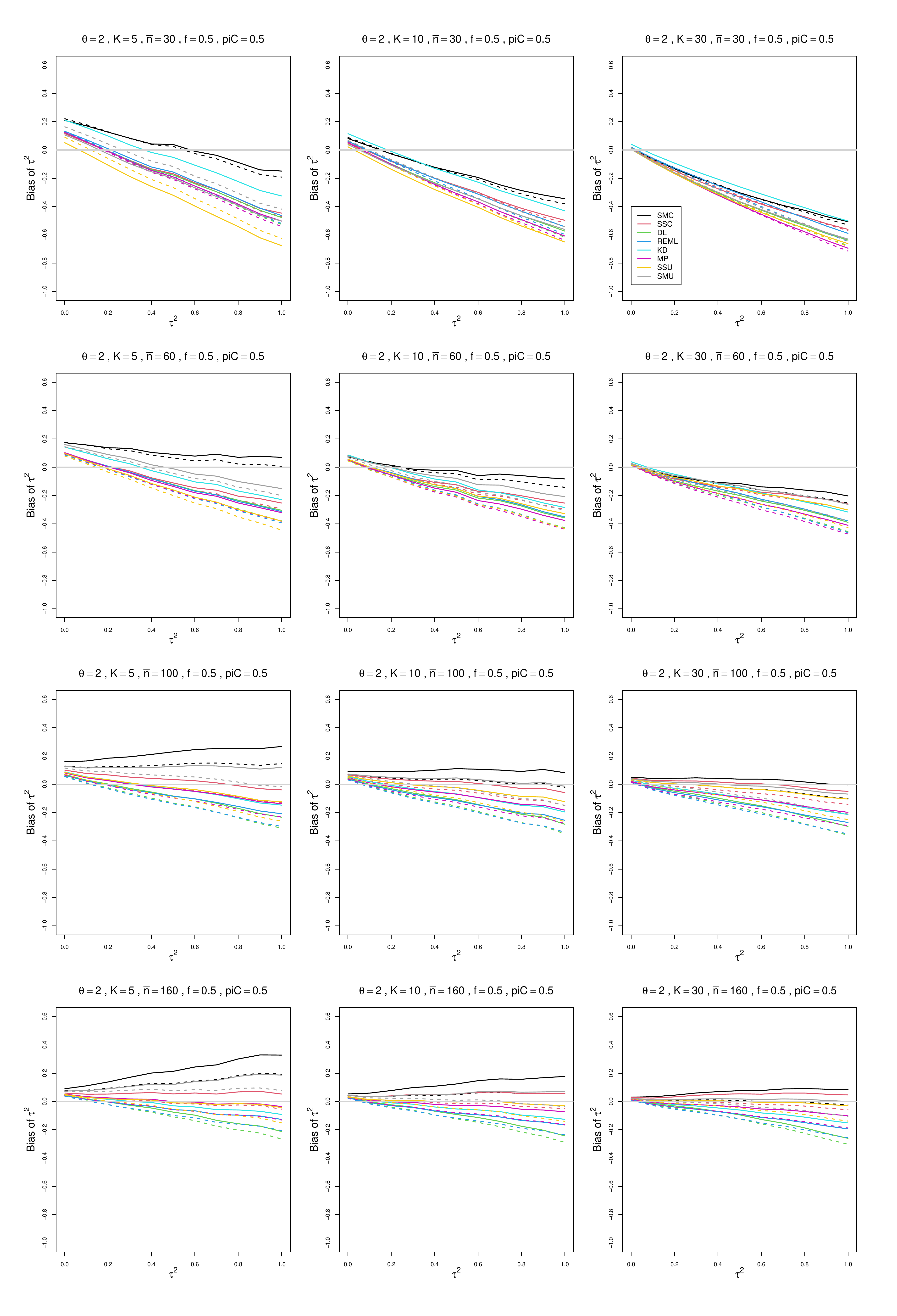}
	\caption{Bias  of estimators of between-study variance of LOR (DL, REML, KD, MP, SMC, SSC, SMU, and SSU) vs $\tau^2$, for unequal sample sizes $\bar{n}=30,\;60,\;100$ and $160$, $p_{iC} = .5$, $\theta=2$ and  $f=0.5$.   Solid lines: DL, REML, MP, SSC, SMC \lq\lq only"; KD; SSU and SMU model-based. Dashed lines: DL, REML, MP, SSC, SMC  \lq\lq always"; SSU and SMU na\"ive. }
	\label{PlotBiasOfTau2_piC_05theta=2_LOR_unequal_sample_sizes}
\end{figure}

\clearpage

\section*{Appendix B: Median bias in point estimators of between-study variance }

Each figure corresponds to a value of the probability of an event in the Control arm $p_{iC}$  (= .1, .2, .5) . \\
The fraction of each study's sample size in the Control arm  ($f$) is held constant at 0.5. For each combination of a value of $n$ (= 20, 40, 100, 250) or  $\bar{n}$ (= 30, 60, 100, 160) and a value of $K$ (= 5, 10, 30), a panel plots median bias versus $\tau^2$ (= 0.0(0.1)1).\\
The point estimators of $\tau^2$ are
\begin{itemize}
\item DL (DerSimonian-Laird method, inverse-variance weights)
\item REML method, inverse-variance weights)
\item MP (Mandel-Paule method, inverse-variance weights)
\item KD (Kulinskaya-Dollinger (2015) approximation, inverse-variance weights)
\item SSC method, effective-sample-size weights, conditional variance of LOR
\item SMC method, median-unbiased, effective-sample-size weights, conditional variance of LOR
\item SSU method, effective-sample-size weights, unconditional variance of LOR
\item SMU method, median-unbiased, effective-sample-size weights, unconditional variance of LOR
\end{itemize}
The plots include two versions of DL, REML, MP, SSC and SMC: adding $1/2$ to all four of $X_{iT},\;X_{iC},\; n_{iT}-X_{iT},\; n_{iC}-X_{iC}$ only when one of these is zero (solid lines) or always (dashed lines).\\
The plots also include two versions of SSU and SMU: model-based estimation of $p_{iT}$ (solid lines) or na\"{i}ve estimation (dashed lines).
%\item 2M SSW Na\"{i}ve (Two-moment gamma approximation, Na\"{i}ve estimation of $p_{iT}$, effective-sample-size weights)
%\item 2M SSW Model (Two-moment gamma approximation, model-based estimation of $p_{iT}$, effective-sample-size weights)
%	\item F SSW  Na\"{i}ve(Farebrother approximation, Na\"{i}ve estimation of $p_{iT}$, effective-sample-size weights)
%	\item F SSW Model  (Farebrother approximation, model-based estimation of $p_{iT}$, effective-sample-size weights)

\clearpage
\setcounter{figure}{0}
\renewcommand{\thefigure}{B.\arabic{figure}}

\begin{figure}[ht]
	\centering
	\includegraphics[scale=0.33]{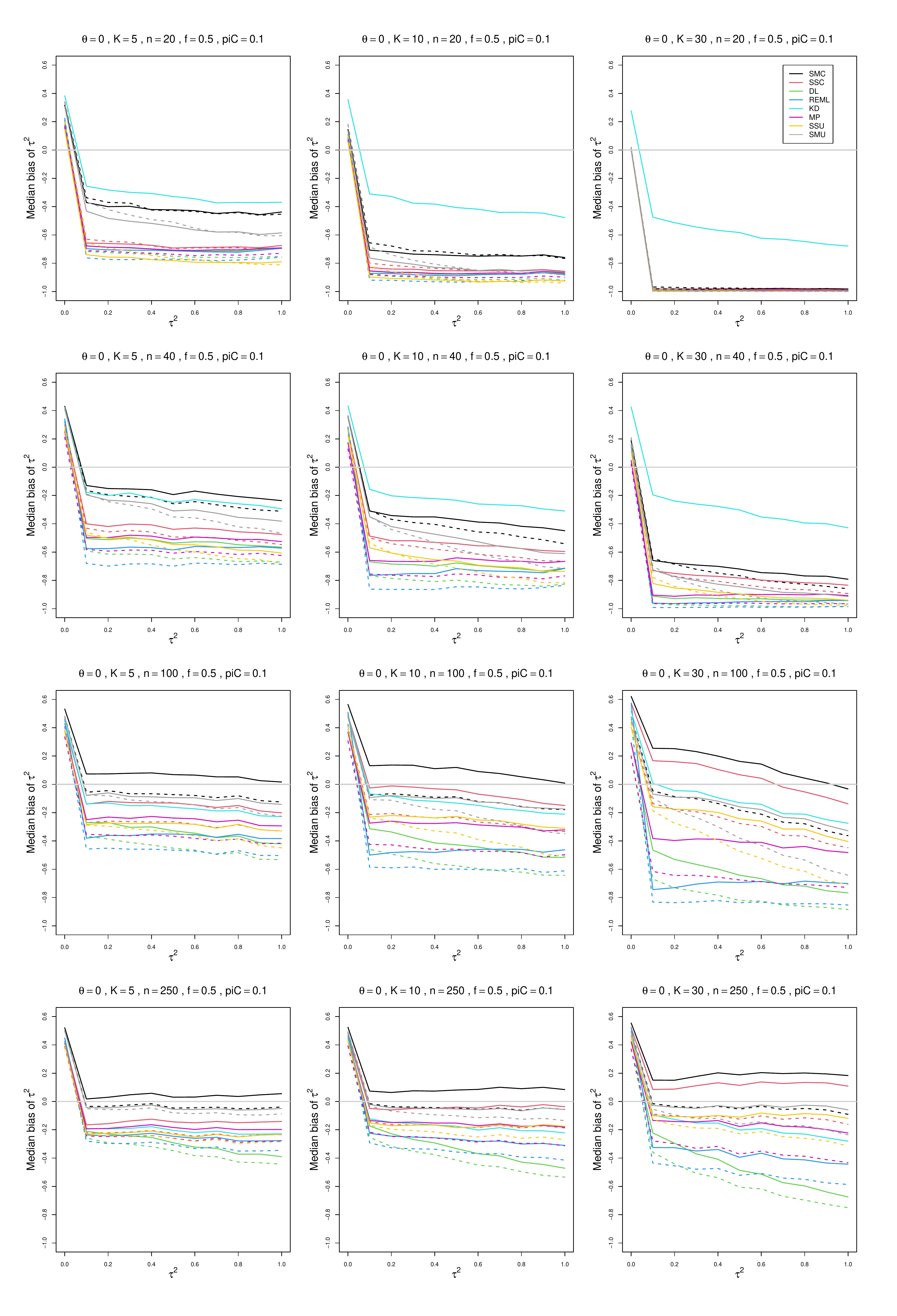}
	\caption{Median bias of estimators of between-study variance of LOR (DL, REML, KD, MP, SMC, SSC, SMU, and SSU) vs $\tau^2$, for equal sample sizes $n = 20,\;40,\;100$ and $250$, $p_{iC} = .1$, $\theta = 0$ and  $f = 0.5$.   Solid lines: DL, REML, MP, SSC, SMC \lq\lq only"; KD; SSU and SMU model-based. Dashed lines: DL, REML, MP, SSC, SMC  \lq\lq always"; SSU and SMU  na\"ive.  }
	\label{PlotMedBiasOfTau2_piC_01theta=0_LOR_equal_sample_sizes}
\end{figure}

\begin{figure}[ht]
	\centering
	\includegraphics[scale=0.33]{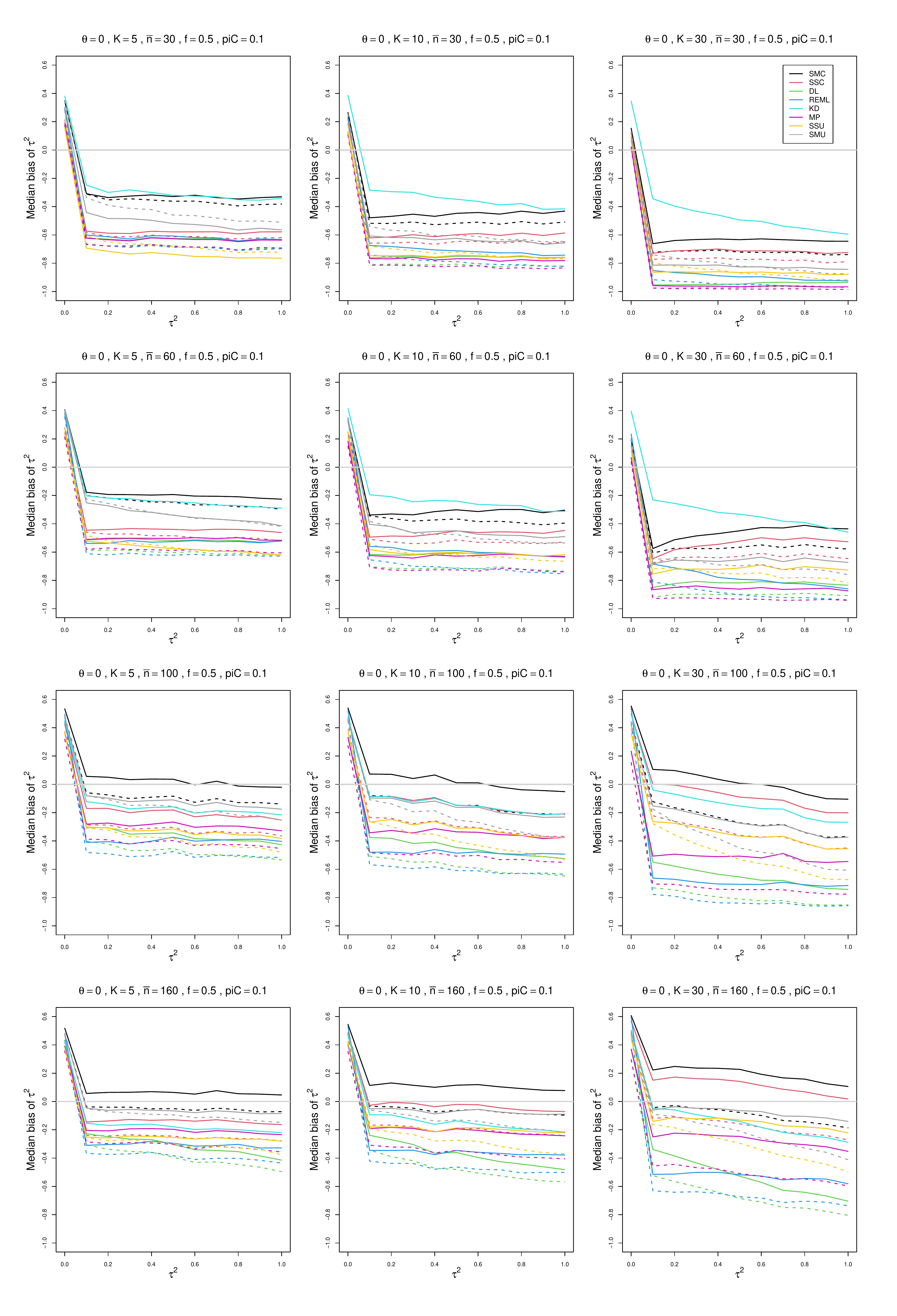}
	\caption{Median bias of estimators of between-study variance of LOR (DL, REML, KD, MP, SMC, SSC, SMU, and SSU) vs $\tau^2$, for unequal sample sizes $\bar{n}=30,\;60,\;100$ and $160$, $p_{iC} = .1$, $\theta = 0$ and  $f = 0.5$.   Solid lines: DL, REML, MP, SSC, SMC \lq\lq only"; KD; SSU and SMU model-based. Dashed lines: DL, REML, MP, SSC, SMC  \lq\lq always"; SSU and SMU  na\"ive.   }
	\label{PlotMedBiasOfTau2_piC_01theta=0_LOR_unequal_sample_sizes}
\end{figure}

\begin{figure}[ht]
	\centering
	\includegraphics[scale=0.33]{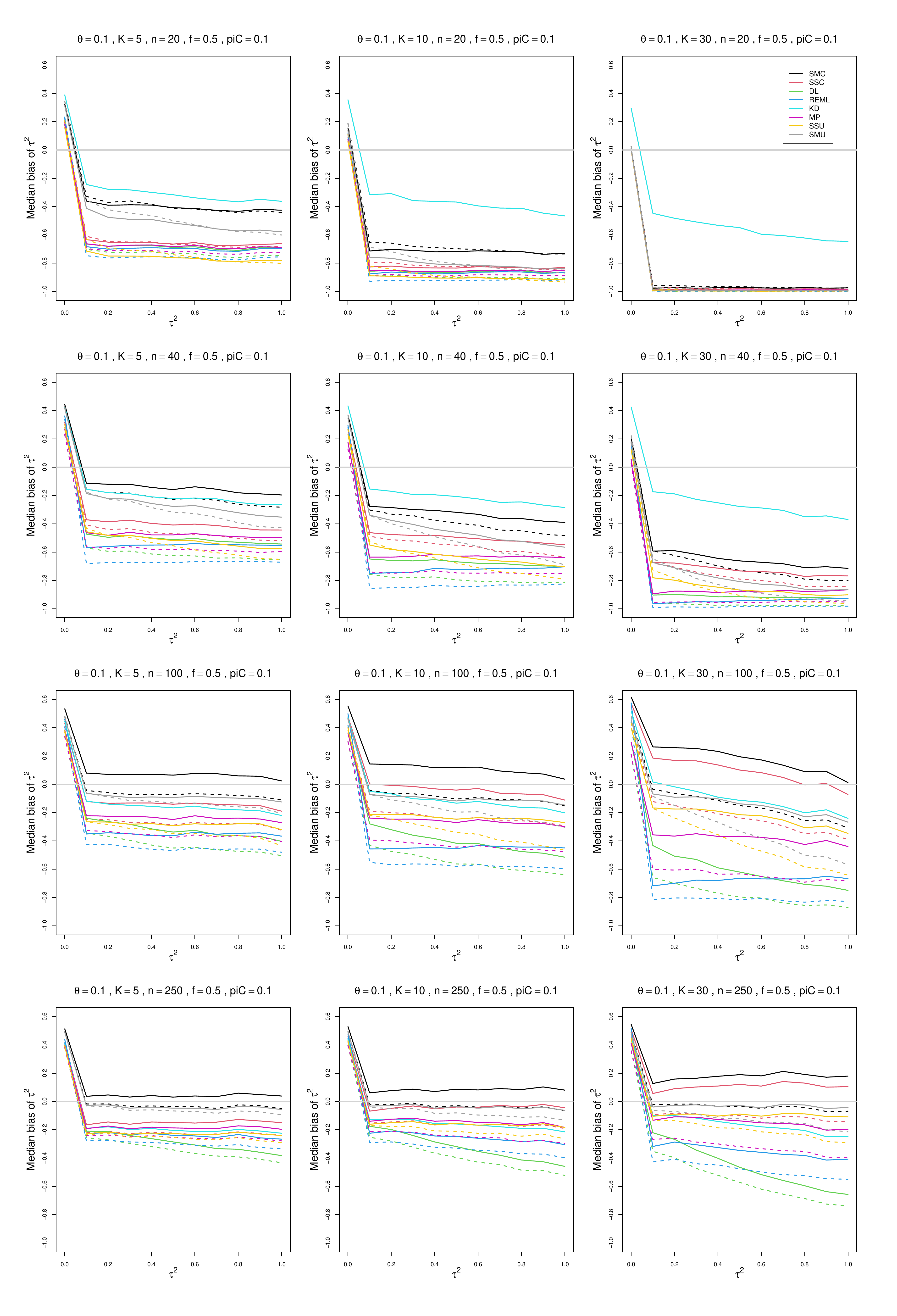}
	\caption{Median bias of estimators of between-study variance of LOR (DL, REML, KD, MP, SMC, SSC, SMU, and SSU) vs $\tau^2$, for equal sample sizes $n = 20,\;40,\;100$ and $250$, $p_{iC} = .1$, $\theta = 0.1$ and  $f = 0.5$.   Solid lines: DL, REML, MP, SSC, SMC \lq\lq only"; KD; SSU and SMU model-based. Dashed lines: DL, REML, MP, SSC, SMC  \lq\lq always"; SSU and SMU  na\"ive.  }
	\label{PlotMedBiasOfTau2_piC_01theta=0.1_LOR_equal_sample_sizes}
\end{figure}

\begin{figure}[ht]
	\centering
	\includegraphics[scale=0.33]{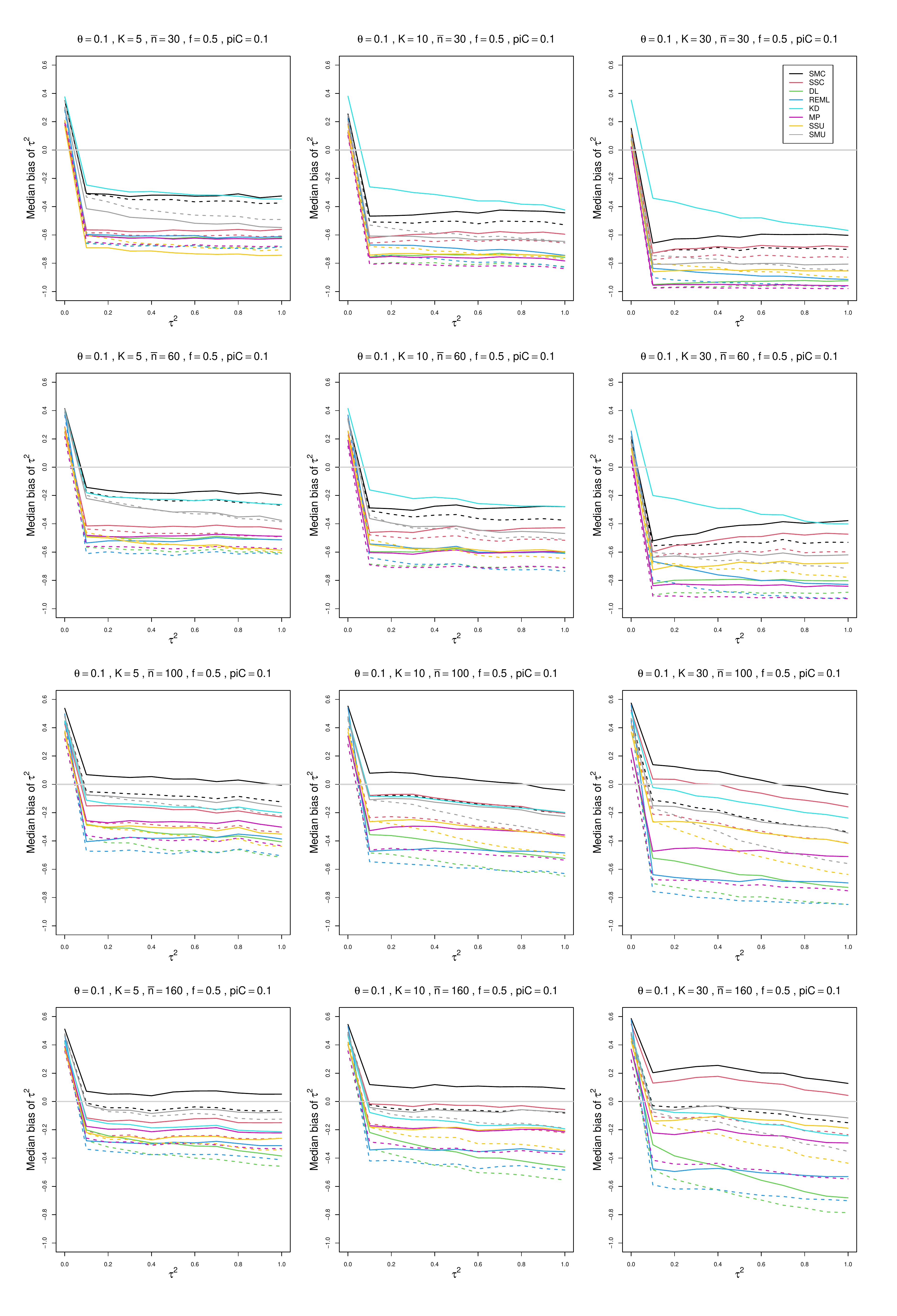}
	\caption{Median bias of estimators of between-study variance of LOR (DL, REML, KD, MP, SMC, SSC, SMU, and SSU) vs $\tau^2$, for unequal sample sizes $\bar{n}=30,\;60,\;100$ and $160$, $p_{iC} = .1$, $\theta = 0.1$ and  $f = 0.5$.   Solid lines: DL, REML, MP, SSC, SMC \lq\lq only"; KD; SSU and SMU model-based. Dashed lines: DL, REML, MP, SSC, SMC  \lq\lq always"; SSU and SMU  na\"ive.   }
	\label{PlotMedBiasOfTau2_piC_01theta=0.1_LOR_unequal_sample_sizes}
\end{figure}

\begin{figure}[ht]
	\centering
	\includegraphics[scale=0.33]{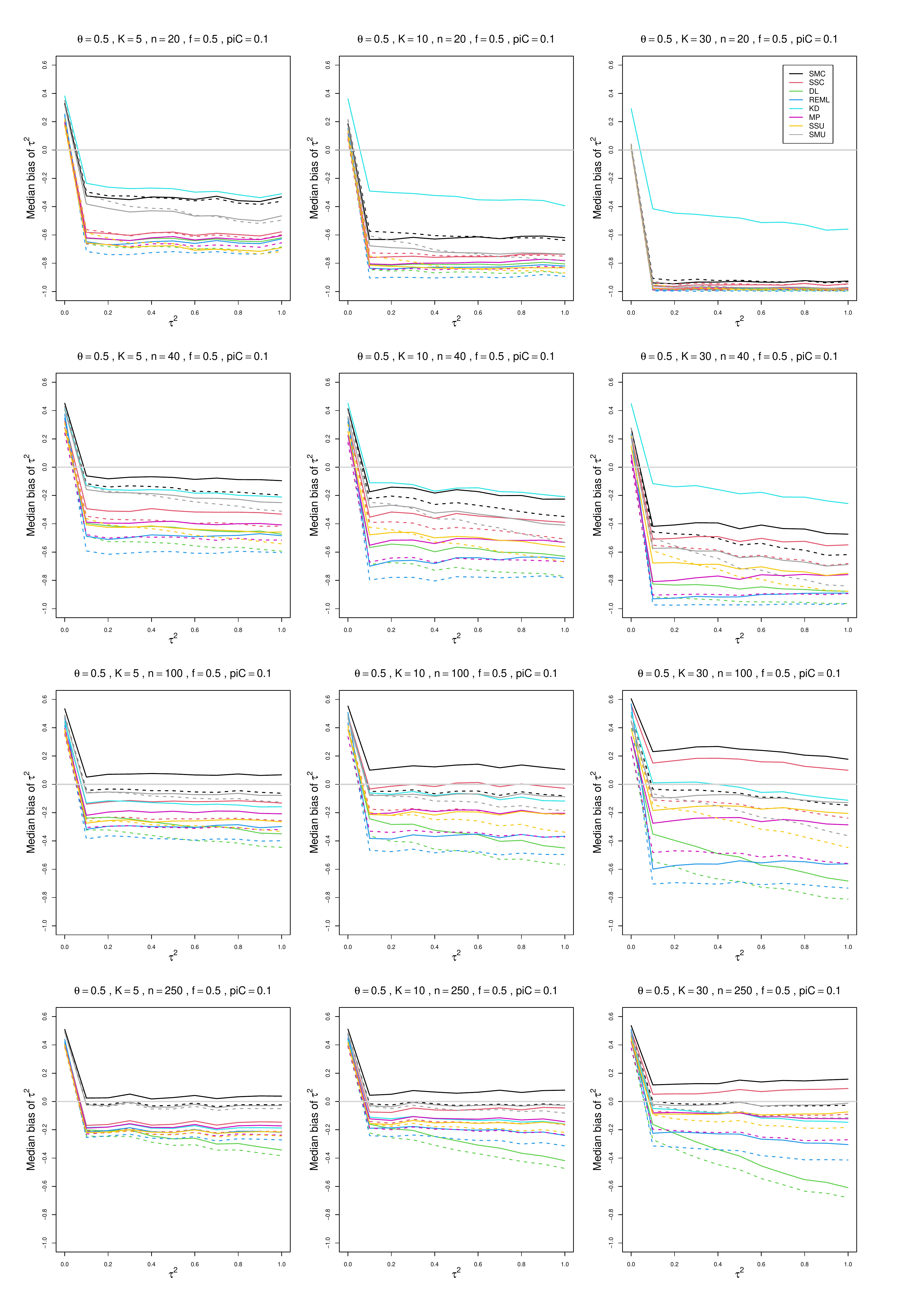}
	\caption{Median bias of estimators of between-study variance of LOR (DL, REML, KD, MP, SMC, SSC, SMU, and SSU) vs $\tau^2$, for equal sample sizes $n = 20,\;40,\;100$ and $250$, $p_{iC} = .1$, $\theta = 0.5$ and  $f = 0.5$.   Solid lines: DL, REML, MP, SSC, SMC \lq\lq only"; KD; SSU and SMU model-based. Dashed lines: DL, REML, MP, SSC, SMC  \lq\lq always"; SSU and SMU  na\"ive.  }
	\label{PlotMedBiasOfTau2_piC_01theta=0.5_LOR_equal_sample_sizes}
\end{figure}

\begin{figure}[ht]
	\centering
	\includegraphics[scale=0.33]{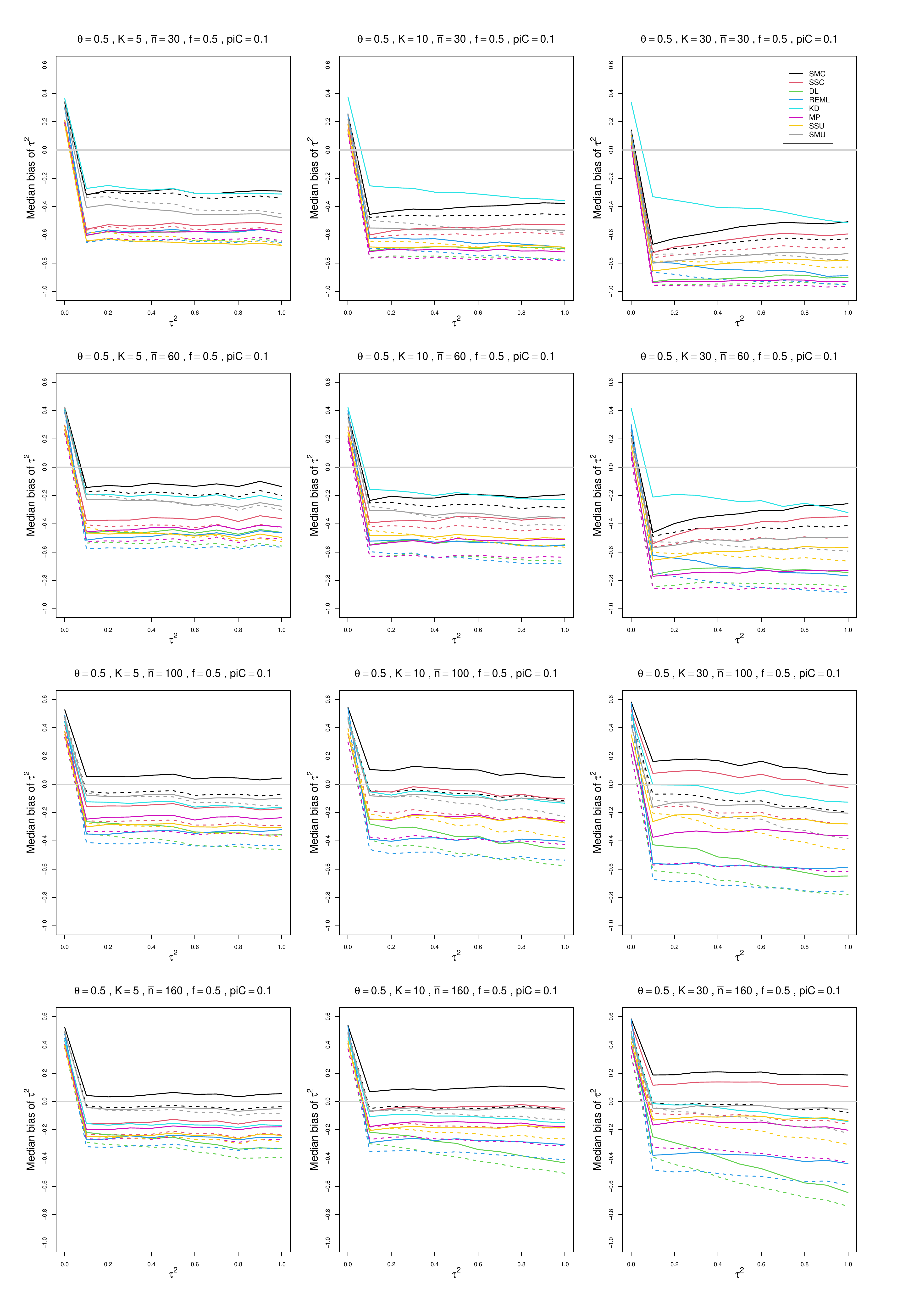}
	\caption{Median bias of estimators of between-study variance of LOR (DL, REML, KD, MP, SMC, SSC, SMU, and SSU) vs $\tau^2$, for unequal sample sizes $\bar{n}=30,\;60,\;100$ and $160$, $p_{iC} = .1$, $\theta = 0.5$ and  $f = 0.5$.   Solid lines: DL, REML, MP, SSC, SMC \lq\lq only"; KD; SSU and SMU model-based. Dashed lines: DL, REML, MP, SSC, SMC  \lq\lq always"; SSU and SMU  na\"ive.  }
	\label{PlotMedBiasOfTau2_piC_01theta=0.5_LOR_unequal_sample_sizes}
\end{figure}

\begin{figure}[ht]
	\centering
	\includegraphics[scale=0.33]{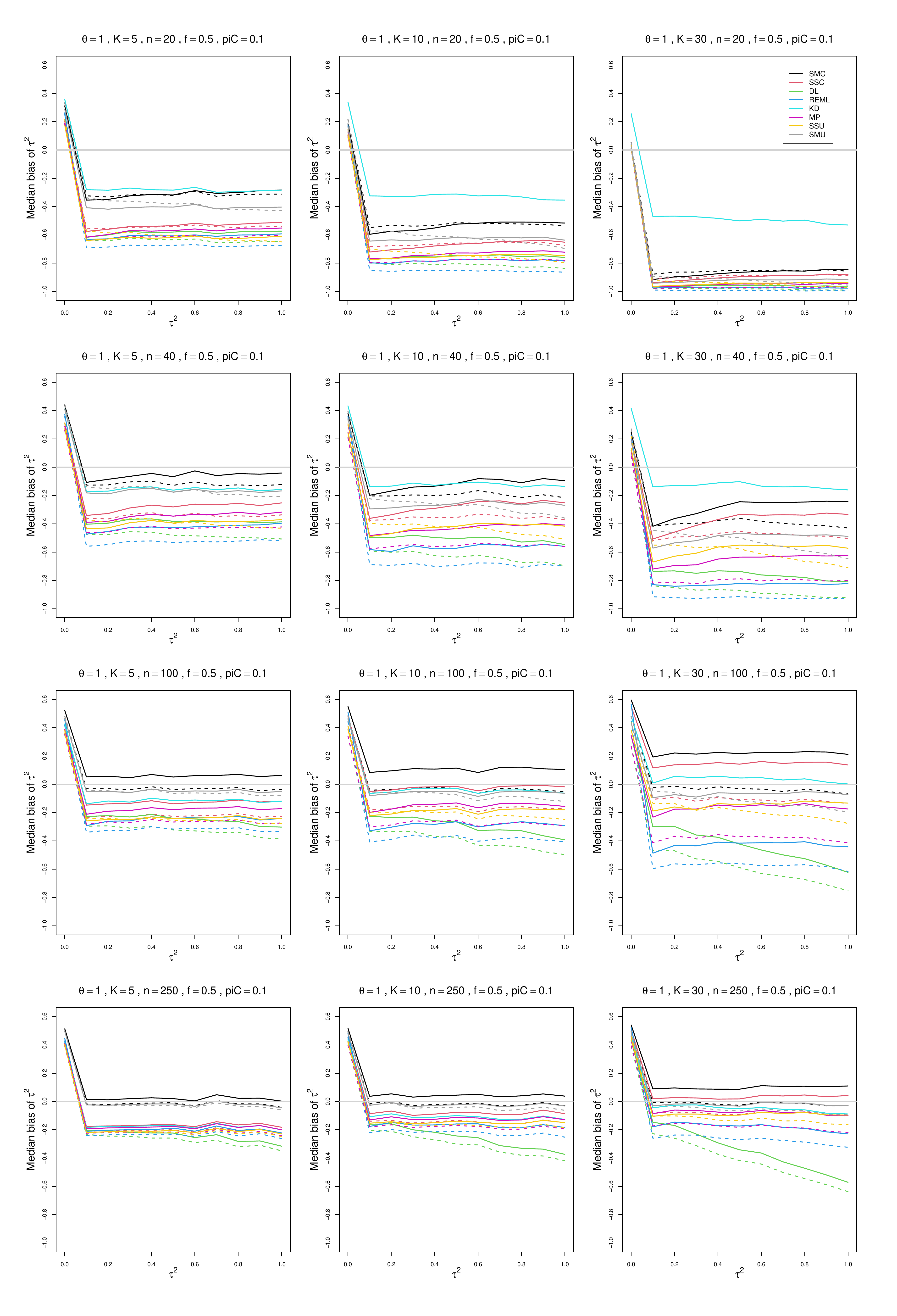}
	\caption{Median bias of estimators of between-study variance of LOR (DL, REML, KD, MP, SMC, SSC, SMU, and SSU) vs $\tau^2$, for equal sample sizes $n = 20,\;40,\;100$ and $250$, $p_{iC} = .1$, $\theta = 1$ and  $f = 0.5$.   Solid lines: DL, REML, MP, SSC, SMC \lq\lq only"; KD; SSU and SMU model-based. Dashed lines: DL, REML, MP, SSC, SMC  \lq\lq always"; SSU and SMU  na\"ive.  }
	\label{PlotMedBiasOfTau2_piC_01theta=1_LOR_equal_sample_sizes}
\end{figure}

\begin{figure}[ht]
	\centering
	\includegraphics[scale=0.33]{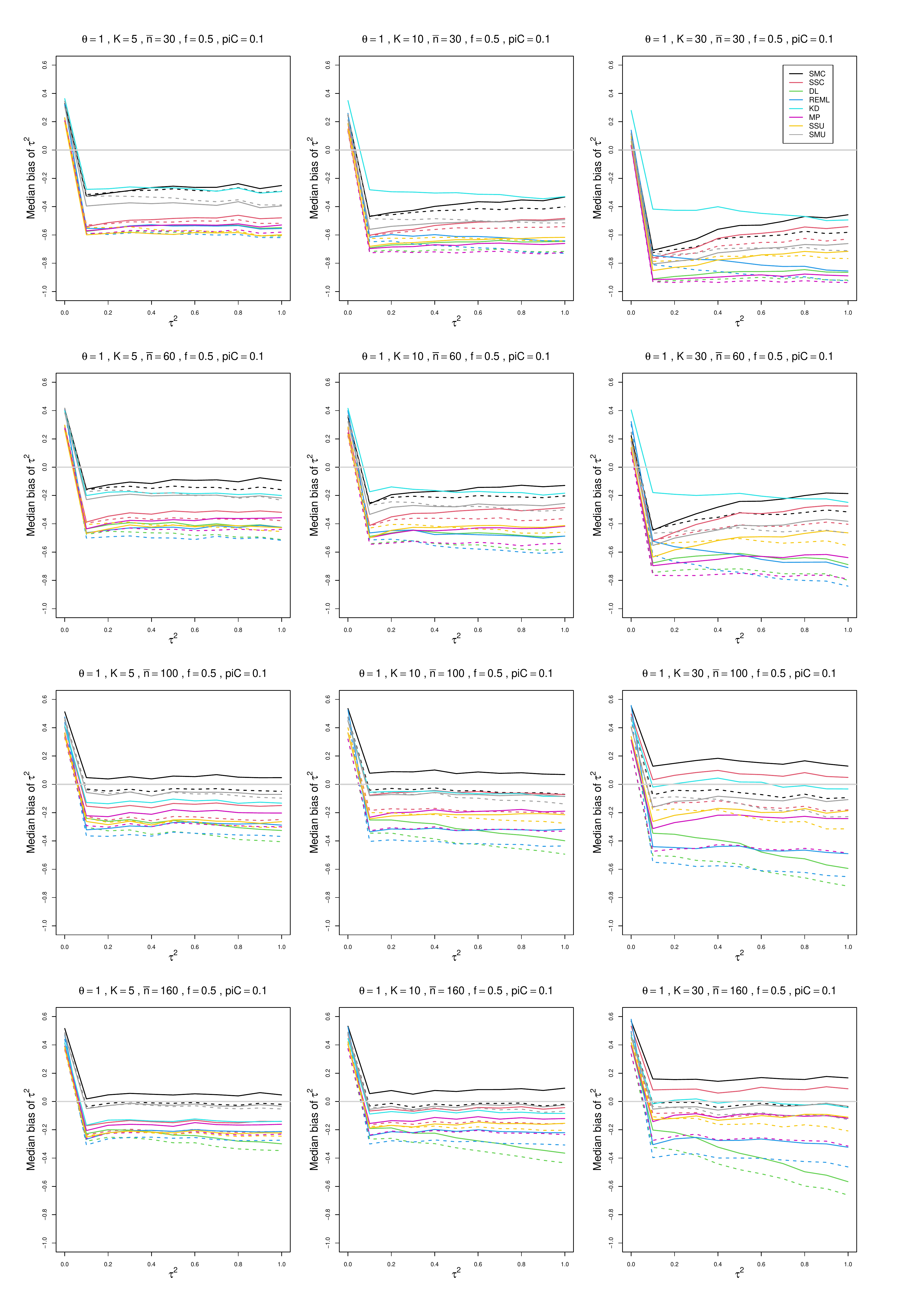}
	\caption{Median bias of estimators of between-study variance of LOR (DL, REML, KD, MP, SMC, SSC, SMU, and SSU) vs $\tau^2$, for unequal sample sizes $\bar{n}=30,\;60,\;100$ and $160$, $p_{iC} = .1$, $\theta = 1$ and  $f = 0.5$.   Solid lines: DL, REML, MP, SSC, SMC \lq\lq only"; KD; SSU and SMU model-based. Dashed lines: DL, REML, MP, SSC, SMC  \lq\lq always"; SSU and SMU  na\"ive.  }
	\label{PlotMedBiasOfTau2_piC_01theta=1_LOR_unequal_sample_sizes}
\end{figure}

\begin{figure}[ht]
	\centering
	\includegraphics[scale=0.33]{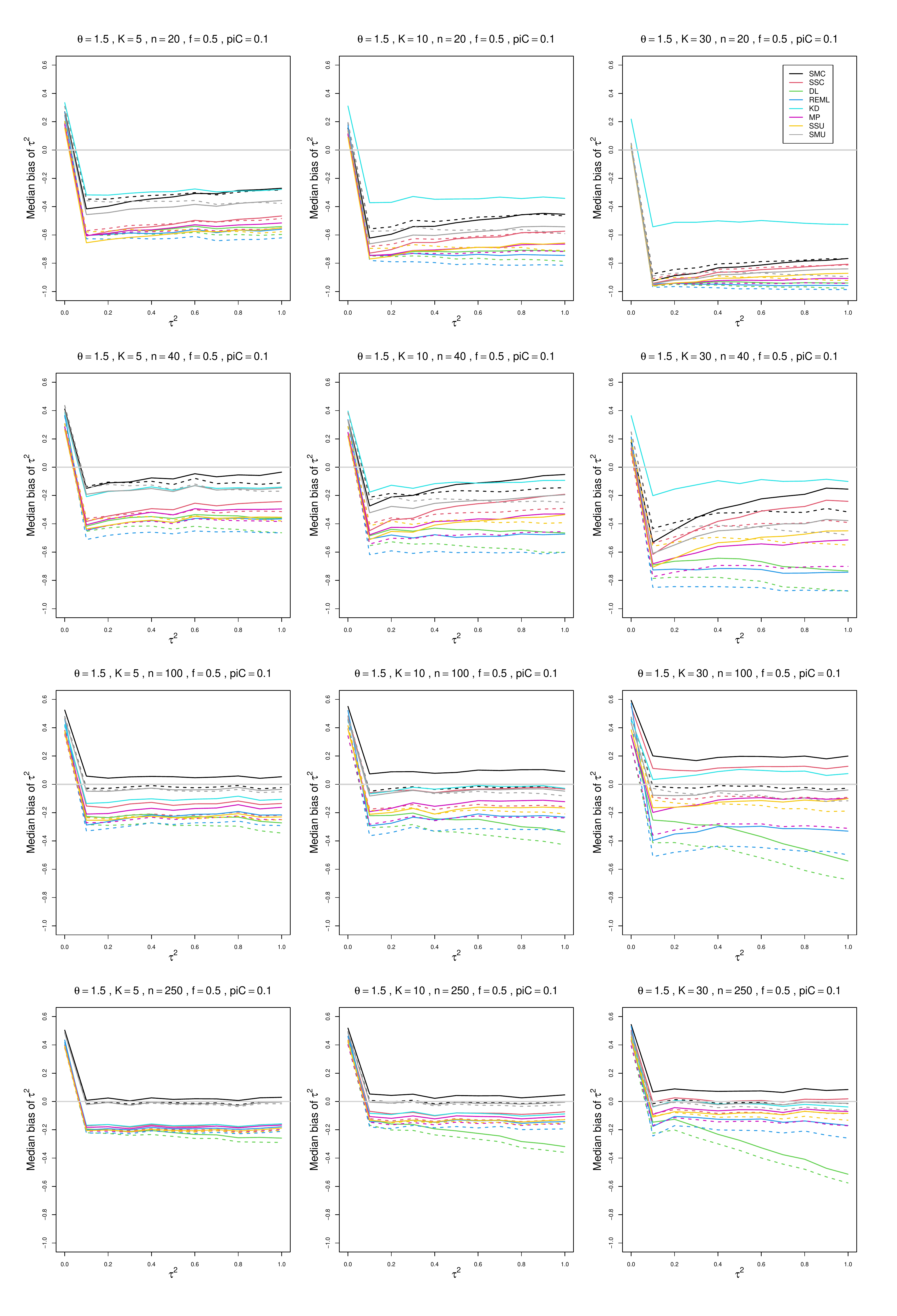}
	\caption{Median bias of estimators of between-study variance of LOR (DL, REML, KD, MP, SMC, SSC, SMU, and SSU) vs $\tau^2$, for equal sample sizes $n = 20,\;40,\;100$ and $250$, $p_{iC} = .1$, $\theta = 1.5$ and  $f = 0.5$.   Solid lines: DL, REML, MP, SSC, SMC \lq\lq only"; KD; SSU and SMU model-based. Dashed lines: DL, REML, MP, SSC, SMC  \lq\lq always"; SSU and SMU  na\"ive.  }
	\label{PlotMedBiasOfTau2_piC_01theta=1.5_LOR_equal_sample_sizes}
\end{figure}

\begin{figure}[ht]
	\centering
	\includegraphics[scale=0.33]{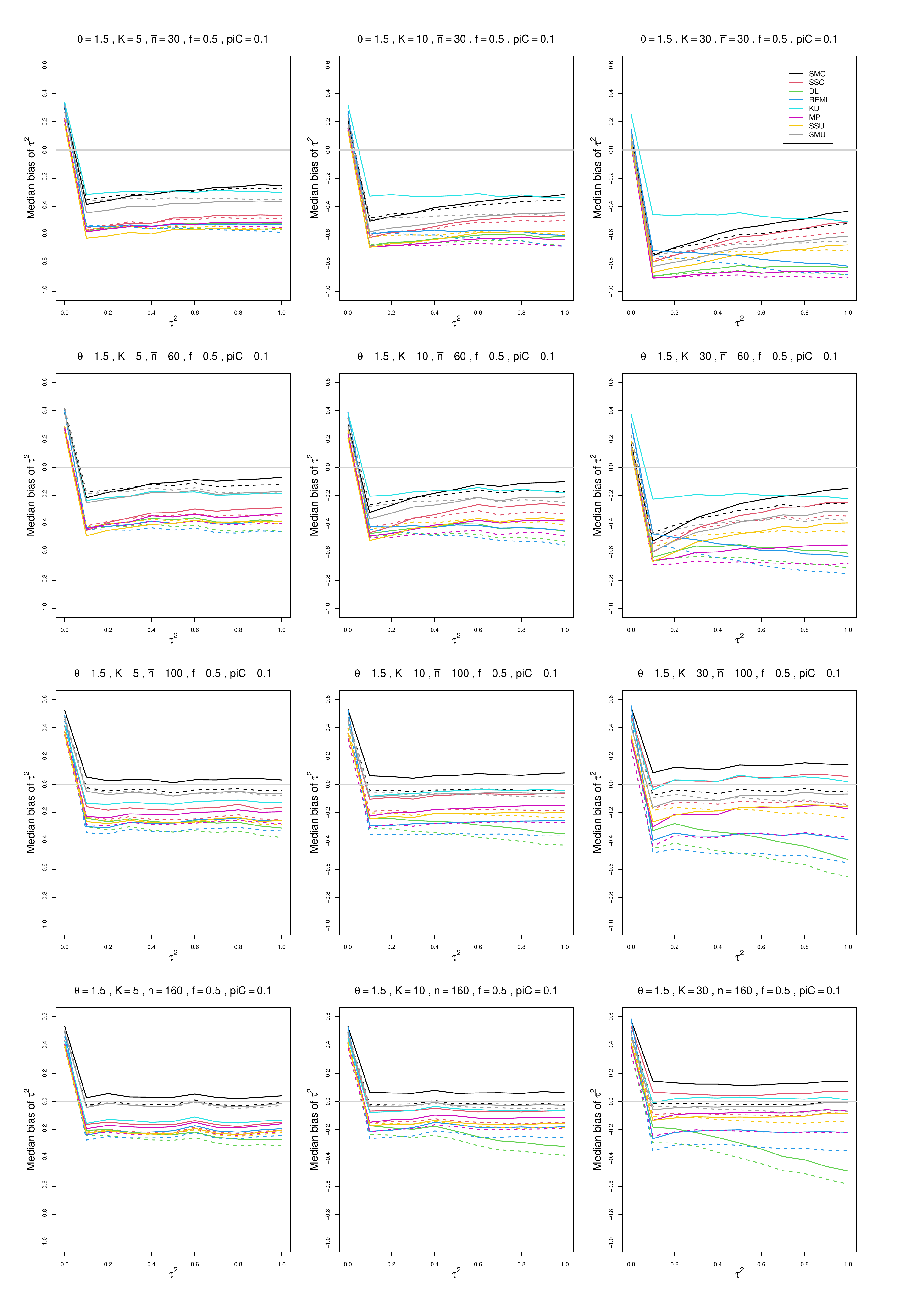}
	\caption{Median bias of estimators of between-study variance of LOR (DL, REML, KD, MP, SMC, SSC, SMU, and SSU) vs $\tau^2$, for unequal sample sizes $\bar{n}=30,\;60,\;100$ and $160$, $p_{iC} = .1$, $\theta = 1.5$ and  $f = 0.5$.   Solid lines: DL, REML, MP, SSC, SMC \lq\lq only"; KD; SSU and SMU model-based. Dashed lines: DL, REML, MP, SSC, SMC  \lq\lq always"; SSU and SMU  na\"ive.  }
	\label{PlotMedBiasOfTau2_piC_01theta=1.5_LOR_unequal_sample_sizes}
\end{figure}

\begin{figure}[ht]
	\centering
	\includegraphics[scale=0.33]{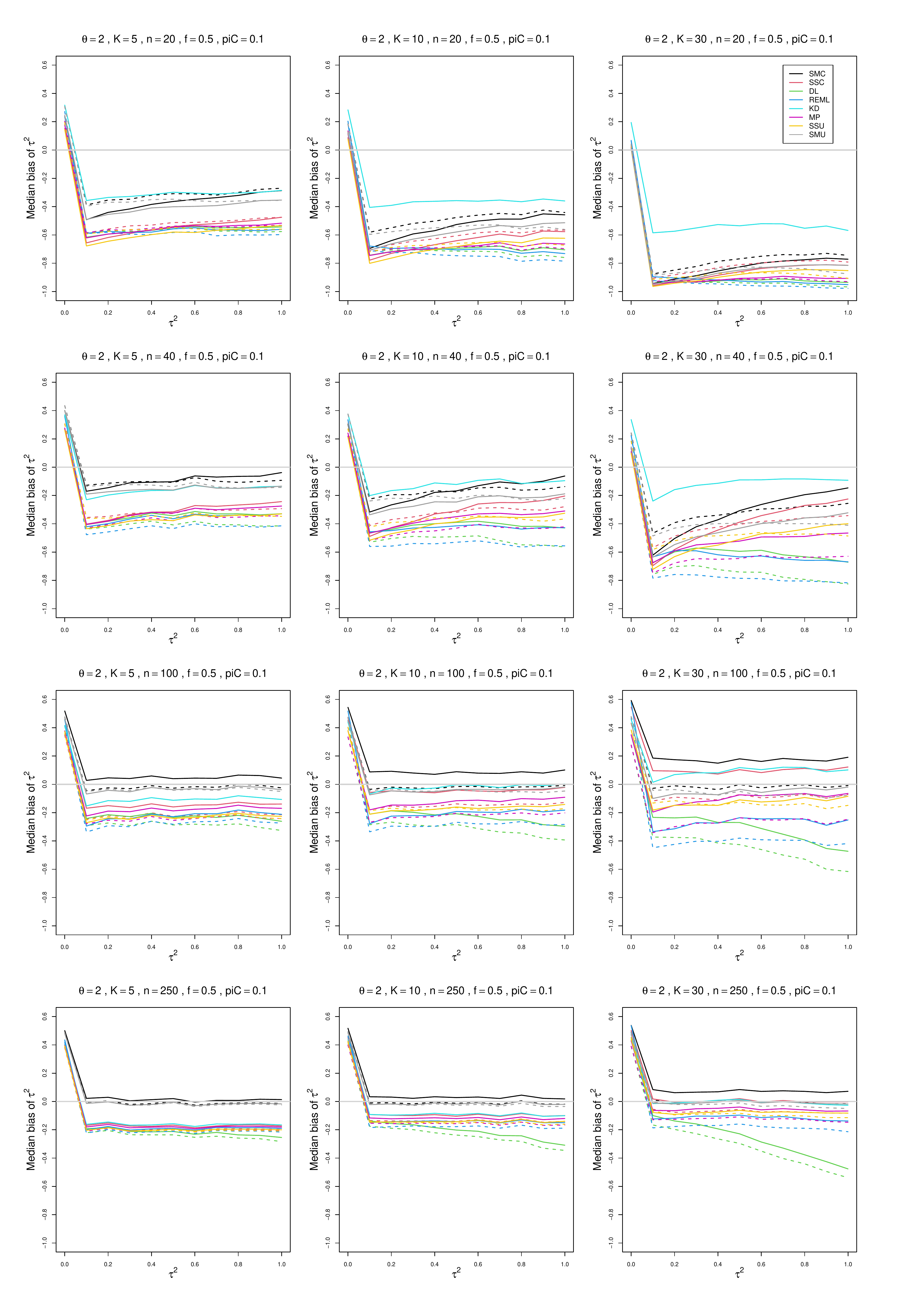}
	\caption{Median bias of estimators of between-study variance of LOR (DL, REML, KD, MP, SMC, SSC, SMU, and SSU) vs $\tau^2$, for equal sample sizes $n = 20,\;40,\;100$ and $250$, $p_{iC} = .1$, $\theta = 2$ and  $f = 0.5$.   Solid lines: DL, REML, MP, SSC, SMC \lq\lq only"; KD; SSU and SMU model-based. Dashed lines: DL, REML, MP, SSC, SMC  \lq\lq always"; SSU and SMU  na\"ive.  }
	\label{PlotMedBiasOfTau2_piC_01theta=2_LOR_equal_sample_sizes}
\end{figure}

\begin{figure}[ht]
	\centering
	\includegraphics[scale=0.33]{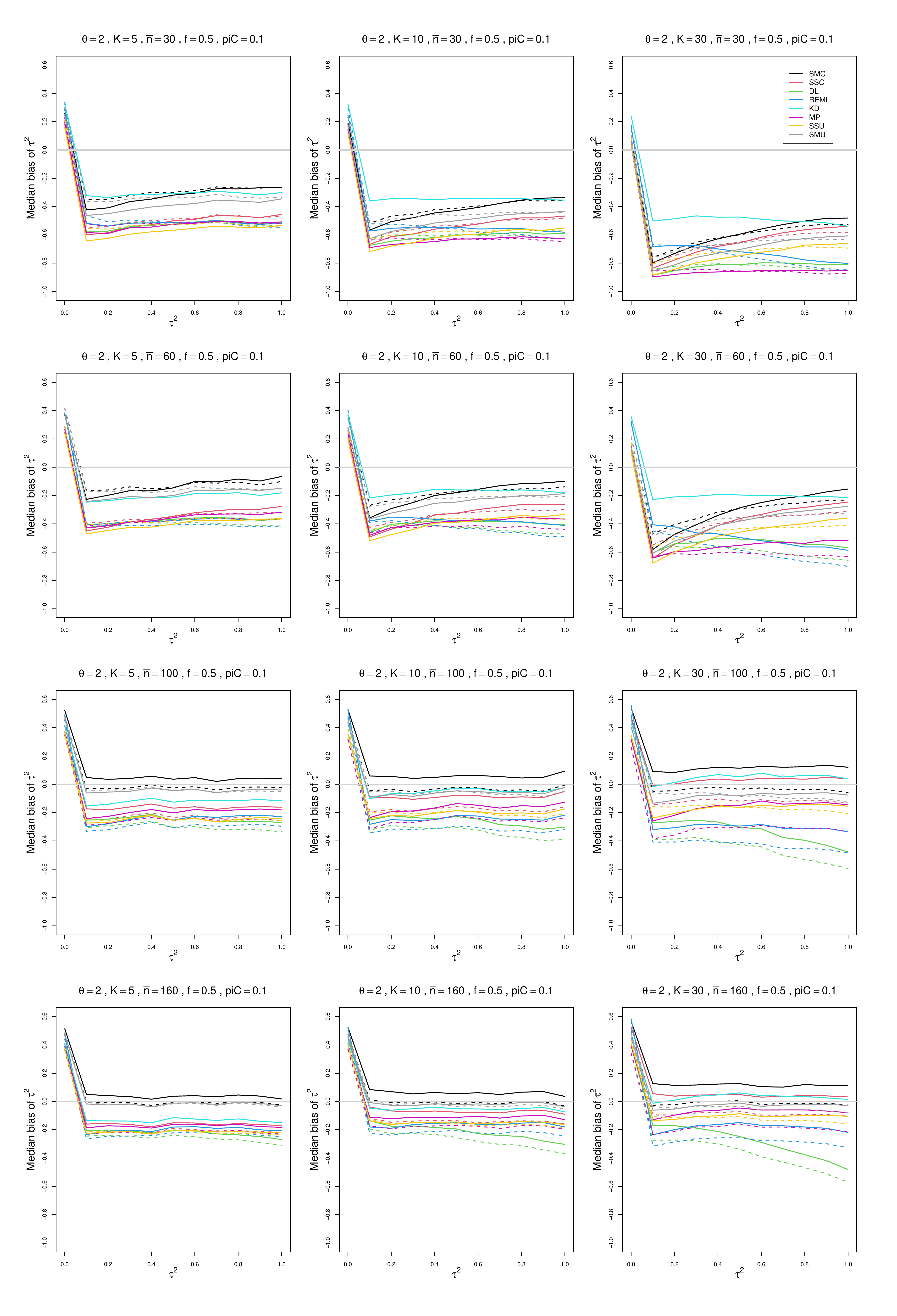}
	\caption{Median bias of estimators of between-study variance of LOR (DL, REML, KD, MP, SMC, SSC, SMU, and SSU) vs $\tau^2$, for unequal sample sizes $\bar{n}=30,\;60,\;100$ and $160$, $p_{iC} = .1$, $\theta = 2$ and  $f = 0.5$.   Solid lines: DL, REML, MP, SSC, SMC \lq\lq only"; KD; SSU and SMU model-based. Dashed lines: DL, REML, MP, SSC, SMC  \lq\lq always"; SSU and SMU  na\"ive.  }
	\label{PlotMedBiasOfTau2_piC_01theta=2_LOR_unequal_sample_sizes}
\end{figure}

\begin{figure}[ht]
	\centering
	\includegraphics[scale=0.33]{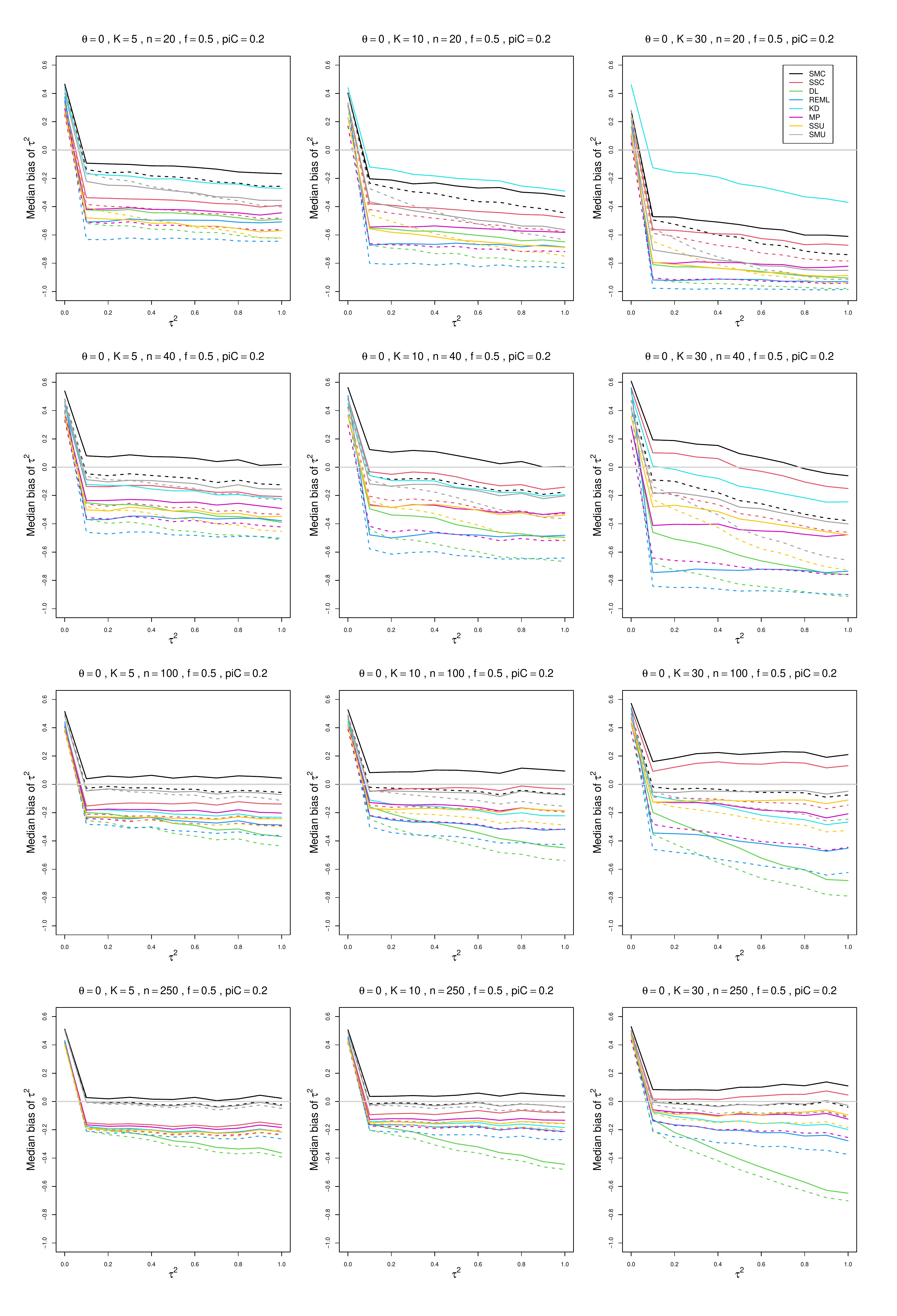}
	\caption{Median bias of estimators of between-study variance of LOR (DL, REML, KD, MP, SMC, SSC, SMU, and SSU) vs $\tau^2$, for equal sample sizes $n = 20,\;40,\;100$ and $250$, $p_{iC} = .2$, $\theta = 0$ and  $f = 0.5$.   Solid lines: DL, REML, MP, SSC, SMC \lq\lq only"; KD; SSU and SMU model-based. Dashed lines: DL, REML, MP, SSC, SMC  \lq\lq always"; SSU and SMU  na\"ive. }
	\label{PlotMedBiasOfTau2_piC_02theta=0_LOR_equal_sample_sizes}
\end{figure}

\begin{figure}[ht]
	\centering
	\includegraphics[scale=0.33]{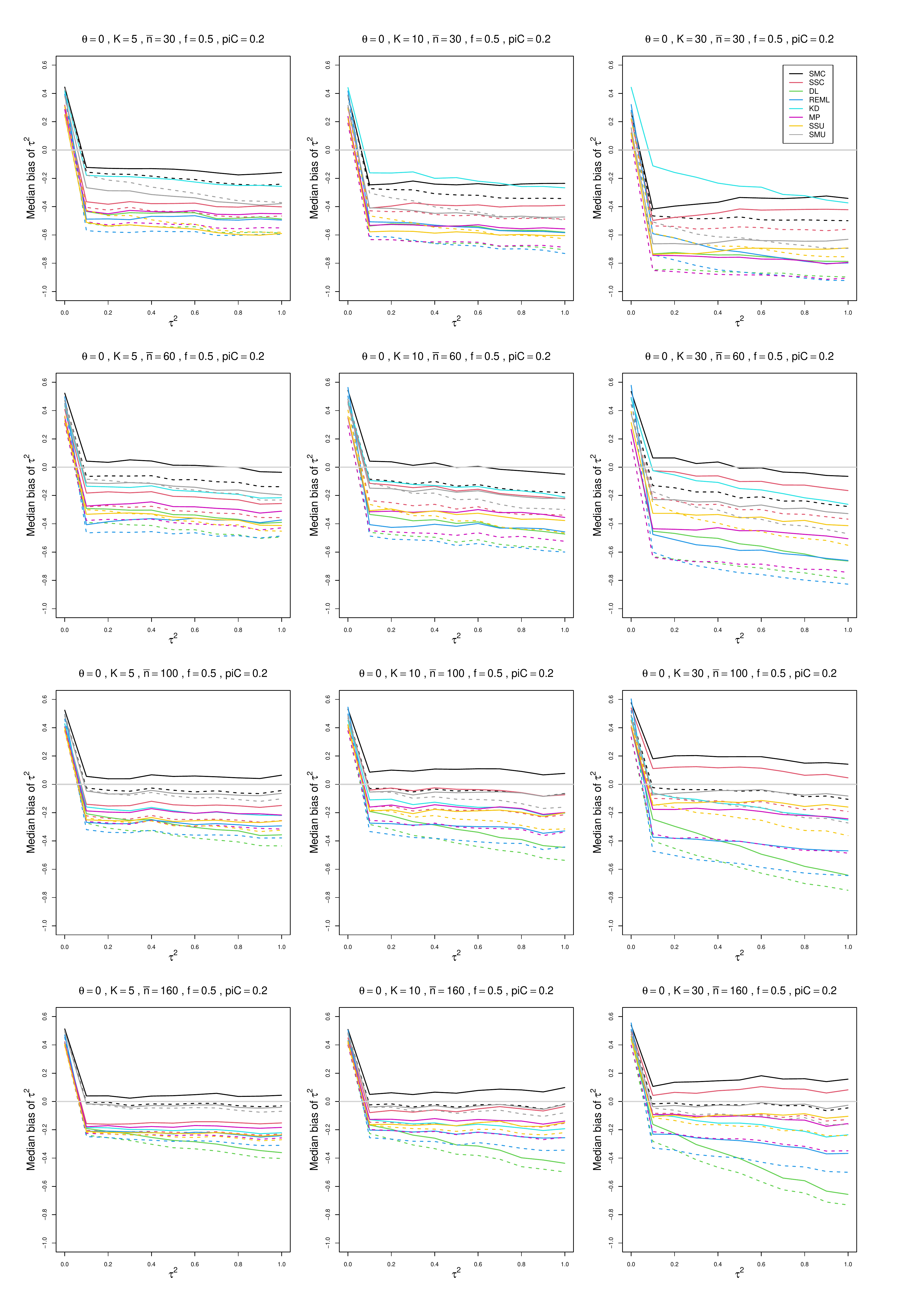}
	\caption{Median bias of estimators of between-study variance of LOR (DL, REML, KD, MP, SMC, SSC, SMU, and SSU) vs $\tau^2$, for unequal sample sizes $\bar{n}=30,\;60,\;100$ and $160$, $p_{iC} = .2$, $\theta = 0$ and  $f = 0.5$.   Solid lines: DL, REML, MP, SSC, SMC \lq\lq only"; KD; SSU and SMU model-based. Dashed lines: DL, REML, MP, SSC, SMC  \lq\lq always"; SSU and SMU  na\"ive.   }
	\label{PlotMedBiasOfTau2_piC_02theta=0_LOR_unequal_sample_sizes}
\end{figure}

\begin{figure}[ht]
	\centering
	\includegraphics[scale=0.33]{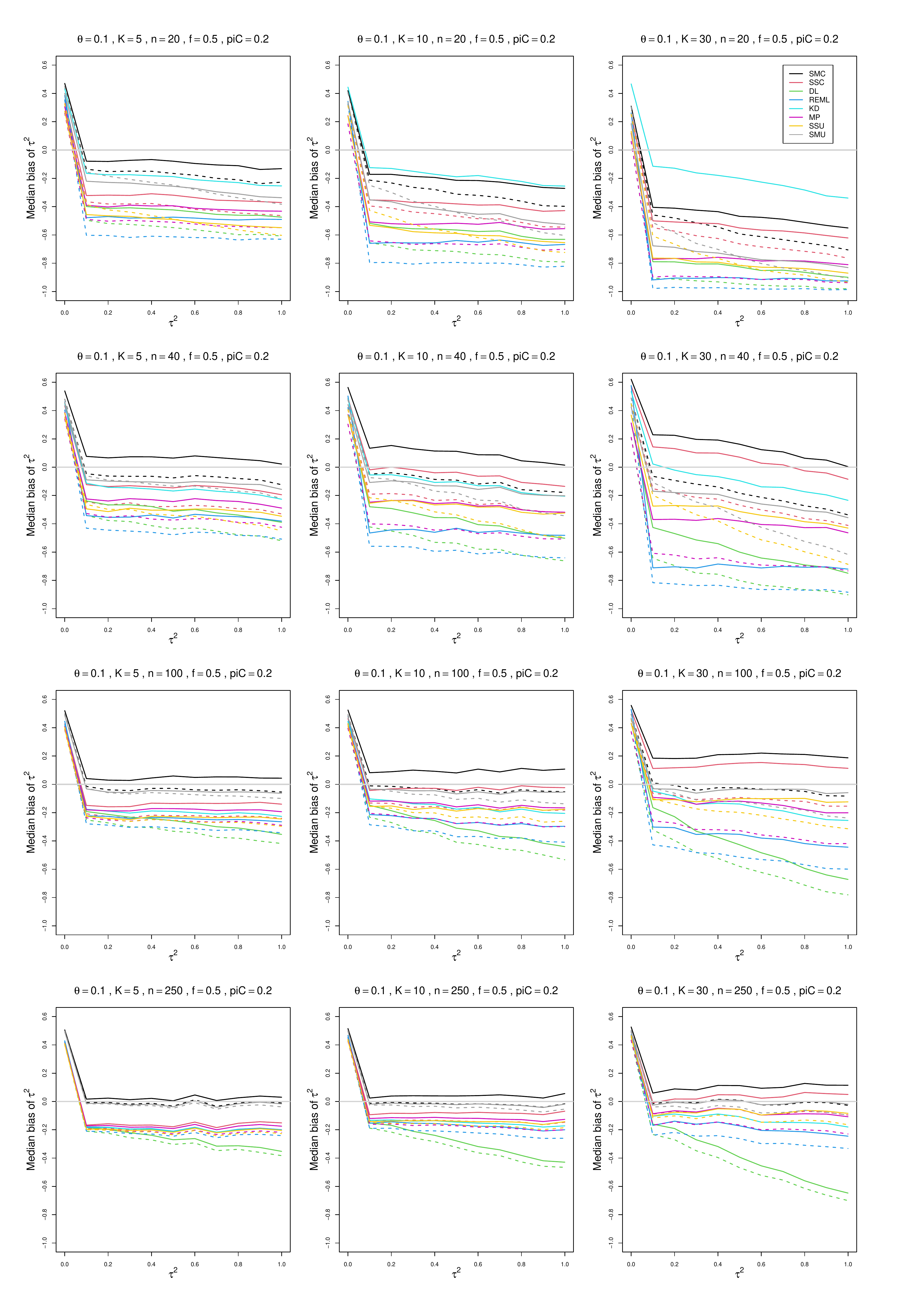}
	\caption{Median bias of estimators of between-study variance of LOR (DL, REML, KD, MP, SMC, SSC, SMU, and SSU) vs $\tau^2$, for equal sample sizes $n = 20,\;40,\;100$ and $250$, $p_{iC} = .2$, $\theta = 0.1$ and  $f = 0.5$.   Solid lines: DL, REML, MP, SSC, SMC \lq\lq only"; KD; SSU and SMU model-based. Dashed lines: DL, REML, MP, SSC, SMC  \lq\lq always"; SSU and SMU  na\"ive.  }
	\label{PlotMedBiasOfTau2_piC_02theta=0.1_LOR_equal_sample_sizes}
\end{figure}

\begin{figure}[ht]
	\centering
	\includegraphics[scale=0.33]{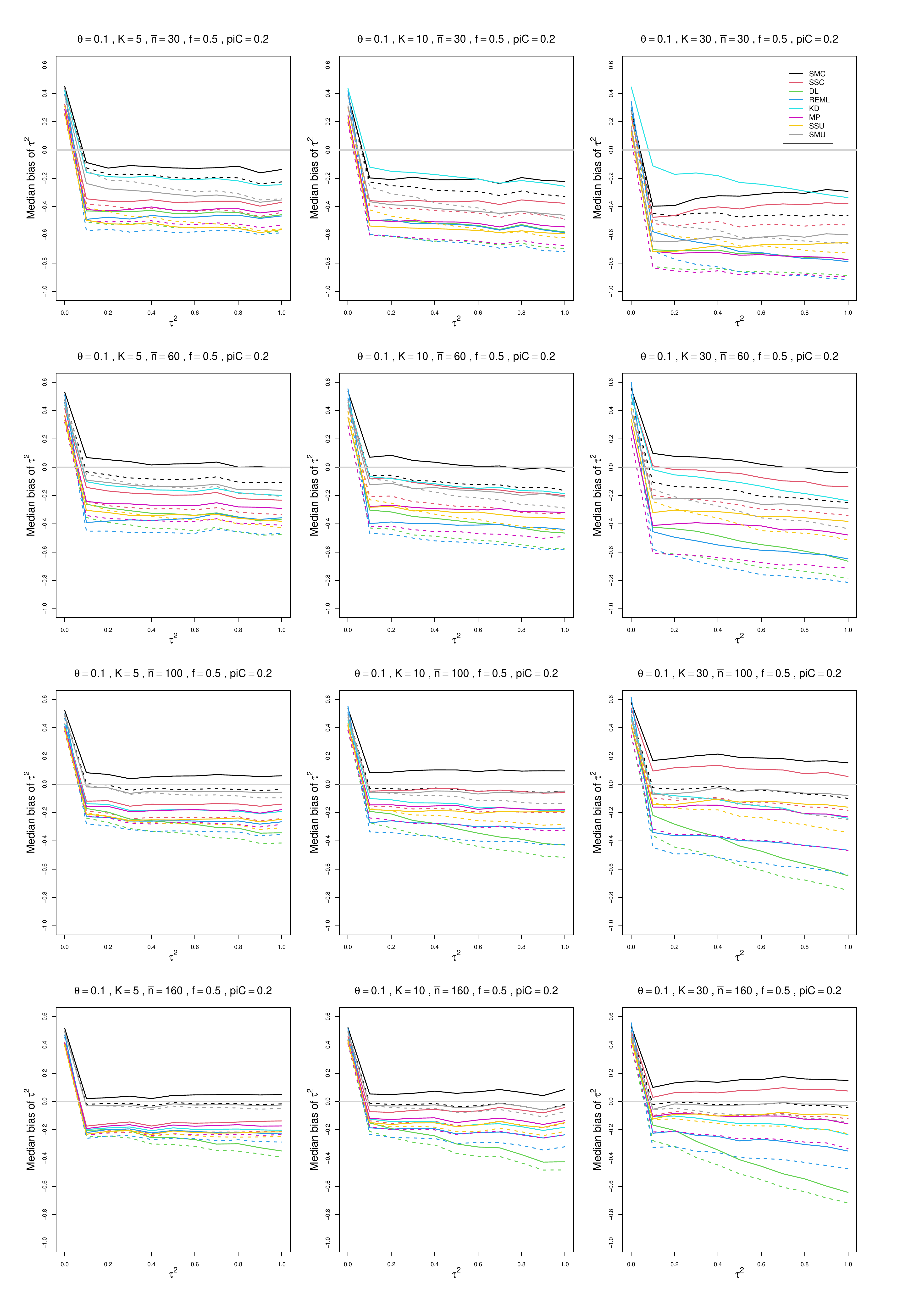}
	\caption{Median bias of estimators of between-study variance of LOR (DL, REML, KD, MP, SMC, SSC, SMU, and SSU) vs $\tau^2$, for unequal sample sizes $\bar{n}=30,\;60,\;100$ and $160$, $p_{iC} = .2$, $\theta = 0.1$ and  $f = 0.5$.   Solid lines: DL, REML, MP, SSC, SMC \lq\lq only"; KD; SSU and SMU model-based. Dashed lines: DL, REML, MP, SSC, SMC  \lq\lq always"; SSU and SMU  na\"ive.  }
	\label{PlotMedBiasOfTau2_piC_02theta=0.1_LOR_unequal_sample_sizes}
\end{figure}

\begin{figure}[ht]
	\centering
	\includegraphics[scale=0.33]{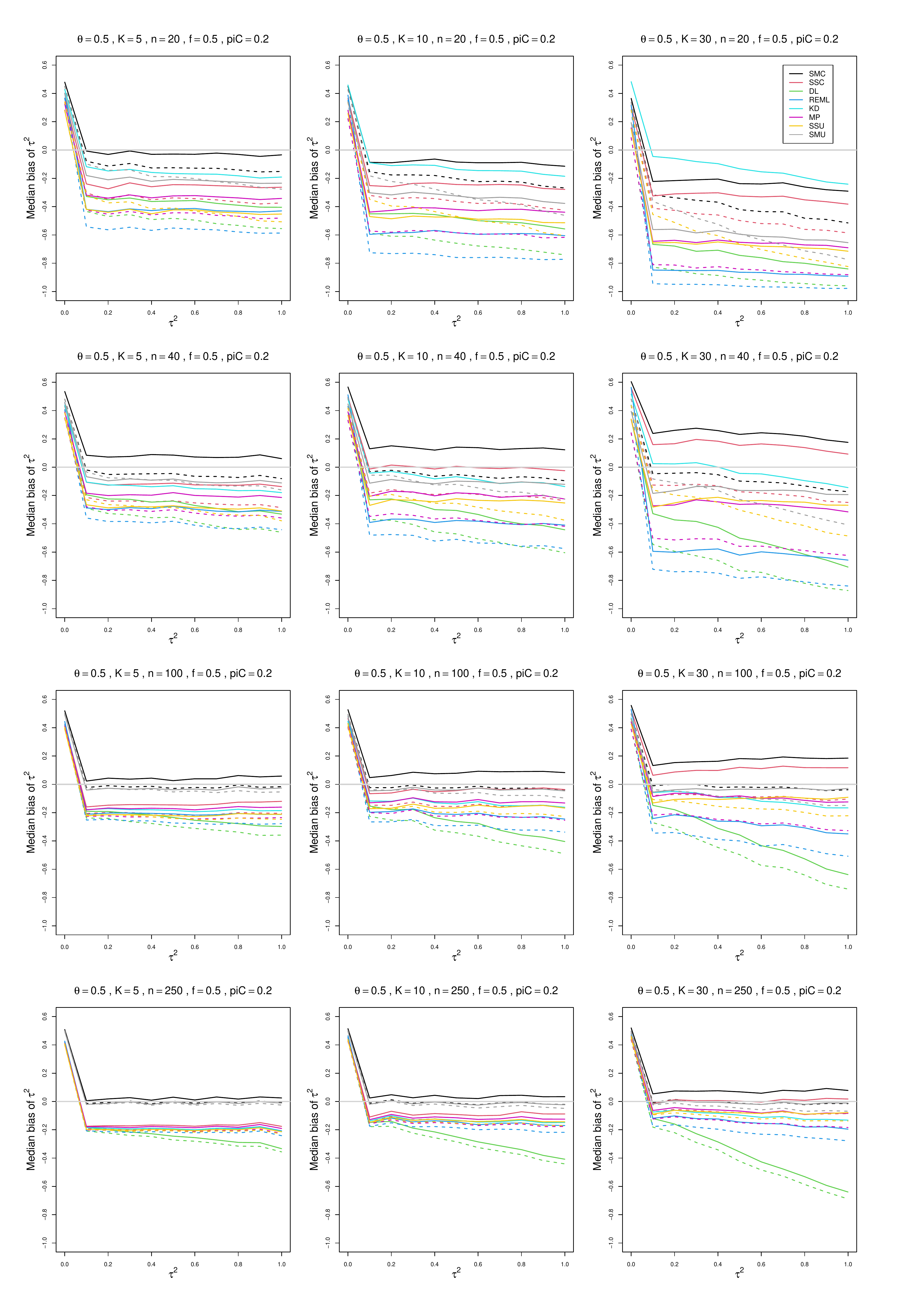}
	\caption{Median bias of estimators of between-study variance of LOR (DL, REML, KD, MP, SMC, SSC, SMU, and SSU) vs $\tau^2$, for equal sample sizes $n = 20,\;40,\;100$ and $250$, $p_{iC} = .2$, $\theta = 0.5$ and  $f = 0.5$.   Solid lines: DL, REML, MP, SSC, SMC \lq\lq only"; KD; SSU and SMU model-based. Dashed lines: DL, REML, MP, SSC, SMC  \lq\lq always"; SSU and SMU  na\"ive.  }
	\label{PlotMedBiasOfTau2_piC_02theta=0.5_LOR_equal_sample_sizes}
\end{figure}

\begin{figure}[ht]
	\centering
	\includegraphics[scale=0.33]{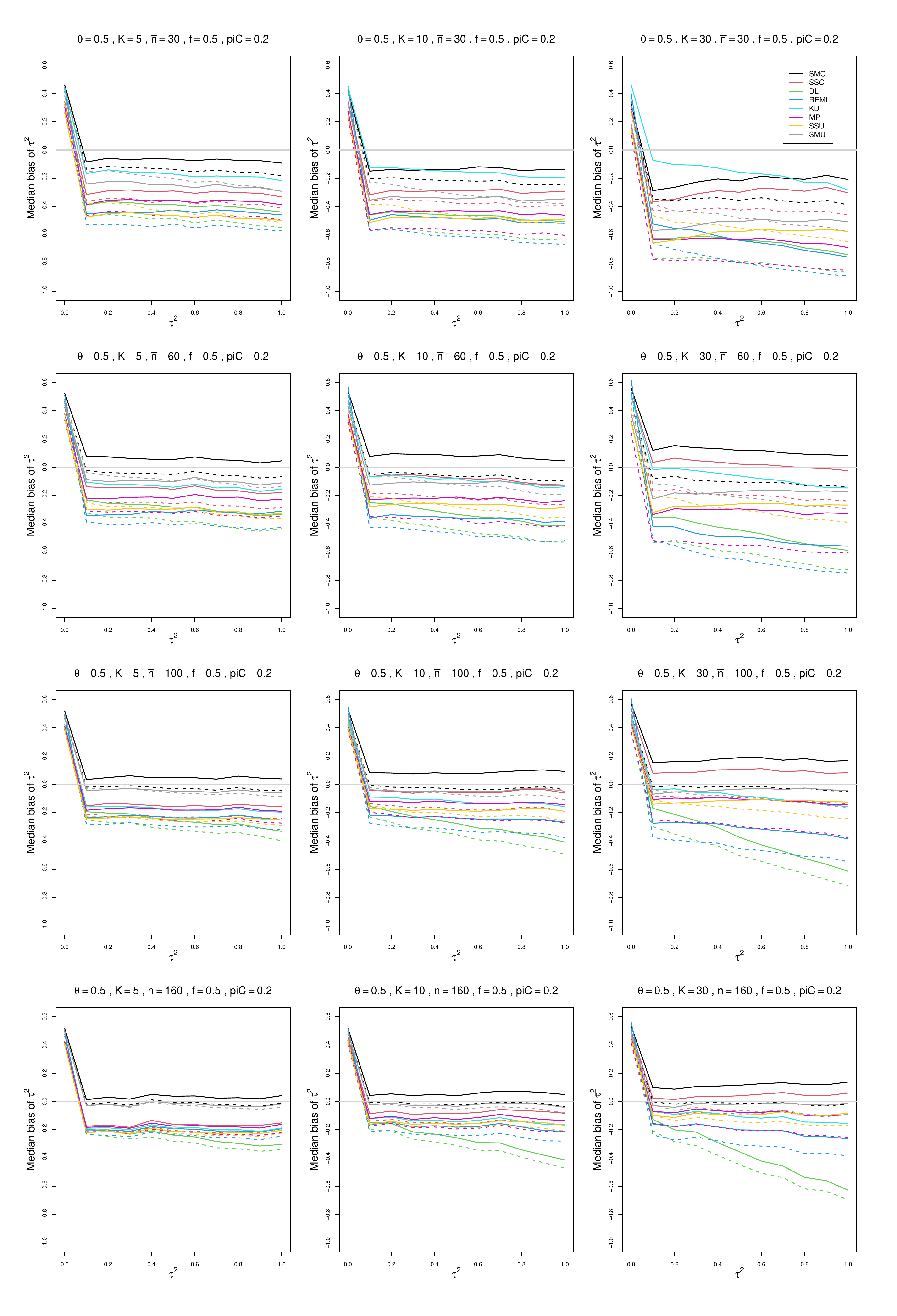}
	\caption{Median bias of estimators of between-study variance of LOR (DL, REML, KD, MP, SMC, SSC, SMU, and SSU) vs $\tau^2$, for unequal sample sizes $\bar{n}=30,\;60,\;100$ and $160$, $p_{iC} = .2$, $\theta = 0.5$ and  $f = 0.5$.   Solid lines: DL, REML, MP, SSC, SMC \lq\lq only"; KD; SSU and SMU model-based. Dashed lines: DL, REML, MP, SSC, SMC  \lq\lq always"; SSU and SMU  na\"ive.   }
	\label{PlotMedBiasOfTau2_piC_02theta=0.5_LOR_unequal_sample_sizes}
\end{figure}

\begin{figure}[ht]
	\centering
	\includegraphics[scale=0.33]{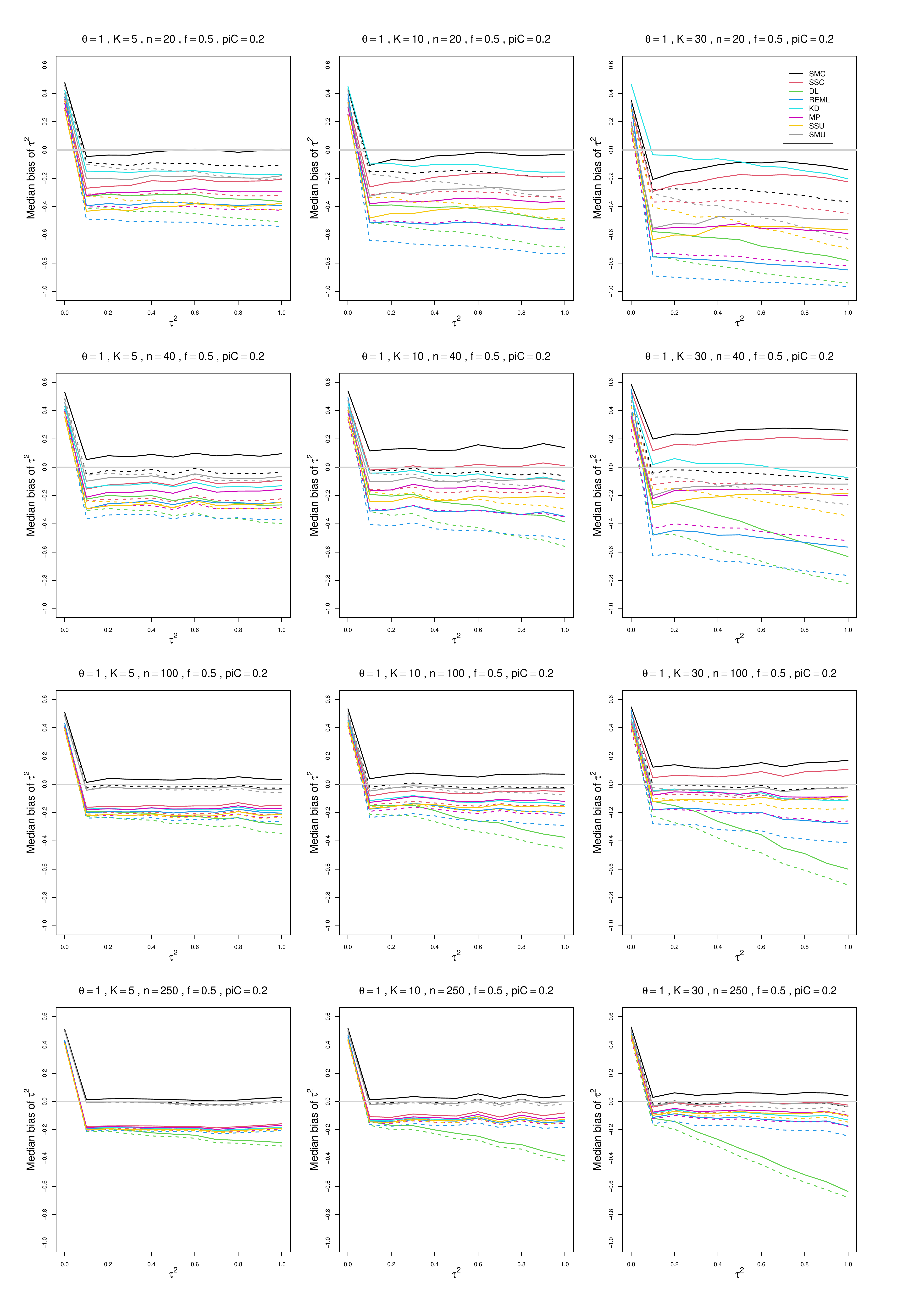}
	\caption{Median bias of estimators of between-study variance of LOR (DL, REML, KD, MP, SMC, SSC, SMU, and SSU) vs $\tau^2$, for equal sample sizes $n = 20,\;40,\;100$ and $250$, $p_{iC} = .2$, $\theta = 1$ and  $f = 0.5$.   Solid lines: DL, REML, MP, SSC, SMC \lq\lq only"; KD; SSU and SMU model-based. Dashed lines: DL, REML, MP, SSC, SMC  \lq\lq always"; SSU and SMU  na\"ive.  }
	\label{PlotMedBiasOfTau2_piC_02theta=1_LOR_equal_sample_sizes}
\end{figure}

\begin{figure}[ht]
	\centering
	\includegraphics[scale=0.33]{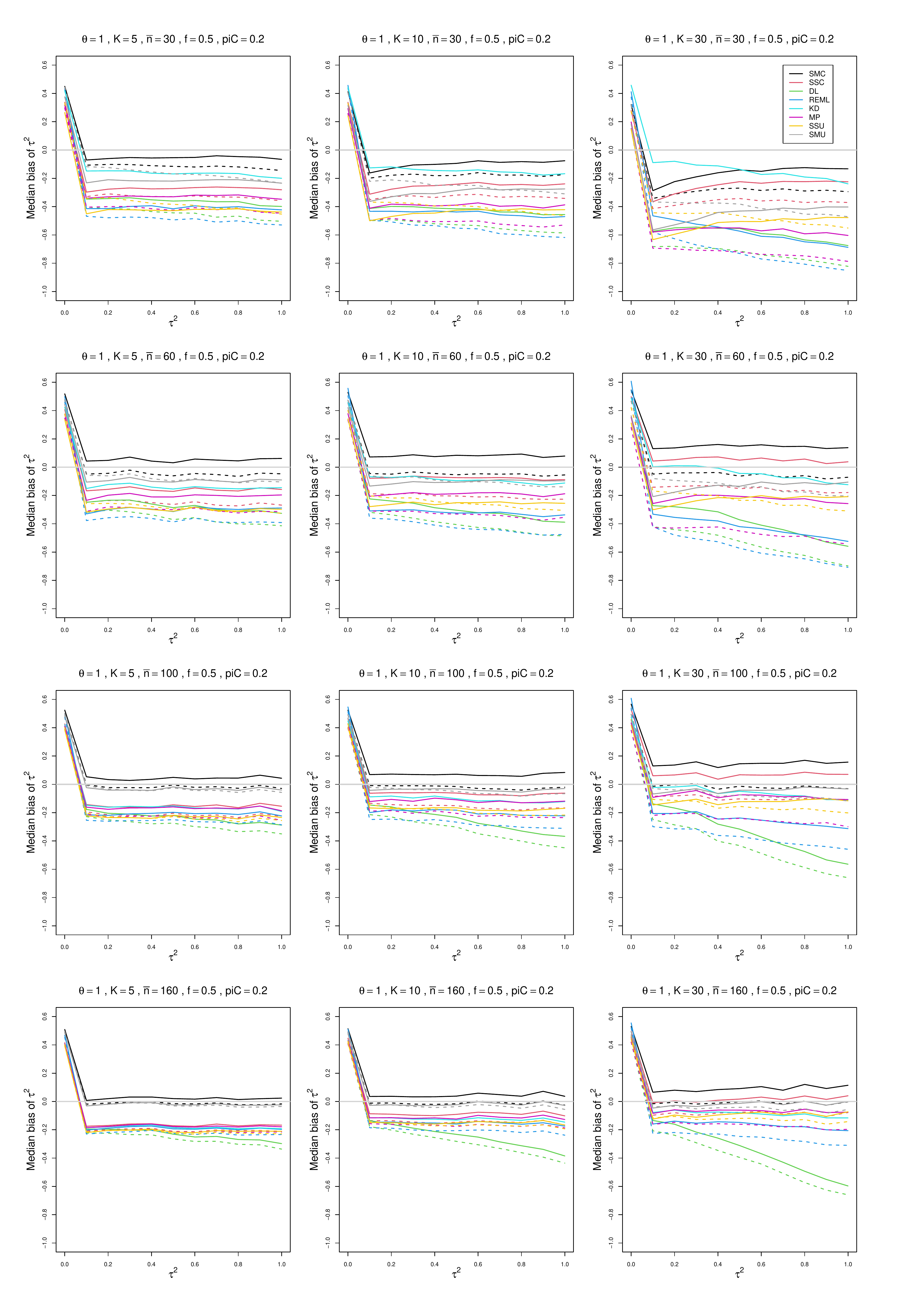}
	\caption{Median bias of estimators of between-study variance of LOR (DL, REML, KD, MP, SMC, SSC, SMU, and SSU) vs $\tau^2$, for unequal sample sizes $\bar{n}=30,\;60,\;100$ and $160$, $p_{iC} = .2$, $\theta = 1$ and  $f = 0.5$.   Solid lines: DL, REML, MP, SSC, SMC \lq\lq only"; KD; SSU and SMU model-based. Dashed lines: DL, REML, MP, SSC, SMC  \lq\lq always"; SSU and SMU  na\"ive.  }
	\label{PlotMedBiasOfTau2_piC_02theta=1_LOR_unequal_sample_sizes}
\end{figure}

\begin{figure}[ht]
	\centering
	\includegraphics[scale=0.33]{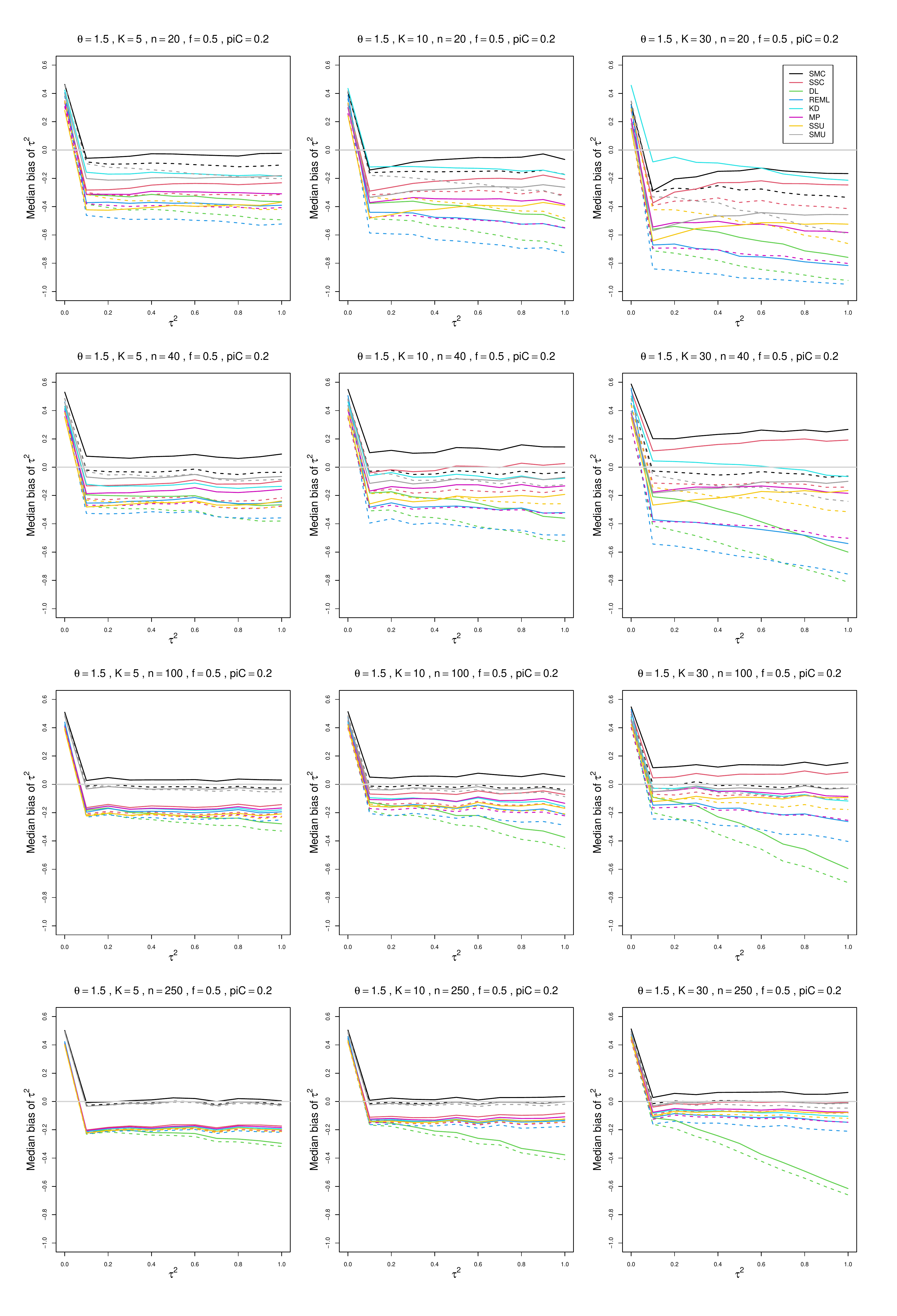}
	\caption{Median bias of estimators of between-study variance of LOR (DL, REML, KD, MP, SMC, SSC, SMU, and SSU) vs $\tau^2$, for equal sample sizes $n = 20,\;40,\;100$ and $250$, $p_{iC} = .2$, $\theta = 1.5$ and  $f = 0.5$.   Solid lines: DL, REML, MP, SSC, SMC \lq\lq only"; KD; SSU and SMU model-based. Dashed lines: DL, REML, MP, SSC, SMC  \lq\lq always"; SSU and SMU  na\"ive.  }
	\label{PlotMedBiasOfTau2_piC_02theta=1.5_LOR_equal_sample_sizes}
\end{figure}

\begin{figure}[ht]
	\centering
	\includegraphics[scale=0.33]{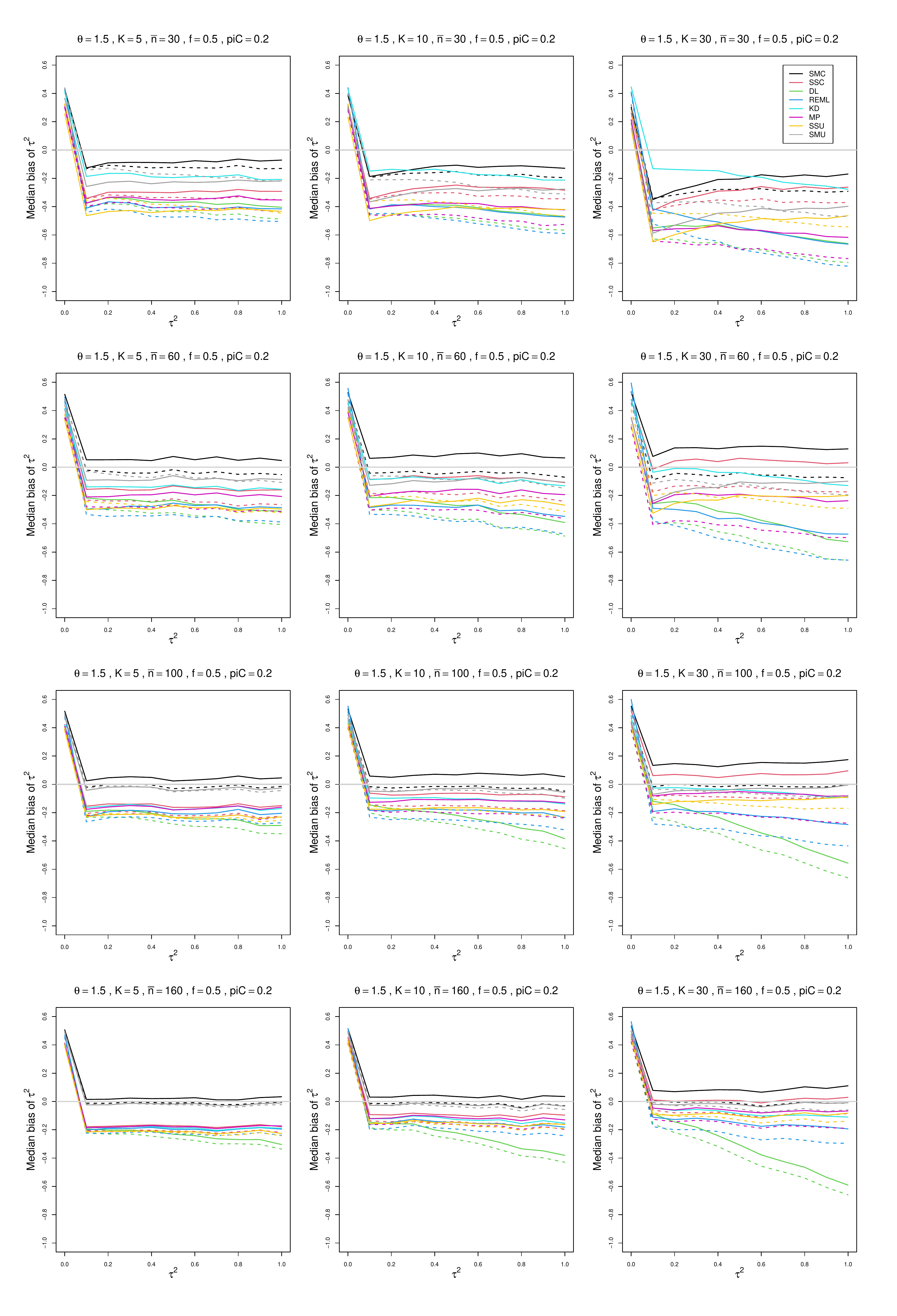}
	\caption{Median bias of estimators of between-study variance of LOR (DL, REML, KD, MP, SMC, SSC, SMU, and SSU) vs $\tau^2$, for unequal sample sizes $\bar{n}=30,\;60,\;100$ and $160$, $p_{iC} = .2$, $\theta = 1.5$ and  $f = 0.5$.   Solid lines: DL, REML, MP, SSC, SMC \lq\lq only"; KD; SSU and SMU model-based. Dashed lines: DL, REML, MP, SSC, SMC  \lq\lq always"; SSU and SMU  na\"ive.  }
	\label{PlotMedBiasOfTau2_piC_02theta=1.5_LOR_unequal_sample_sizes}
\end{figure}

\begin{figure}[ht]
	\centering
	\includegraphics[scale=0.33]{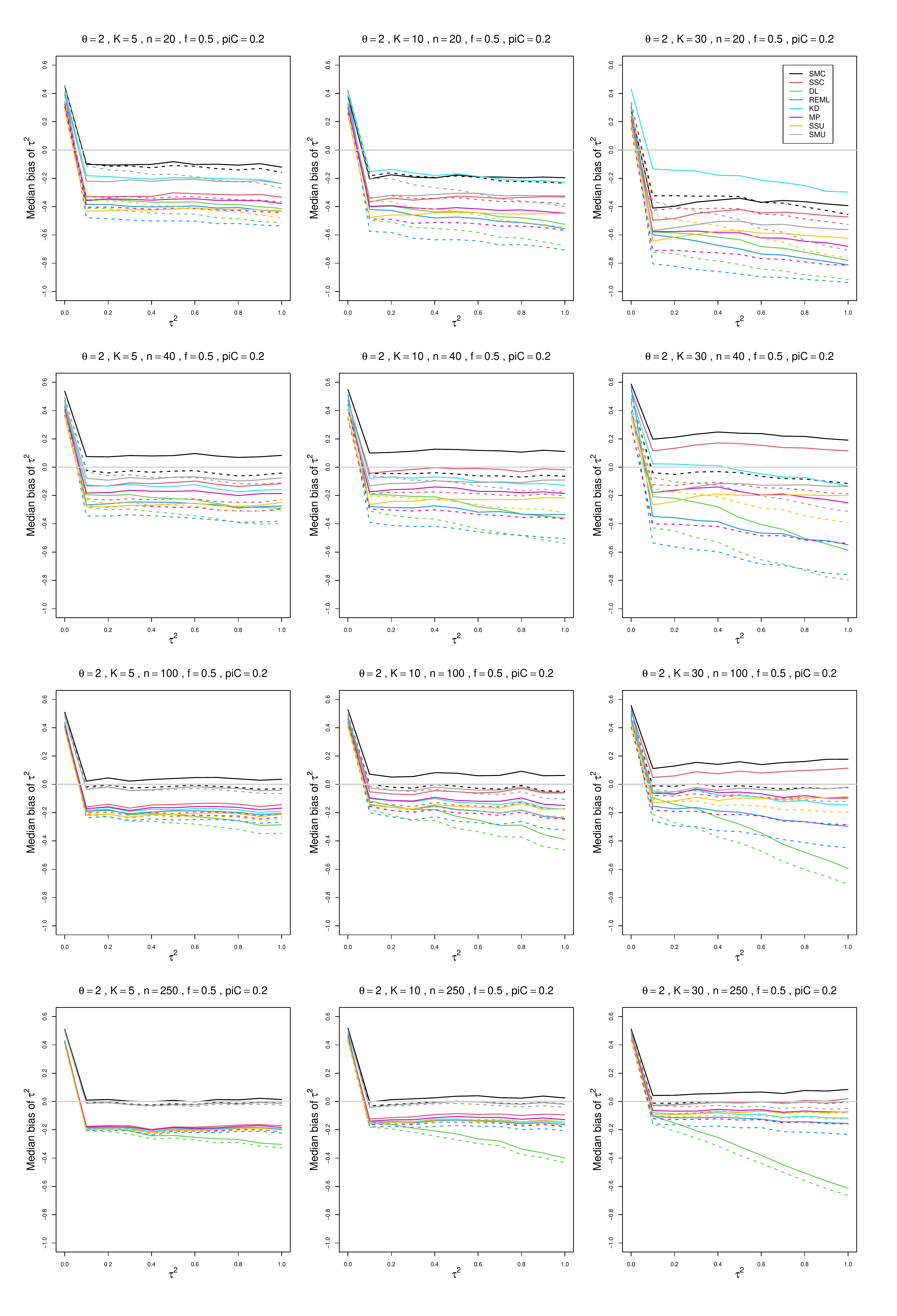}
	\caption{Median bias of estimators of between-study variance of LOR (DL, REML, KD, MP, SMC, SSC, SMU, and SSU) vs $\tau^2$, for equal sample sizes $n = 20,\;40,\;100$ and $250$, $p_{iC} = .2$, $\theta = 2$ and  $f = 0.5$.   Solid lines: DL, REML, MP, SSC, SMC \lq\lq only"; KD; SSU and SMU model-based. Dashed lines: DL, REML, MP, SSC, SMC  \lq\lq always"; SSU and SMU  na\"ive.  }
	\label{PlotMedBiasOfTau2_piC_02theta=2_LOR_equal_sample_sizes}
\end{figure}

\begin{figure}[ht]
	\centering
	\includegraphics[scale=0.33]{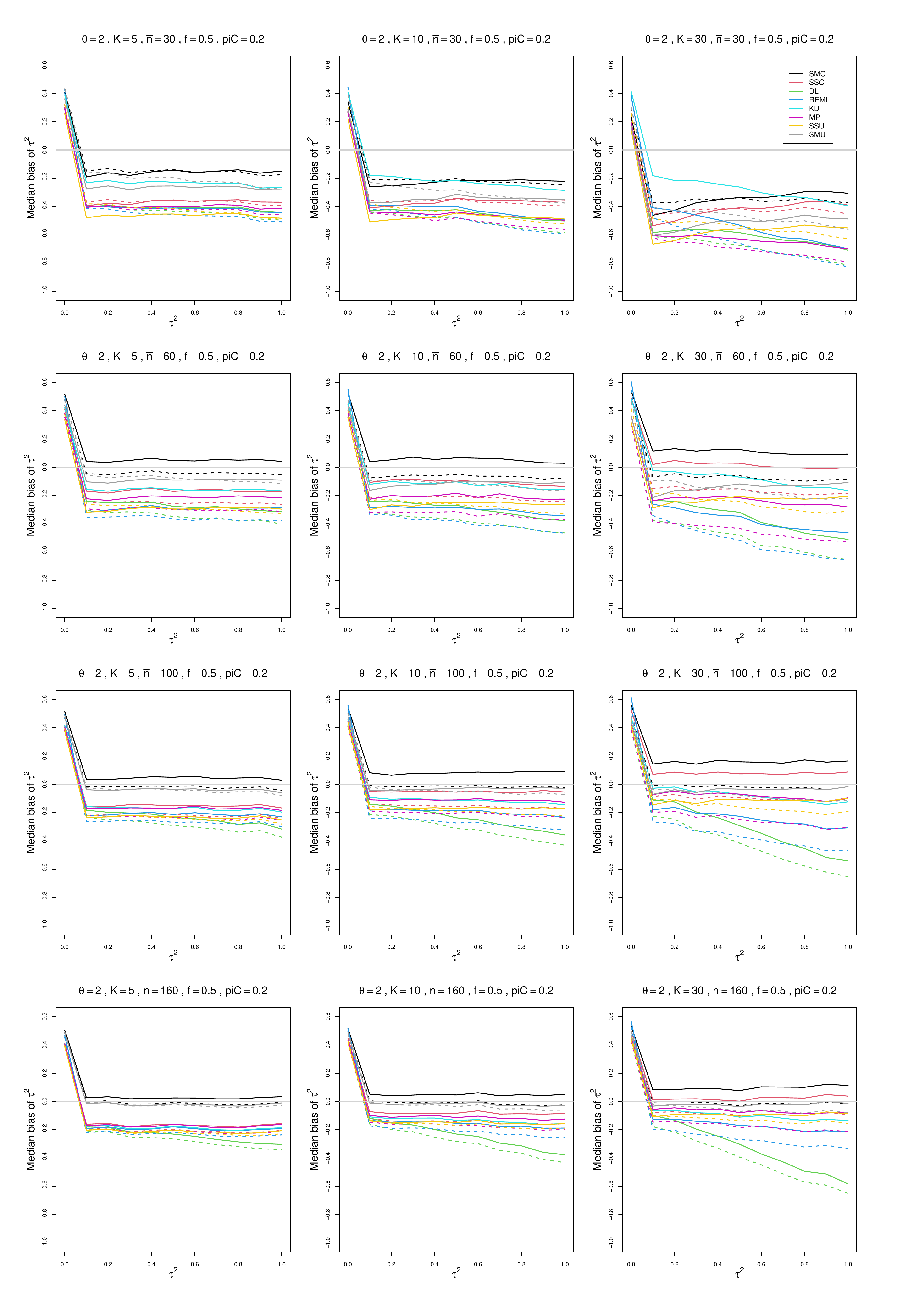}
	\caption{Median bias of estimators of between-study variance of LOR (DL, REML, KD, MP, SMC, SSC, SMU, and SSU) vs $\tau^2$, for unequal sample sizes $\bar{n}=30,\;60,\;100$ and $160$, $p_{iC} = .2$, $\theta = 2$ and  $f = 0.5$.   Solid lines: DL, REML, MP, SSC, SMC \lq\lq only"; KD; SSU and SMU model-based. Dashed lines: DL, REML, MP, SSC, SMC  \lq\lq always"; SSU and SMU  na\"ive.   }
	\label{PlotMedBiasOfTau2_piC_02theta=2_LOR_unequal_sample_sizes}
\end{figure}

\begin{figure}[ht]
	\centering
	\includegraphics[scale=0.33]{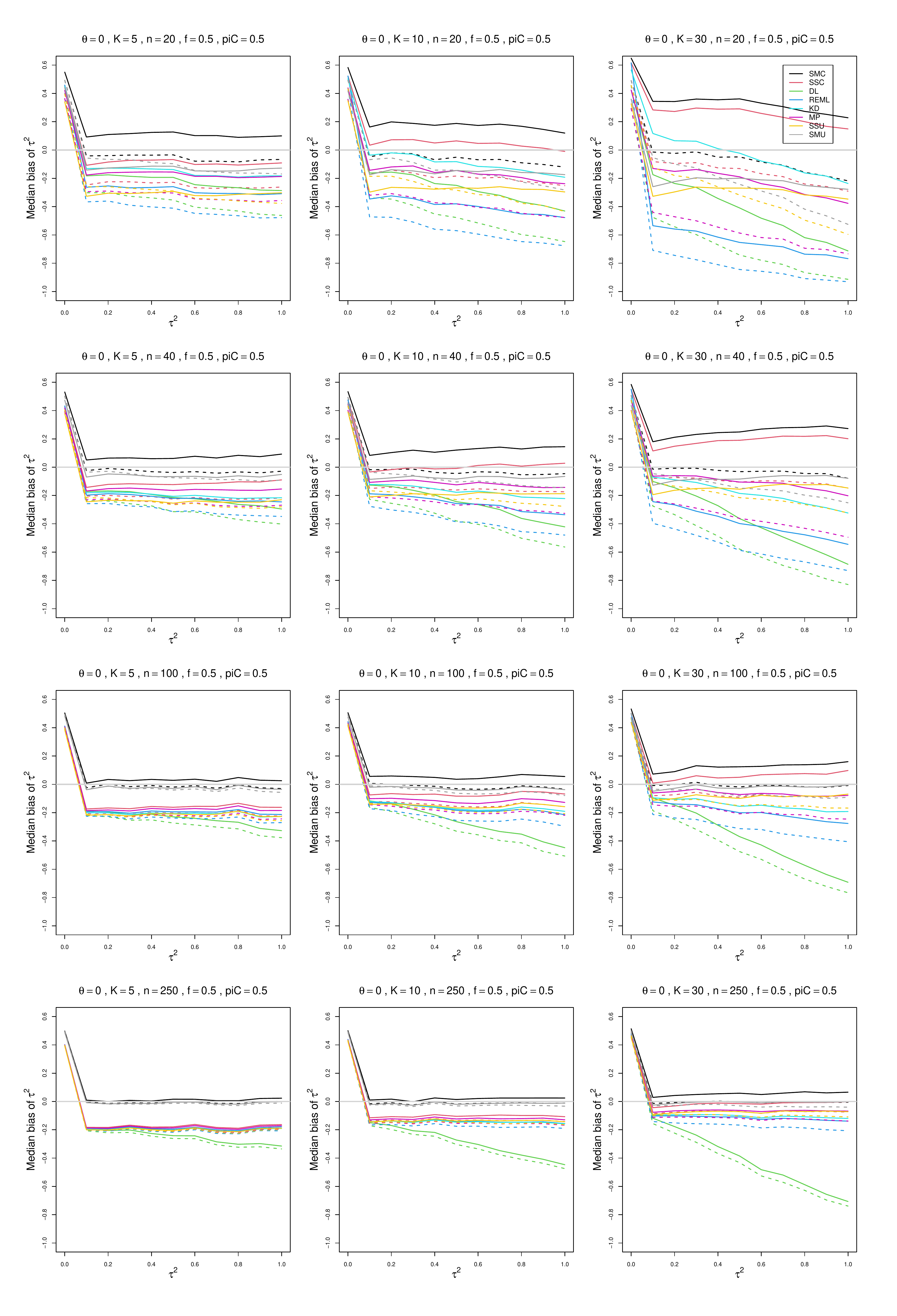}
	\caption{Median bias of estimators of between-study variance of LOR (DL, REML, KD, MP, SMC, SSC, SMU, and SSU) vs $\tau^2$, for equal sample sizes $n = 20,\;40,\;100$ and $250$, $p_{iC} = .5$, $\theta = 0$ and  $f = 0.5$.   Solid lines: DL, REML, MP, SSC, SMC \lq\lq only"; KD; SSU and SMU model-based. Dashed lines: DL, REML, MP, SSC, SMC  \lq\lq always"; SSU and SMU  na\"ive.  }
	\label{PlotMedBiasOfTau2_piC_05theta=0_LOR_equal_sample_sizes}
\end{figure}

\begin{figure}[ht]
	\centering
	\includegraphics[scale=0.33]{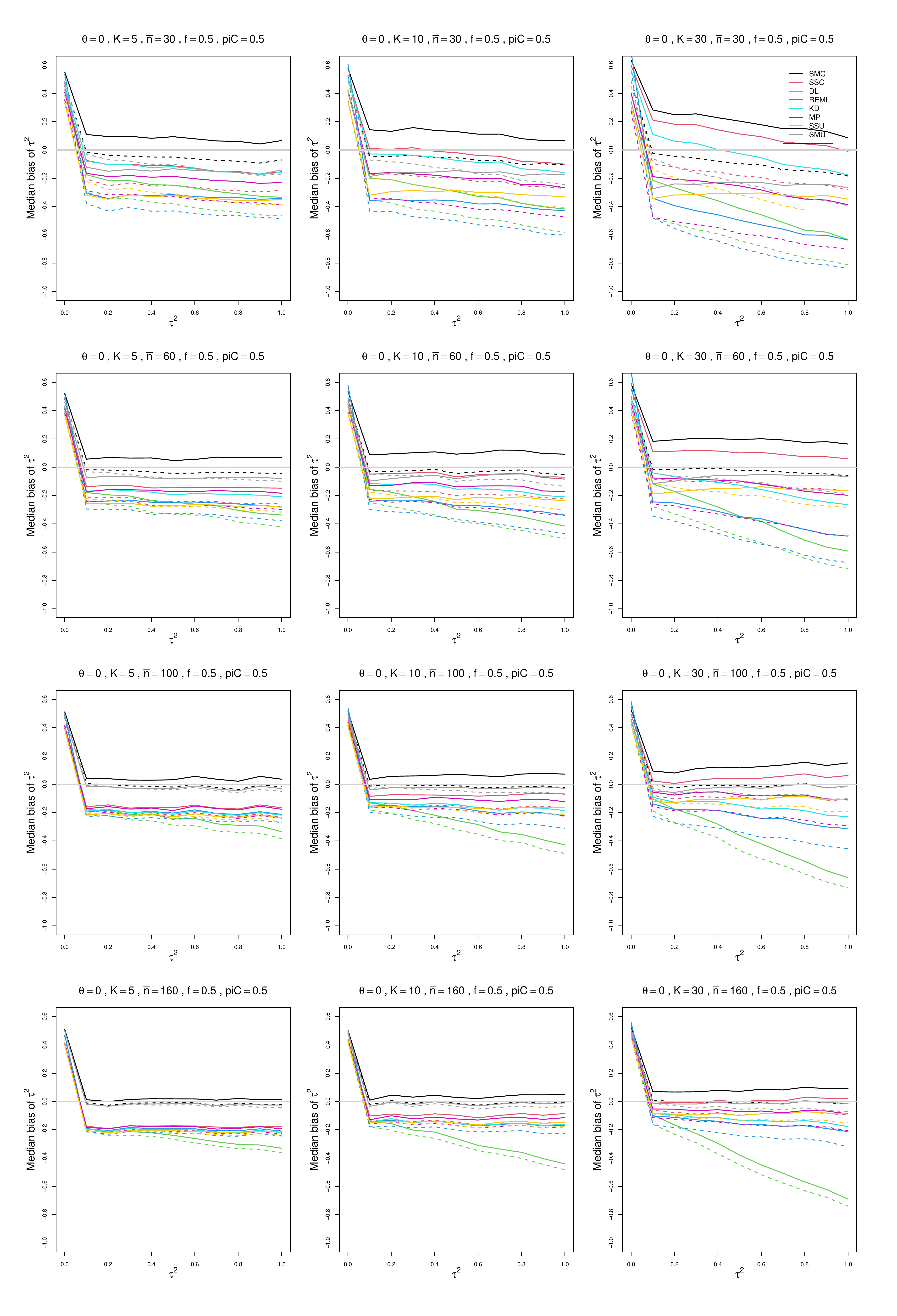}
	\caption{Median bias of estimators of between-study variance of LOR (DL, REML, KD, MP, SMC, SSC, SMU, and SSU) vs $\tau^2$, for unequal sample sizes $\bar{n}=30,\;60,\;100$ and $160$, $p_{iC} = .5$, $\theta = 0$ and  $f = 0.5$.   Solid lines: DL, REML, MP, SSC, SMC \lq\lq only"; KD; SSU and SMU model-based. Dashed lines: DL, REML, MP, SSC, SMC  \lq\lq always"; SSU and SMU  na\"ive.  }
	\label{PlotMedBiasOfTau2_piC_05theta=0_LOR_unequal_sample_sizes}
\end{figure}

\begin{figure}[ht]
	\centering
	\includegraphics[scale=0.33]{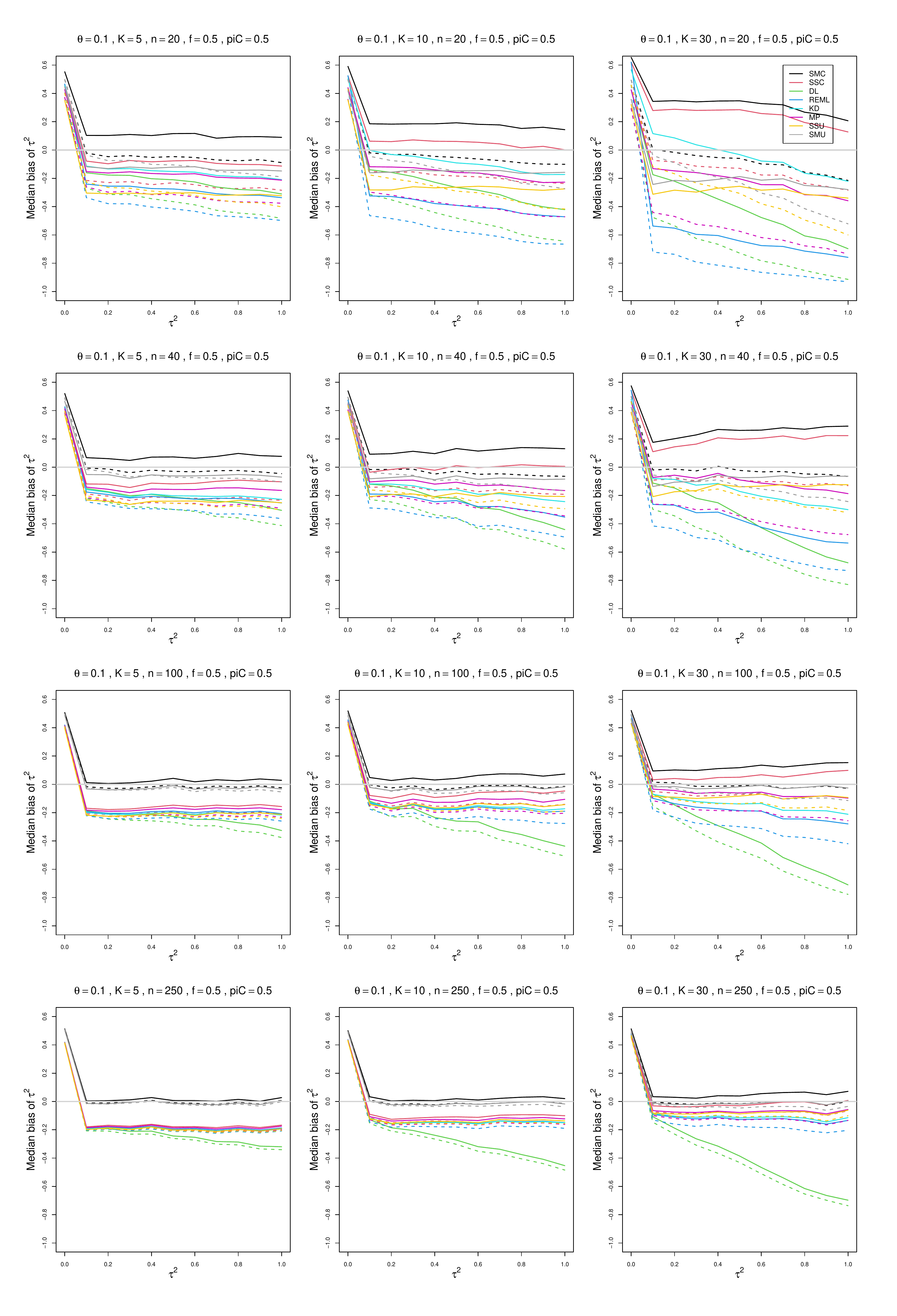}
	\caption{Median bias of estimators of between-study variance of LOR (DL, REML, KD, MP, SMC, SSC, SMU, and SSU) vs $\tau^2$, for equal sample sizes $n = 20,\;40,\;100$ and $250$, $p_{iC} = .5$, $\theta = 0.1$ and  $f = 0.5$.   Solid lines: DL, REML, MP, SSC, SMC \lq\lq only"; KD; SSU and SMU model-based. Dashed lines: DL, REML, MP, SSC, SMC  \lq\lq always"; SSU and SMU  na\"ive.    }
	\label{PlotMedBiasOfTau2_piC_05theta=0.1_LOR_equal_sample_sizes}
\end{figure}

\begin{figure}[ht]
	\centering
	\includegraphics[scale=0.33]{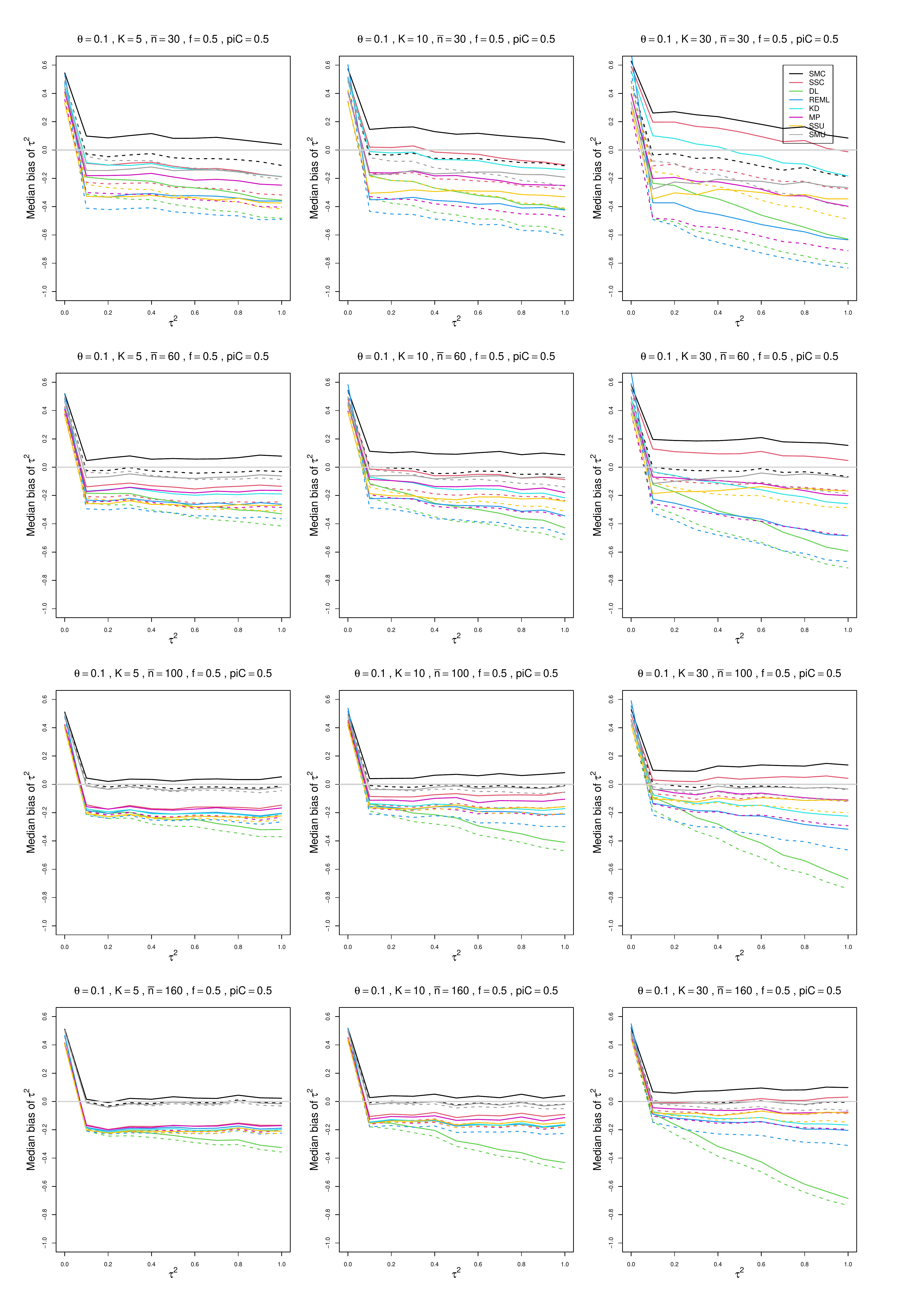}
	\caption{Median bias of estimators of between-study variance of LOR (DL, REML, KD, MP, SMC, SSC, SMU, and SSU) vs $\tau^2$, for unequal sample sizes $\bar{n}=30,\;60,\;100$ and $160$, $p_{iC} = .5$, $\theta = 0.1$ and  $f = 0.5$.   Solid lines: DL, REML, MP, SSC, SMC \lq\lq only"; KD; SSU and SMU model-based. Dashed lines: DL, REML, MP, SSC, SMC  \lq\lq always"; SSU and SMU  na\"ive.  }
	\label{PlotMedBiasOfTau2_piC_05theta=0.1_LOR_unequal_sample_sizes}
\end{figure}

\begin{figure}[ht]
	\centering
	\includegraphics[scale=0.33]{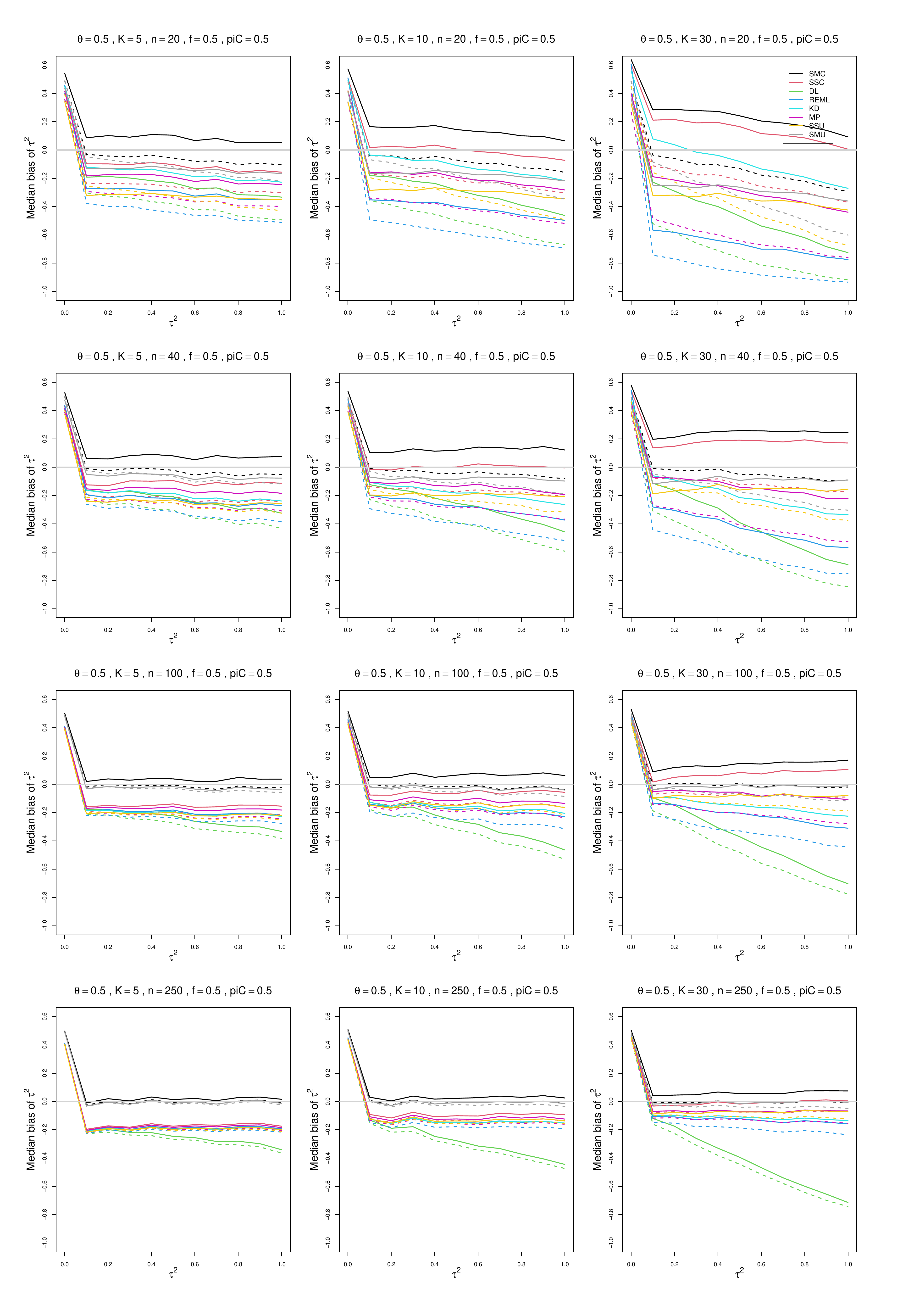}
	\caption{Median bias of estimators of between-study variance of LOR (DL, REML, KD, MP, SMC, SSC, SMU, and SSU) vs $\tau^2$, for equal sample sizes $n = 20,\;40,\;100$ and $250$, $p_{iC} = .5$, $\theta = 0.5$ and  $f = 0.5$.   Solid lines: DL, REML, MP, SSC, SMC \lq\lq only"; KD; SSU and SMU model-based. Dashed lines: DL, REML, MP, SSC, SMC  \lq\lq always"; SSU and SMU  na\"ive.  }
	\label{PlotMedBiasOfTau2_piC_05theta=0.5_LOR_equal_sample_sizes}
\end{figure}

\begin{figure}[ht]
	\centering
	\includegraphics[scale=0.33]{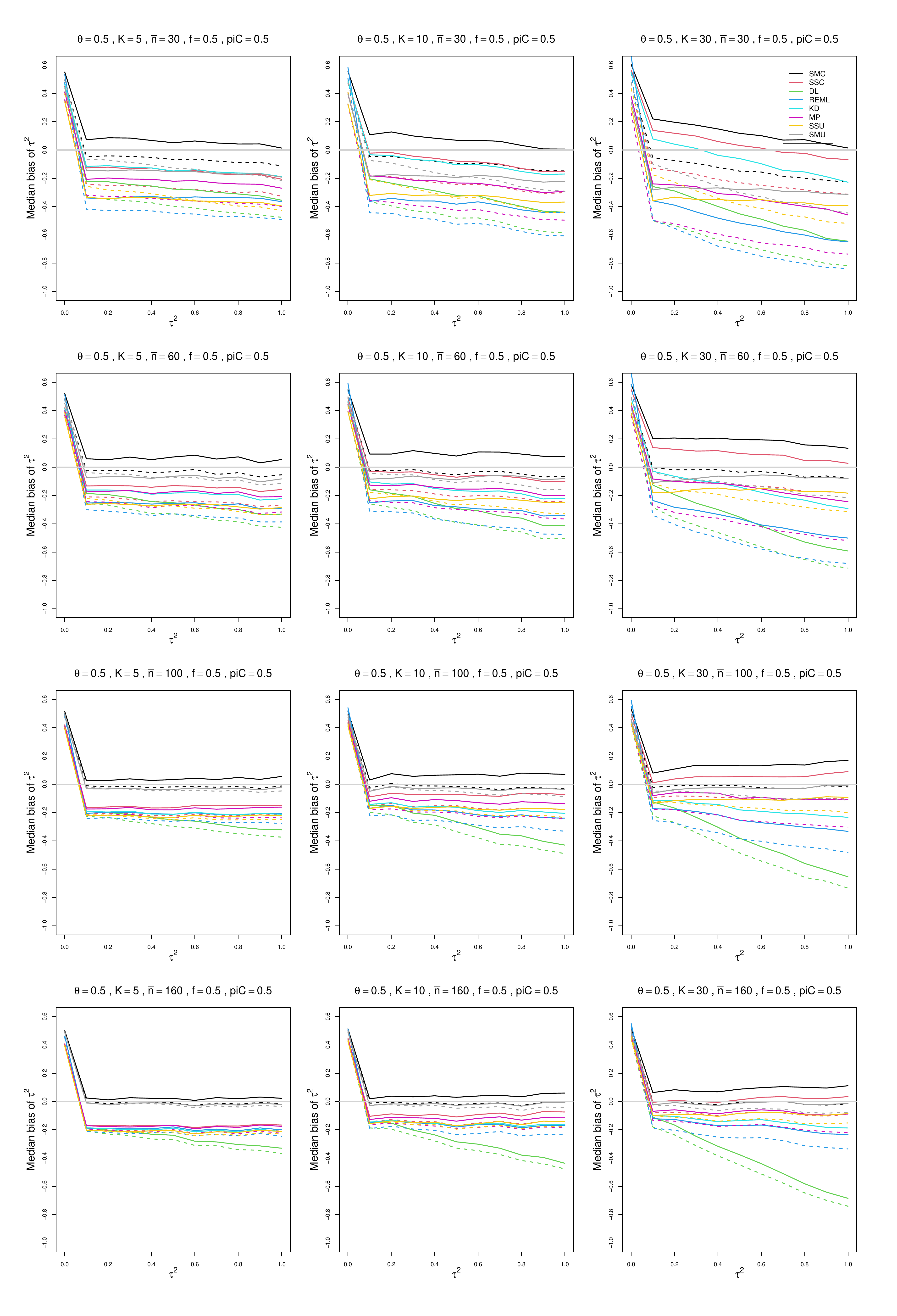}
	\caption{Median bias of estimators of between-study variance of LOR (DL, REML, KD, MP, SMC, SSC, SMU, and SSU) vs $\tau^2$, for unequal sample sizes $\bar{n}=30,\;60,\;100$ and $160$, $p_{iC} = .5$, $\theta = 0.5$ and  $f = 0.5$.   Solid lines: DL, REML, MP, SSC, SMC \lq\lq only"; KD; SSU and SMU model-based. Dashed lines: DL, REML, MP, SSC, SMC  \lq\lq always"; SSU and SMU  na\"ive.  }
	\label{PlotMedBiasOfTau2_piC_05theta=0.5_LOR_unequal_sample_sizes}
\end{figure}

\begin{figure}[ht]
	\centering
	\includegraphics[scale=0.33]{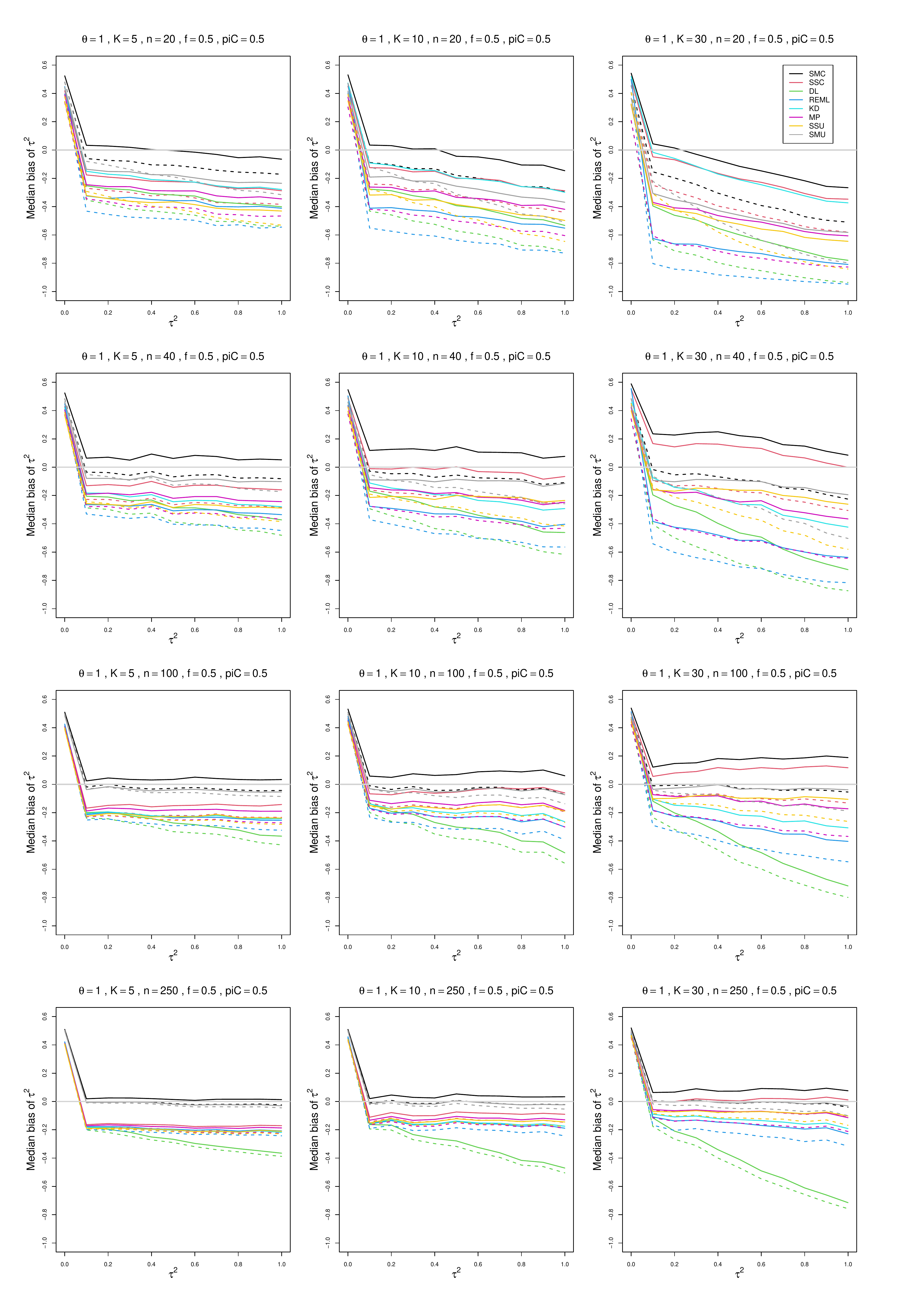}
	\caption{Median bias of estimators of between-study variance of LOR (DL, REML, KD, MP, SMC, SSC, SMU, and SSU) vs $\tau^2$, for equal sample sizes $n = 20,\;40,\;100$ and $250$, $p_{iC} = .5$, $\theta = 1$ and  $f = 0.5$.   Solid lines: DL, REML, MP, SSC, SMC \lq\lq only"; KD; SSU and SMU model-based. Dashed lines: DL, REML, MP, SSC, SMC  \lq\lq always"; SSU and SMU  na\"ive.   }
	\label{PlotMedBiasOfTau2_piC_05theta=1_LOR_equal_sample_sizes}
\end{figure}

\begin{figure}[ht]
	\centering
	\includegraphics[scale=0.33]{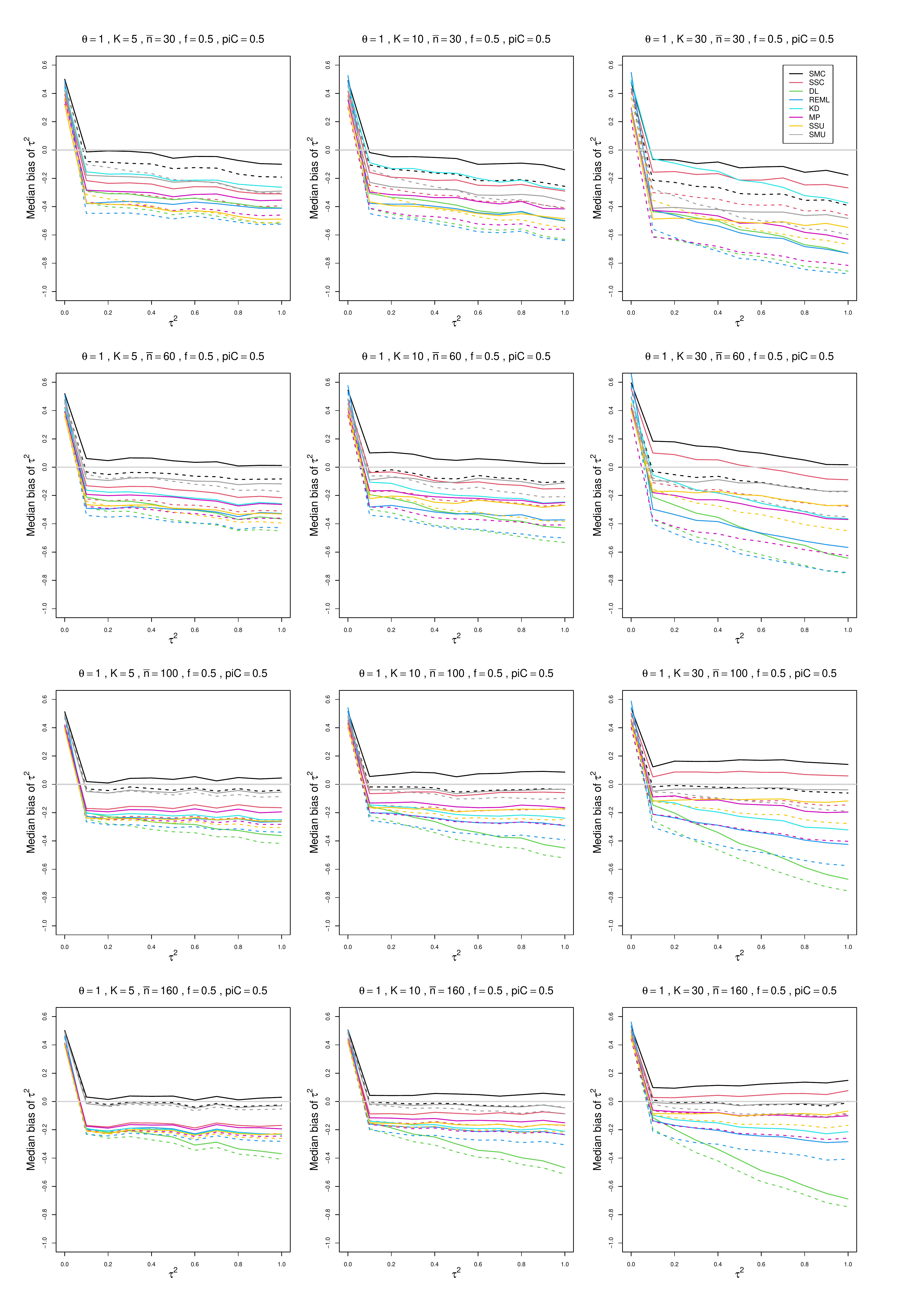}
	\caption{Median bias of estimators of between-study variance of LOR (DL, REML, KD, MP, SMC, SSC, SMU, and SSU) vs $\tau^2$, for unequal sample sizes $\bar{n}=30,\;60,\;100$ and $160$, $p_{iC} = .5$, $\theta = 1$ and  $f = 0.5$.   Solid lines: DL, REML, MP, SSC, SMC \lq\lq only"; KD; SSU and SMU model-based. Dashed lines: DL, REML, MP, SSC, SMC  \lq\lq always"; SSU and SMU  na\"ive.  }
	\label{PlotMedBiasOfTau2_piC_05theta=1_LOR_unequal_sample_sizes}
\end{figure}

\begin{figure}[ht]
	\centering
	\includegraphics[scale=0.33]{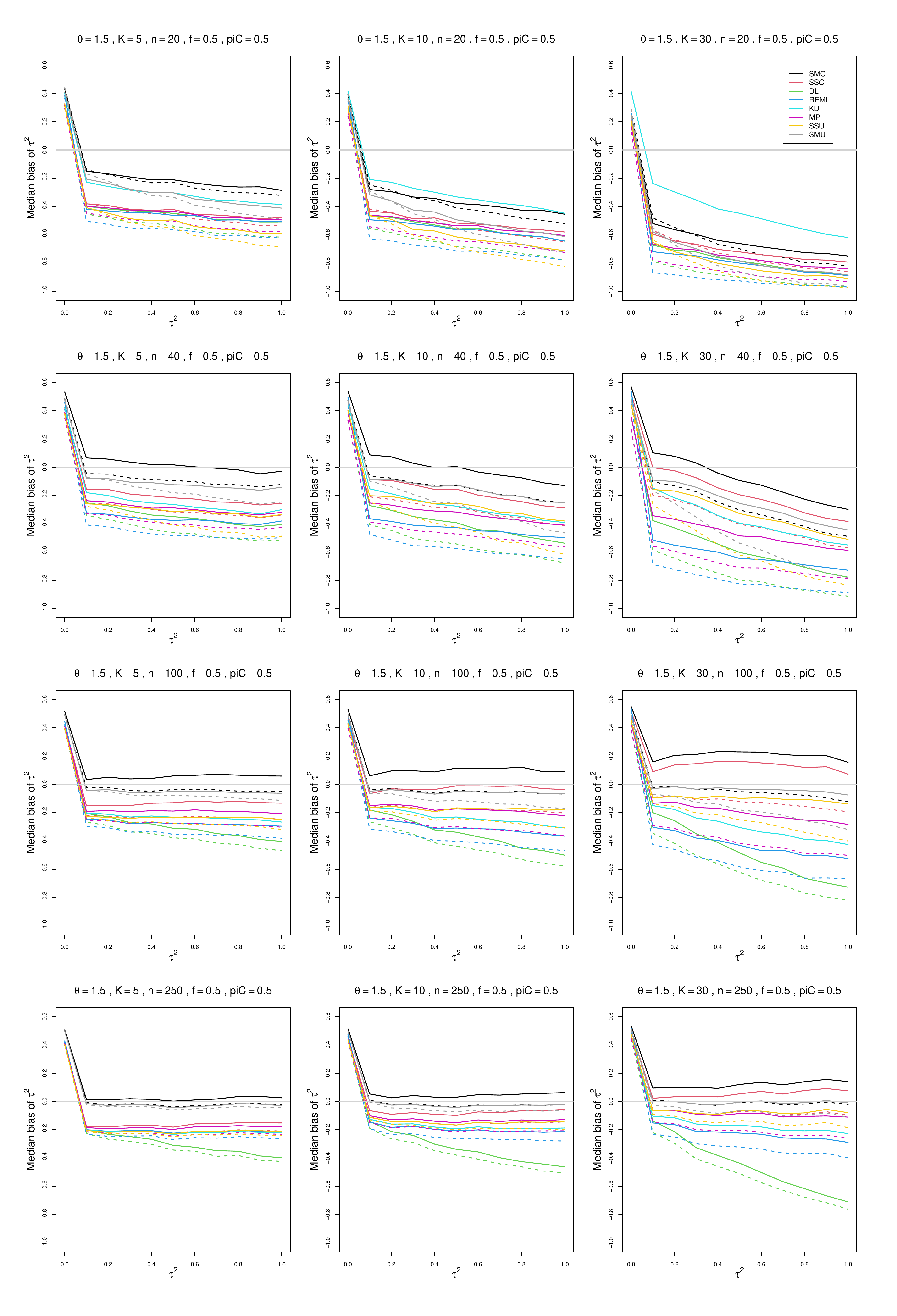}
	\caption{Median bias of estimators of between-study variance of LOR (DL, REML, KD, MP, SMC, SSC, SMU, and SSU) vs $\tau^2$, for equal sample sizes $n = 20,\;40,\;100$ and $250$, $p_{iC} = .5$, $\theta = 1.5$ and  $f = 0.5$.   Solid lines: DL, REML, MP, SSC, SMC \lq\lq only"; KD; SSU and SMU model-based. Dashed lines: DL, REML, MP, SSC, SMC  \lq\lq always"; SSU and SMU  na\"ive.  }
	\label{PlotMedBiasOfTau2_piC_05theta=1.5_LOR_equal_sample_sizes}
\end{figure}

\begin{figure}[ht]
	\centering
	\includegraphics[scale=0.33]{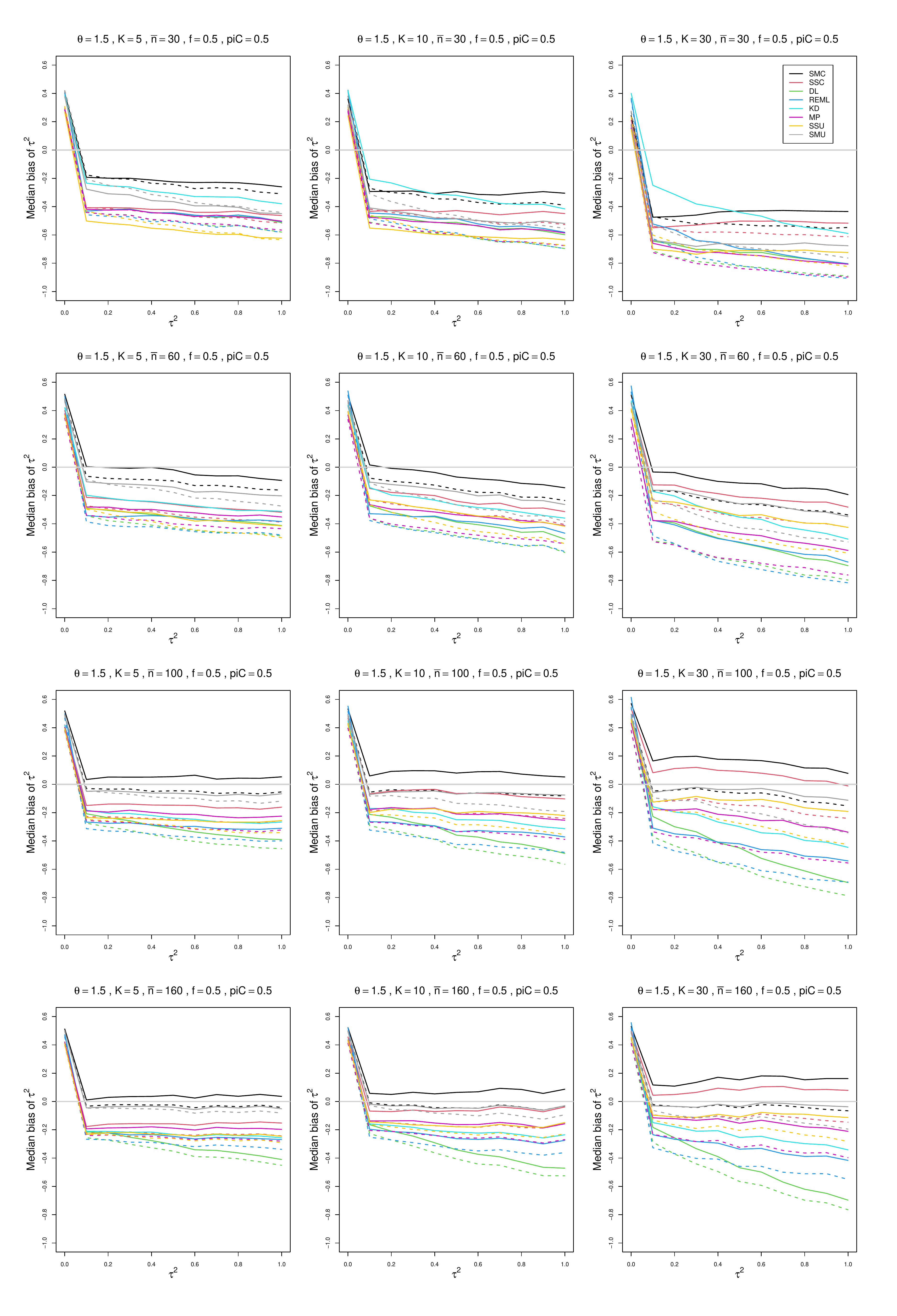}
	\caption{Median bias of estimators of between-study variance of LOR (DL, REML, KD, MP, SMC, SSC, SMU, and SSU) vs $\tau^2$, for unequal sample sizes $\bar{n}=30,\;60,\;100$ and $160$, $p_{iC} = .5$, $\theta = 1.5$ and  $f = 0.5$.   Solid lines: DL, REML, MP, SSC, SMC \lq\lq only"; KD; SSU and SMU model-based. Dashed lines: DL, REML, MP, SSC, SMC  \lq\lq always"; SSU and SMU  na\"ive.  }
	\label{PlotMedBiasOfTau2_piC_05theta=1.5_LOR_unequal_sample_sizes}
\end{figure}

\begin{figure}[ht]
	\centering
	\includegraphics[scale=0.33]{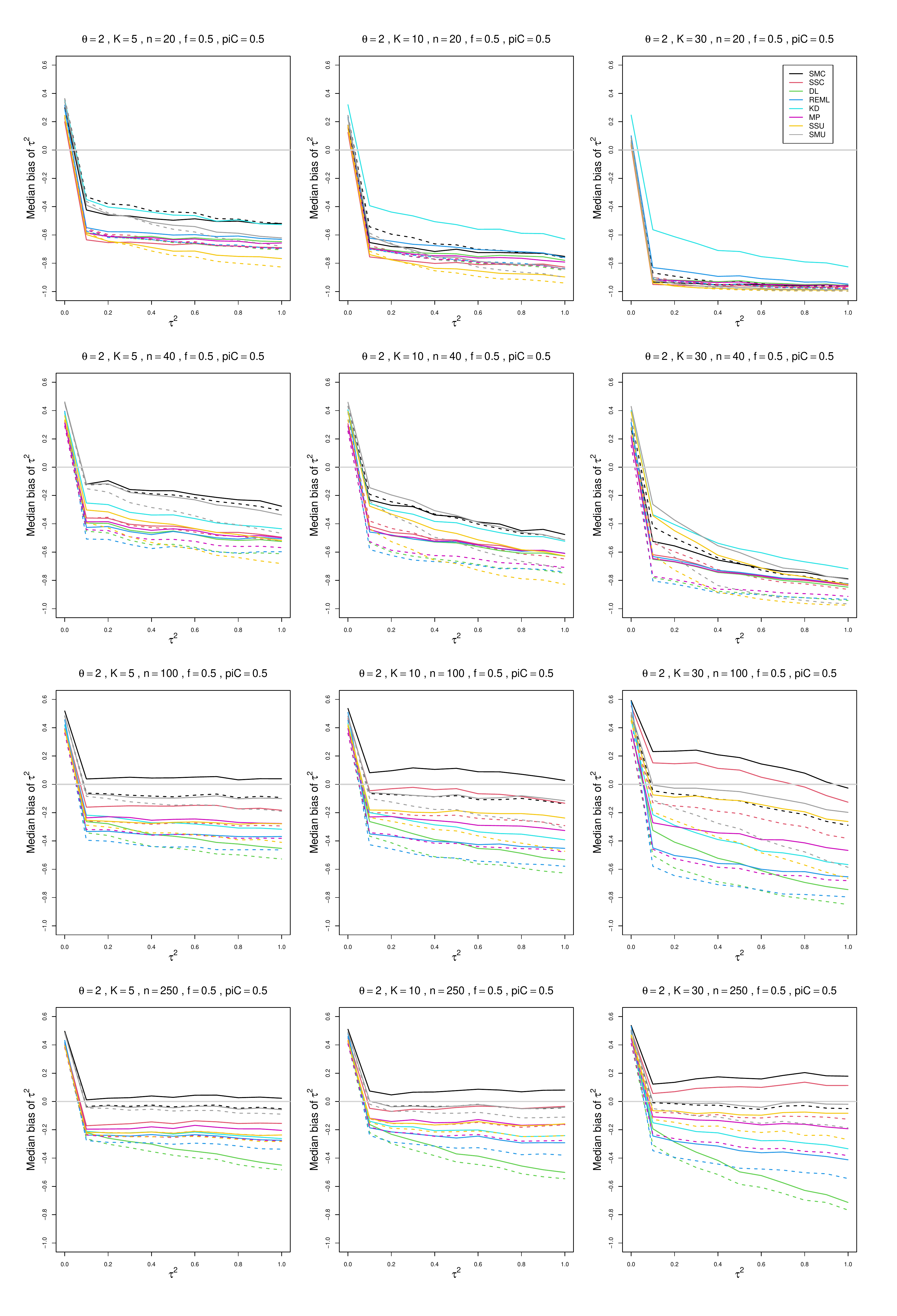}
	\caption{Median bias of estimators of between-study variance of LOR (DL, REML, KD, MP, SMC, SSC, SMU, and SSU) vs $\tau^2$, for equal sample sizes $n = 20,\;40,\;100$ and $250$, $p_{iC} = .5$, $\theta = 2$ and  $f = 0.5$.   Solid lines: DL, REML, MP, SSC, SMC \lq\lq only"; KD; SSU and SMU model-based. Dashed lines: DL, REML, MP, SSC, SMC  \lq\lq always"; SSU and SMU  na\"ive.  }
	\label{PlotMedBiasOfTau2_piC_05theta=2_LOR_equal_sample_sizes}
\end{figure}

\begin{figure}[ht]
	\centering
	\includegraphics[scale=0.33]{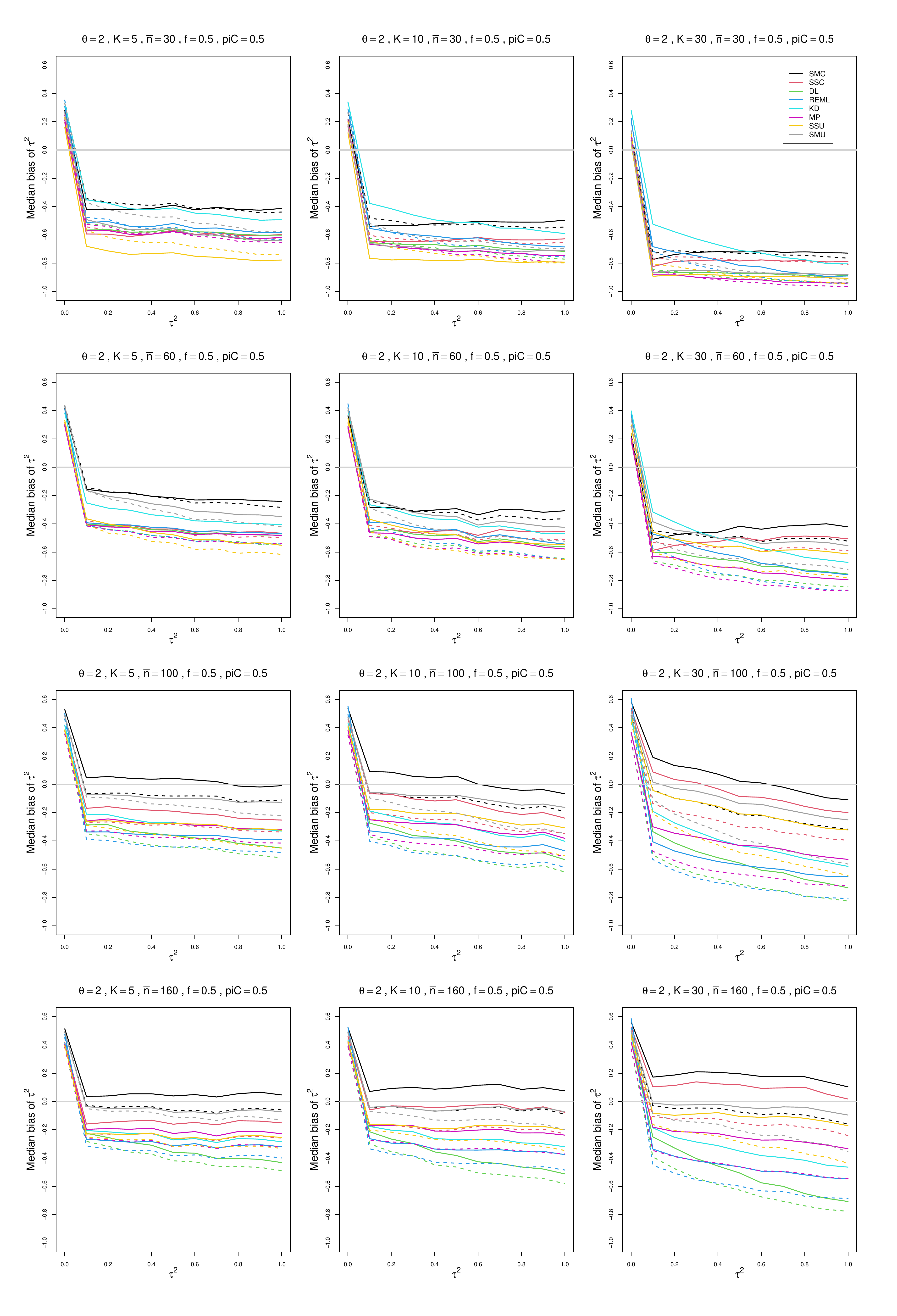}
	\caption{Median bias of estimators of between-study variance of LOR (DL, REML, KD, MP, SMC, SSC, SMU, and SSU) vs $\tau^2$, for unequal sample sizes $\bar{n}=30,\;60,\;100$ and $160$, $p_{iC} = .5$, $\theta = 2$ and  $f = 0.5$.   Solid lines: DL, REML, MP, SSC, SMC \lq\lq only"; KD; SSU and SMU model-based. Dashed lines: DL, REML, MP, SSC, SMC  \lq\lq always"; SSU and SMU  na\"ive.  }
	\label{PlotMedBiasOfTau2_piC_05theta=2_LOR_unequal_sample_sizes}
\end{figure}

\clearpage
\section*{Appendix C: Coverage of 95\% confidence intervals for between-study variance}

Each figure corresponds to a value of the probability of an event in the Control arm $p_{iC}$  (= .1, .2, .5) . \\
The fraction of each study's sample size in the Control arm  ($f$) is held constant at 0.5. For each combination of a value of $n$ (= 20, 40, 100, 250) or $\bar{n}$=(= 30, 60, 100, 160) and a value of $K$ (= 5, 10, 30), a panel plots coverage of 95\% confidence intervals versus $\tau^2$ (= 0.0(0.1)1).\\
The confidence intervals for $\tau^2$ are
\begin{itemize}
\item PL (Profile likelihood), inverse-variance weights)
\item QP (Q-profile, inverse-variance weights)
\item KD (based on  Kulinskaya-Dollinger (2015) approximation, inverse-variance weights)
\item FPC (based on Farebrother approximation, effective-sample-size weights, conditional variance of LOR)
\item FPU (based on Farebrother approximation, effective-sample-size weights, unconditional variance of LOR)
\end{itemize}
The plots include two versions of PL, QP, and FPC: adding $1/2$ to all four of $X_{iT},\;X_{iC},\; n_{iT}-X_{iT},\; n_{iC}-X_{iC}$ only when one of these is zero (solid lines) or always (dashed lines).\\
The plots also include two versions of FPU: model-based estimation of $p_{iT}$ (solid lines) or na\"{i}ve estimation (dashed lines).

\clearpage
\setcounter{figure}{0}
\renewcommand{\thefigure}{C.\arabic{figure}}

\begin{figure}[ht]
	\centering
	\includegraphics[scale=0.33]{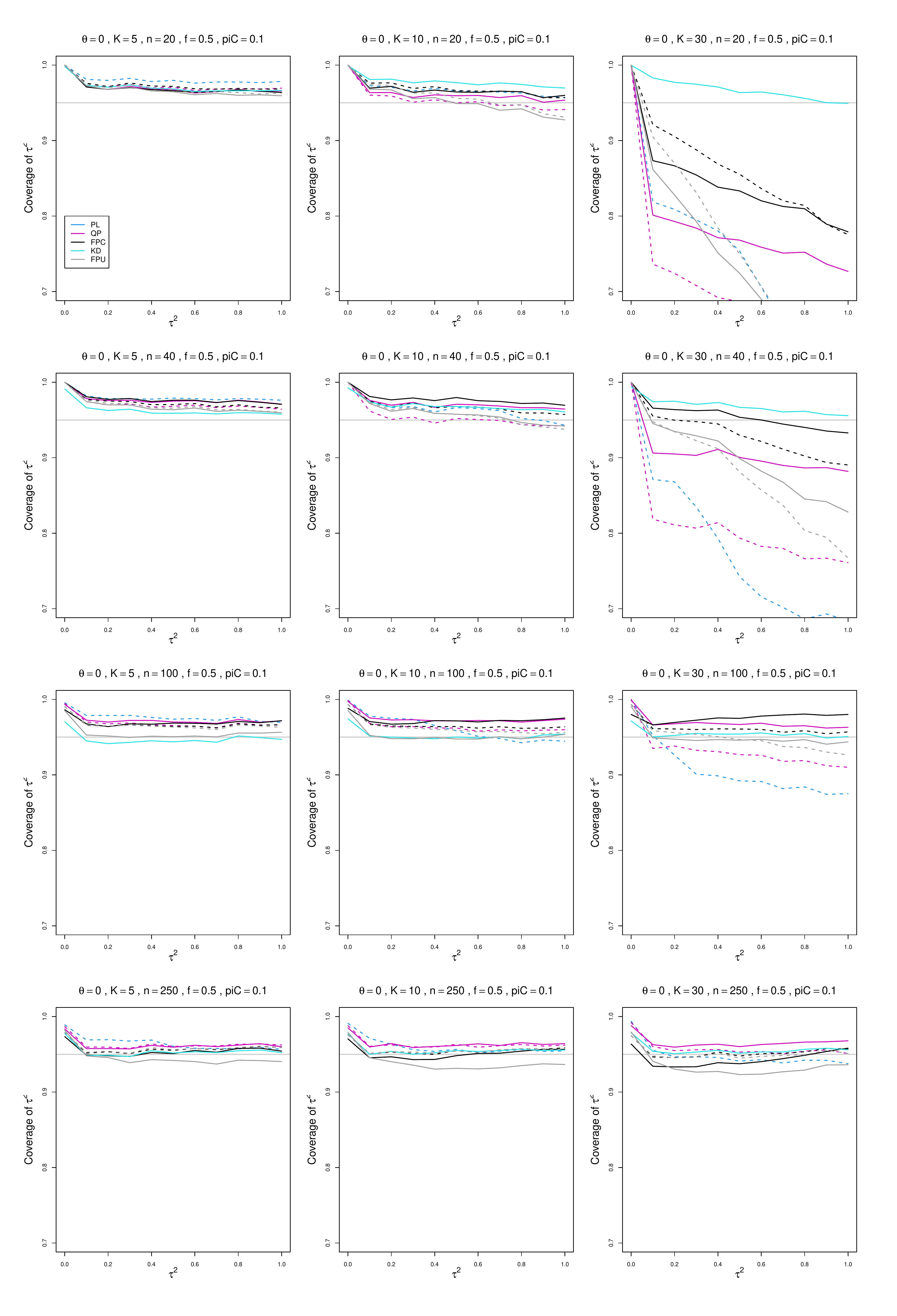}
	\caption{Coverage of PL, QP, KD, FPC, and FPU 95\% confidence intervals for between-study variance of LOR vs $\tau^2$, for equal sample sizes $n = 20,\;40,\;100$ and $250$, $p_{iC} = .1$, $\theta = 0$ and  $f = 0.5$.  Solid lines: PL, QP, and FPC \lq\lq only", FPU model-based, and KD. Dashed lines: PL, QP, and FPC \lq\lq always" and FPU na\"{i}ve.  }
	\label{PlotCovOfTau2_piC_01theta=0_LOR_equal_sample_sizes}
\end{figure}

\begin{figure}[ht]
	\centering
	\includegraphics[scale=0.33]{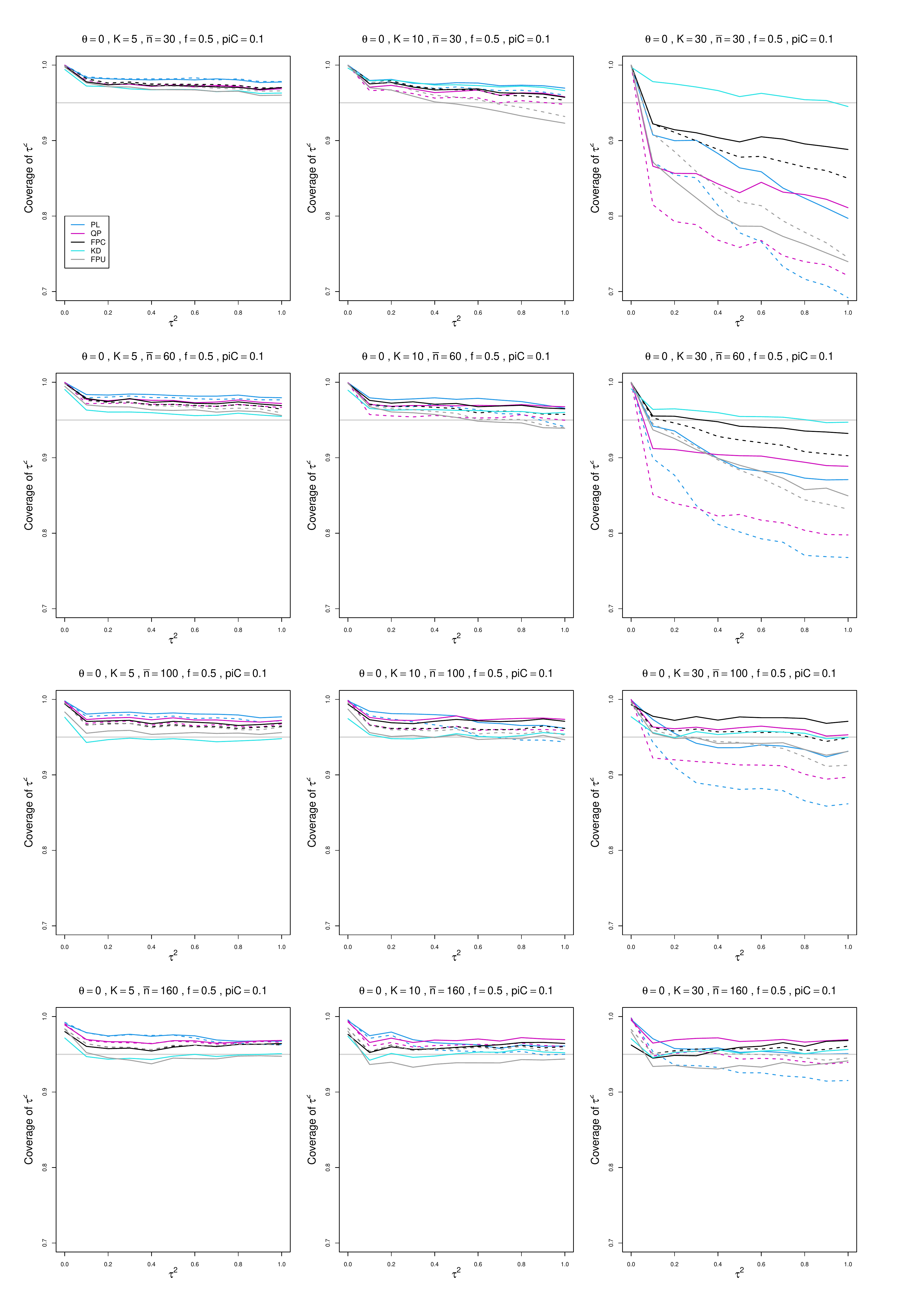}
	\caption{Coverage of PL, QP, KD, FPC, and FPU 95\% confidence intervals for between-study variance of LOR vs $\tau^2$, for unequal sample sizes $\bar{n}=30,\;60,\;100$ and $160$, $p_{iC} = .1$, $\theta = 0$ and  $f = 0.5$.  Solid lines: PL, QP, and FPC \lq\lq only", FPU model-based, and KD. Dashed lines: PL, QP, and FPC \lq\lq always" and FPU na\"{i}ve.    }
	\label{PlotCovOfTau2_piC_01theta=0_LOR_unequal_sample_sizes}
\end{figure}

\begin{figure}[ht]
	\centering
	\includegraphics[scale=0.33]{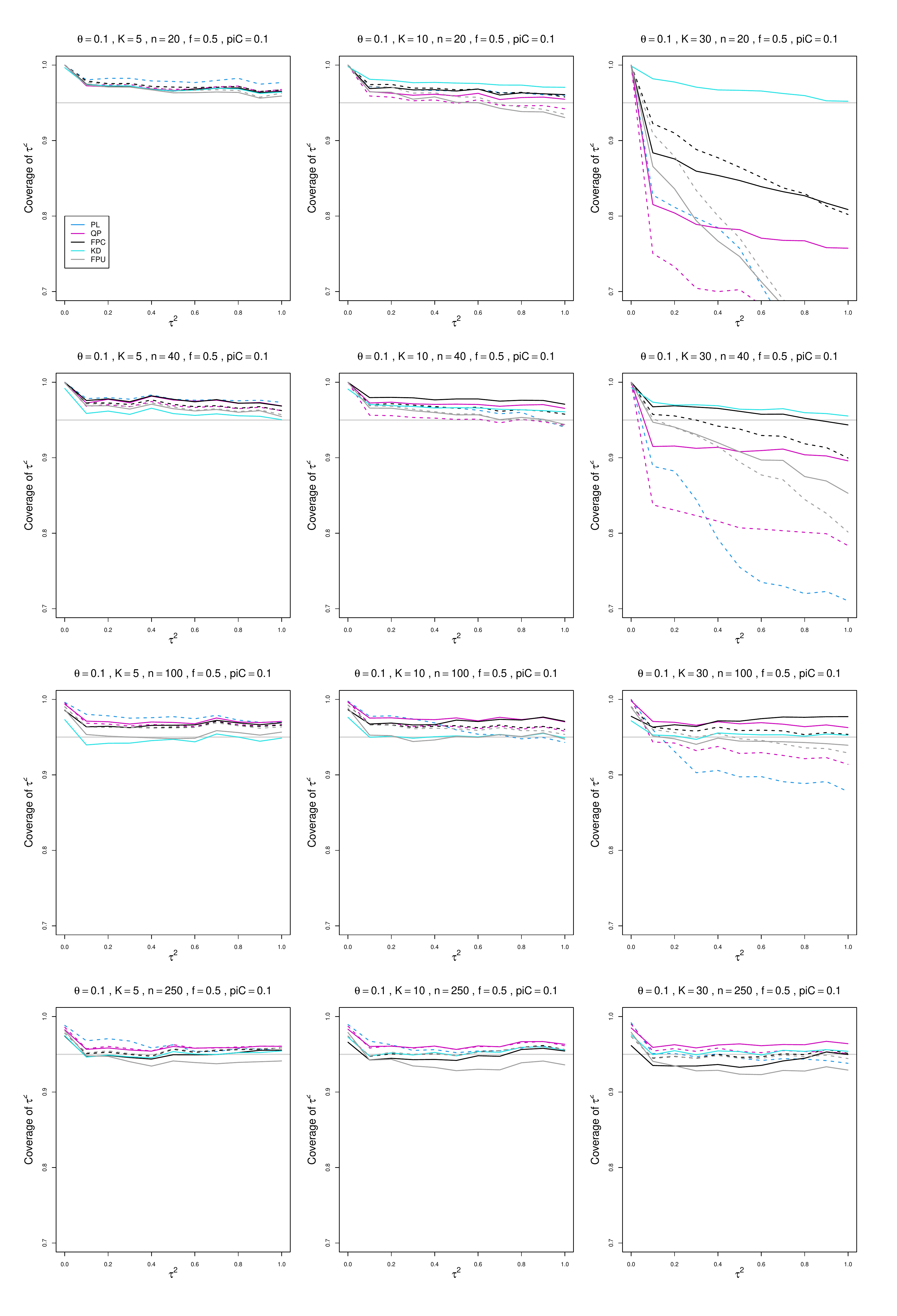}
	\caption{Coverage of PL, QP, KD, FPC, and FPU 95\% confidence intervals for between-study variance of LOR vs $\tau^2$, for equal sample sizes $n = 20,\;40,\;100$ and $250$, $p_{iC} = .1$, $\theta = 0.1$ and  $f = 0.5$.  Solid lines: PL, QP, and FPC \lq\lq only", FPU model-based, and KD. Dashed lines: PL, QP, and FPC \lq\lq always" and FPU na\"{i}ve.   }
	\label{PlotCovOfTau2_piC_01theta=0.1_LOR_equal_sample_sizes}
\end{figure}

\begin{figure}[ht]
	\centering
	\includegraphics[scale=0.33]{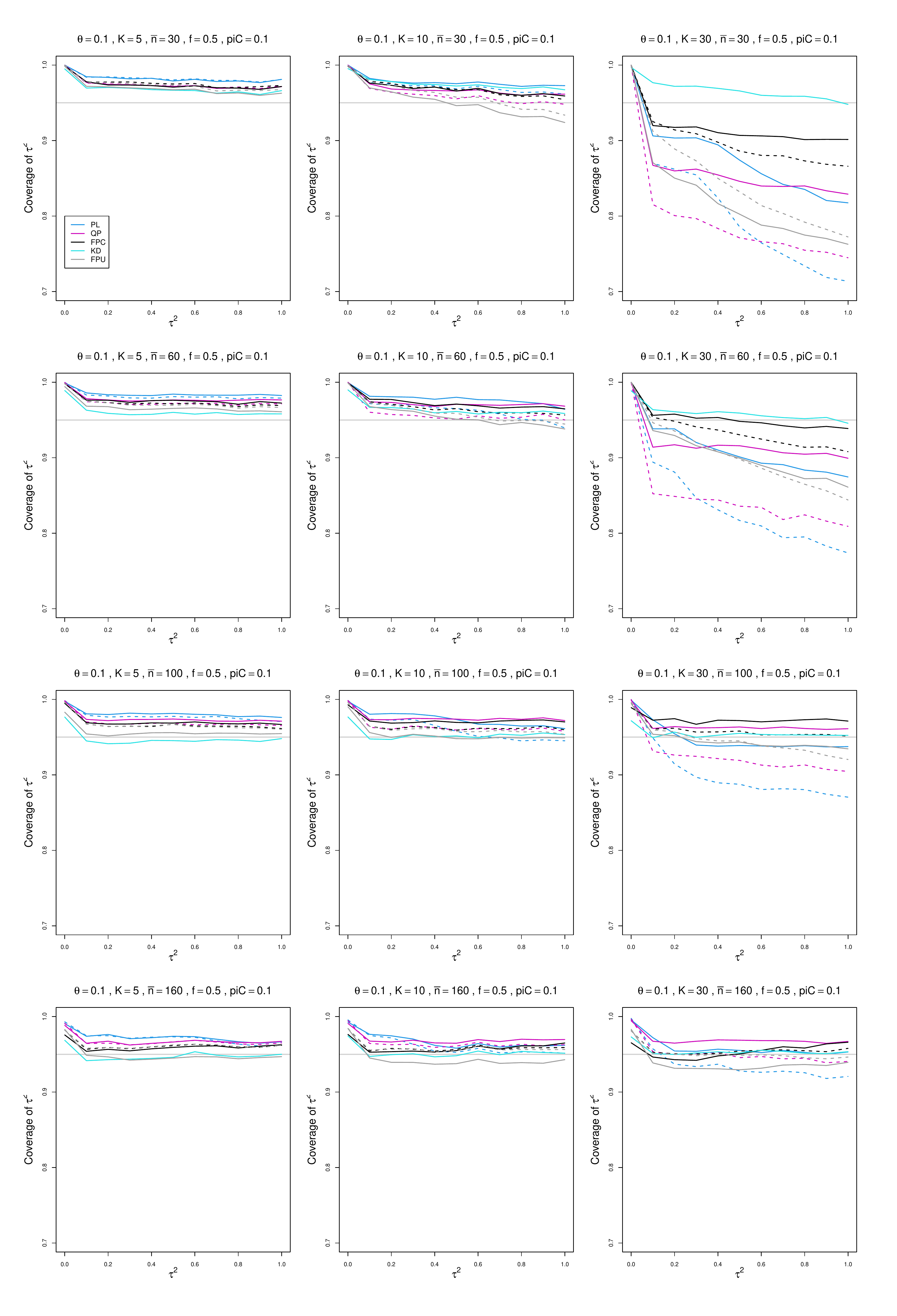}
	\caption{Coverage of PL, QP, KD, FPC, and FPU 95\% confidence intervals for between-study variance of LOR vs $\tau^2$, for unequal sample sizes $\bar{n}=30,\;60,\;100$ and $160$, $p_{iC} = .1$, $\theta = 0.1$ and  $f = 0.5$.  Solid lines: PL, QP, and FPC \lq\lq only", FPU model-based, and KD. Dashed lines: PL, QP, and FPC \lq\lq always" and FPU na\"{i}ve.    }
	\label{PlotCovOfTau2_piC_01theta=0.1_LOR_unequal_sample_sizes}
\end{figure}

\begin{figure}[ht]
	\centering
	\includegraphics[scale=0.33]{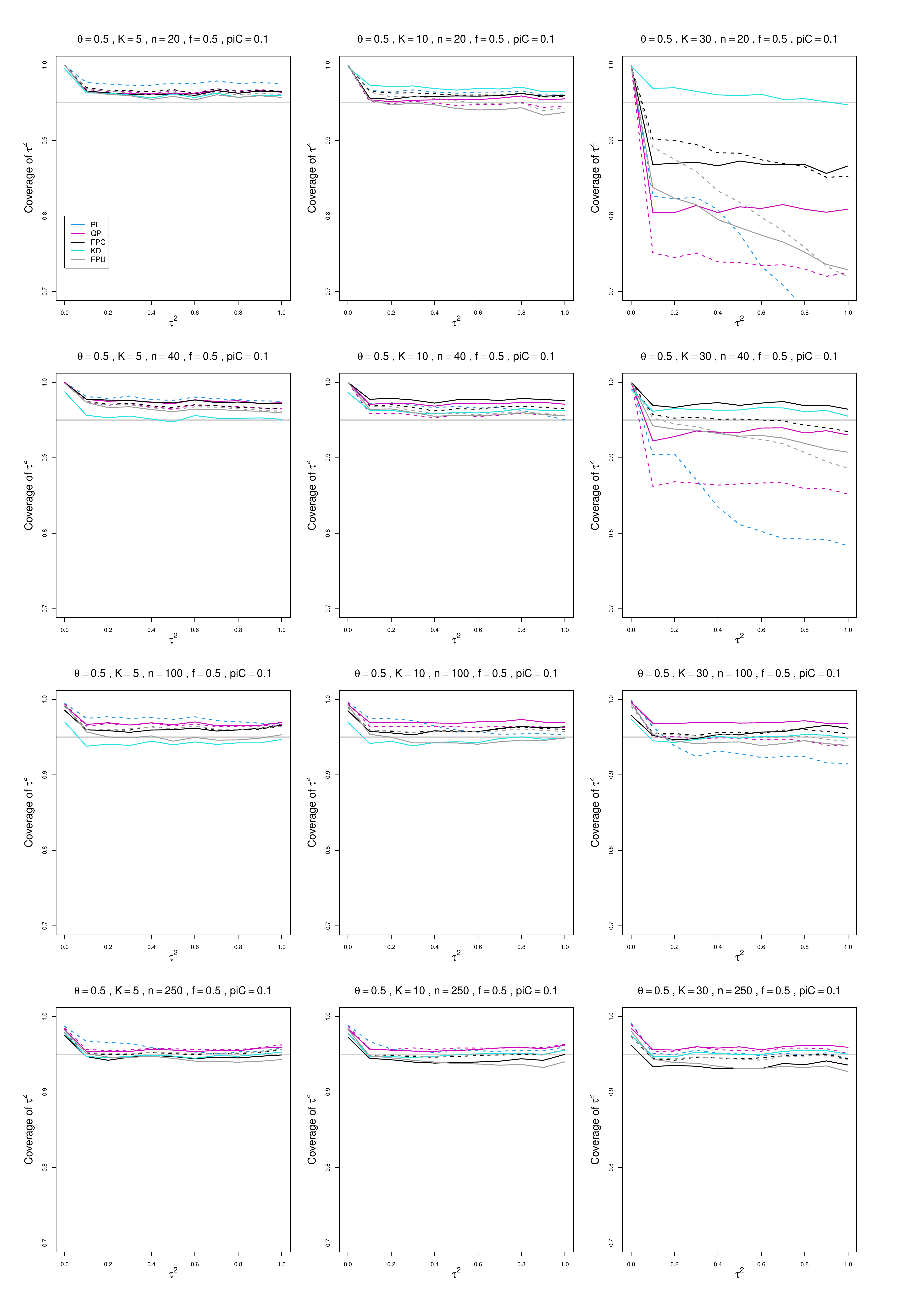}
	\caption{Coverage of PL, QP, KD, FPC, and FPU 95\% confidence intervals for between-study variance of LOR vs $\tau^2$, for equal sample sizes $n = 20,\;40,\;100$ and $250$, $p_{iC} = .1$, $\theta = 0.5$ and  $f = 0.5$.  Solid lines: PL, QP, and FPC \lq\lq only", FPU model-based, and KD. Dashed lines: PL, QP, and FPC \lq\lq always" and FPU na\"{i}ve.   }
	\label{PlotCovOfTau2_piC_01theta=0.5_LOR_equal_sample_sizes}
\end{figure}

\begin{figure}[ht]
	\centering
	\includegraphics[scale=0.33]{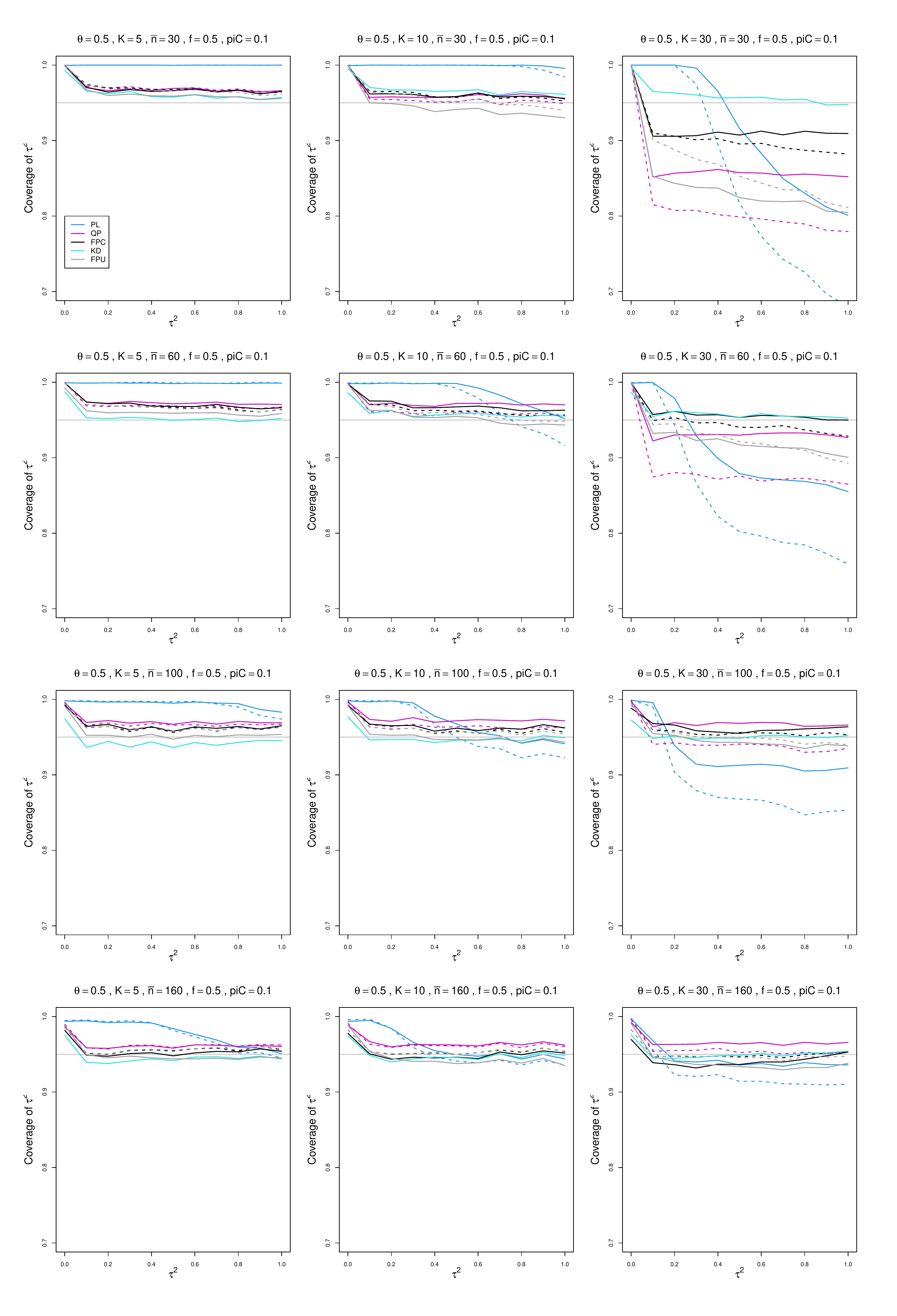}
	\caption{Coverage of PL, QP, KD, FPC, and FPU 95\% confidence intervals for between-study variance of LOR vs $\tau^2$, for unequal sample sizes $\bar{n}=30,\;60,\;100$ and $160$, $p_{iC} = .1$, $\theta = 0.5$ and  $f = 0.5$.  Solid lines: PL, QP, and FPC \lq\lq only", FPU model-based, and KD. Dashed lines: PL, QP, and FPC \lq\lq always" and FPU na\"{i}ve.   }
	\label{PlotCovOfTau2_piC_01theta=0.5_LOR_unequal_sample_sizes}
\end{figure}

\begin{figure}[ht]
	\centering
	\includegraphics[scale=0.33]{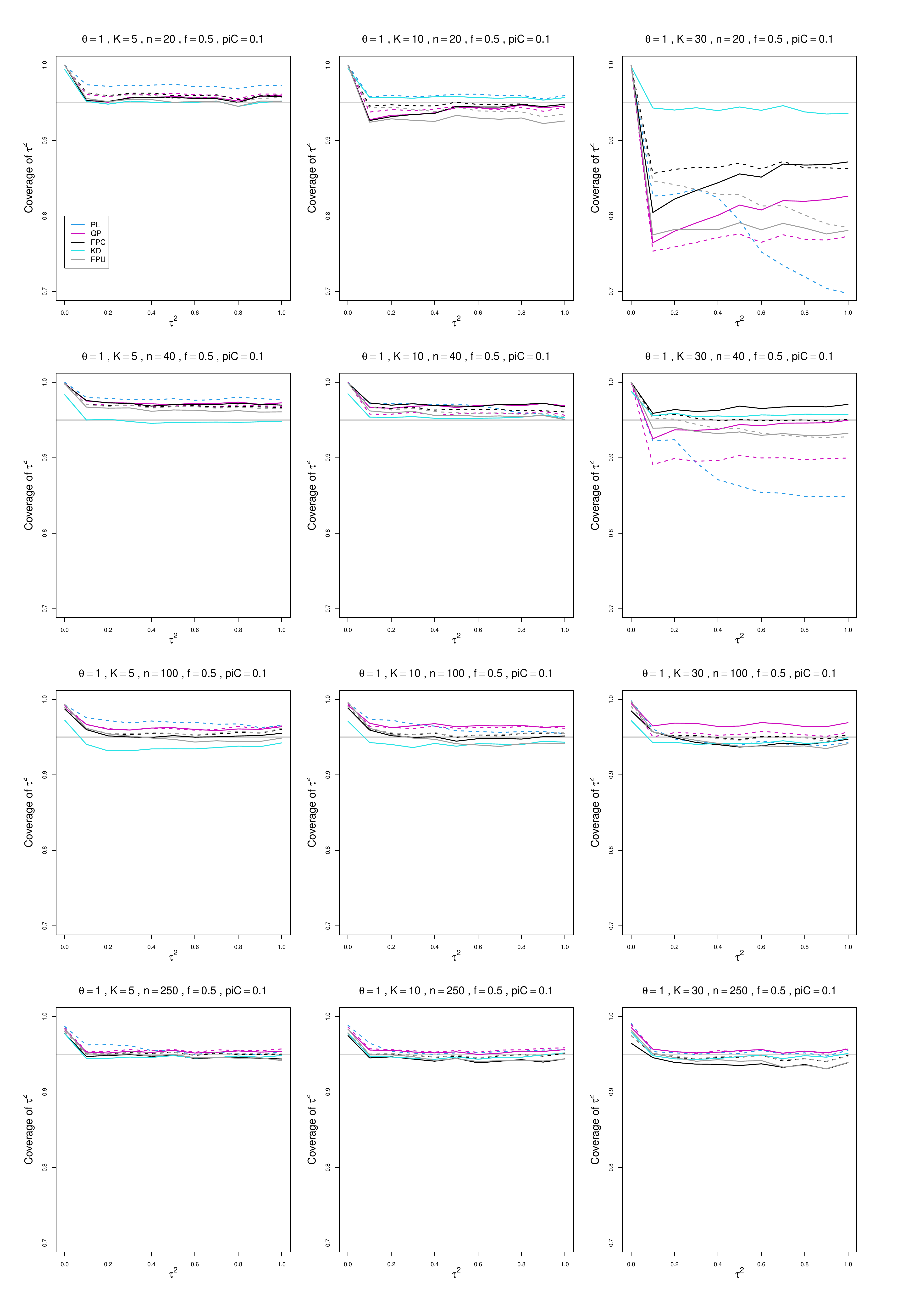}
	\caption{Coverage of PL, QP, KD, FPC, and FPU 95\% confidence intervals for between-study variance of LOR vs $\tau^2$, for equal sample sizes $n = 20,\;40,\;100$ and $250$, $p_{iC} = .1$, $\theta = 1$ and  $f = 0.5$.  Solid lines: PL, QP, and FPC \lq\lq only", FPU model-based, and KD. Dashed lines: PL, QP, and FPC \lq\lq always" and FPU na\"{i}ve.   }
	\label{PlotCovOfTau2_piC_01theta=1_LOR_equal_sample_sizes}
\end{figure}

\begin{figure}[ht]
	\centering
	\includegraphics[scale=0.33]{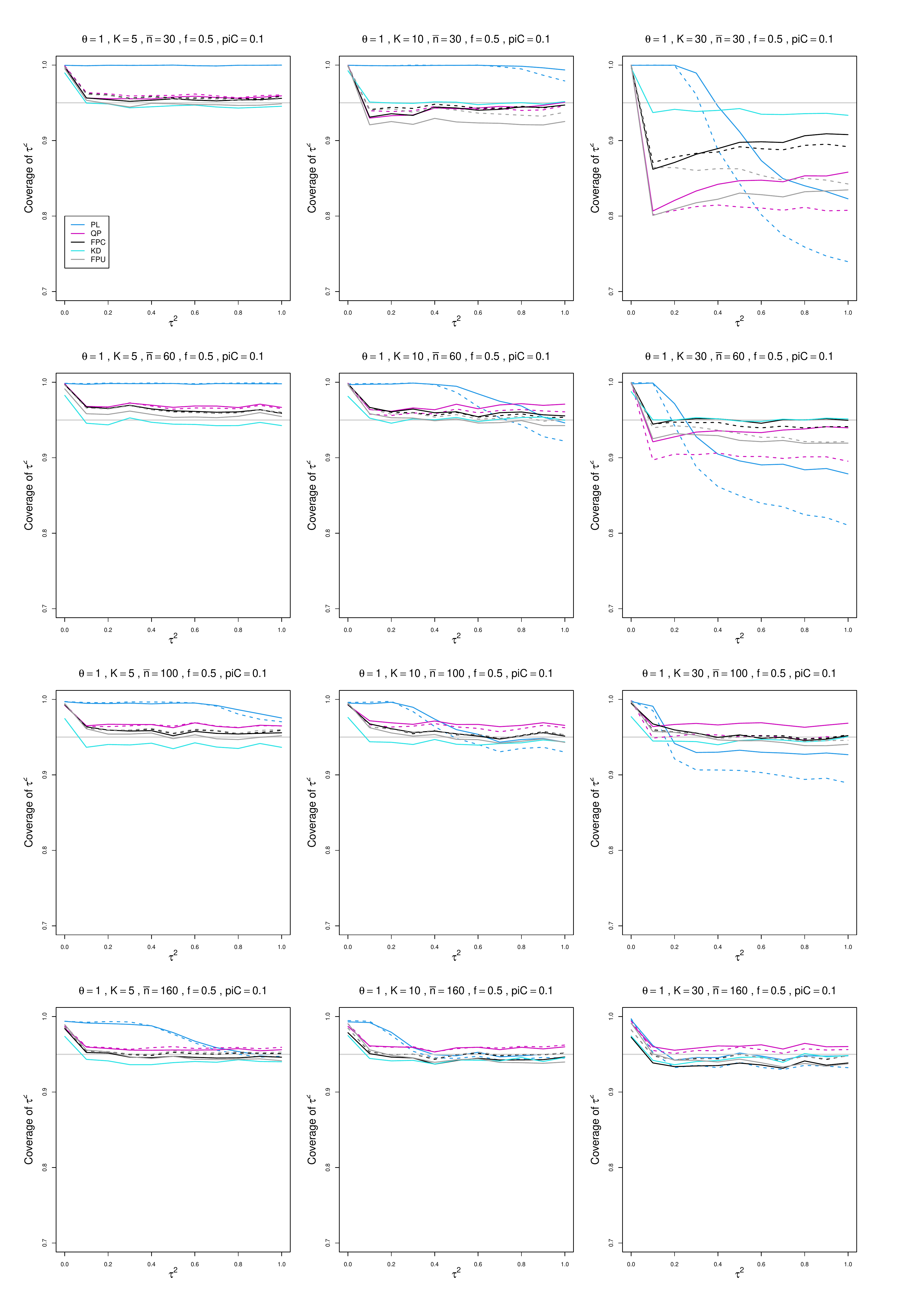}
	\caption{Coverage of PL, QP, KD, FPC, and FPU 95\% confidence intervals for between-study variance of LOR vs $\tau^2$, for unequal sample sizes $\bar{n}=30,\;60,\;100$ and $160$, $p_{iC} = .1$, $\theta = 1$ and  $f = 0.5$.  Solid lines: PL, QP, and FPC \lq\lq only", FPU model-based, and KD. Dashed lines: PL, QP, and FPC \lq\lq always" and FPU na\"{i}ve.   }
	\label{PlotCovOfTau2_piC_01theta=1_LOR_unequal_sample_sizes}
\end{figure}

\begin{figure}[ht]
	\centering
	\includegraphics[scale=0.33]{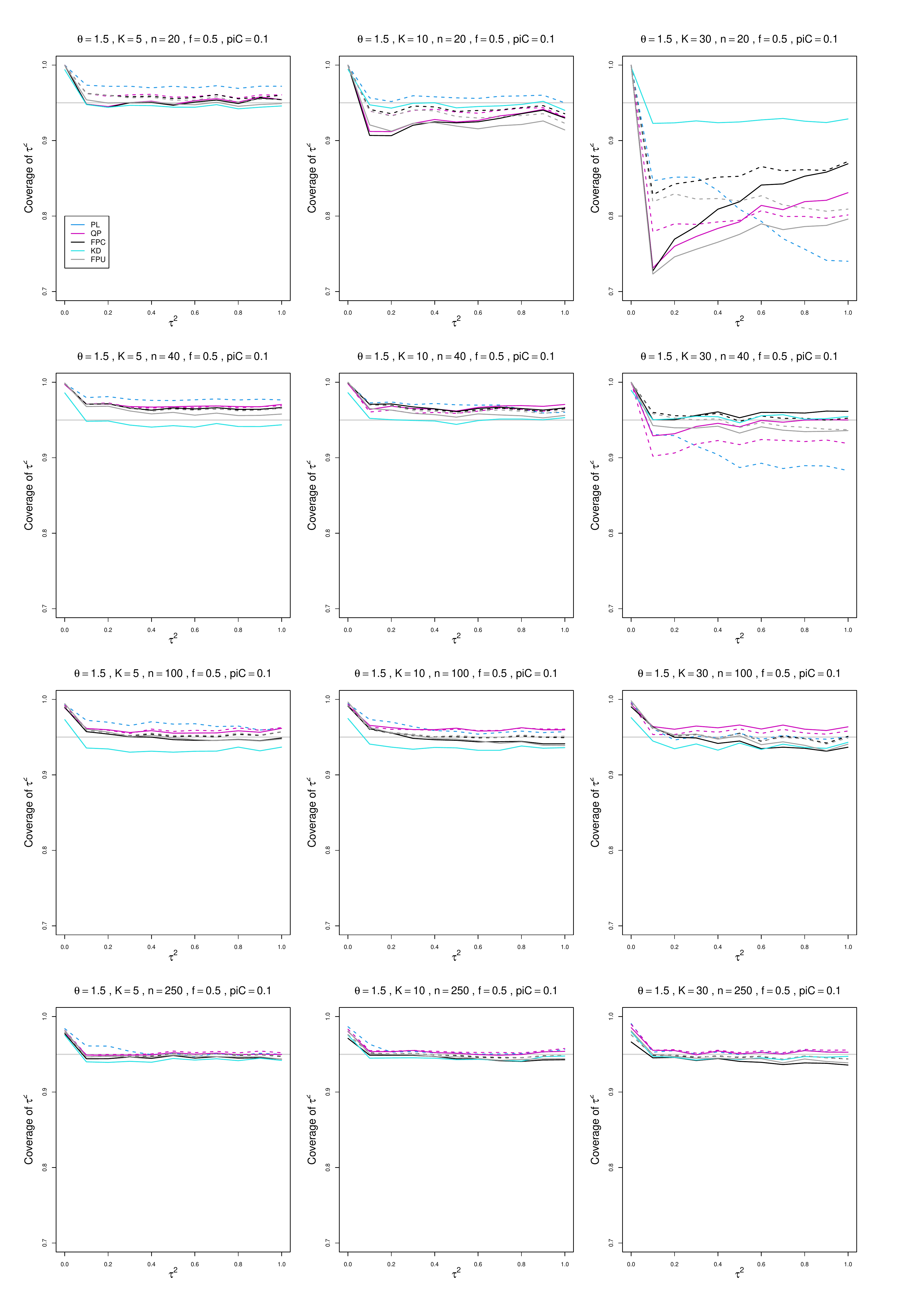}
	\caption{Coverage of PL, QP, KD, FPC, and FPU 95\% confidence intervals for between-study variance of LOR vs $\tau^2$, for equal sample sizes $n = 20,\;40,\;100$ and $250$, $p_{iC} = .1$, $\theta = 1.5$ and  $f = 0.5$.  Solid lines: PL, QP, and FPC \lq\lq only", FPU model-based, and KD. Dashed lines: PL, QP, and FPC \lq\lq always" and FPU na\"{i}ve.  }
	\label{PlotCovOfTau2_piC_01theta=1.5_LOR_equal_sample_sizes}
\end{figure}

\begin{figure}[ht]
	\centering
	\includegraphics[scale=0.33]{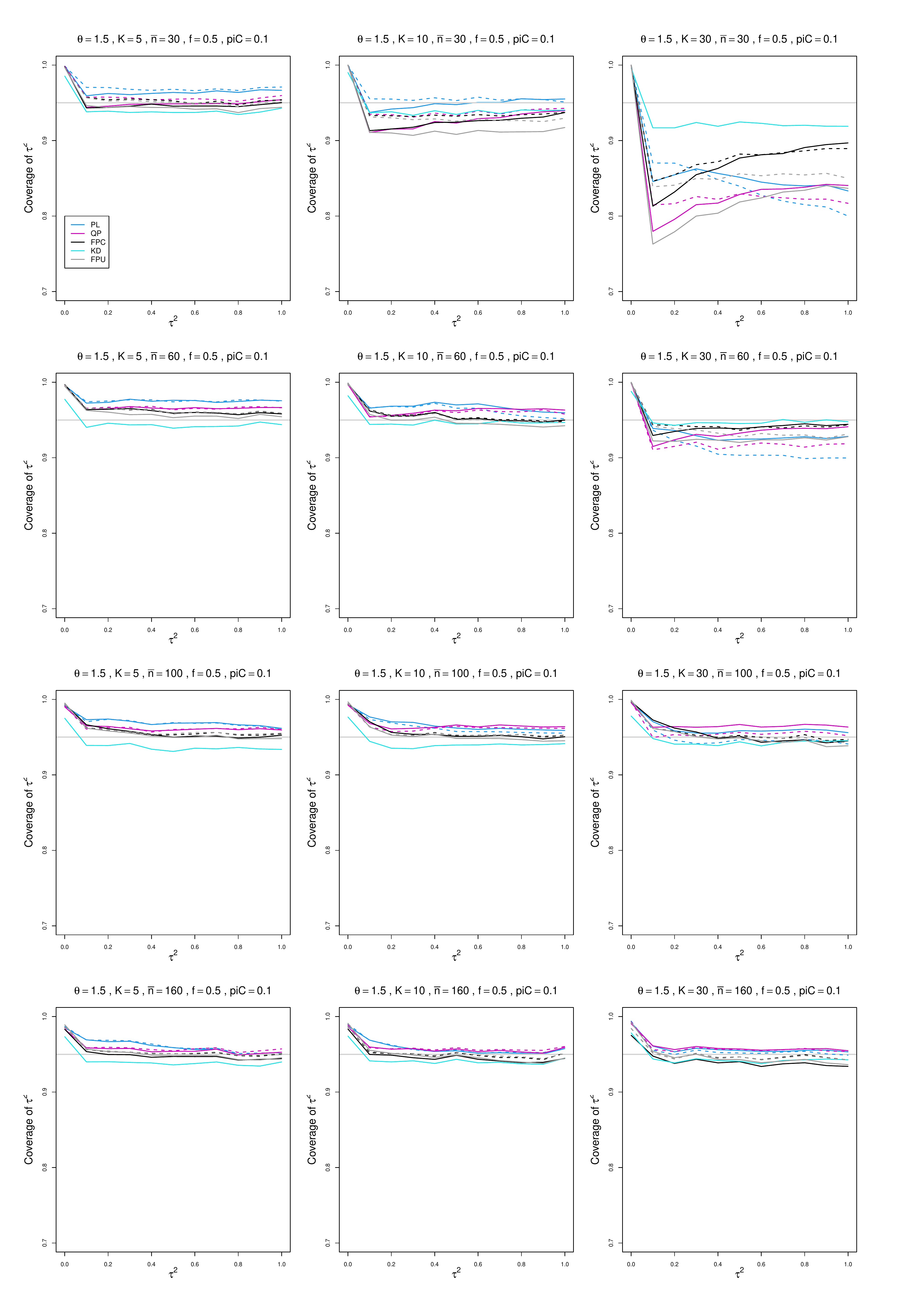}
	\caption{Coverage of PL, QP, KD, FPC, and FPU 95\% confidence intervals for between-study variance of LOR vs $\tau^2$, for unequal sample sizes $\bar{n}=30,\;60,\;100$ and $160$, $p_{iC} = .1$, $\theta = 1.5$ and  $f = 0.5$.  Solid lines: PL, QP, and FPC \lq\lq only", FPU model-based, and KD. Dashed lines: PL, QP, and FPC \lq\lq always" and FPU na\"{i}ve.    }
	\label{PlotCovOfTau2_piC_01theta=1.5_LOR_unequal_sample_sizes}
\end{figure}

\begin{figure}[ht]
	\centering
	\includegraphics[scale=0.33]{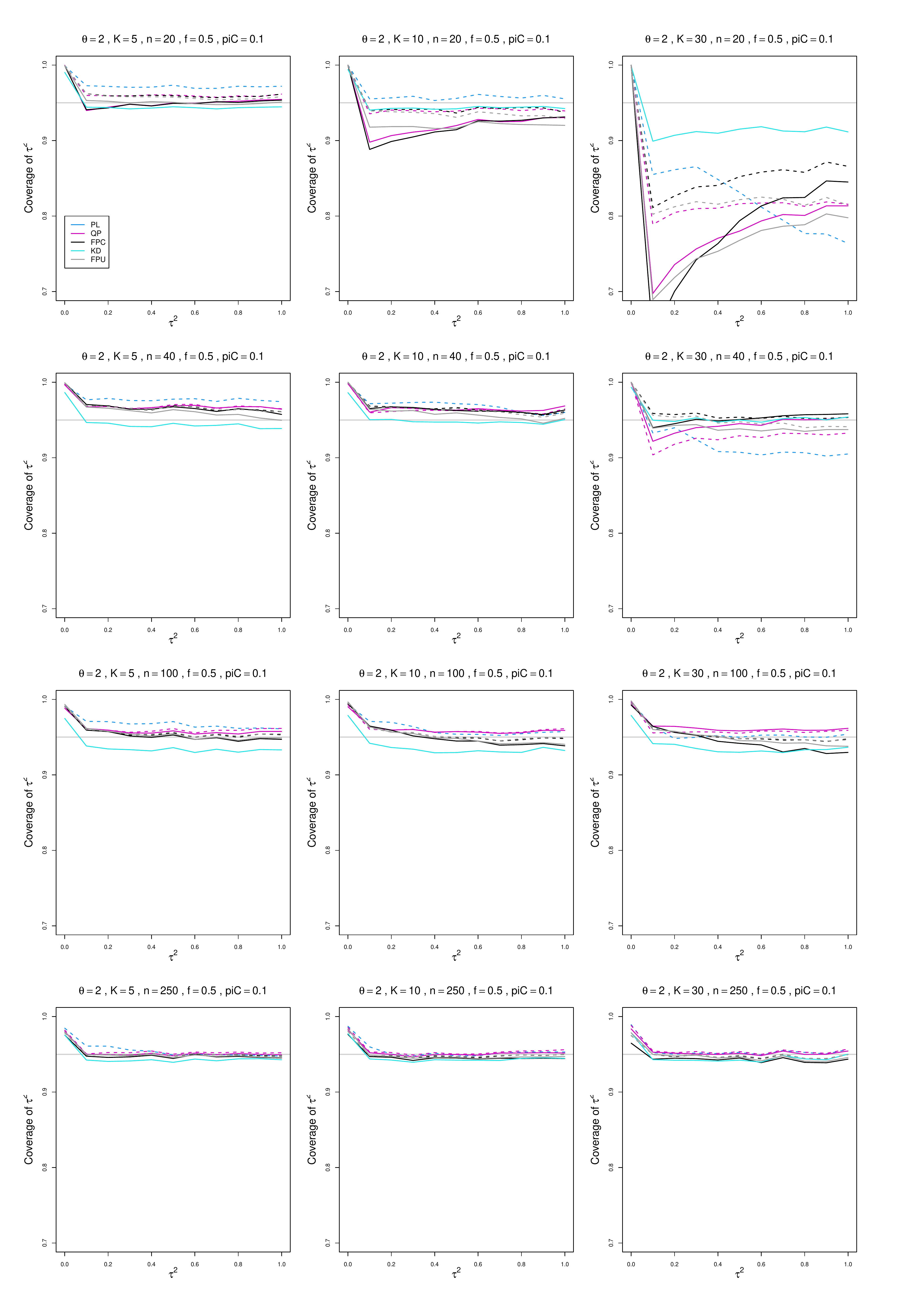}
	\caption{Coverage of PL, QP, KD, FPC, and FPU 95\% confidence intervals for between-study variance of LOR vs $\tau^2$, for equal sample sizes $n = 20,\;40,\;100$ and $250$, $p_{iC} = .1$, $\theta = 2$ and  $f = 0.5$.  Solid lines: PL, QP, and FPC \lq\lq only", FPU model-based, and KD. Dashed lines: PL, QP, and FPC \lq\lq always" and FPU na\"{i}ve.   }
	\label{PlotCovOfTau2_piC_01theta=2_LOR_equal_sample_sizes}
\end{figure}

\begin{figure}[ht]
	\centering
	\includegraphics[scale=0.33]{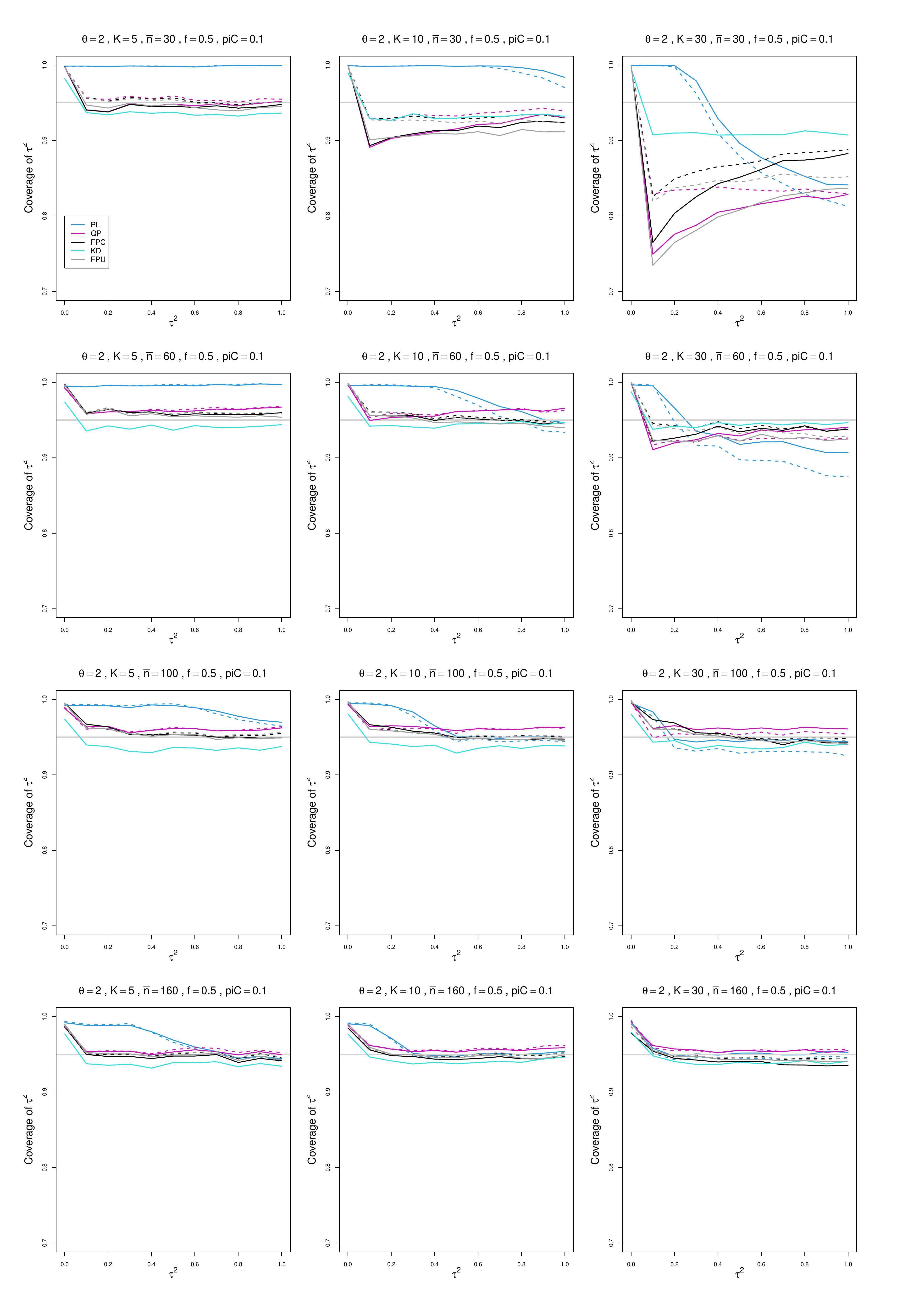}
	\caption{Coverage of PL, QP, KD, FPC, and FPU 95\% confidence intervals for between-study variance of LOR vs $\tau^2$, for unequal sample sizes $\bar{n}=30,\;60,\;100$ and $160$, $p_{iC} = .1$, $\theta = 2$ and  $f = 0.5$.  Solid lines: PL, QP, and FPC \lq\lq only", FPU model-based, and KD. Dashed lines: PL, QP, and FPC \lq\lq always" and FPU na\"{i}ve.   }
	\label{PlotCovOfTau2_piC_01theta=2_LOR_unequal_sample_sizes}
\end{figure}

\begin{figure}[ht]
	\centering
	\includegraphics[scale=0.33]{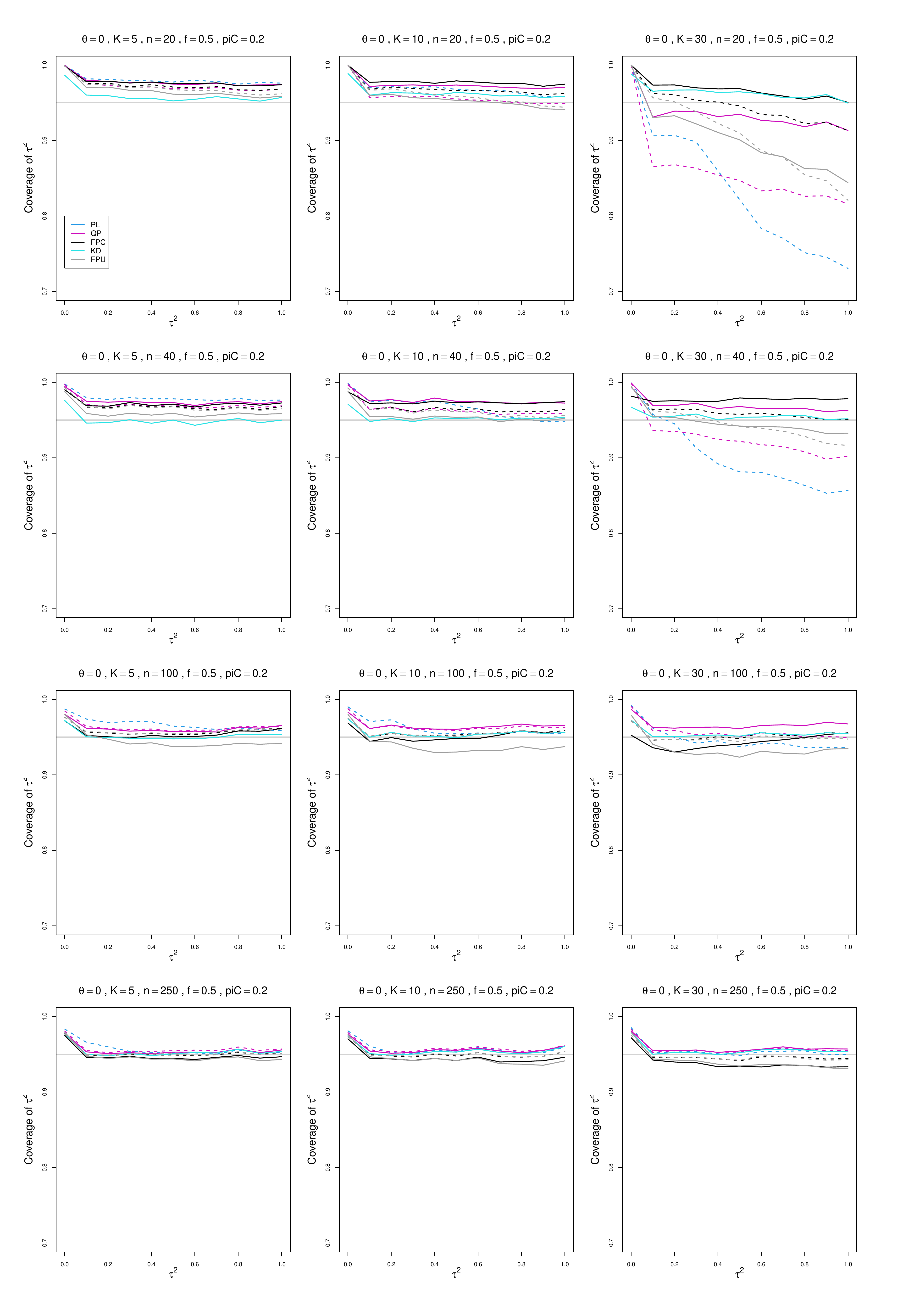}
	\caption{Coverage of PL, QP, KD, FPC, and FPU 95\% confidence intervals for between-study variance of LOR vs $\tau^2$, for equal sample sizes $n = 20,\;40,\;100$ and $250$, $p_{iC} = .2$, $\theta = 0$ and  $f = 0.5$.  Solid lines: PL, QP, and FPC \lq\lq only", FPU model-based, and KD. Dashed lines: PL, QP, and FPC \lq\lq always" and FPU na\"{i}ve.  }
	\label{PlotCovOfTau2_piC_02theta=0_LOR_equal_sample_sizes}
\end{figure}

\begin{figure}[ht]
	\centering
	\includegraphics[scale=0.33]{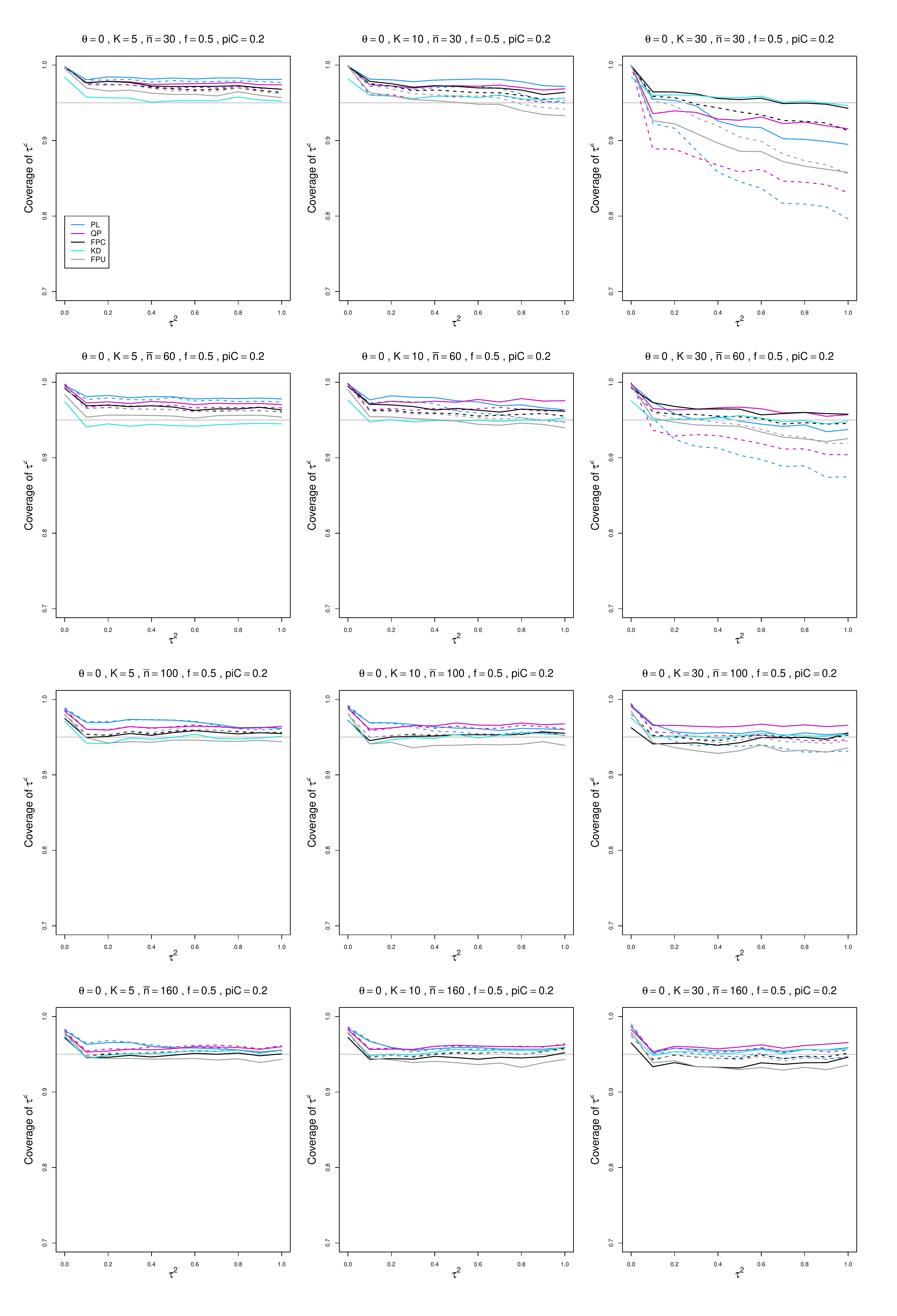}
	\caption{Coverage of PL, QP, KD, FPC, and FPU 95\% confidence intervals for between-study variance of LOR vs $\tau^2$, for unequal sample sizes $\bar{n}=30,\;60,\;100$ and $160$, $p_{iC} = .2$, $\theta = 0$ and  $f = 0.5$.  Solid lines: PL, QP, and FPC \lq\lq only", FPU model-based, and KD. Dashed lines: PL, QP, and FPC \lq\lq always" and FPU na\"{i}ve.   }
	\label{PlotCovOfTau2_piC_02theta=0_LOR_unequal_sample_sizes}
\end{figure}

\begin{figure}[ht]
	\centering
	\includegraphics[scale=0.33]{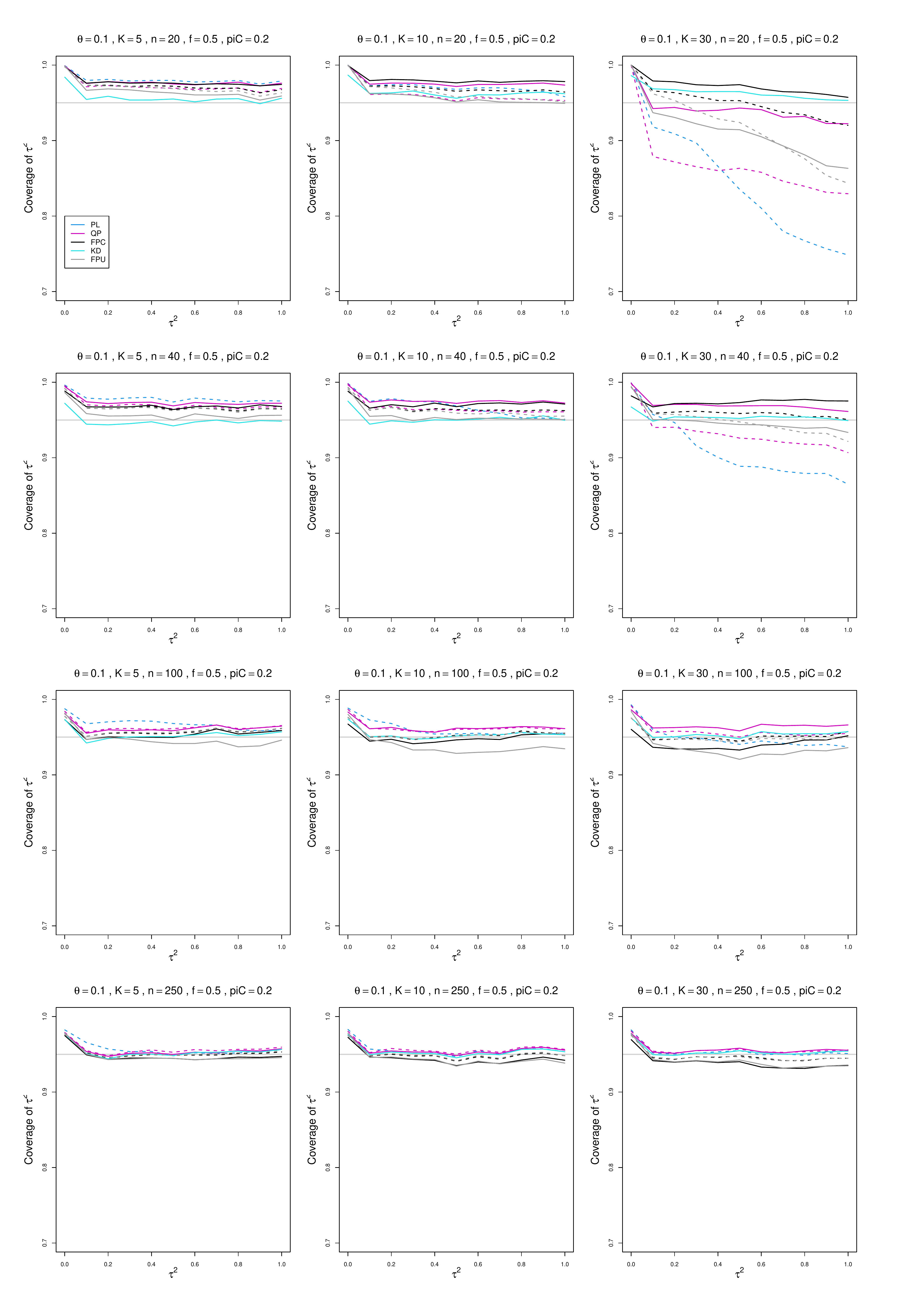}
	\caption{Coverage of PL, QP, KD, FPC, and FPU 95\% confidence intervals for between-study variance of LOR vs $\tau^2$, for equal sample sizes $n = 20,\;40,\;100$ and $250$, $p_{iC} = .2$, $\theta = 0.1$ and  $f = 0.5$.  Solid lines: PL, QP, and FPC \lq\lq only", FPU model-based, and KD. Dashed lines: PL, QP, and FPC \lq\lq always" and FPU na\"{i}ve.  }
	\label{PlotCovOfTau2_piC_02theta=0.1_LOR_equal_sample_sizes}
\end{figure}

\begin{figure}[ht]
	\centering
	\includegraphics[scale=0.33]{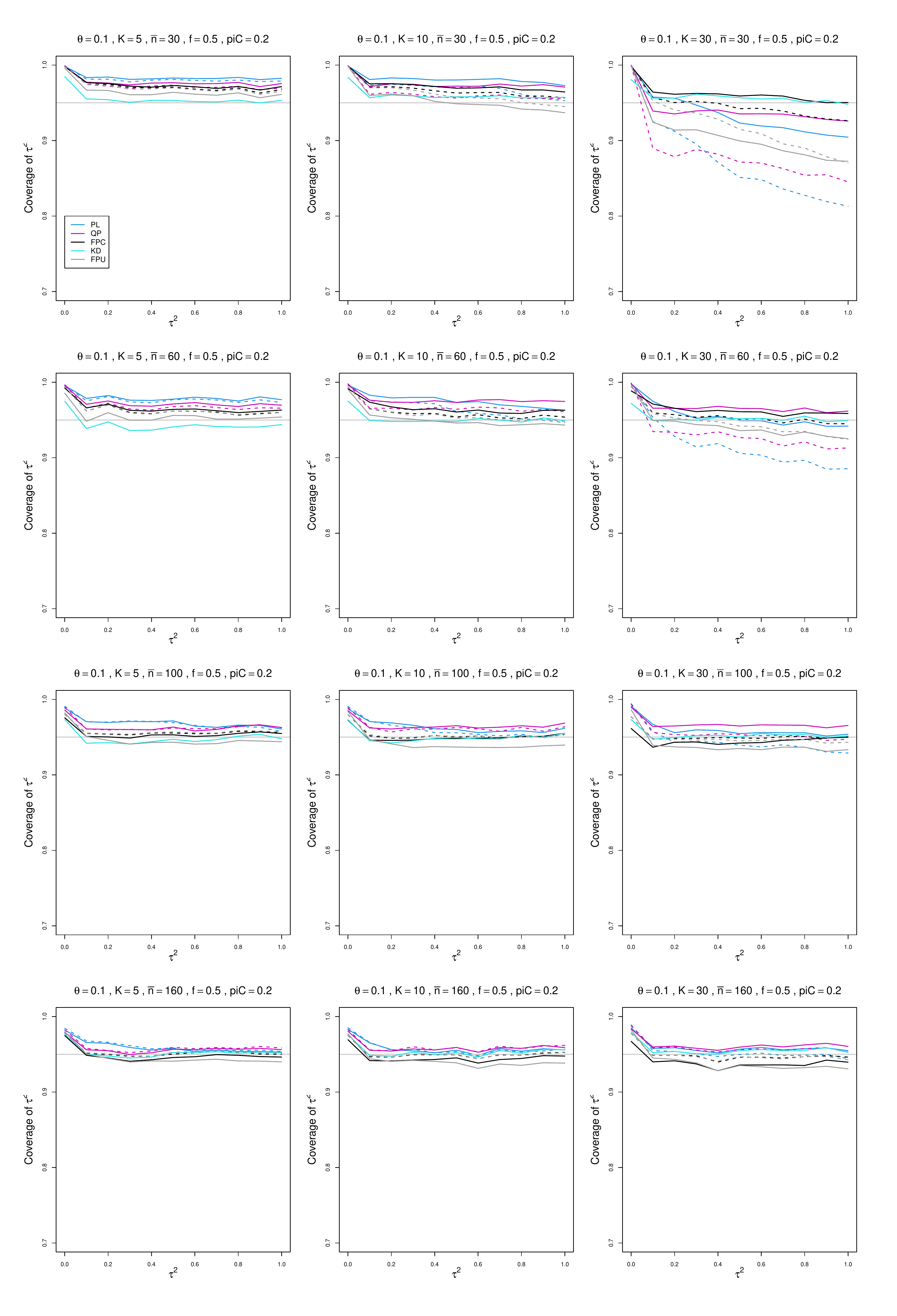}
	\caption{Coverage of PL, QP, KD, FPC, and FPU 95\% confidence intervals for between-study variance of LOR vs $\tau^2$, for unequal sample sizes $\bar{n}=30,\;60,\;100$ and $160$, $p_{iC} = .2$, $\theta = 0.1$ and  $f = 0.5$.  Solid lines: PL, QP, and FPC \lq\lq only", FPU model-based, and KD. Dashed lines: PL, QP, and FPC \lq\lq always" and FPU na\"{i}ve.   }
	\label{PlotCovOfTau2_piC_02theta=0.1_LOR_unequal_sample_sizes}
\end{figure}

\begin{figure}[ht]
	\centering
	\includegraphics[scale=0.33]{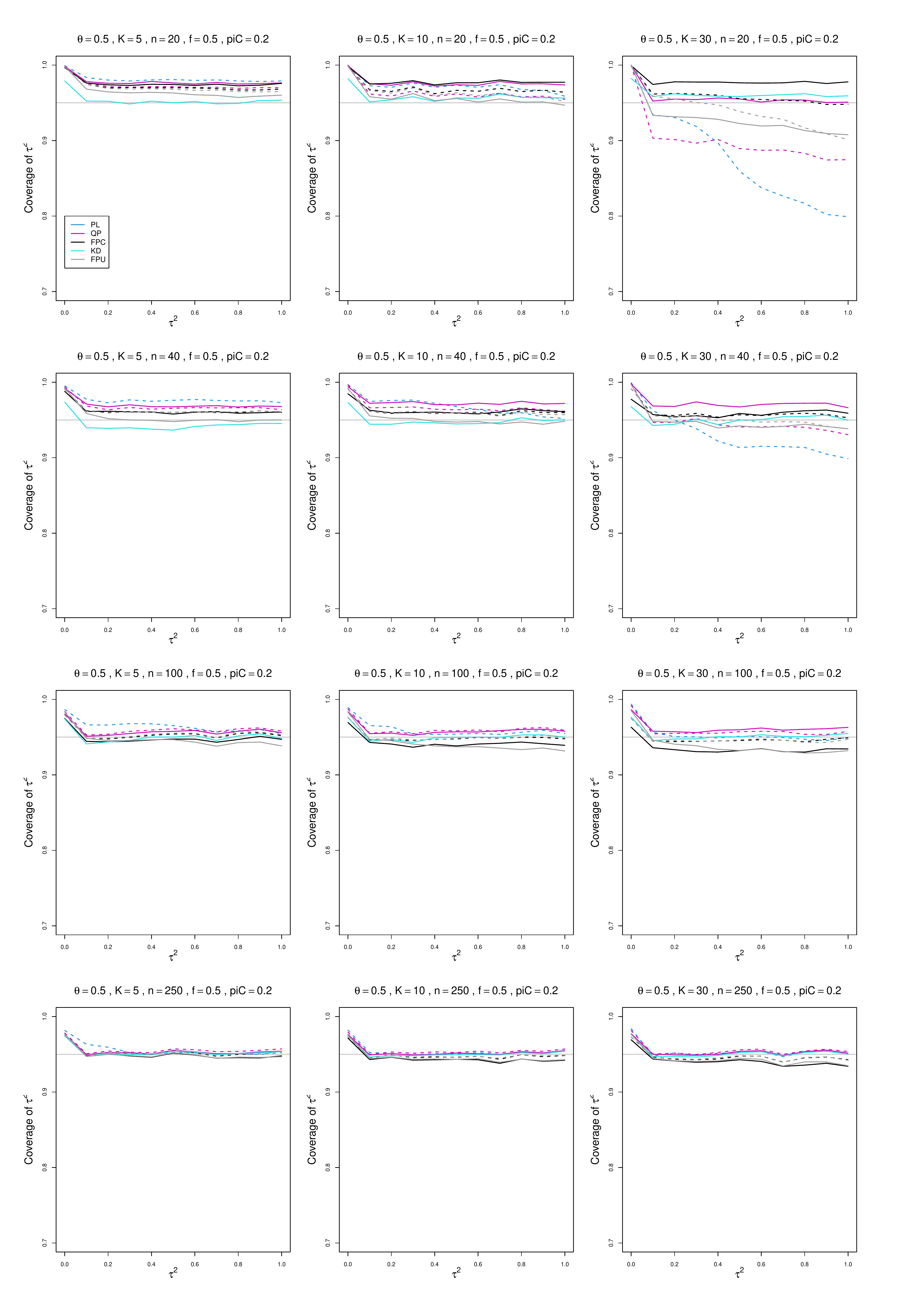}
	\caption{Coverage of PL, QP, KD, FPC, and FPU 95\% confidence intervals for between-study variance of LOR vs $\tau^2$, for equal sample sizes $n = 20,\;40,\;100$ and $250$, $p_{iC} = .2$, $\theta = 0.5$ and  $f = 0.5$.  Solid lines: PL, QP, and FPC \lq\lq only", FPU model-based, and KD. Dashed lines: PL, QP, and FPC \lq\lq always" and FPU na\"{i}ve.  }
	\label{PlotCovOfTau2_piC_02theta=0.5_LOR_equal_sample_sizes}
\end{figure}

\begin{figure}[ht]
	\centering
	\includegraphics[scale=0.33]{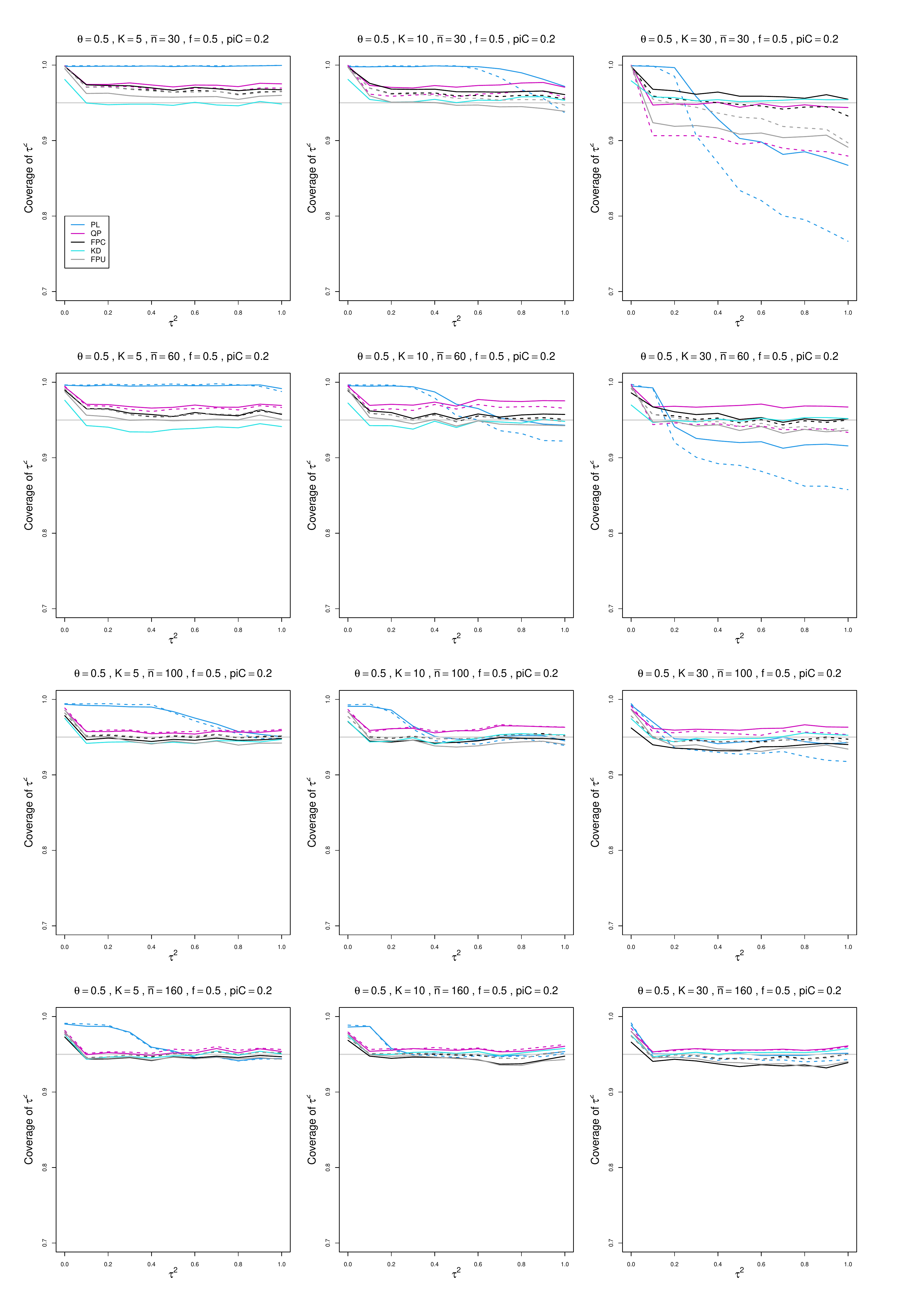}
	\caption{Coverage of PL, QP, KD, FPC, and FPU 95\% confidence intervals for between-study variance of LOR vs $\tau^2$, for unequal sample sizes $\bar{n}=30,\;60,\;100$ and $160$, $p_{iC} = .2$, $\theta = 0.5$ and  $f = 0.5$.  Solid lines: PL, QP, and FPC \lq\lq only", FPU model-based, and KD. Dashed lines: PL, QP, and FPC \lq\lq always" and FPU na\"{i}ve.  }
	\label{PlotCovOfTau2_piC_02theta=0.5_LOR_unequal_sample_sizes}
\end{figure}

\begin{figure}[ht]
	\centering
	\includegraphics[scale=0.33]{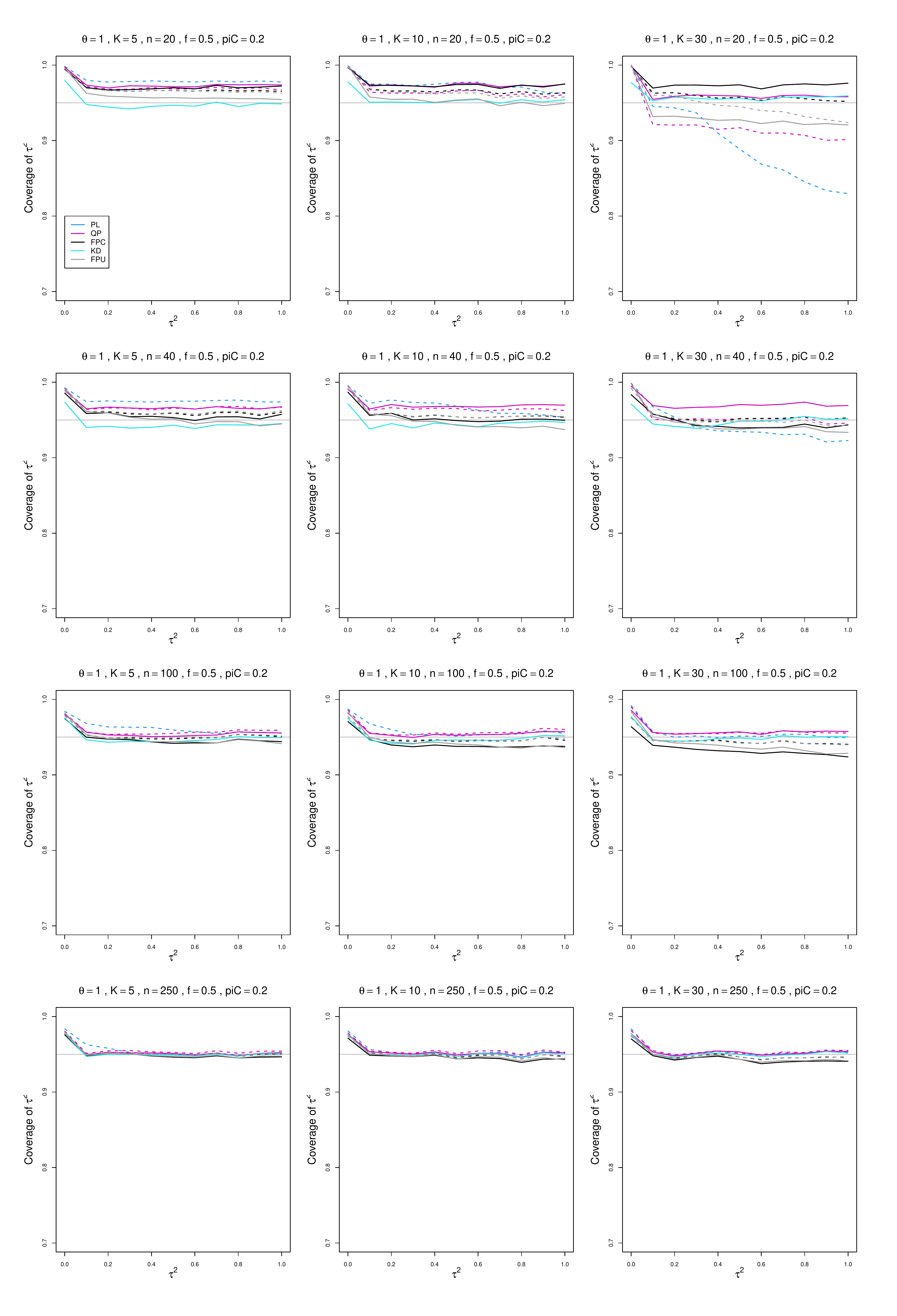}
	\caption{Coverage of PL, QP, KD, FPC, and FPU 95\% confidence intervals for between-study variance of LOR vs $\tau^2$, for equal sample sizes $n = 20,\;40,\;100$ and $250$, $p_{iC} = .2$, $\theta = 1$ and  $f = 0.5$.  Solid lines: PL, QP, and FPC \lq\lq only", FPU model-based, and KD. Dashed lines: PL, QP, and FPC \lq\lq always" and FPU na\"{i}ve.  }
	\label{PlotCovOfTau2_piC_02theta=1_LOR_equal_sample_sizes}
\end{figure}

\begin{figure}[ht]
	\centering
	\includegraphics[scale=0.33]{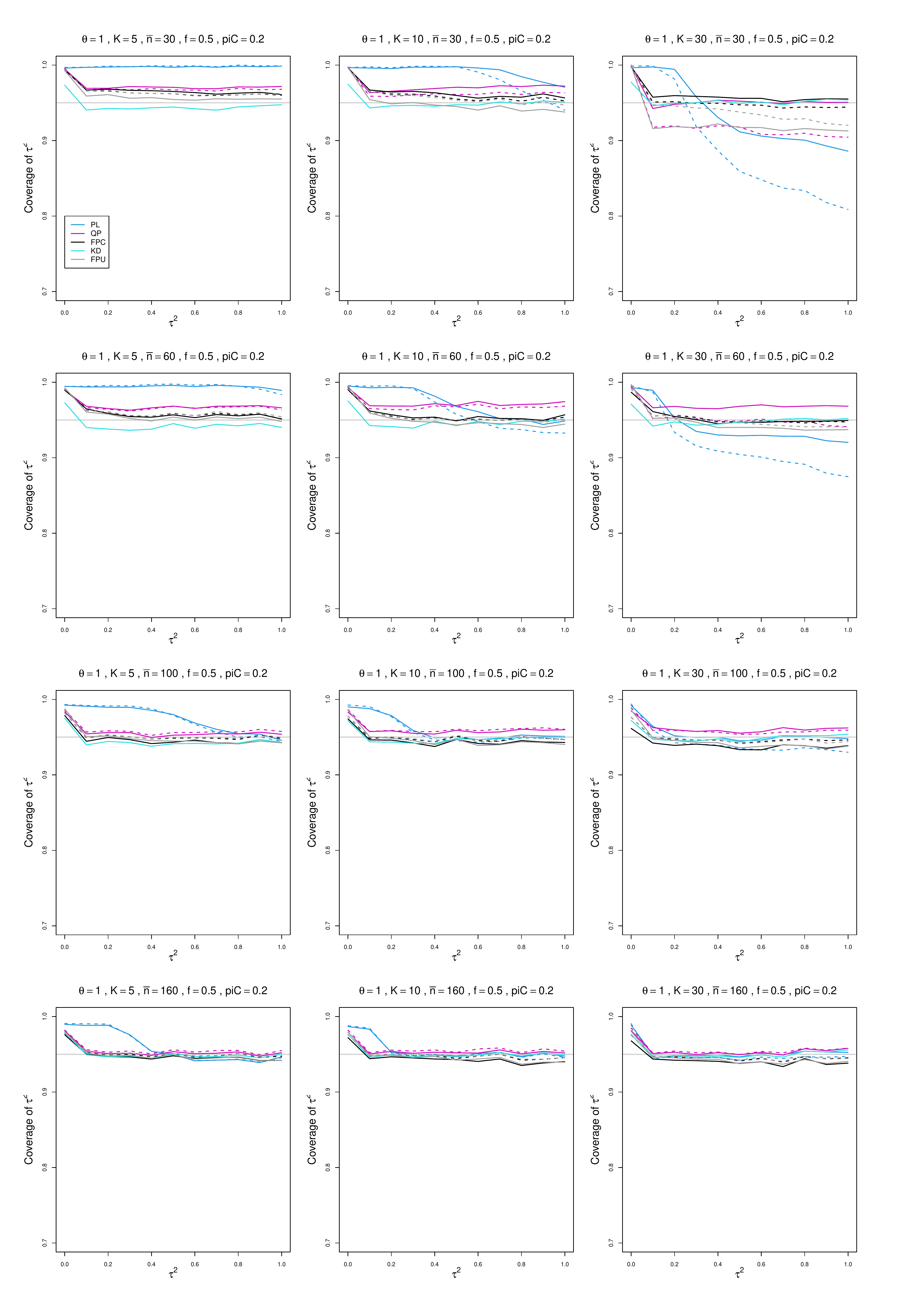}
	\caption{Coverage of PL, QP, KD, FPC, and FPU 95\% confidence intervals for between-study variance of LOR vs $\tau^2$, for unequal sample sizes $\bar{n}=30,\;60,\;100$ and $160$, $p_{iC} = .2$, $\theta = 1$ and  $f = 0.5$.  Solid lines: PL, QP, and FPC \lq\lq only", FPU model-based, and KD. Dashed lines: PL, QP, and FPC \lq\lq always" and FPU na\"{i}ve.  }
	\label{PlotCovOfTau2_piC_02theta=1_LOR_unequal_sample_sizes}
\end{figure}

\begin{figure}[ht]
	\centering
	\includegraphics[scale=0.33]{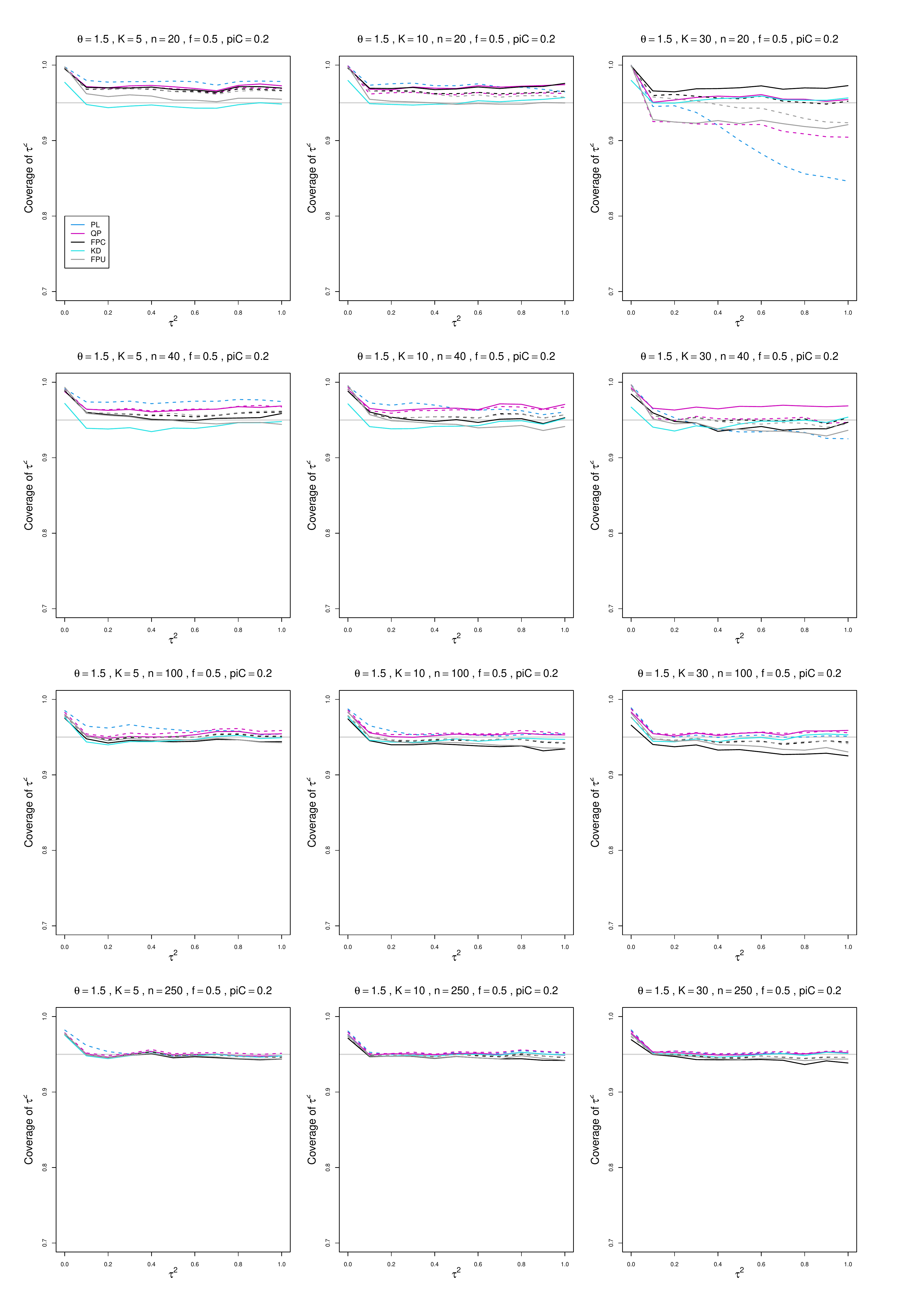}
	\caption{Coverage of PL, QP, KD, FPC, and FPU 95\% confidence intervals for between-study variance of LOR vs $\tau^2$, for equal sample sizes $n = 20,\;40,\;100$ and $250$, $p_{iC} = .2$, $\theta = 1.5$ and  $f = 0.5$.  Solid lines: PL, QP, and FPC \lq\lq only", FPU model-based, and KD. Dashed lines: PL, QP, and FPC \lq\lq always" and FPU na\"{i}ve.  }
	\label{PlotCovOfTau2_piC_02theta=1.5_LOR_equal_sample_sizes}
\end{figure}

\begin{figure}[ht]
	\centering
	\includegraphics[scale=0.33]{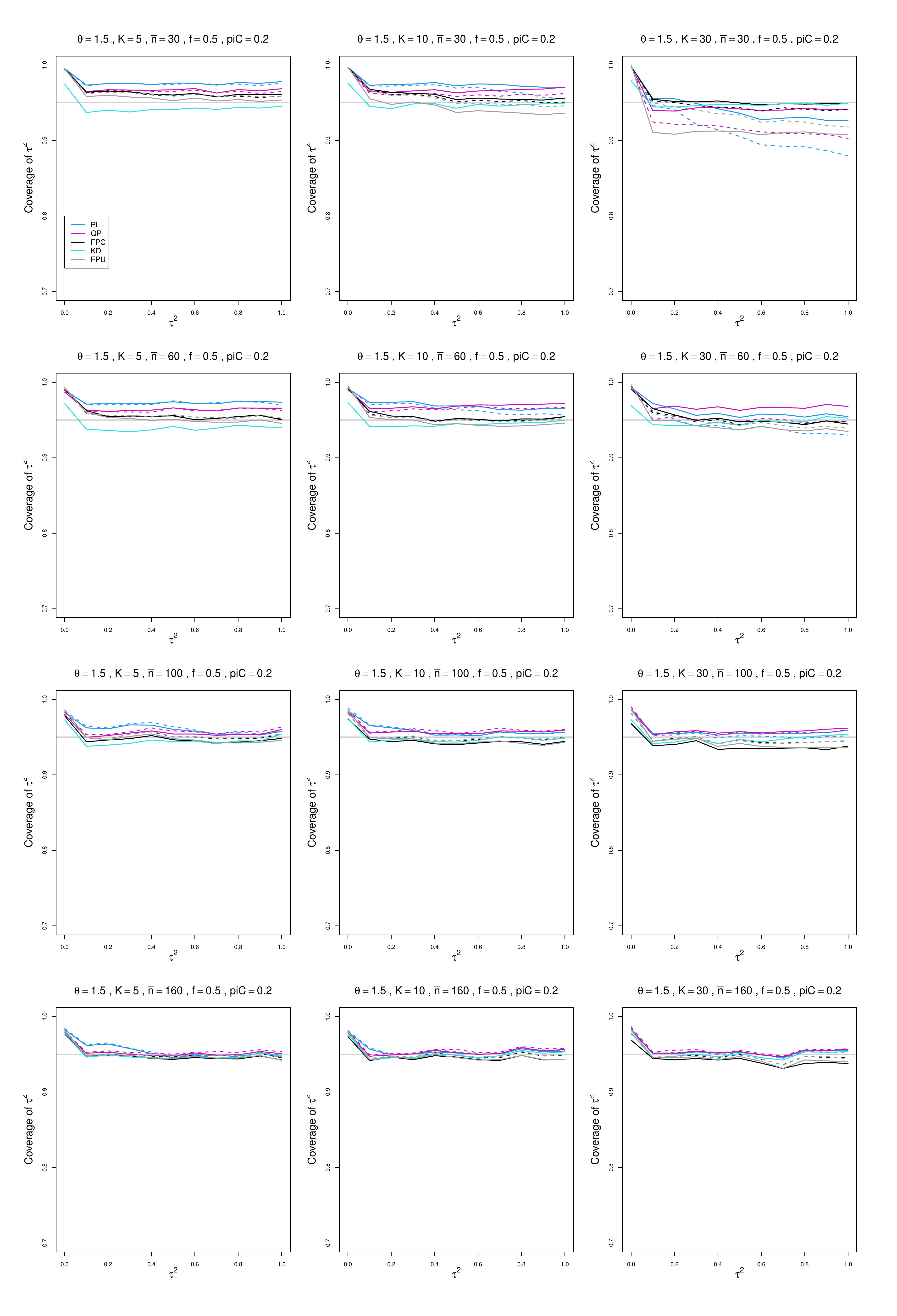}
	\caption{Coverage of PL, QP, KD, FPC, and FPU 95\% confidence intervals for between-study variance of LOR vs $\tau^2$, for unequal sample sizes $\bar{n}=30,\;60,\;100$ and $160$, $p_{iC} = .2$, $\theta = 1.5$ and  $f = 0.5$.  Solid lines: PL, QP, and FPC \lq\lq only", FPU model-based, and KD. Dashed lines: PL, QP, and FPC \lq\lq always" and FPU na\"{i}ve.   }
	\label{PlotCovOfTau2_piC_02theta=1.5_LOR_unequal_sample_sizes}
\end{figure}

\begin{figure}[ht]
	\centering
	\includegraphics[scale=0.33]{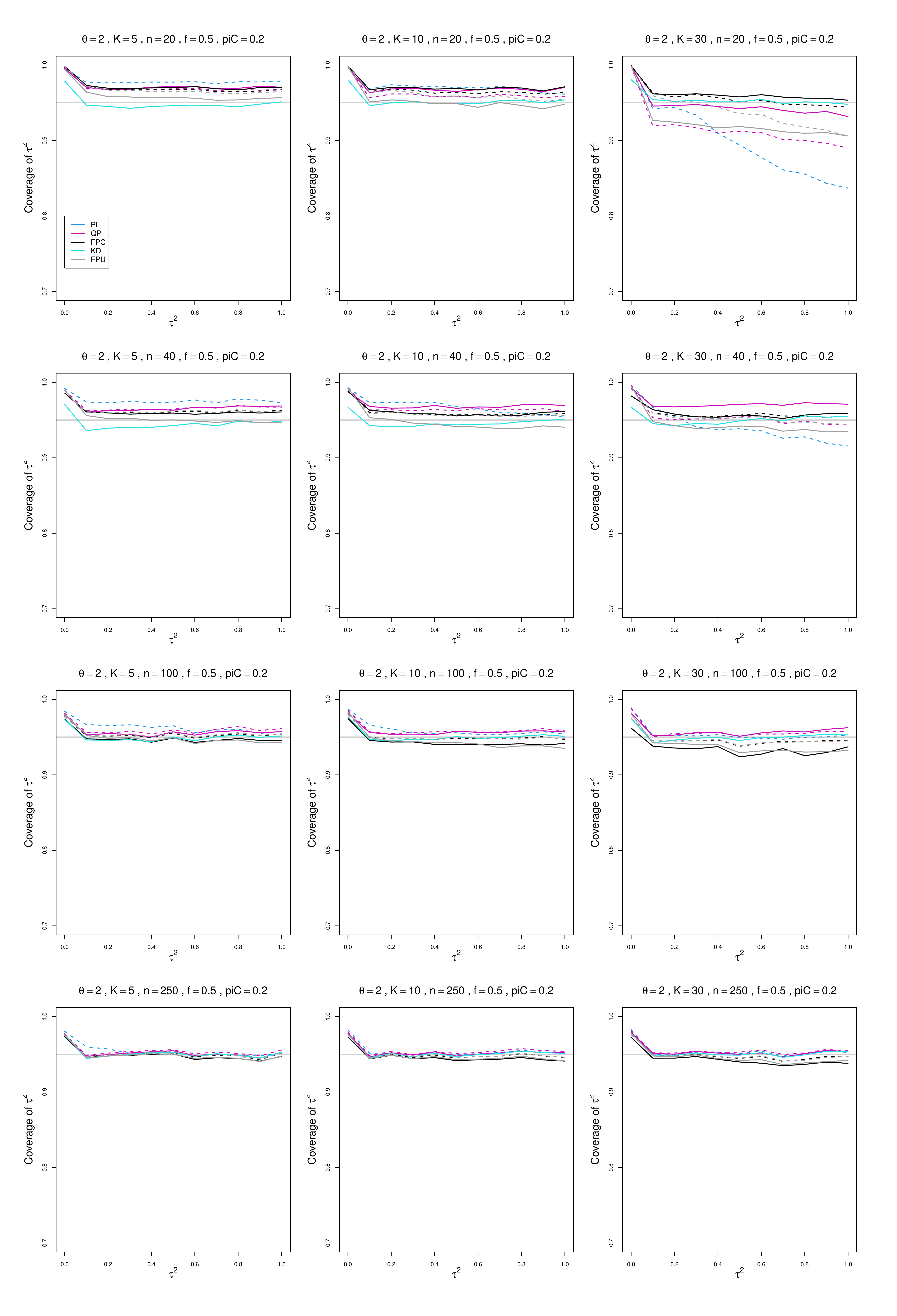}
	\caption{Coverage of PL, QP, KD, FPC, and FPU 95\% confidence intervals for between-study variance of LOR vs $\tau^2$, for equal sample sizes $n = 20,\;40,\;100$ and $250$, $p_{iC} = .2$, $\theta = 2$ and  $f = 0.5$.  Solid lines: PL, QP, and FPC \lq\lq only", FPU model-based, and KD. Dashed lines: PL, QP, and FPC \lq\lq always" and FPU na\"{i}ve.  }
	\label{PlotCpvOfTau2_piC_02theta=2_LOR_equal_sample_sizes}
\end{figure}

\begin{figure}[ht]
	\centering
	\includegraphics[scale=0.33]{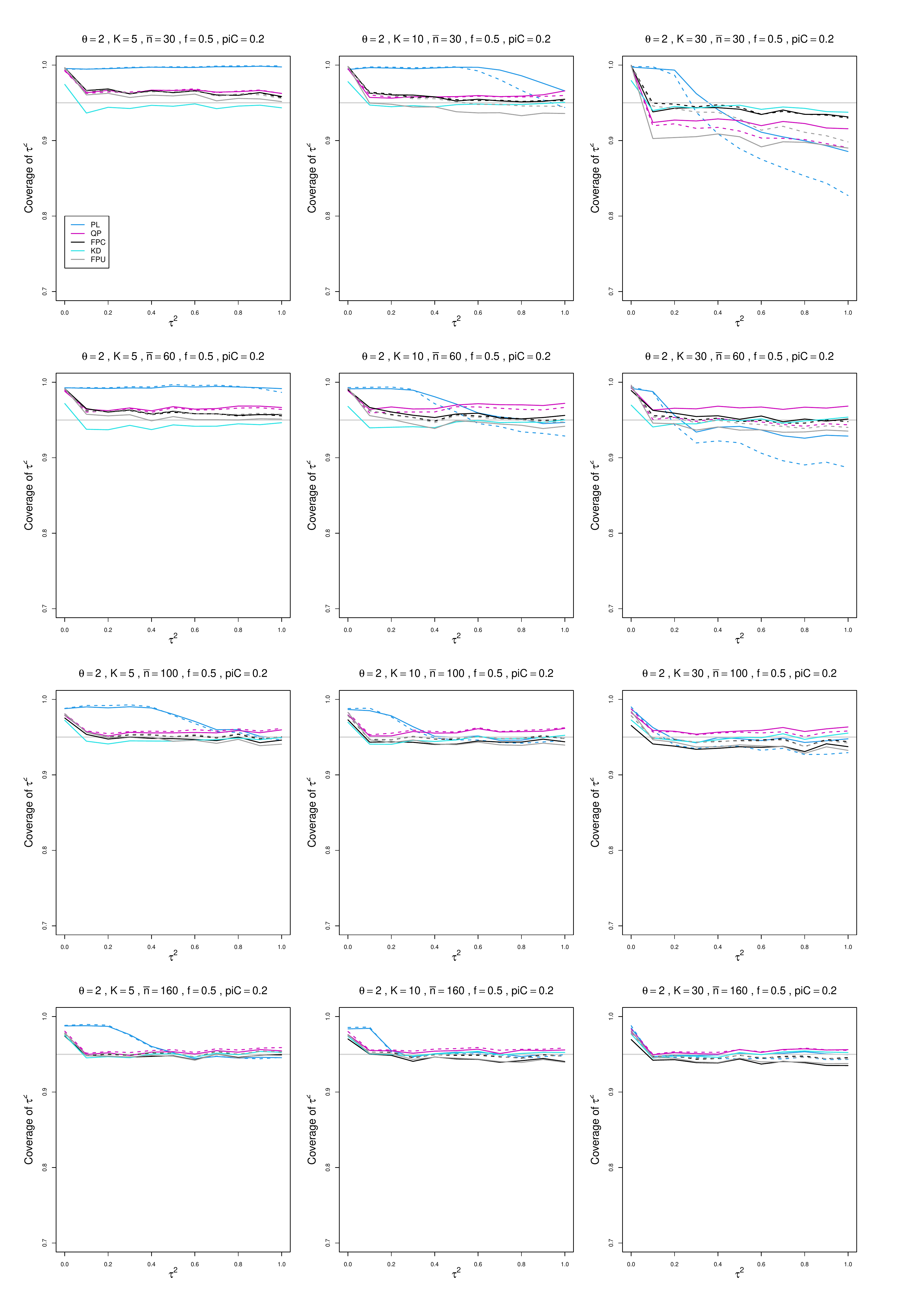}
	\caption{Coverage of PL, QP, KD, FPC, and FPU 95\% confidence intervals for between-study variance of LOR vs $\tau^2$, for unequal sample sizes $\bar{n}=30,\;60,\;100$ and $160$, $p_{iC} = .2$, $\theta = 2$ and  $f = 0.5$.  Solid lines: PL, QP, and FPC \lq\lq only", FPU model-based, and KD. Dashed lines: PL, QP, and FPC \lq\lq always" and FPU na\"{i}ve.  }
	\label{PlotCovOfTau2_piC_02theta=2_LOR_unequal_sample_sizes}
\end{figure}

\begin{figure}[ht]
	\centering
	\includegraphics[scale=0.33]{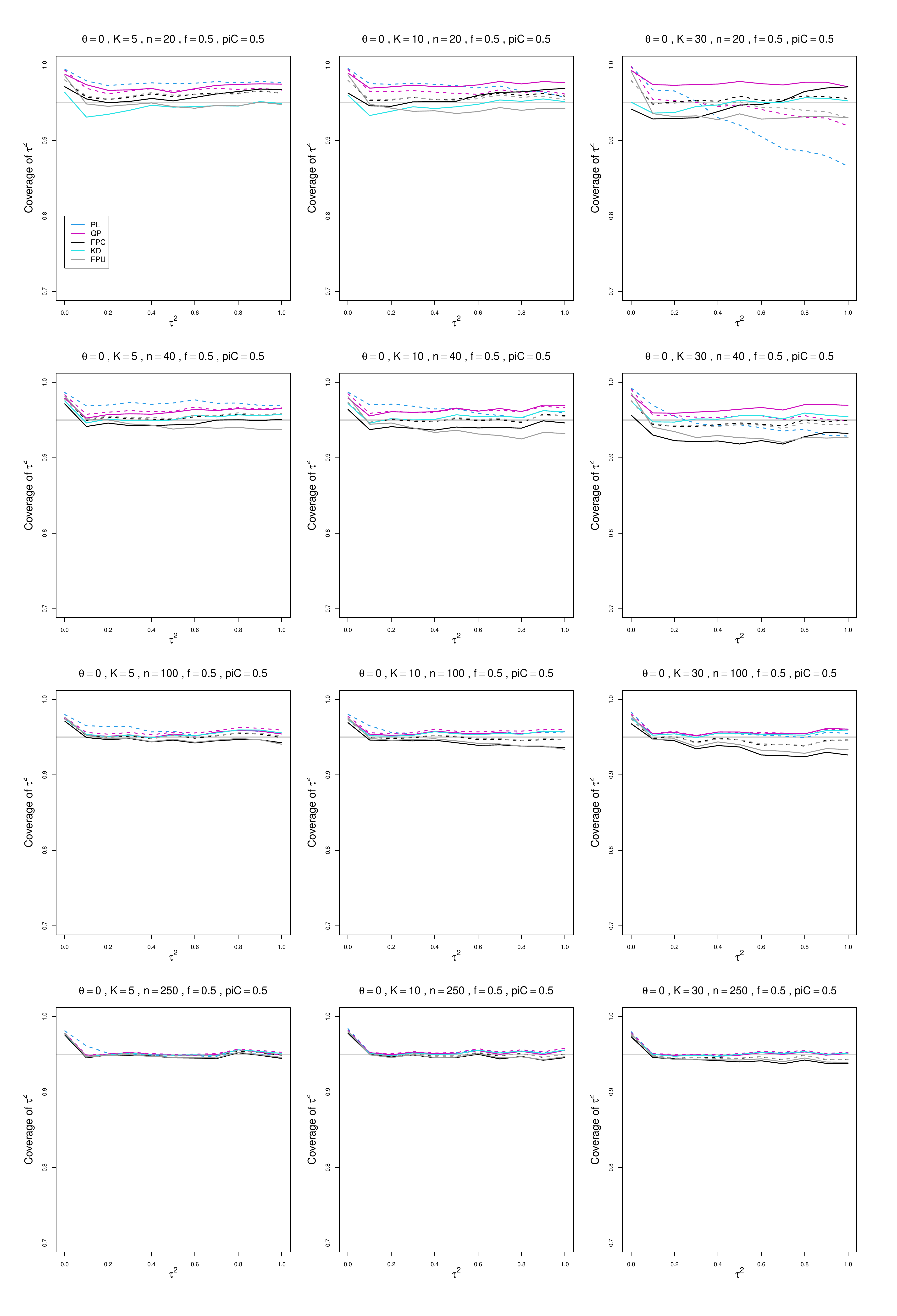}
	\caption{Coverage of PL, QP, KD, FPC, and FPU 95\% confidence intervals for between-study variance of LOR vs $\tau^2$, for equal sample sizes $n = 20,\;40,\;100$ and $250$, $p_{iC} = .5$, $\theta = 0$ and  $f = 0.5$.  Solid lines: PL, QP, and FPC \lq\lq only", FPU model-based, and KD. Dashed lines: PL, QP, and FPC \lq\lq always" and FPU na\"{i}ve.  }
	\label{PlotCovOfTau2_piC_05theta=0_LOR_equal_sample_sizes}
\end{figure}

\begin{figure}[ht]
	\centering
	\includegraphics[scale=0.33]{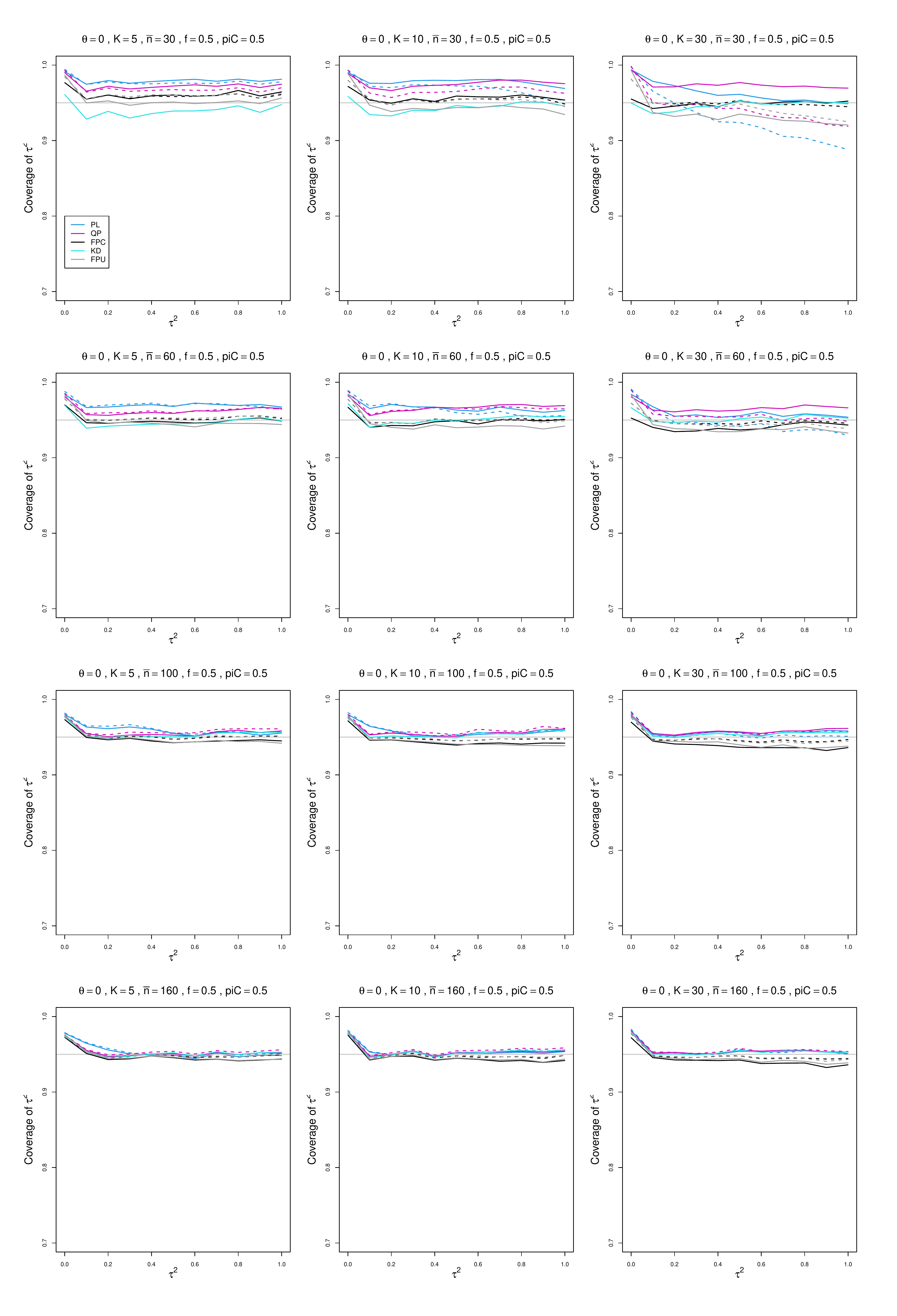}
	\caption{Coverage of PL, QP, KD, FPC, and FPU 95\% confidence intervals for between-study variance of LOR vs $\tau^2$, for unequal sample sizes $\bar{n}=30,\;60,\;100$ and $160$, $p_{iC} = .5$, $\theta = 0$ and  $f = 0.5$.  Solid lines: PL, QP, and FPC \lq\lq only", FPU model-based, and KD. Dashed lines: PL, QP, and FPC \lq\lq always" and FPU na\"{i}ve.   }
	\label{PlotCovOfTau2_piC_05theta=0_LOR_unequal_sample_sizes}
\end{figure}

\begin{figure}[ht]
	\centering
	\includegraphics[scale=0.33]{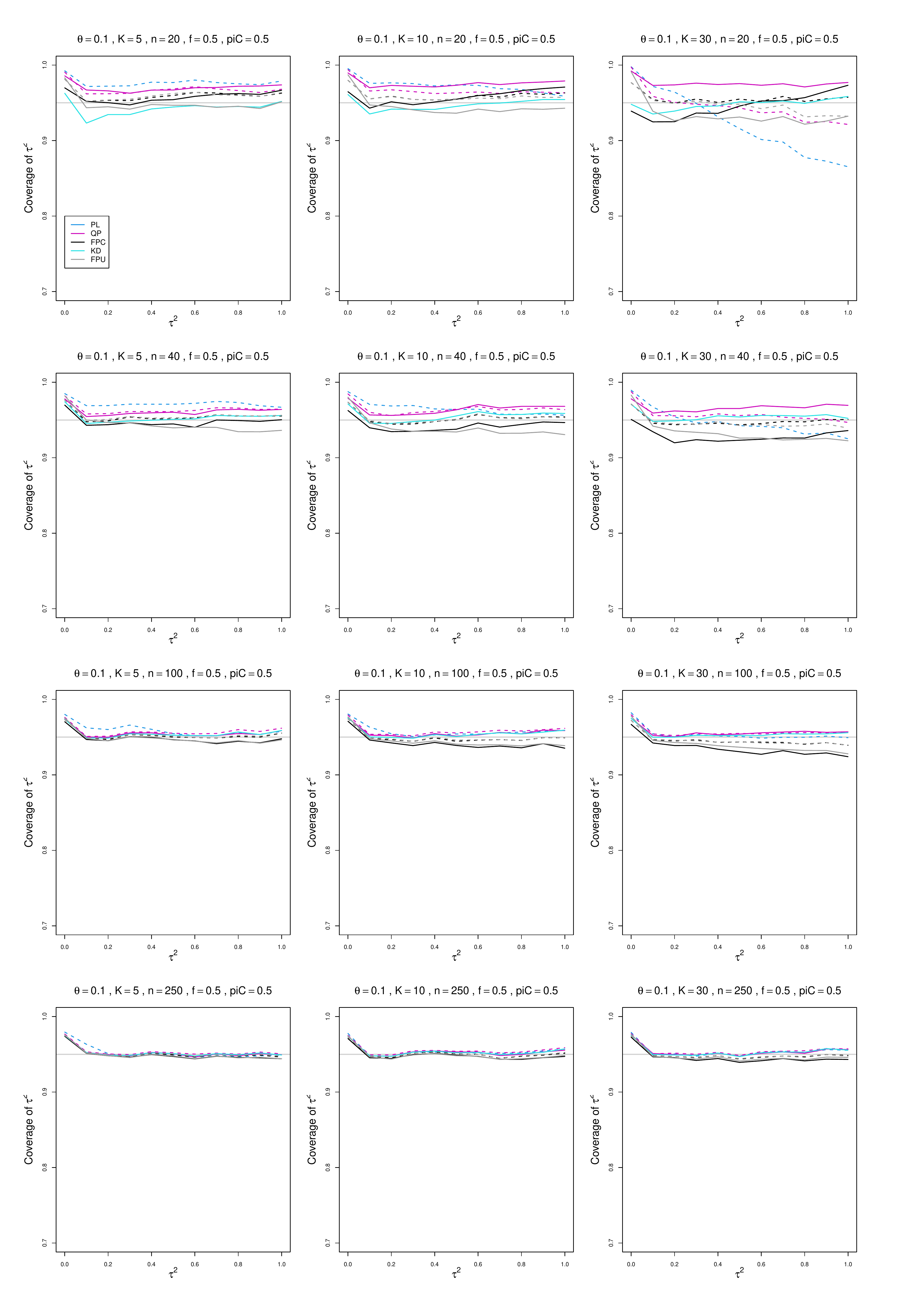}
	\caption{Coverage of PL, QP, KD, FPC, and FPU 95\% confidence intervals for between-study variance of LOR vs $\tau^2$, for equal sample sizes $n = 20,\;40,\;100$ and $250$, $p_{iC} = .5$, $\theta = 0.1$ and  $f = 0.5$.  Solid lines: PL, QP, and FPC \lq\lq only", FPU model-based, and KD. Dashed lines: PL, QP, and FPC \lq\lq always" and FPU na\"{i}ve.   }
	\label{PlotCovOfTau2_piC_05theta=0.1_LOR_equal_sample_sizes}
\end{figure}

\begin{figure}[ht]
	\centering
	\includegraphics[scale=0.33]{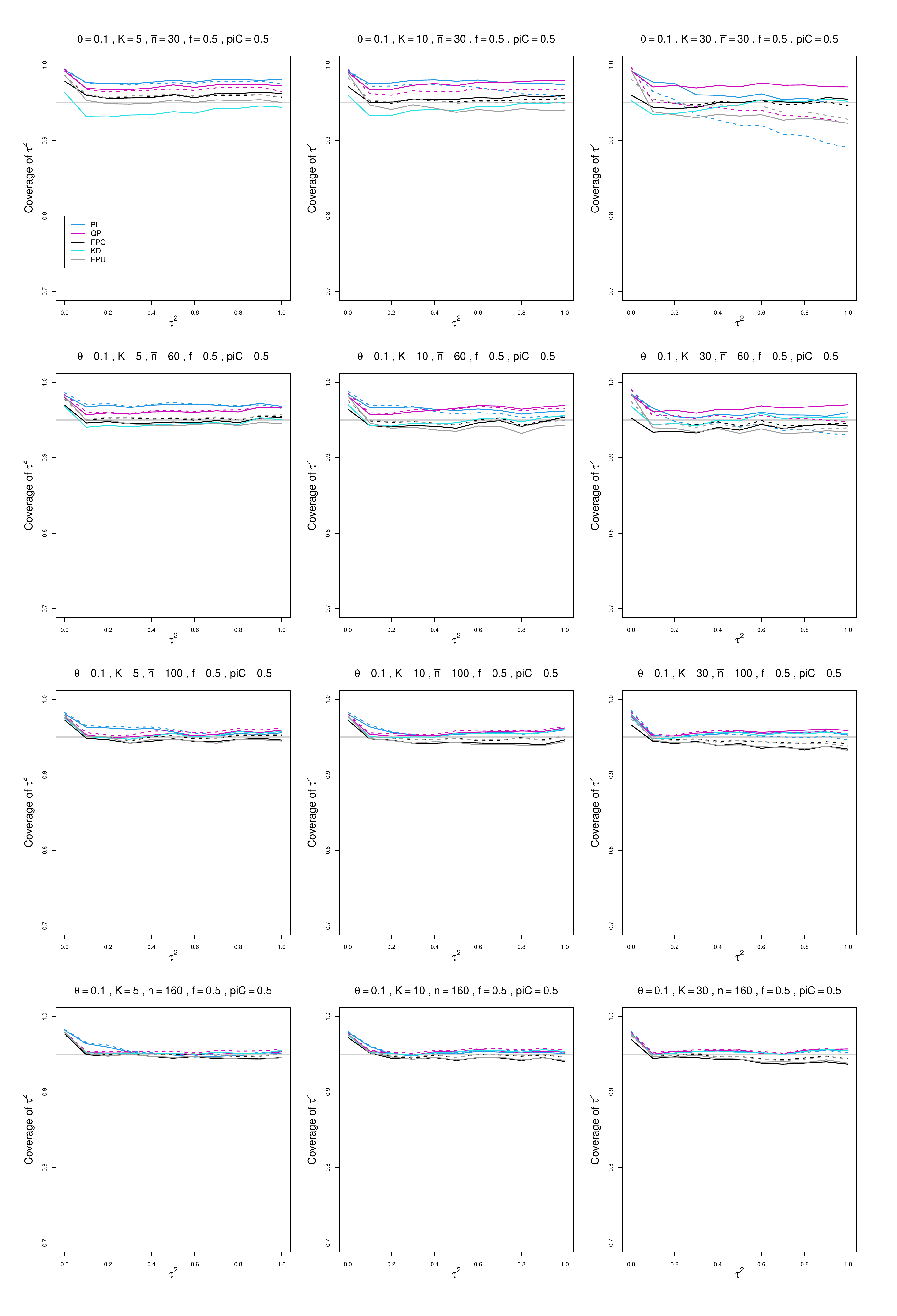}
	\caption{Coverage of PL, QP, KD, FPC, and FPU 95\% confidence intervals for between-study variance of LOR vs $\tau^2$, for unequal sample sizes $\bar{n}=30,\;60,\;100$ and $160$, $p_{iC} = .5$, $\theta = 0.1$ and  $f = 0.5$.  Solid lines: PL, QP, and FPC \lq\lq only", FPU model-based, and KD. Dashed lines: PL, QP, and FPC \lq\lq always" and FPU na\"{i}ve.  }
	\label{PlotCovOfTau2_piC_05theta=0.1_LOR_unequal_sample_sizes}
\end{figure}

\begin{figure}[ht]
	\centering
	\includegraphics[scale=0.33]{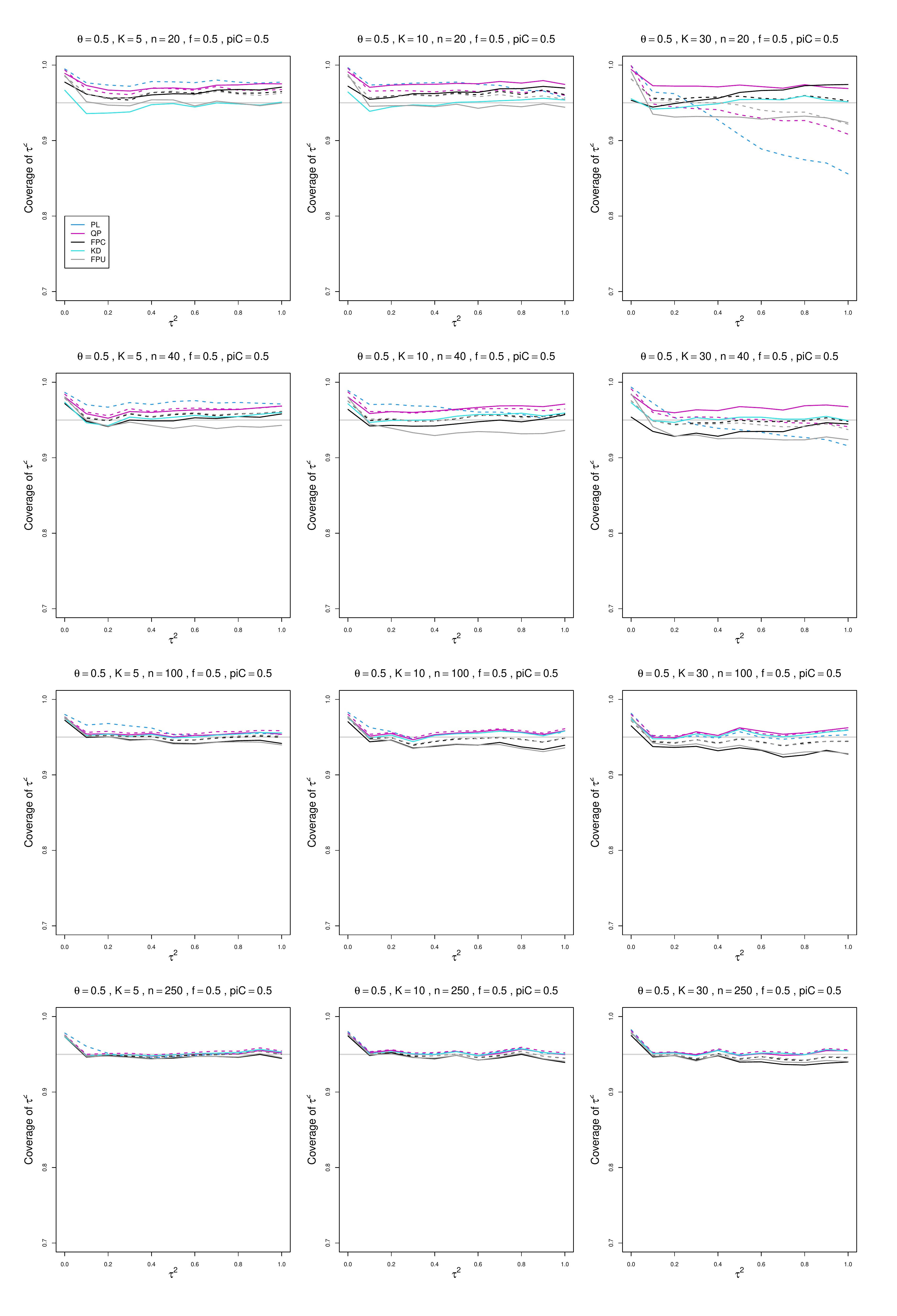}
	\caption{Coverage of PL, QP, KD, FPC, and FPU 95\% confidence intervals for between-study variance of LOR vs $\tau^2$, for equal sample sizes $n = 20,\;40,\;100$ and $250$, $p_{iC} = .5$, $\theta = 0.5$ and  $f = 0.5$.  Solid lines: PL, QP, and FPC \lq\lq only", FPU model-based, and KD. Dashed lines: PL, QP, and FPC \lq\lq always" and FPU na\"{i}ve.  }
	\label{PlotCovOfTau2_piC_05theta=0.5_LOR_equal_sample_sizes}
\end{figure}

\begin{figure}[ht]
	\centering
	\includegraphics[scale=0.33]{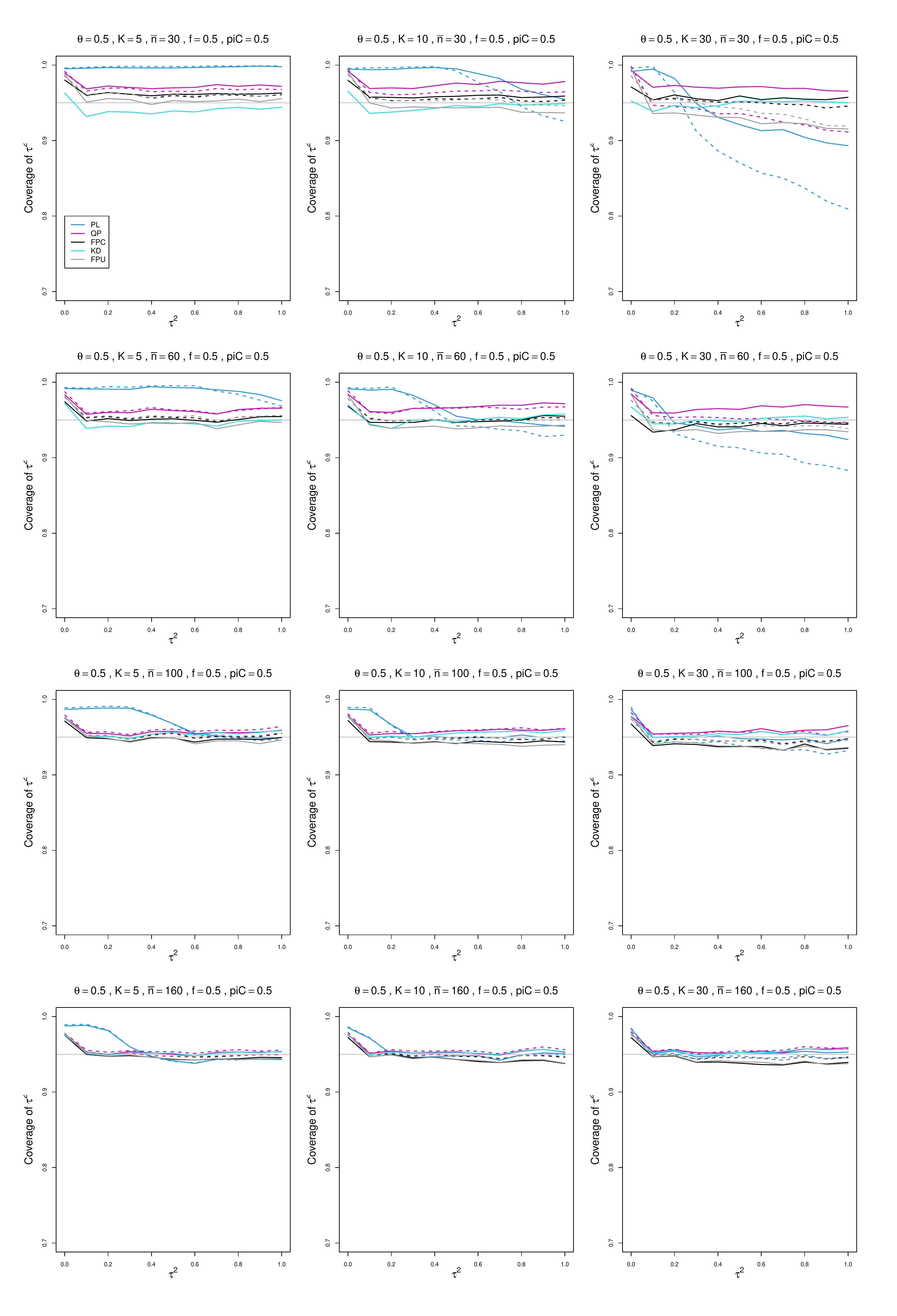}
	\caption{Coverage of PL, QP, KD, FPC, and FPU 95\% confidence intervals for between-study variance of LOR vs $\tau^2$, for unequal sample sizes $\bar{n}=30,\;60,\;100$ and $160$, $p_{iC} = .5$, $\theta = 0.5$ and  $f = 0.5$.  Solid lines: PL, QP, and FPC \lq\lq only", FPU model-based, and KD. Dashed lines: PL, QP, and FPC \lq\lq always" and FPU na\"{i}ve.  }
	\label{PlotCovOfTau2_piC_05theta=0.5_LOR_unequal_sample_sizes}
\end{figure}

\begin{figure}[ht]
	\centering
	\includegraphics[scale=0.33]{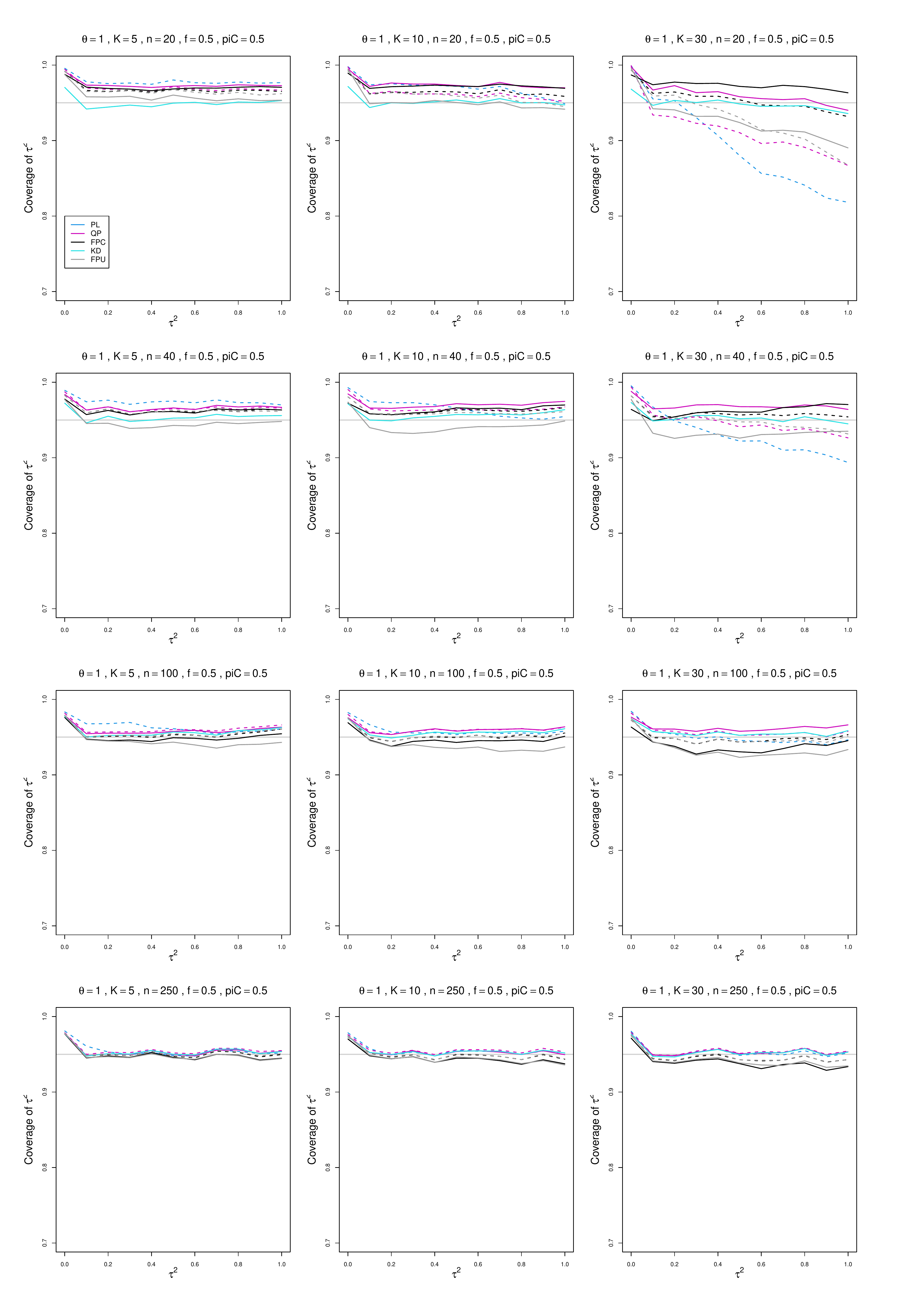}
	\caption{Coverage of PL, QP, KD, FPC, and FPU 95\% confidence intervals for between-study variance of LOR vs $\tau^2$, for equal sample sizes $n = 20,\;40,\;100$ and $250$, $p_{iC} = .5$, $\theta = 1$ and  $f = 0.5$.  Solid lines: PL, QP, and FPC \lq\lq only", FPU model-based, and KD. Dashed lines: PL, QP, and FPC \lq\lq always" and FPU na\"{i}ve.  }
	\label{PlotCovOfTau2_piC_05theta=1_LOR_equal_sample_sizes}
\end{figure}

\begin{figure}[ht]
	\centering
	\includegraphics[scale=0.33]{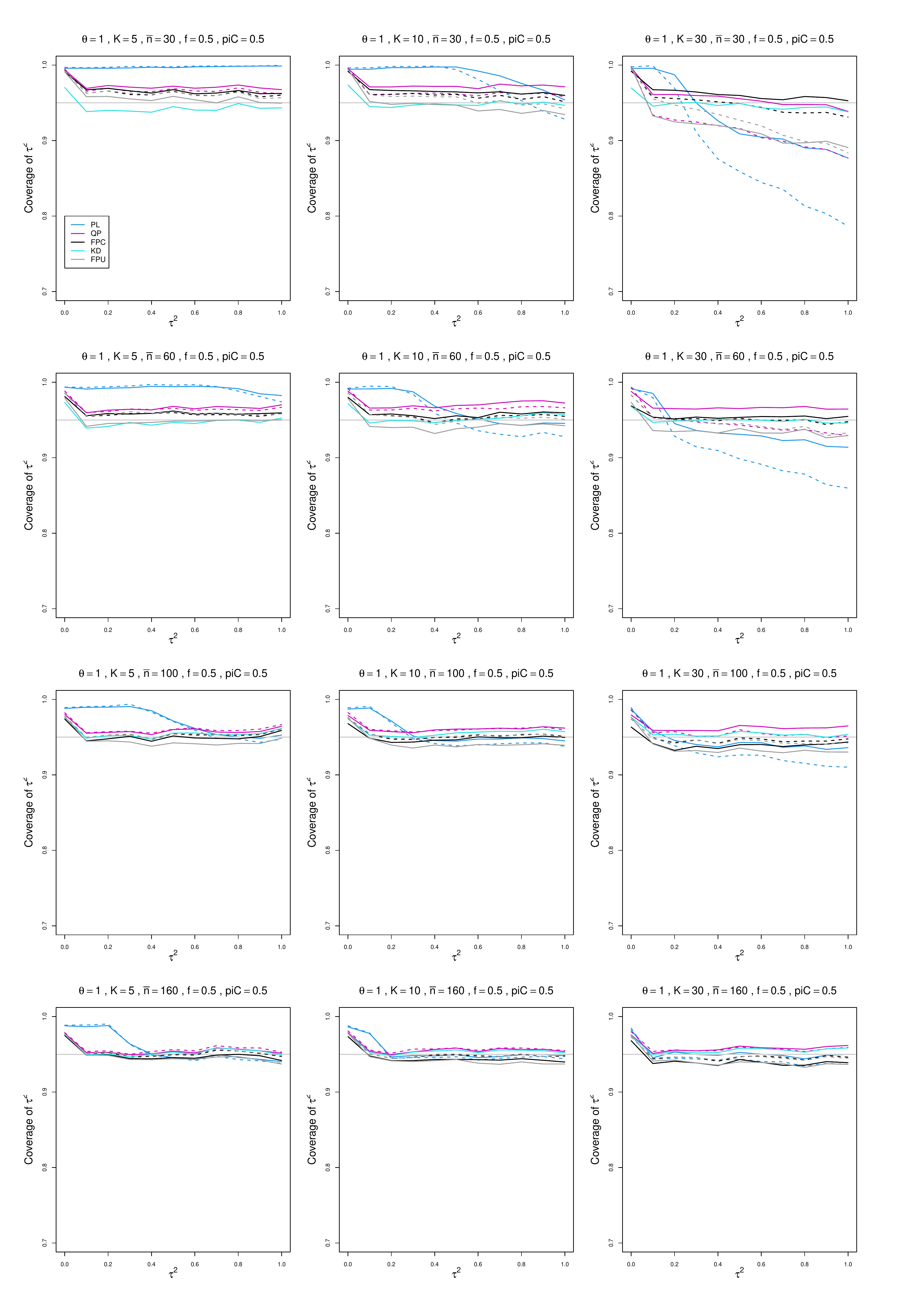}
	\caption{Coverage of PL, QP, KD, FPC, and FPU 95\% confidence intervals for between-study variance of LOR vs $\tau^2$, for unequal sample sizes $\bar{n}=30,\;60,\;100$ and $160$, $p_{iC} = .5$, $\theta = 1$ and  $f = 0.5$.  Solid lines: PL, QP, and FPC \lq\lq only", FPU model-based, and KD. Dashed lines: PL, QP, and FPC \lq\lq always" and FPU na\"{i}ve.  }
	\label{PlotCovOfTau2_piC_05theta=1_LOR_unequal_sample_sizes}
\end{figure}

\begin{figure}[ht]
	\centering
	\includegraphics[scale=0.33]{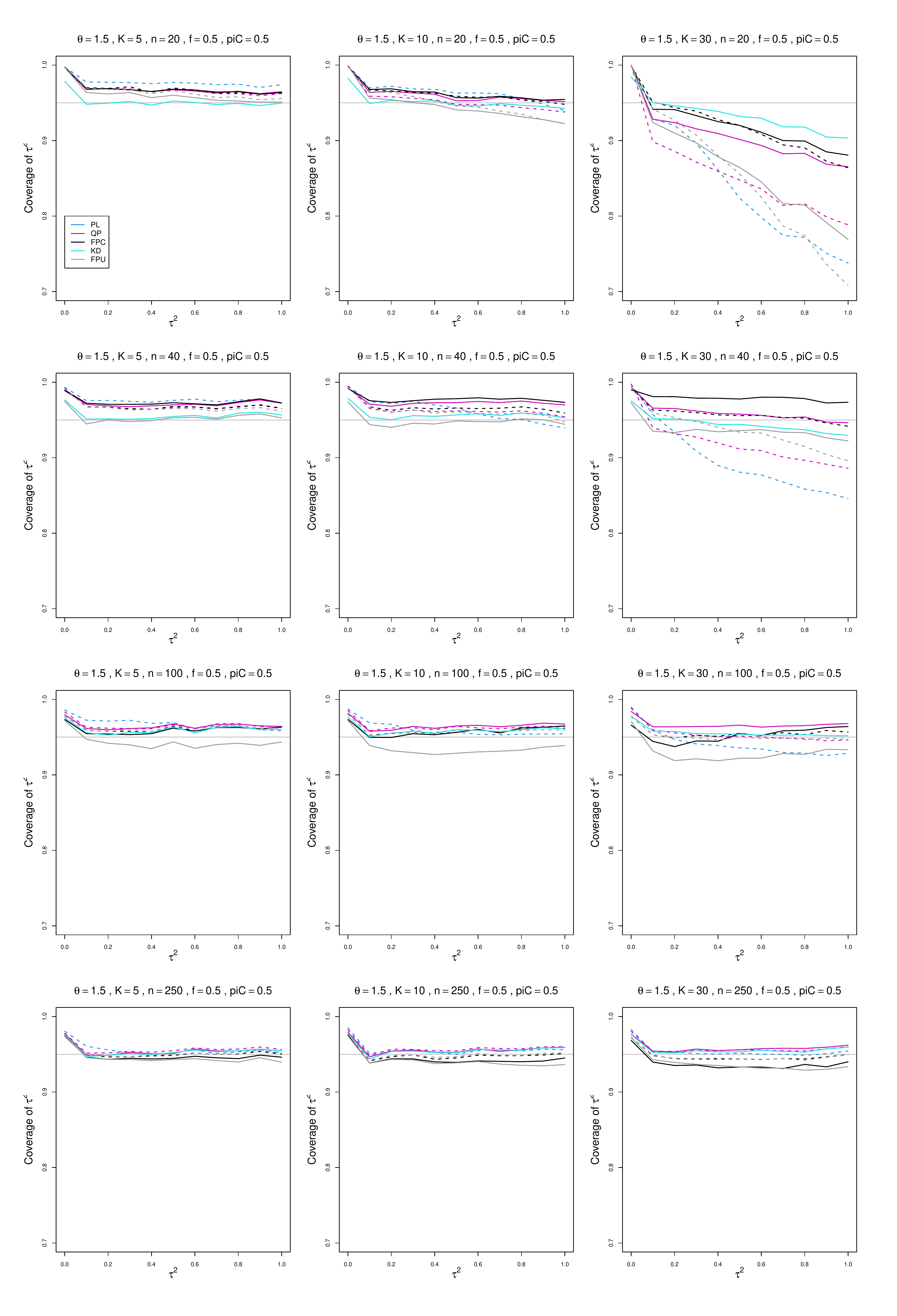}
	\caption{Coverage of PL, QP, KD, FPC, and FPU 95\% confidence intervals for between-study variance of LOR vs $\tau^2$, for equal sample sizes $n = 20,\;40,\;100$ and $250$, $p_{iC} = .5$, $\theta = 1.5$ and  $f = 0.5$.  Solid lines: PL, QP, and FPC \lq\lq only", FPU model-based, and KD. Dashed lines: PL, QP, and FPC \lq\lq always" and FPU na\"{i}ve.  }
	\label{PlotCovOfTau2_piC_05theta=1.5_LOR_equal_sample_sizes}
\end{figure}

\begin{figure}[ht]
	\centering
	\includegraphics[scale=0.33]{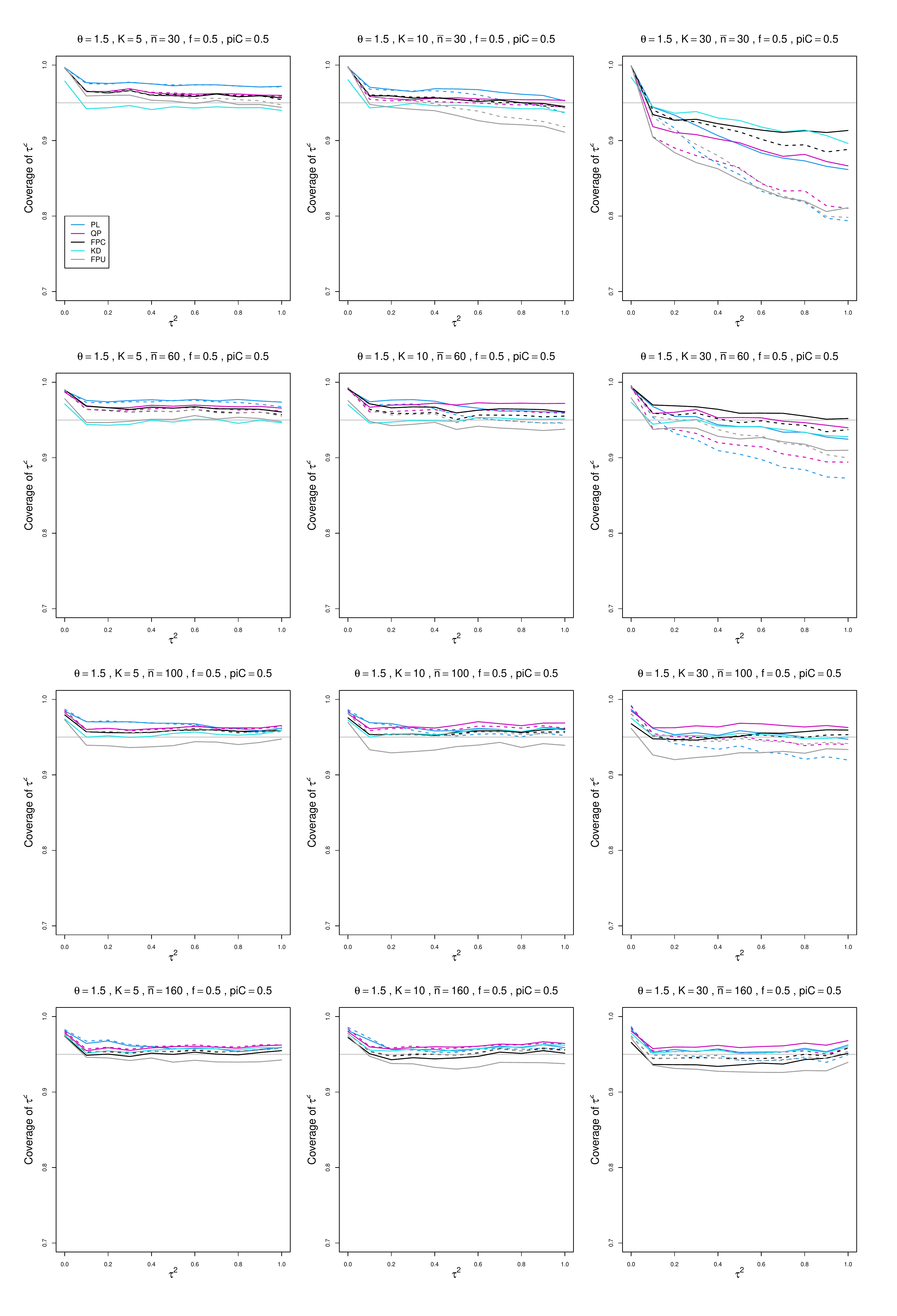}
	\caption{Coverage of PL, QP, KD, FPC, and FPU 95\% confidence intervals for between-study variance of LOR vs $\tau^2$, for unequal sample sizes $\bar{n}=30,\;60,\;100$ and $160$, $p_{iC} = .5$, $\theta = 1.5$ and  $f = 0.5$.  Solid lines: PL, QP, and FPC \lq\lq only", FPU model-based, and KD. Dashed lines: PL, QP, and FPC \lq\lq always" and FPU na\"{i}ve.  }
	\label{PlotCovOfTau2_piC_05theta=1.5_LOR_unequal_sample_sizes}
\end{figure}

\begin{figure}[ht]
	\centering
	\includegraphics[scale=0.33]{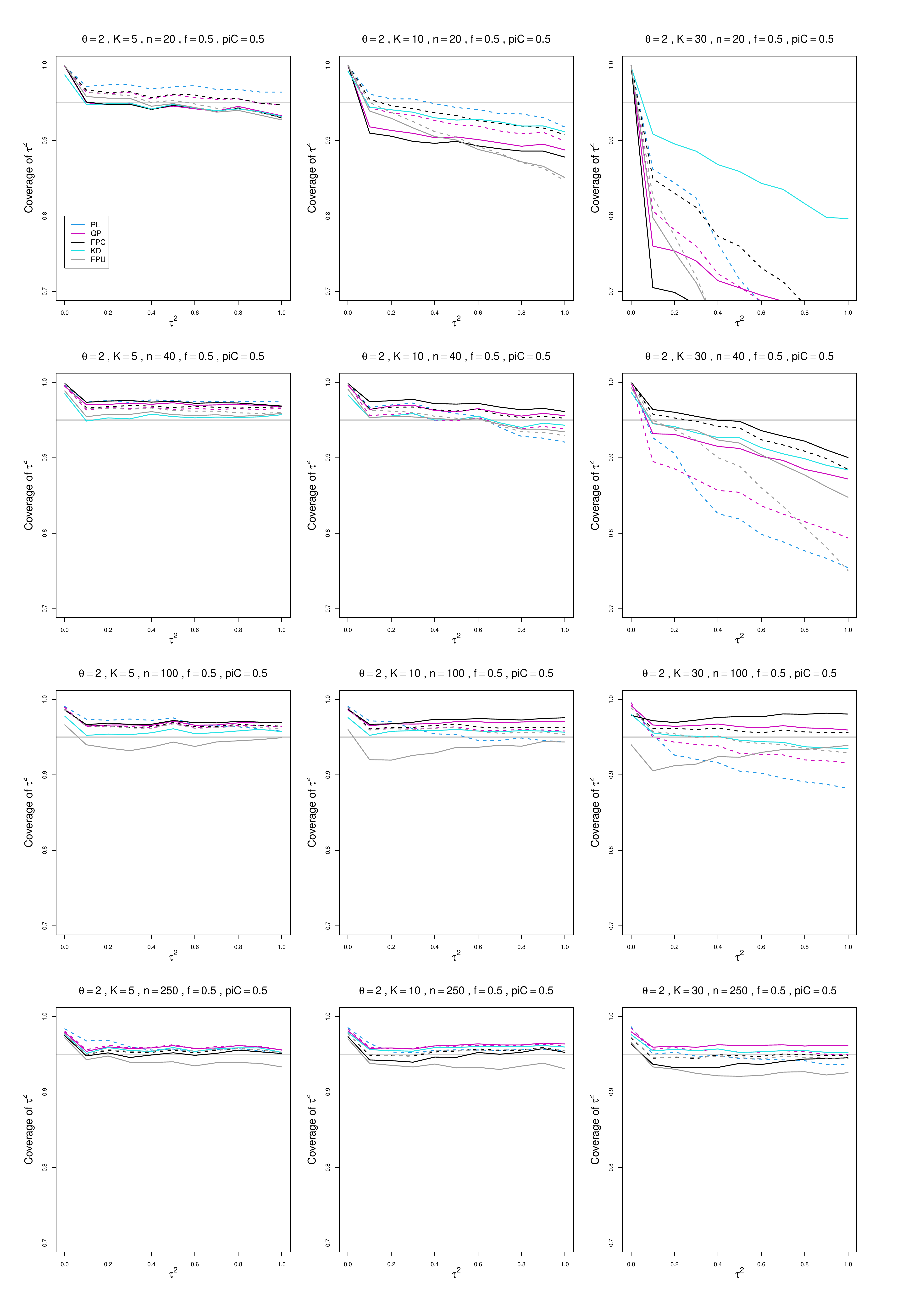}
	\caption{Coverage of PL, QP, KD, FPC, and FPU 95\% confidence intervals for between-study variance of LOR vs $\tau^2$, for equal sample sizes $n = 20,\;40,\;100$ and $250$, $p_{iC} = .5$, $\theta = 2$ and  $f = 0.5$.  Solid lines: PL, QP, and FPC \lq\lq only", FPU model-based, and KD. Dashed lines: PL, QP, and FPC \lq\lq always" and FPU na\"{i}ve.   }
	\label{PlotCovOfTau2_piC_05theta=2_LOR_equal_sample_sizes}
\end{figure}

\begin{figure}[ht]
	\centering
	\includegraphics[scale=0.33]{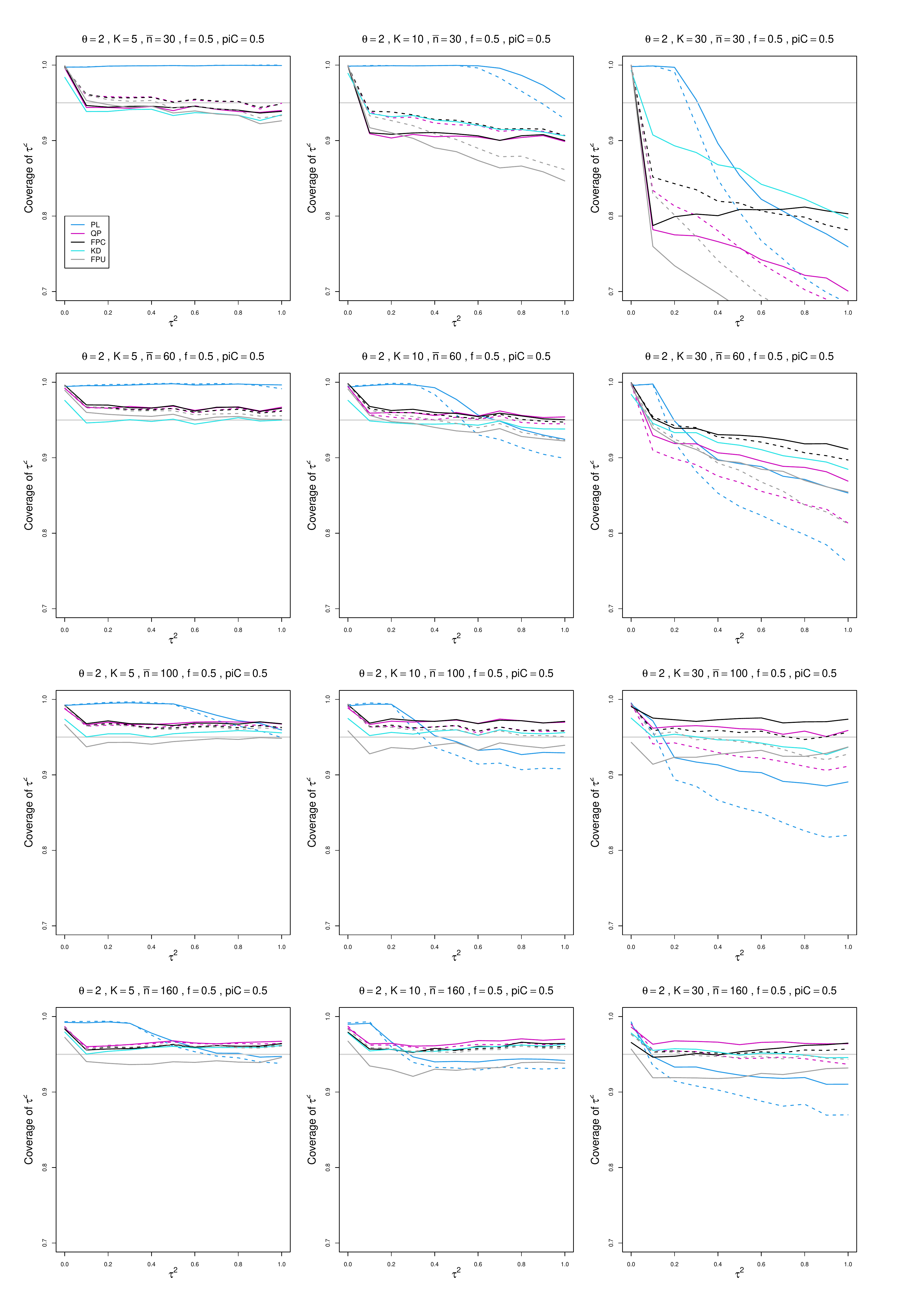}
	\caption{Coverage of PL, QP, KD, FPC, and FPU 95\% confidence intervals for between-study variance of LOR vs $\tau^2$, for unequal sample sizes $\bar{n}=30,\;60,\;100$ and $160$, $p_{iC} = .5$, $\theta = 2$ and  $f = 0.5$.  Solid lines: PL, QP, and FPC \lq\lq only", FPU model-based, and KD. Dashed lines: PL, QP, and FPC \lq\lq always" and FPU na\"{i}ve.  }
	\label{PlotCovOfTau2_piC_05theta=2_LOR_unequal_sample_sizes}
\end{figure}

\clearpage

\section*{Appendix D: Sample miss-left probability for 95\% confidence intervals for between-study variance}

Each figure corresponds to a value of the probability of an event in the Control arm $p_{iC}$  (= .1, .2, .5) . \\
The fraction of each study's sample size in the Control arm  $f$ is held constant at 0.5. For each combination of a value of $n$ (= 20, 40, 100, 250) or  $\bar{n}$=(= 30, 60, 100, 160) and a value of $K$ (= 5, 10, 30), a panel plots the probability that the parameter is to the left of the lower confidence limit of the confidence interval versus $\tau^2$ (= 0.0(0.1)1).\\
The confidence intervals for $\tau^2$ are
\begin{itemize}
\item PL (Profile Likelihood), inverse-variance weights)
\item QP (Q profile, inverse-variance weights)
\item KD (based on  Kulinskaya-Dollinger (2015) approximation, inverse-variance weights)
\item FPC (based on Farebrother approximation, effective-sample-size weights, conditional variance of LOR)
\item FPU (based on Farebrother approximation, effective-sample-size weights, unconditional variance of LOR)
\end{itemize}
The plots include two versions of PL, QP, and FPC: adding $1/2$ to all four of $X_{iT},\;X_{iC},\; n_{iT}-X_{iT},\; n_{iC}-X_{iC}$ only when one of these is zero (solid lines) or always (dashed lines).\\
The plots also include two versions of FPU: model-based estimation of $p_{iT}$ (solid lines) or na\"{i}ve estimation (dashed lines).

\clearpage
\setcounter{figure}{0}
\renewcommand{\thefigure}{D.\arabic{figure}}

\begin{figure}[ht]
	\centering
	\includegraphics[scale=0.33]{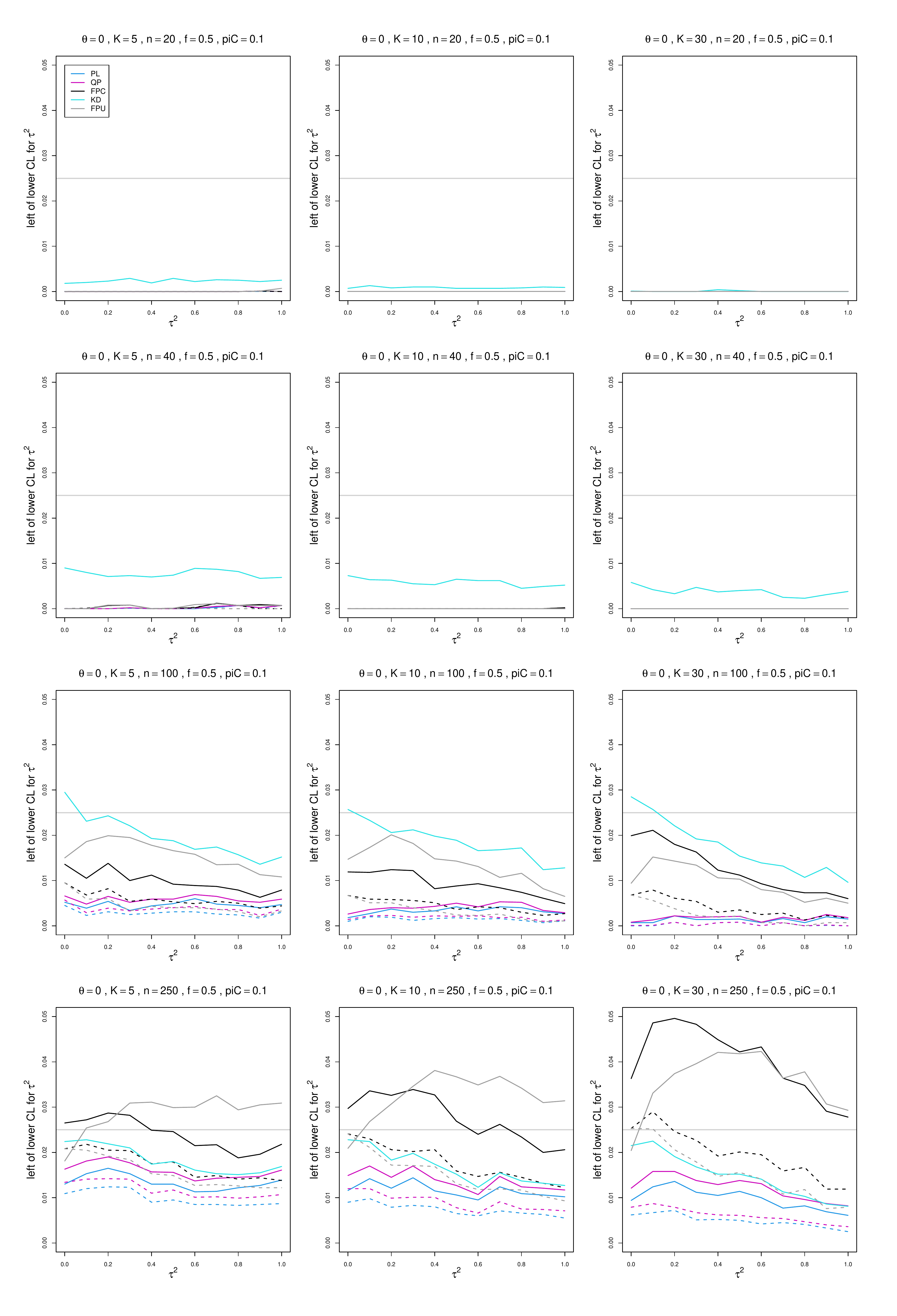}
	\caption{Miss-left probability  of  PL, QP, KD, FPC, and FPU 95\% confidence intervals for between-study variance of LOR vs $\tau^2$, for equal sample sizes $n=20,\;40,\;100$ and $250$, $p_{iC} = .1$, $\theta=0$ and  $f=0.5$.   Solid lines: PL, QP, and FPC \lq\lq only", FPU model-based, and KD. Dashed lines: PL, QP, and FPC \lq\lq always" and FPU na\"{i}ve.  }
	\label{PlotCovLeftOfTau2_piC_01theta=0_LOR_equal_sample_sizes}
\end{figure}

\begin{figure}[ht]
	\centering
	\includegraphics[scale=0.33]{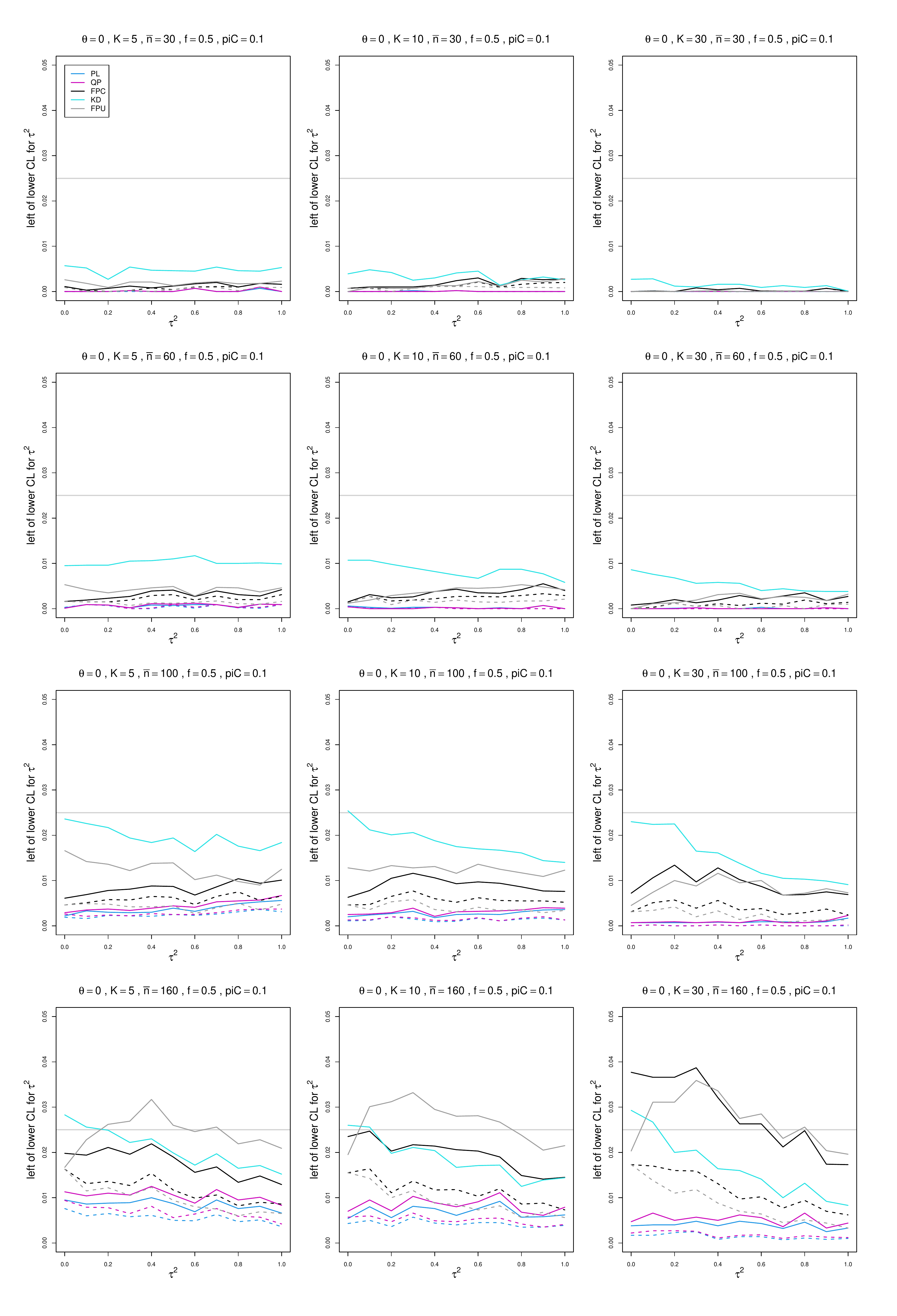}
	\caption{Miss-left probability  of  PL, QP, KD, FPC, and FPU 95\% confidence intervals for between-study variance of LOR vs $\tau^2$, for unequal sample sizes $\bar{n}=30,\;60,\;100$ and $160$, $p_{iC} = .1$, $\theta=0$ and  $f=0.5$.   Solid lines: PL, QP, and FPC \lq\lq only", FPU model-based, and KD. Dashed lines: PL, QP, and FPC \lq\lq always" and FPU na\"{i}ve. }
	\label{PlotCovLeftOfTau2_piC_01theta=0_LOR_unequal_sample_sizes}
\end{figure}

\begin{figure}[ht]
	\centering
	\includegraphics[scale=0.33]{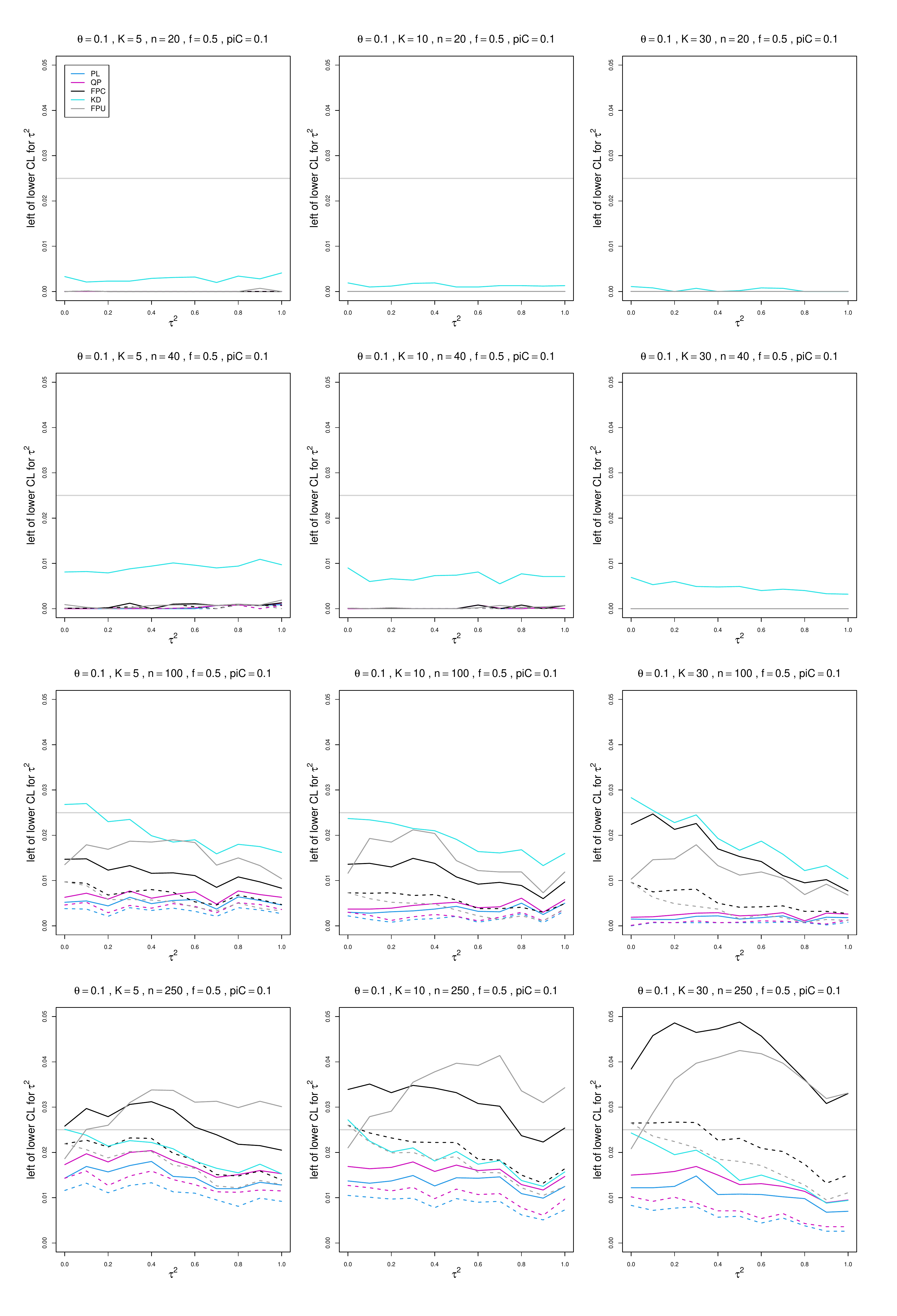}
	\caption{Miss-left probability  of  PL, QP, KD, FPC, and FPU 95\% confidence intervals for between-study variance of LOR vs $\tau^2$, for equal sample sizes $n=20,\;40,\;100$ and $250$, $p_{iC} = .1$, $\theta=0.1$ and  $f=0.5$.   Solid lines: PL, QP, and FPC \lq\lq only", FPU model-based, and KD. Dashed lines: PL, QP, and FPC \lq\lq always" and FPU na\"{i}ve.   }
	\label{PlotCovLeftOfTau2_piC_01theta=0.1_LOR_equal_sample_sizes}
\end{figure}

\begin{figure}[ht]
	\centering
	\includegraphics[scale=0.33]{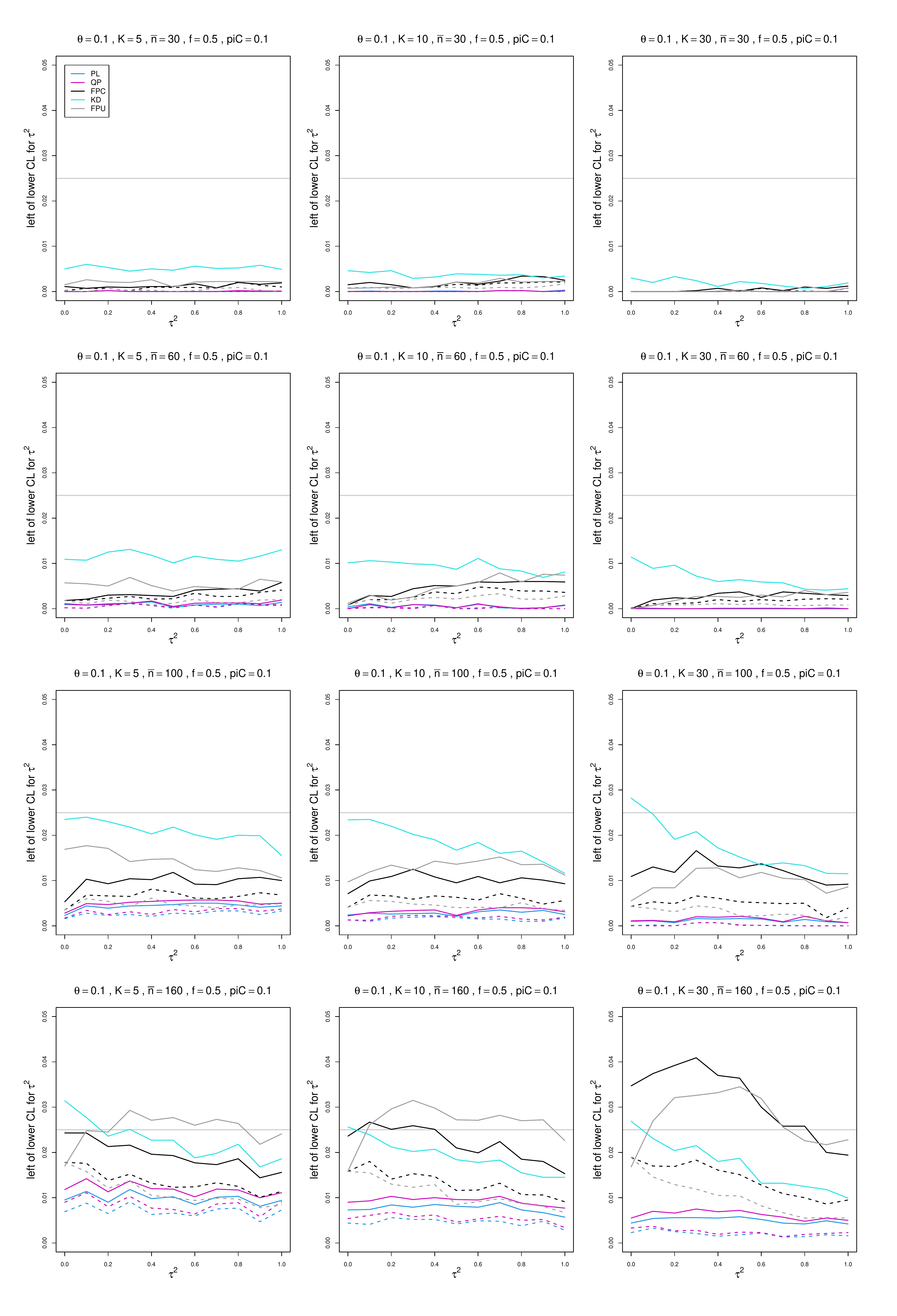}
	\caption{Miss-left probability  of  PL, QP, KD, FPC, and FPU 95\% confidence intervals for between-study variance of LOR vs $\tau^2$, for unequal sample sizes $\bar{n}=30,\;60,\;100$ and $160$, $p_{iC} = .1$, $\theta=0.1$ and  $f=0.5$.   Solid lines: PL, QP, and FPC \lq\lq only", FPU model-based, and KD. Dashed lines: PL, QP, and FPC \lq\lq always" and FPU na\"{i}ve.  }
	\label{PlotCovLeftOfTau2_piC_01theta=0.1_LOR_unequal_sample_sizes}
\end{figure}

\begin{figure}[ht]
	\centering
	\includegraphics[scale=0.33]{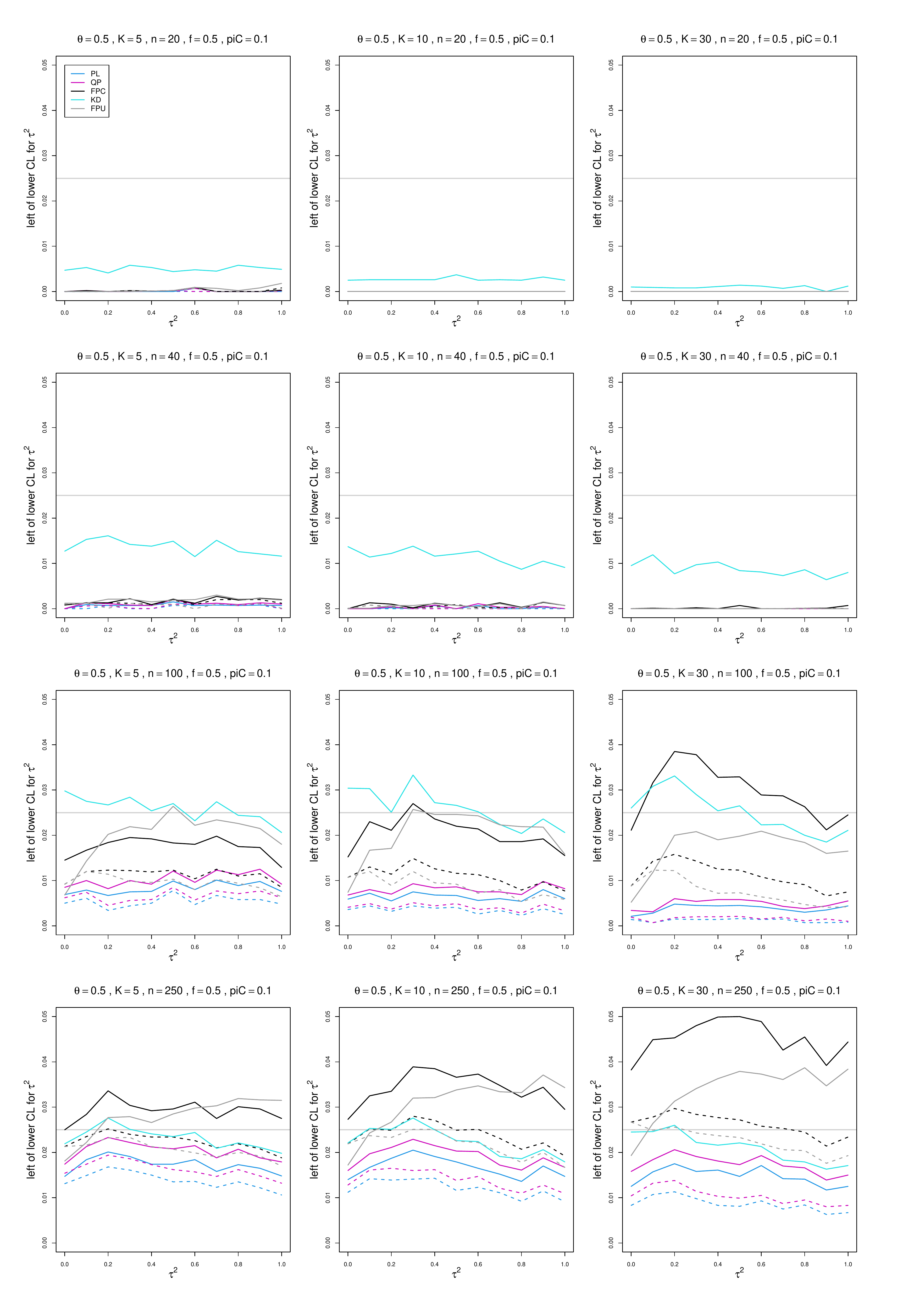}
	\caption{Miss-left probability  of  PL, QP, KD, FPC, and FPU 95\% confidence intervals for between-study variance of LOR vs $\tau^2$, for equal sample sizes $n=20,\;40,\;100$ and $250$, $p_{iC} = .1$, $\theta=0.5$ and  $f=0.5$.   Solid lines: PL, QP, and FPC \lq\lq only", FPU model-based, and KD. Dashed lines: PL, QP, and FPC \lq\lq always" and FPU na\"{i}ve.   }
	\label{PlotLeftOfTau2_piC_01theta=0.5_LOR_equal_sample_sizes}
\end{figure}

\begin{figure}[ht]
	\centering
	\includegraphics[scale=0.33]{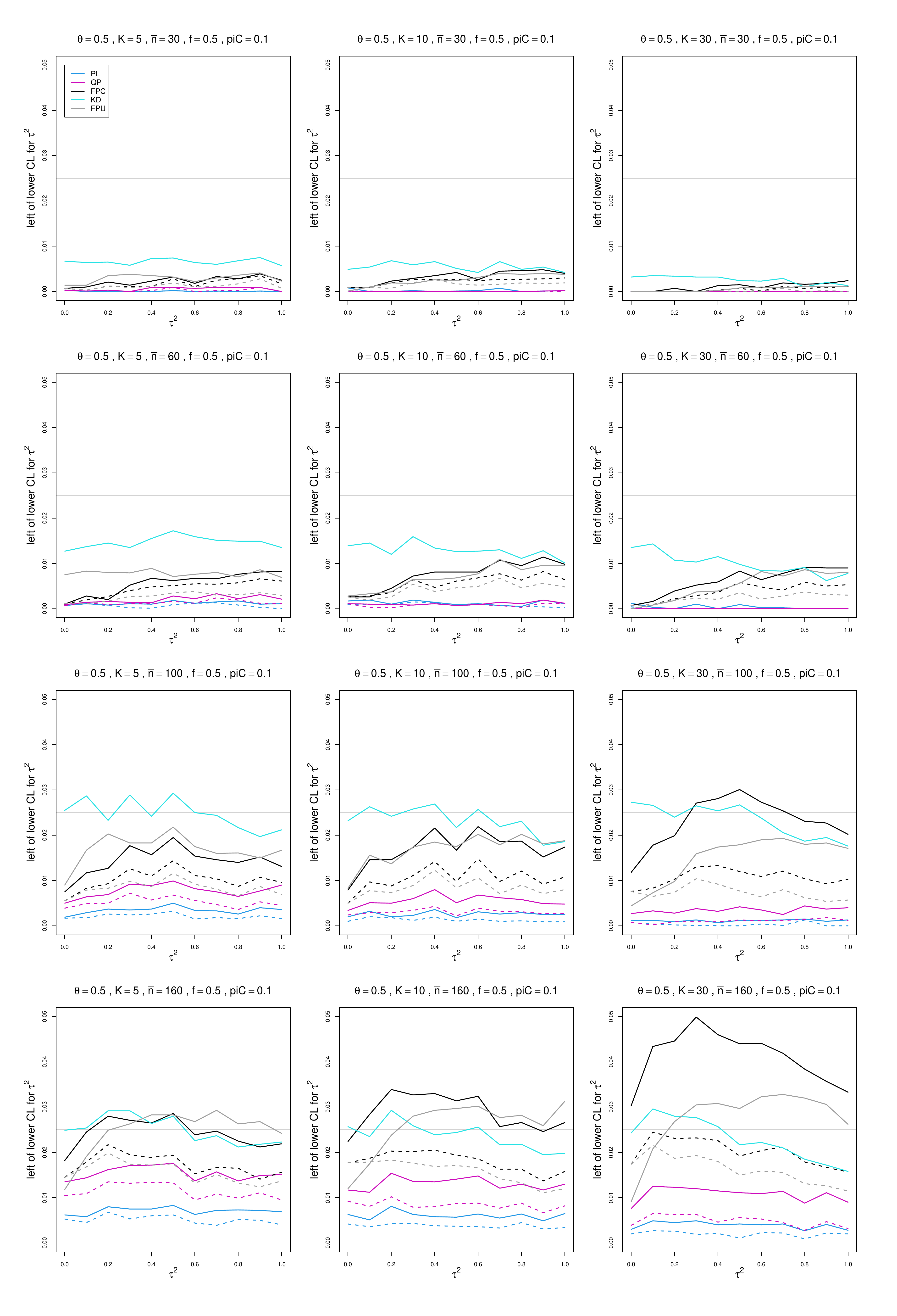}
	\caption{Miss-left probability  of  PL, QP, KD, FPC, and FPU 95\% confidence intervals for between-study variance of LOR vs $\tau^2$, for unequal sample sizes $\bar{n}=30,\;60,\;100$ and $160$, $p_{iC} = .1$, $\theta=0.5$ and  $f=0.5$.   Solid lines: PL, QP, and FPC \lq\lq only", FPU model-based, and KD. Dashed lines: PL, QP, and FPC \lq\lq always" and FPU na\"{i}ve.  }
	\label{PlotCovLeftOfTau2_piC_01theta=0.5_LOR_unequal_sample_sizes}
\end{figure}

\begin{figure}[ht]
	\centering
	\includegraphics[scale=0.33]{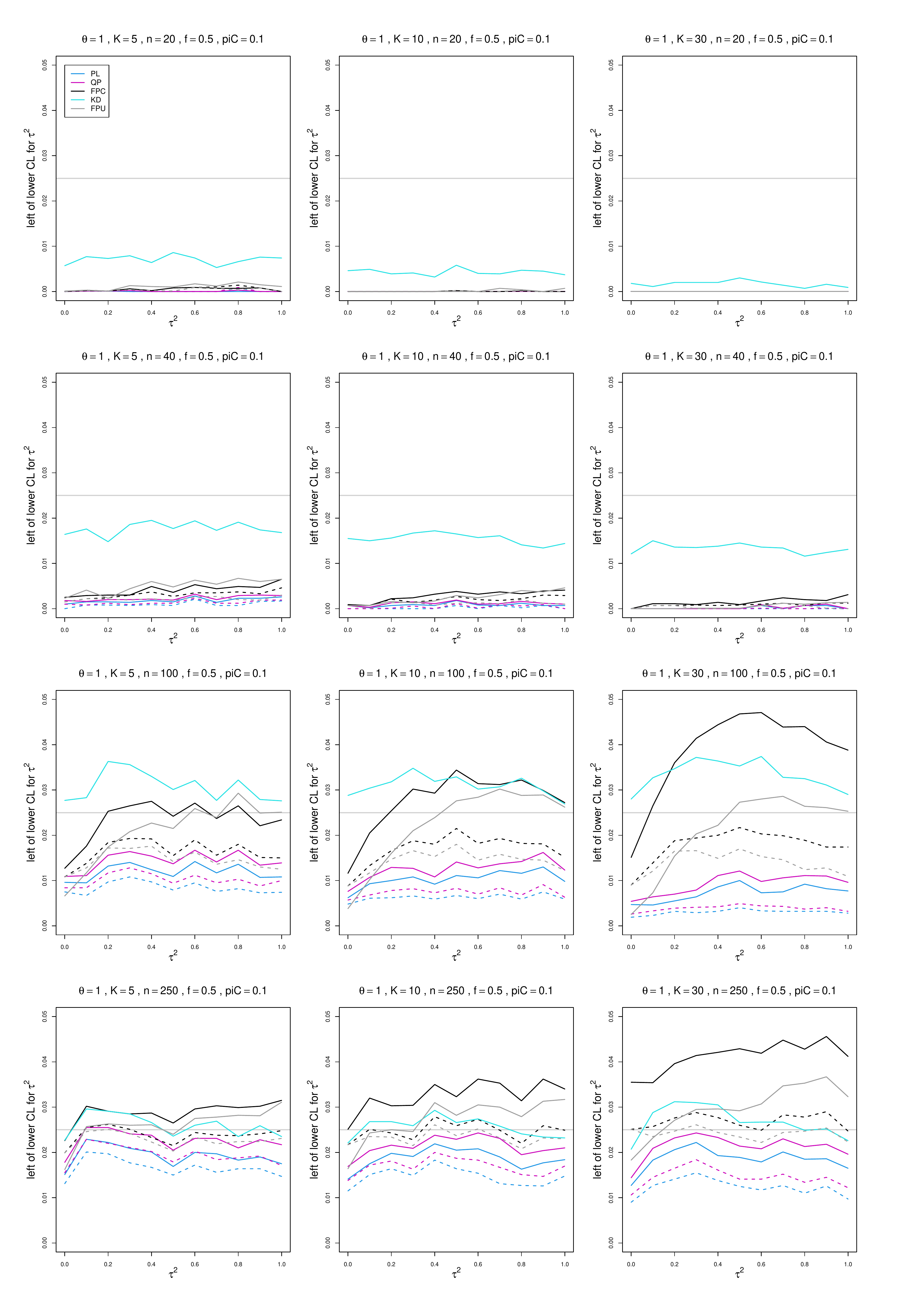}
	\caption{Miss-left probability  of  PL, QP, KD, FPC, and FPU 95\% confidence intervals for between-study variance of LOR vs $\tau^2$, for equal sample sizes $n=20,\;40,\;100$ and $250$, $p_{iC} = .1$, $\theta=1$ and  $f=0.5$.   Solid lines: PL, QP, and FPC \lq\lq only", FPU model-based, and KD. Dashed lines: PL, QP, and FPC \lq\lq always" and FPU na\"{i}ve.  }
	\label{PlotCovLeftOfTau2_piC_01theta=1_LOR_equal_sample_sizes}
\end{figure}

\begin{figure}[ht]
	\centering
	\includegraphics[scale=0.33]{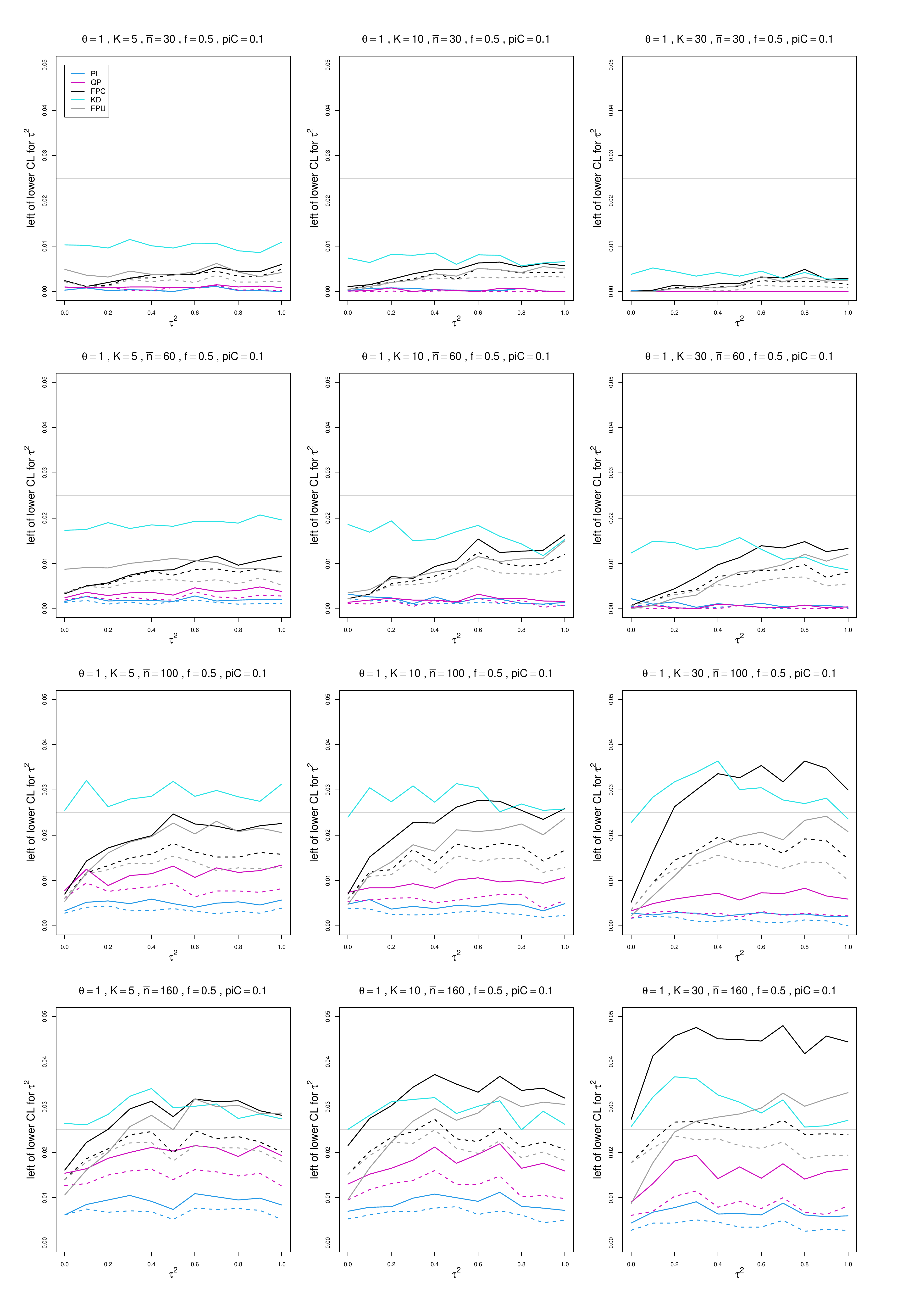}
	\caption{Miss-left probability  of  PL, QP, KD, FPC, and FPU 95\% confidence intervals for between-study variance of LOR vs $\tau^2$, for unequal sample sizes $\bar{n}=30,\;60,\;100$ and $160$, $p_{iC} = .1$, $\theta=1$ and  $f=0.5$.   Solid lines: PL, QP, and FPC \lq\lq only", FPU model-based, and KD. Dashed lines: PL, QP, and FPC \lq\lq always" and FPU na\"{i}ve.  }
	\label{PlotCovLeftOfTau2_piC_01theta=1_LOR_unequal_sample_sizes}
\end{figure}

\begin{figure}[ht]
	\centering
	\includegraphics[scale=0.33]{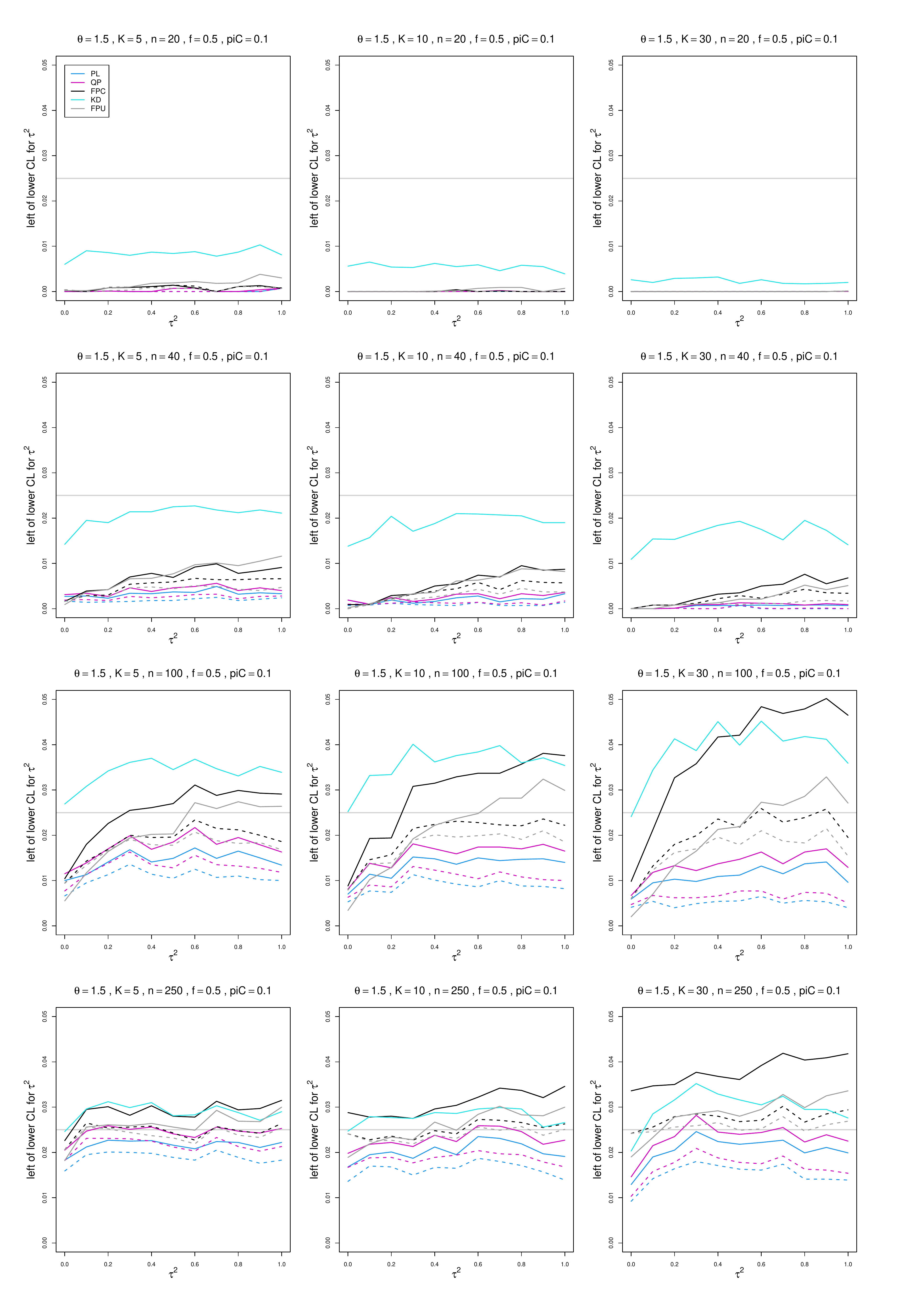}
	\caption{Miss-left probability  of  PL, QP, KD, FPC, and FPU 95\% confidence intervals for between-study variance of LOR vs $\tau^2$, for equal sample sizes $n=20,\;40,\;100$ and $250$, $p_{iC} = .1$, $\theta=1.5$ and  $f=0.5$.   Solid lines: PL, QP, and FPC \lq\lq only", FPU model-based, and KD. Dashed lines: PL, QP, and FPC \lq\lq always" and FPU na\"{i}ve.  }
	\label{PlotCovLeftOfTau2_piC_01theta=1.5_LOR_equal_sample_sizes}
\end{figure}

\begin{figure}[ht]
	\centering
	\includegraphics[scale=0.33]{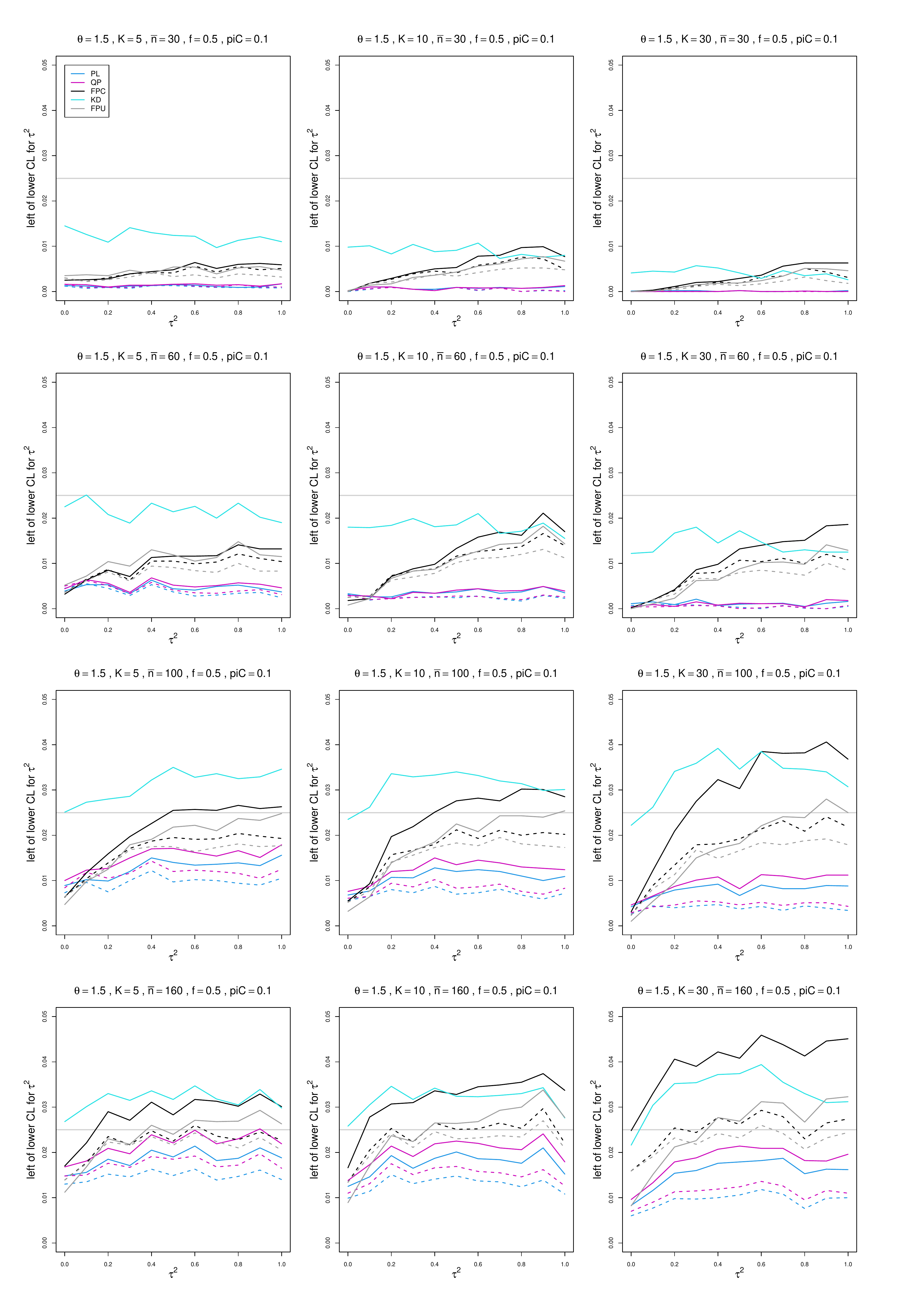}
	\caption{Miss-left probability  of  PL, QP, KD, FPC, and FPU 95\% confidence intervals for between-study variance of LOR vs $\tau^2$, for unequal sample sizes $\bar{n}=30,\;60,\;100$ and $160$, $p_{iC} = .1$, $\theta=1.5$ and  $f=0.5$.   Solid lines: PL, QP, and FPC \lq\lq only", FPU model-based, and KD. Dashed lines: PL, QP, and FPC \lq\lq always" and FPU na\"{i}ve.  }
	\label{PlotCovLeftOfTau2_piC_01theta=1.5_LOR_unequal_sample_sizes}
\end{figure}

\begin{figure}[ht]
	\centering
	\includegraphics[scale=0.33]{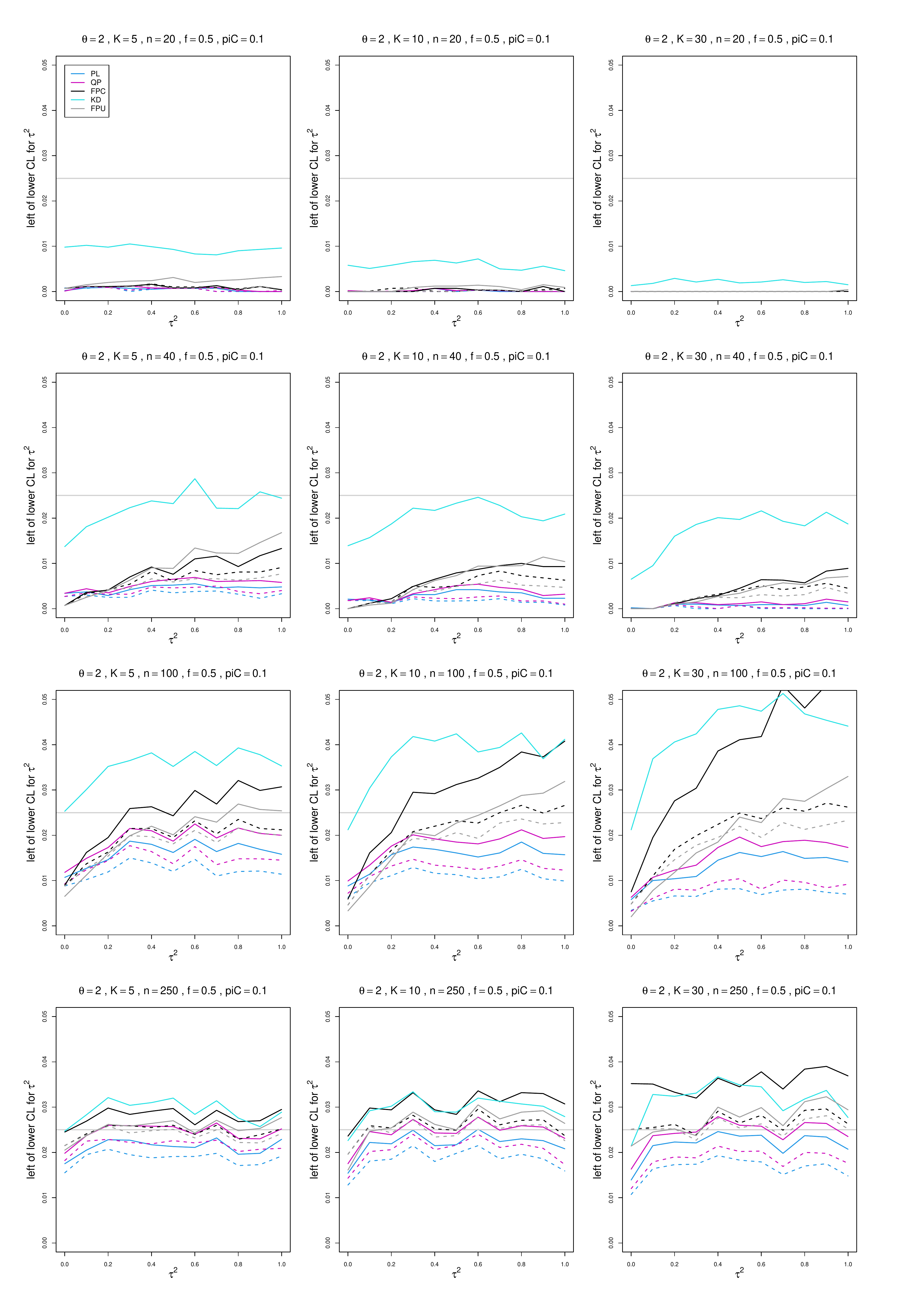}
	\caption{Miss-left probability  of  PL, QP, KD, FPC, and FPU 95\% confidence intervals for between-study variance of LOR vs $\tau^2$, for equal sample sizes $n=20,\;40,\;100$ and $250$, $p_{iC} = .1$, $\theta=2$ and  $f=0.5$.   Solid lines: PL, QP, and FPC \lq\lq only", FPU model-based, and KD. Dashed lines: PL, QP, and FPC \lq\lq always" and FPU na\"{i}ve.  }
	\label{PlotCovLeftOfTau2_piC_01theta=2_LOR_equal_sample_sizes}
\end{figure}

\begin{figure}[ht]
	\centering
	\includegraphics[scale=0.33]{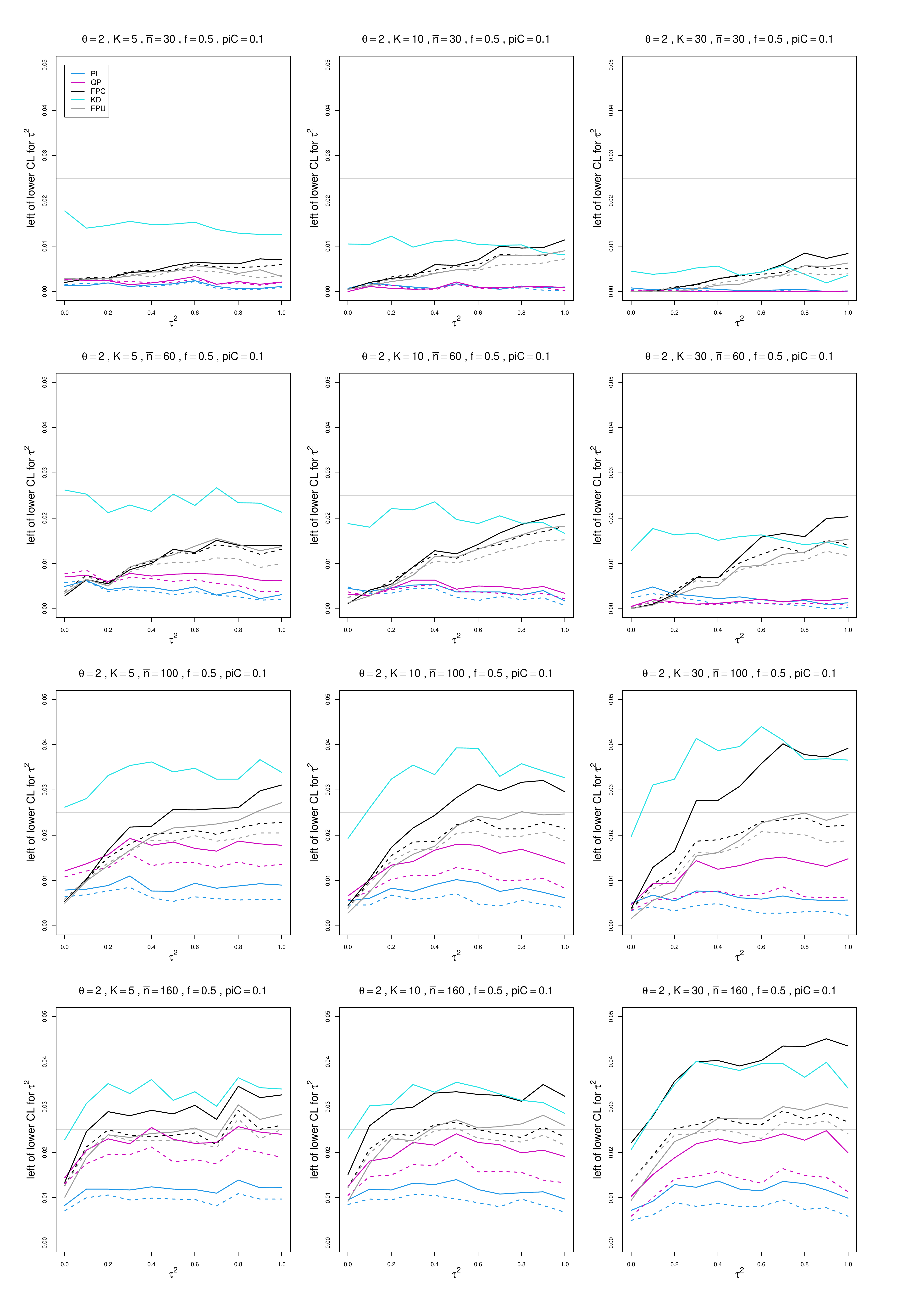}
	\caption{Miss-left probability  of  PL, QP, KD, FPC, and FPU 95\% confidence intervals for between-study variance of LOR vs $\tau^2$, for unequal sample sizes $\bar{n}=30,\;60,\;100$ and $160$, $p_{iC} = .1$, $\theta=2$ and  $f=0.5$.   Solid lines: PL, QP, and FPC \lq\lq only", FPU model-based, and KD. Dashed lines: PL, QP, and FPC \lq\lq always" and FPU na\"{i}ve.  }
	\label{PlotCovLeftOfTau2_piC_01theta=2_LOR_unequal_sample_sizes}
\end{figure}

\begin{figure}[ht]
	\centering
	\includegraphics[scale=0.33]{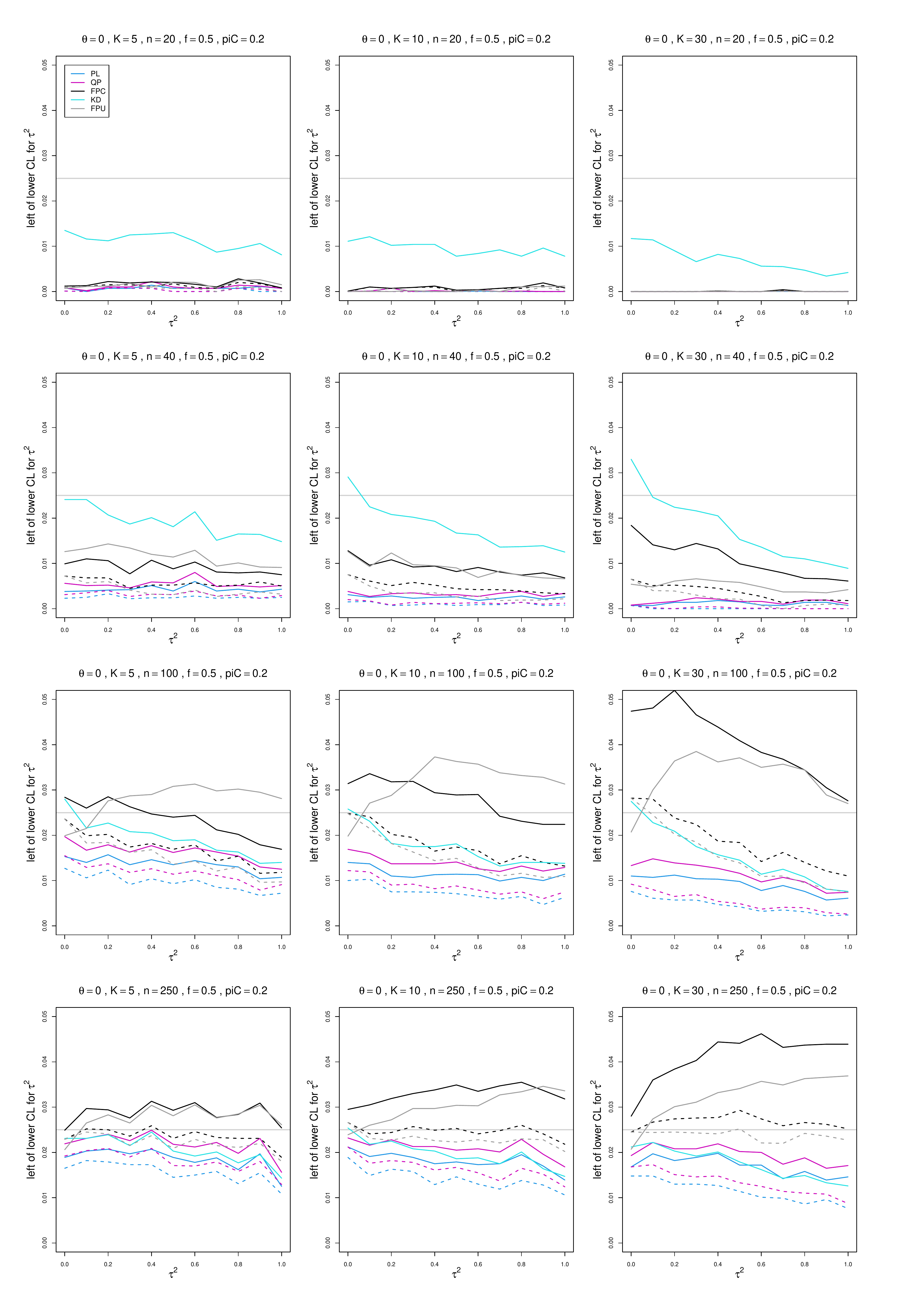}
	\caption{Miss-left probability  of  PL, QP, KD, FPC, and FPU 95\% confidence intervals for between-study variance of LOR vs $\tau^2$, for equal sample sizes $n=20,\;40,\;100$ and $250$, $p_{iC} = .2$, $\theta=0$ and  $f=0.5$.   Solid lines: PL, QP, and FPC \lq\lq only", FPU model-based, and KD. Dashed lines: PL, QP, and FPC \lq\lq always" and FPU na\"{i}ve.  }
	\label{PlotCovLeftOfTau2_piC_02theta=0_LOR_equal_sample_sizes}
\end{figure}

\begin{figure}[ht]
	\centering
	\includegraphics[scale=0.33]{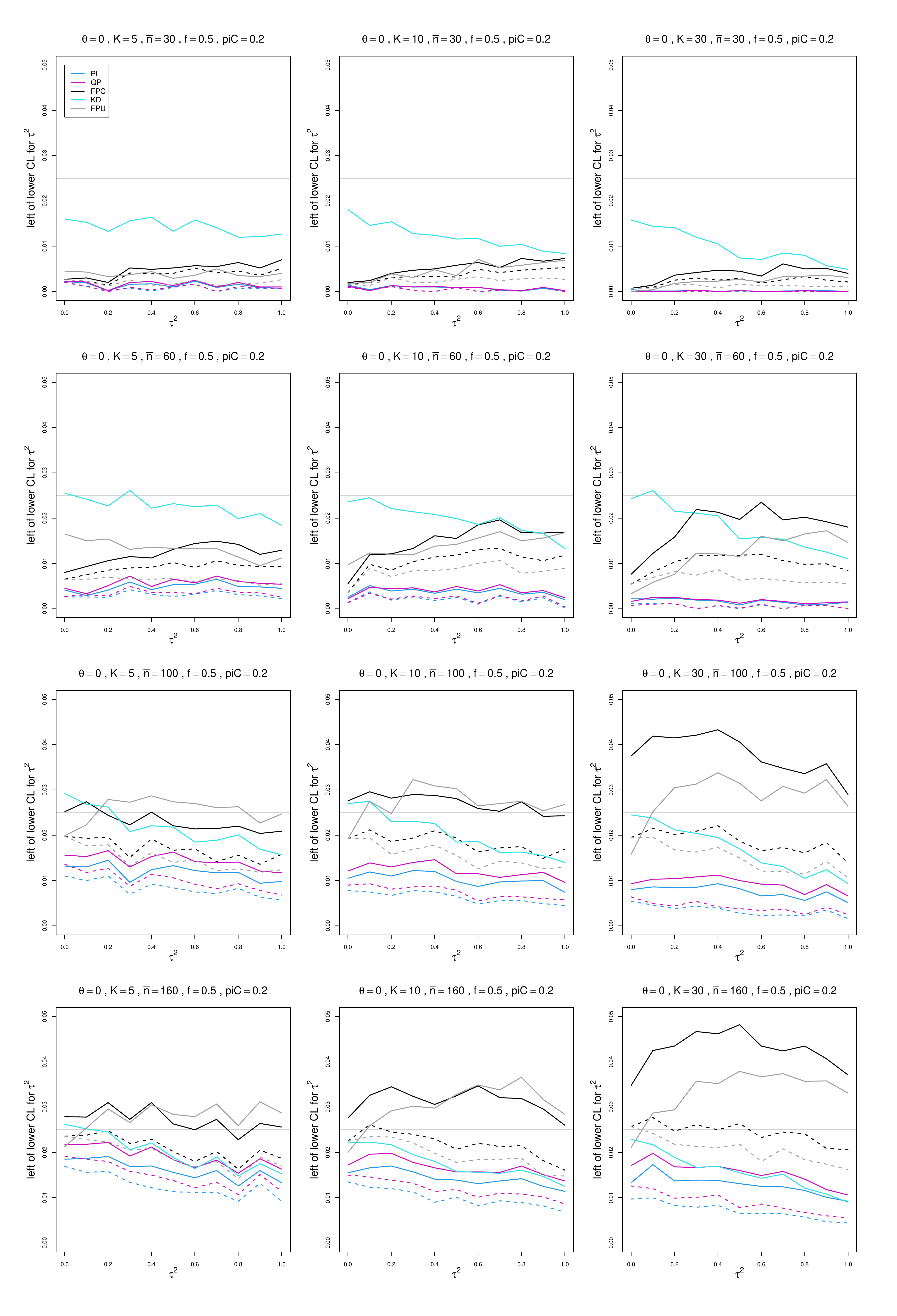}
	\caption{Miss-left probability  of  PL, QP, KD, FPC, and FPU 95\% confidence intervals for between-study variance of LOR vs $\tau^2$, for unequal sample sizes $\bar{n}=30,\;60,\;100$ and $160$, $p_{iC} = .2$, $\theta=0$ and  $f=0.5$.   Solid lines: PL, QP, and FPC \lq\lq only", FPU model-based, and KD. Dashed lines: PL, QP, and FPC \lq\lq always" and FPU na\"{i}ve.  }
	\label{PlotCovLeftOfTau2_piC_02theta=0_LOR_unequal_sample_sizes}
\end{figure}

\begin{figure}[ht]
	\centering
	\includegraphics[scale=0.33]{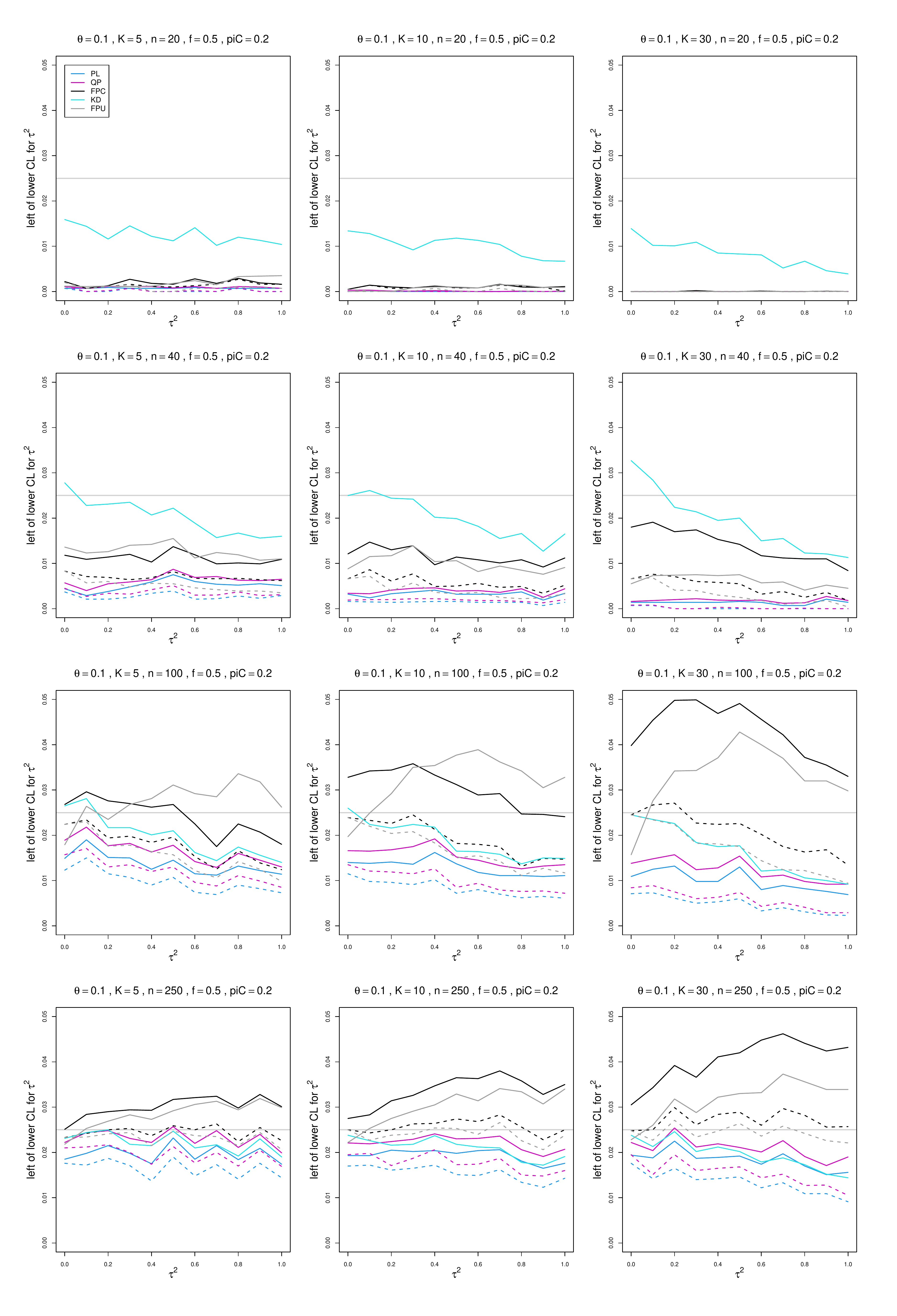}
	\caption{Miss-left probability  of  PL, QP, KD, FPC, and FPU 95\% confidence intervals for between-study variance of LOR vs $\tau^2$, for equal sample sizes $n=20,\;40,\;100$ and $250$, $p_{iC} = .2$, $\theta=0.1$ and  $f=0.5$.   Solid lines: PL, QP, and FPC \lq\lq only", FPU model-based, and KD. Dashed lines: PL, QP, and FPC \lq\lq always" and FPU na\"{i}ve.  }
	\label{PlotCovLeftOfTau2_piC_02theta=0.1_LOR_equal_sample_sizes}
\end{figure}

\begin{figure}[ht]
	\centering
	\includegraphics[scale=0.33]{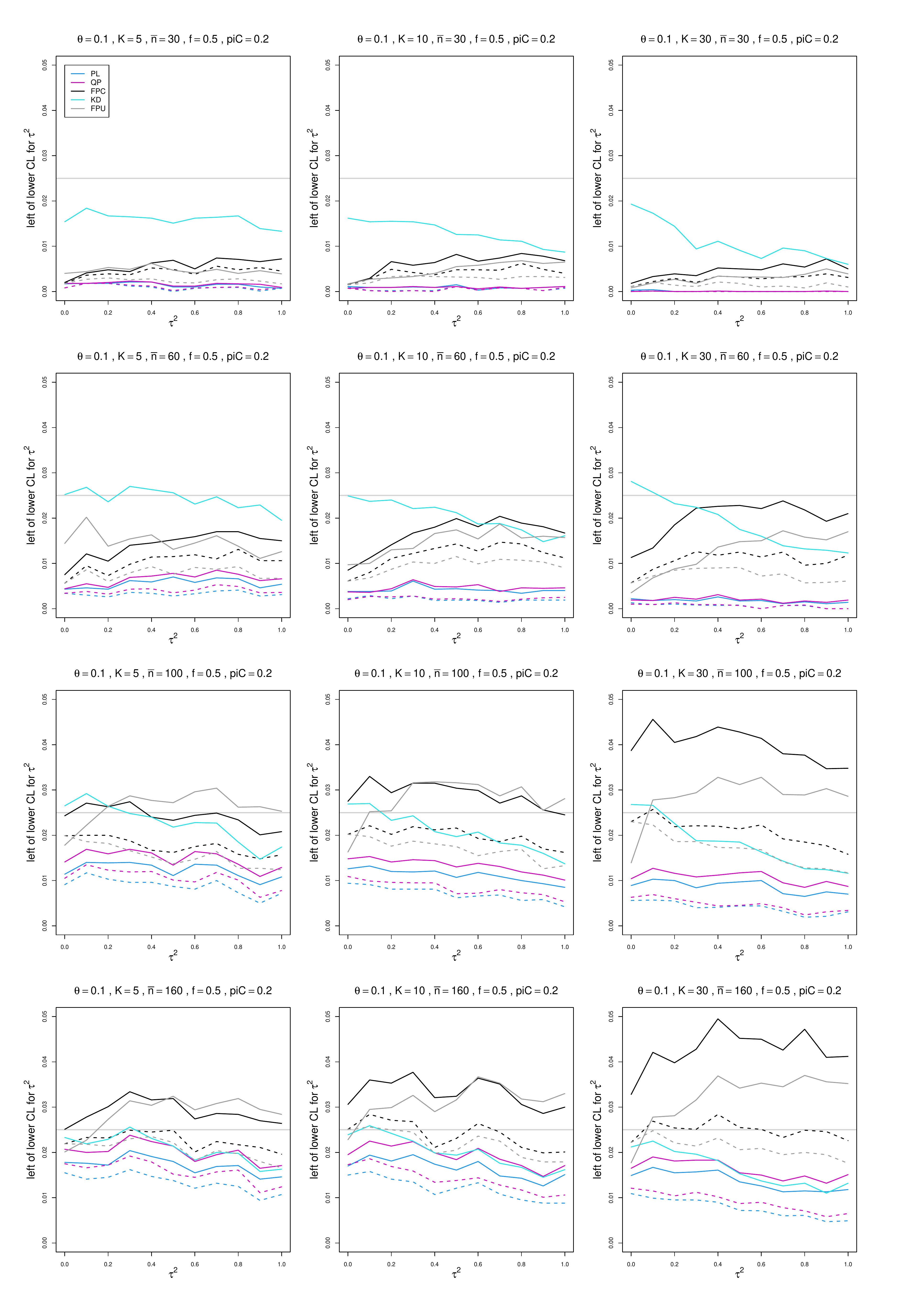}
	\caption{Miss-left probability  of  PL, QP, KD, FPC, and FPU 95\% confidence intervals for between-study variance of LOR vs $\tau^2$, for unequal sample sizes $\bar{n}=30,\;60,\;100$ and $160$, $p_{iC} = .2$, $\theta=0.1$ and  $f=0.5$.   Solid lines: PL, QP, and FPC \lq\lq only", FPU model-based, and KD. Dashed lines: PL, QP, and FPC \lq\lq always" and FPU na\"{i}ve.  }
	\label{PlotCovLeftOfTau2_piC_02theta=0.1_LOR_unequal_sample_sizes}
\end{figure}

\begin{figure}[ht]
	\centering
	\includegraphics[scale=0.33]{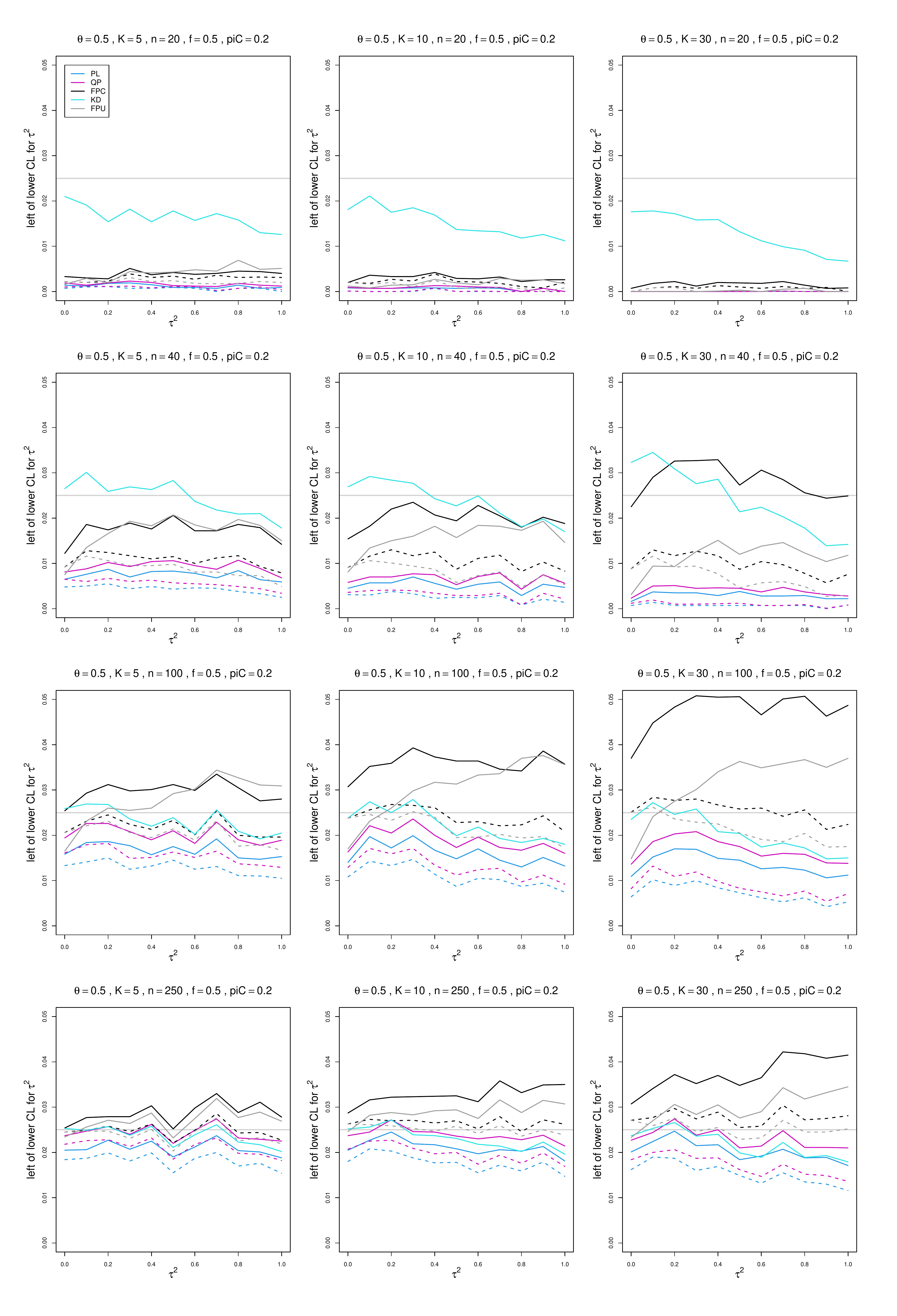}
	\caption{Miss-left probability  of  PL, QP, KD, FPC, and FPU 95\% confidence intervals for between-study variance of LOR vs $\tau^2$, for equal sample sizes $n=20,\;40,\;100$ and $250$, $p_{iC} = .2$, $\theta=0.5$ and  $f=0.5$.   Solid lines: PL, QP, and FPC \lq\lq only", FPU model-based, and KD. Dashed lines: PL, QP, and FPC \lq\lq always" and FPU na\"{i}ve.   }
	\label{PlotCovLeftOfTau2_piC_02theta=0.5_LOR_equal_sample_sizes}
\end{figure}

\begin{figure}[ht]
	\centering
	\includegraphics[scale=0.33]{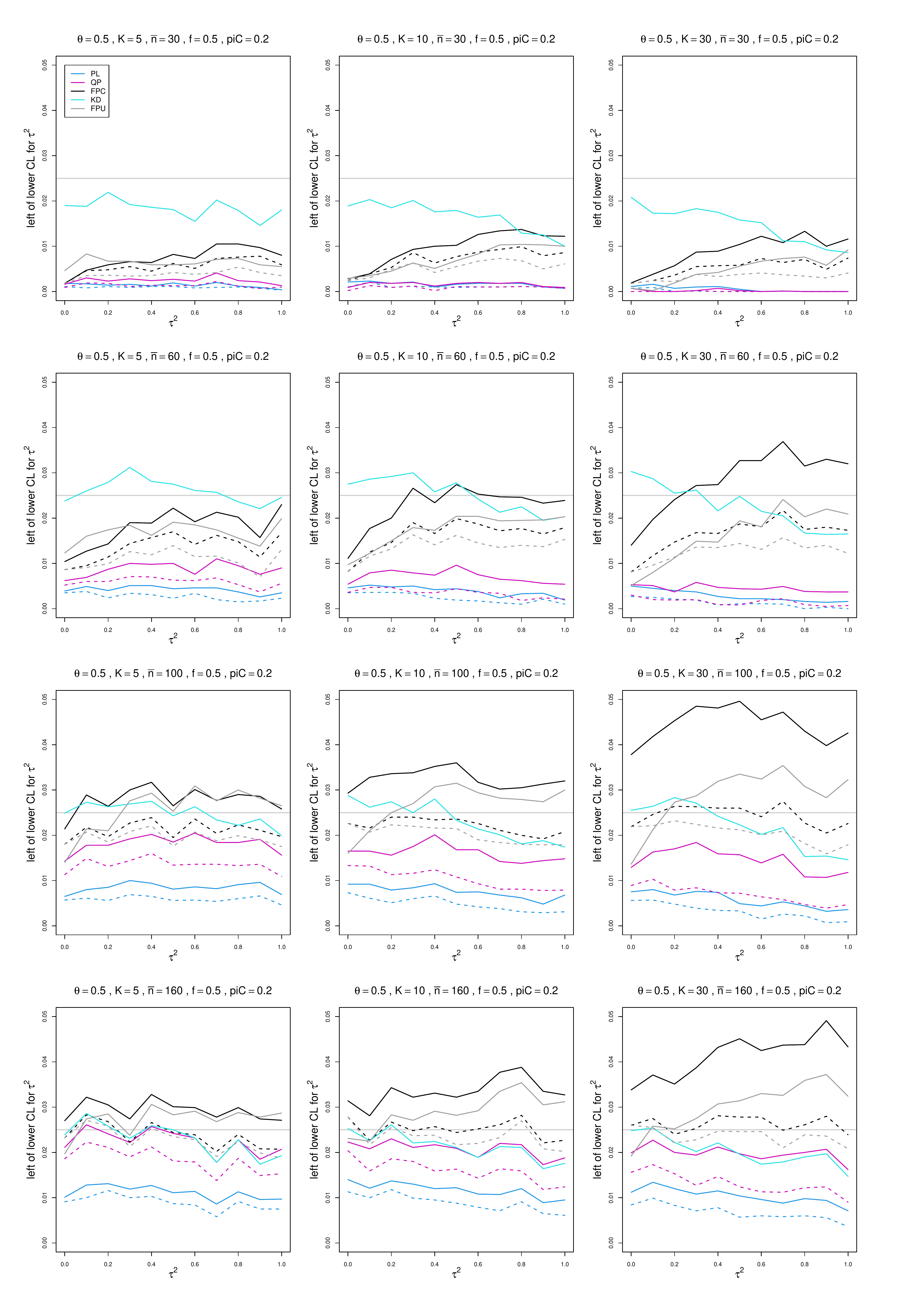}
	\caption{Miss-left probability  of  PL, QP, KD, FPC, and FPU 95\% confidence intervals for between-study variance of LOR vs $\tau^2$, for unequal sample sizes $\bar{n}=30,\;60,\;100$ and $160$, $p_{iC} = .2$, $\theta=0.5$ and  $f=0.5$.   Solid lines: PL, QP, and FPC \lq\lq only", FPU model-based, and KD. Dashed lines: PL, QP, and FPC \lq\lq always" and FPU na\"{i}ve.  }
	\label{PlotCovLeftOfTau2_piC_02theta=0.5_LOR_unequal_sample_sizes}
\end{figure}

\begin{figure}[ht]
	\centering
	\includegraphics[scale=0.33]{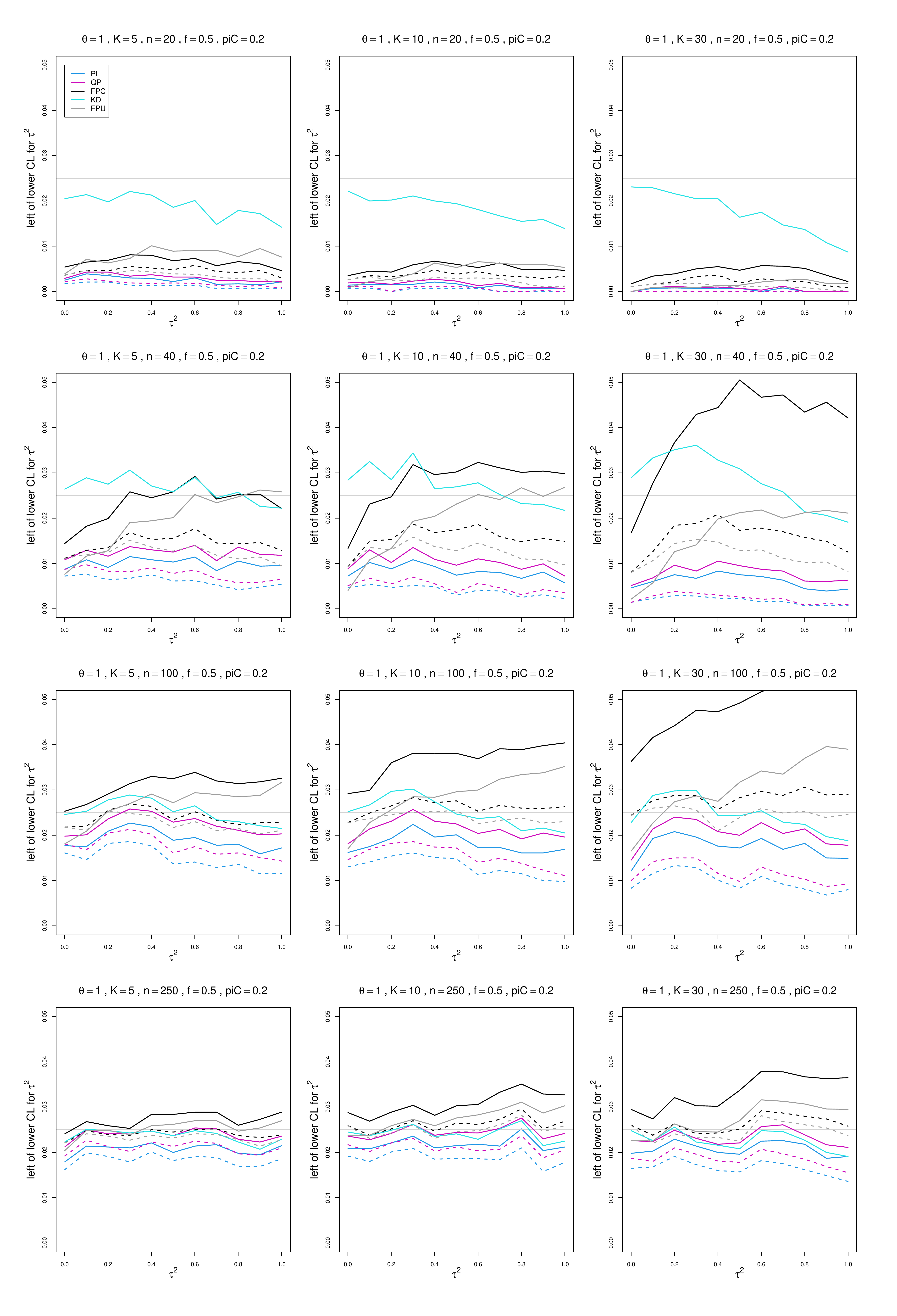}
	\caption{Miss-left probability  of  PL, QP, KD, FPC, and FPU 95\% confidence intervals for between-study variance of LOR vs $\tau^2$, for equal sample sizes $n=20,\;40,\;100$ and $250$, $p_{iC} = .2$, $\theta=1$ and  $f=0.5$.   Solid lines: PL, QP, and FPC \lq\lq only", FPU model-based, and KD. Dashed lines: PL, QP, and FPC \lq\lq always" and FPU na\"{i}ve.   }
	\label{PlotCovLeftOfTau2_piC_02theta=1_LOR_equal_sample_sizes}
\end{figure}

\begin{figure}[ht]
	\centering
	\includegraphics[scale=0.33]{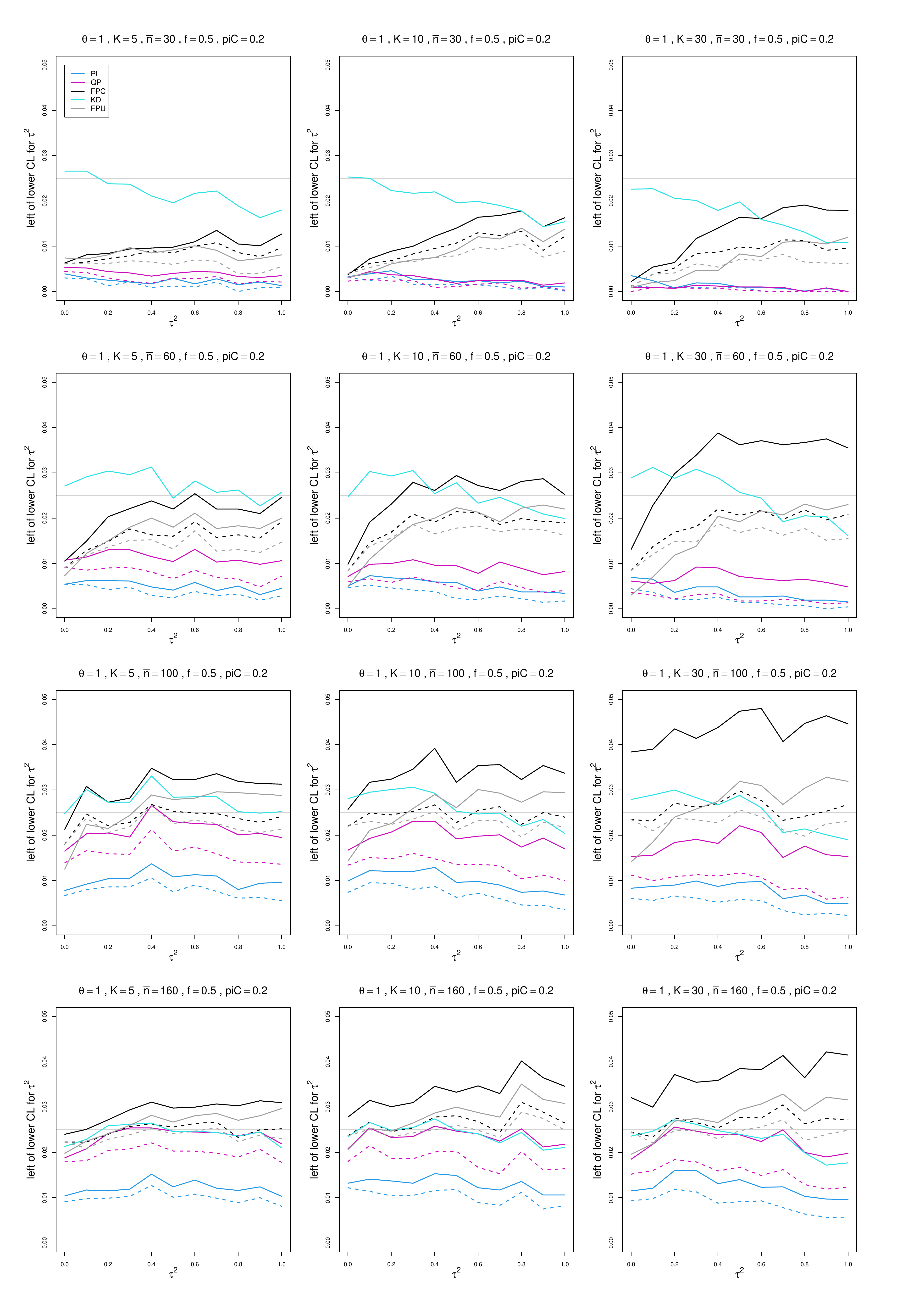}
	\caption{Miss-left probability  of  PL, QP, KD, FPC, and FPU 95\% confidence intervals for between-study variance of LOR vs $\tau^2$, for unequal sample sizes $\bar{n}=30,\;60,\;100$ and $160$, $p_{iC} = .2$, $\theta=1$ and  $f=0.5$.   Solid lines: PL, QP, and FPC \lq\lq only", FPU model-based, and KD. Dashed lines: PL, QP, and FPC \lq\lq always" and FPU na\"{i}ve.  }
	\label{PlotCovLeftOfTau2_piC_02theta=1_LOR_unequal_sample_sizes}
\end{figure}

\begin{figure}[ht]
	\centering
	\includegraphics[scale=0.33]{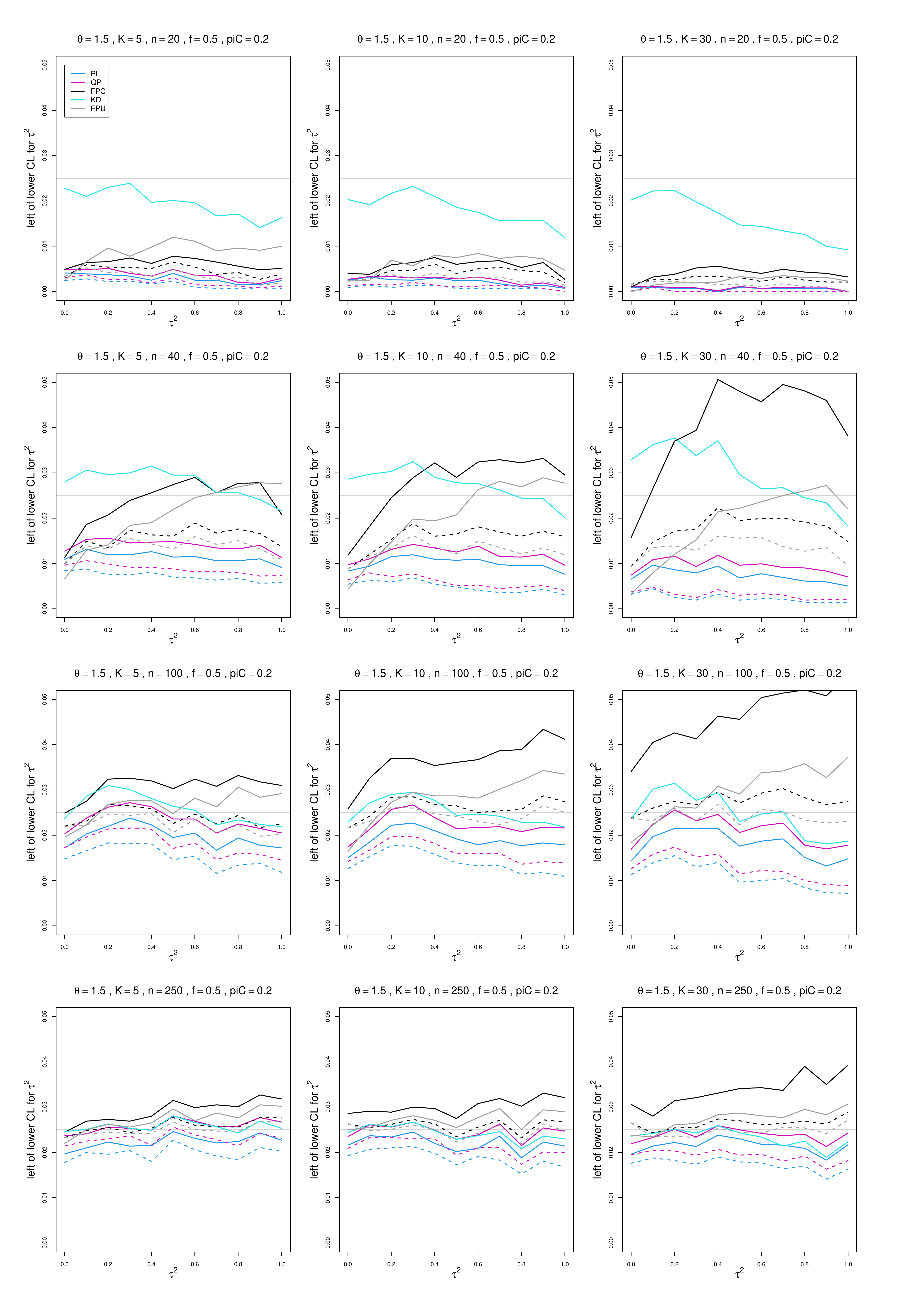}
	\caption{Miss-left probability  of  PL, QP, KD, FPC, and FPU 95\% confidence intervals for between-study variance of LOR vs $\tau^2$, for equal sample sizes $n=20,\;40,\;100$ and $250$, $p_{iC} = .2$, $\theta=1.5$ and  $f=0.5$.   Solid lines: PL, QP, and FPC \lq\lq only", FPU model-based, and KD. Dashed lines: PL, QP, and FPC \lq\lq always" and FPU na\"{i}ve.   }
	\label{PlotCovLeftOfTau2_piC_02theta=1.5_LOR_equal_sample_sizes}
\end{figure}

\begin{figure}[ht]
	\centering
	\includegraphics[scale=0.33]{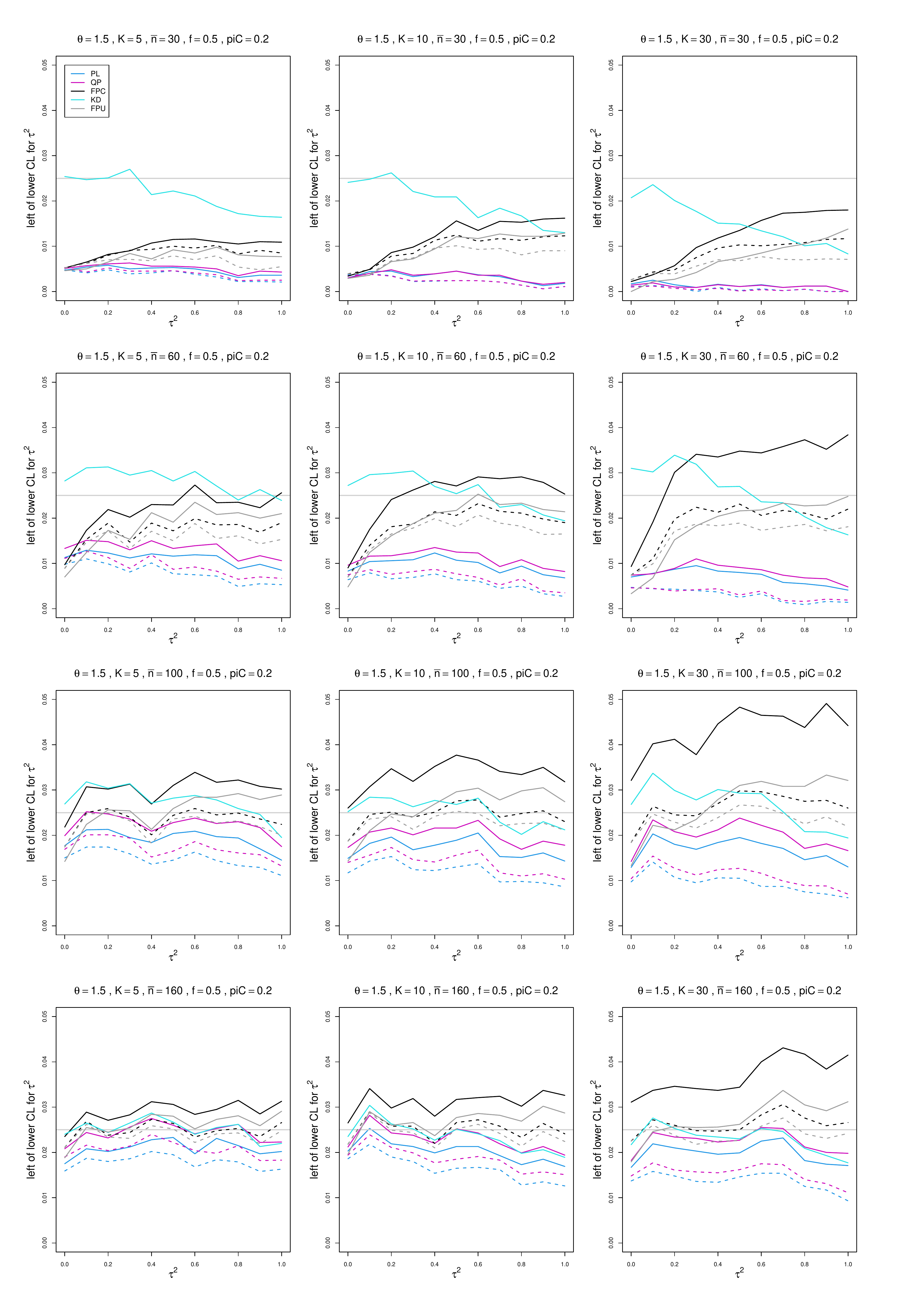}
	\caption{Miss-left probability  of  PL, QP, KD, FPC, and FPU 95\% confidence intervals for between-study variance of LOR vs $\tau^2$, for unequal sample sizes $\bar{n}=30,\;60,\;100$ and $160$, $p_{iC} = .2$, $\theta=1.5$ and  $f=0.5$.   Solid lines: PL, QP, and FPC \lq\lq only", FPU model-based, and KD. Dashed lines: PL, QP, and FPC \lq\lq always" and FPU na\"{i}ve.  }
	\label{PlotCovLeftOfTau2_piC_02theta=1.5_LOR_unequal_sample_sizes}
\end{figure}

\begin{figure}[ht]
	\centering
	\includegraphics[scale=0.33]{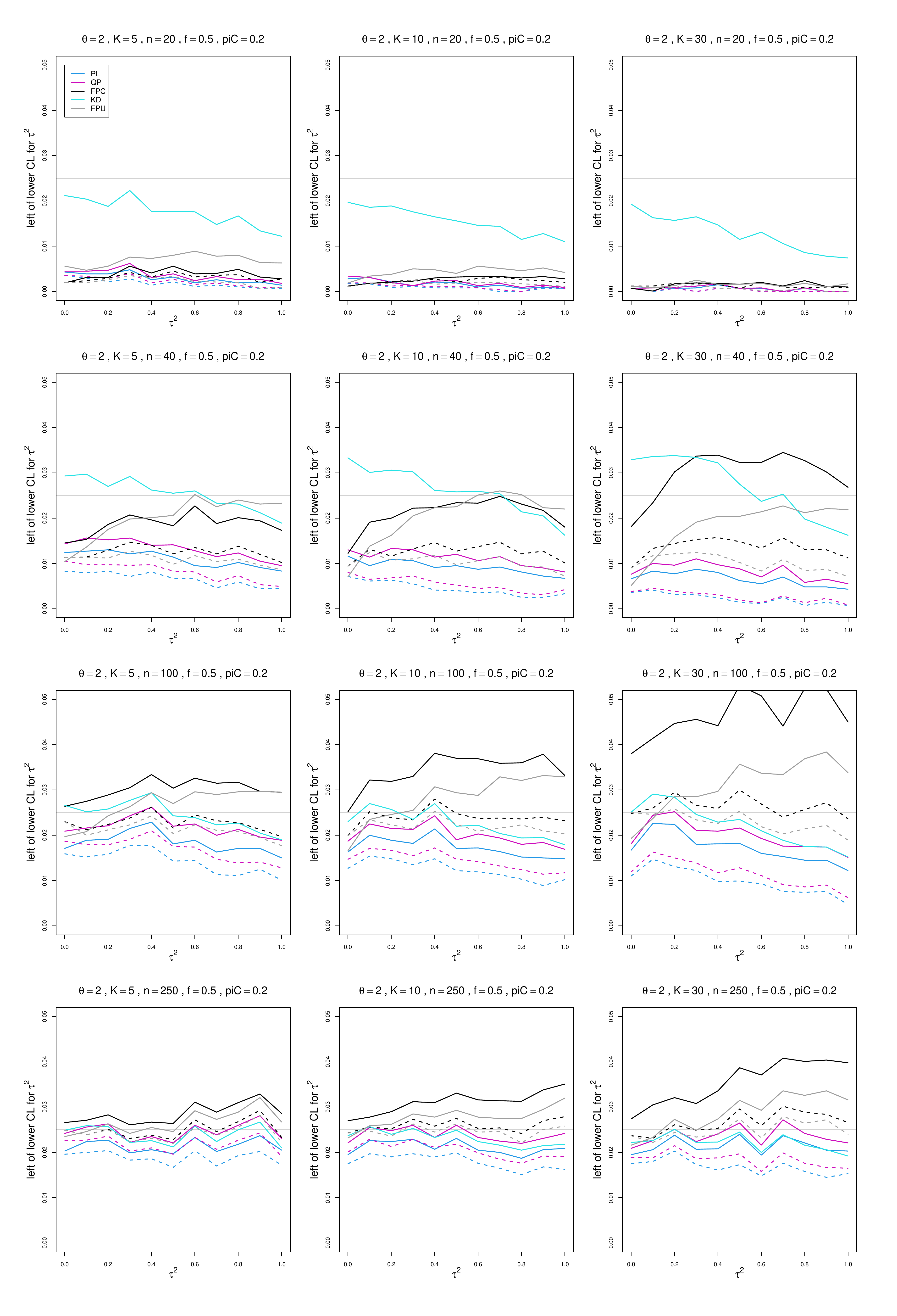}
	\caption{Miss-left probability  of  PL, QP, KD, FPC, and FPU 95\% confidence intervals for between-study variance of LOR vs $\tau^2$, for equal sample sizes $n=20,\;40,\;100$ and $250$, $p_{iC} = .2$, $\theta=2$ and  $f=0.5$.   Solid lines: PL, QP, and FPC \lq\lq only", FPU model-based, and KD. Dashed lines: PL, QP, and FPC \lq\lq always" and FPU na\"{i}ve.   }
	\label{PlotCovLeftOfTau2_piC_02theta=2_LOR_equal_sample_sizes}
\end{figure}

\begin{figure}[ht]
	\centering
	\includegraphics[scale=0.33]{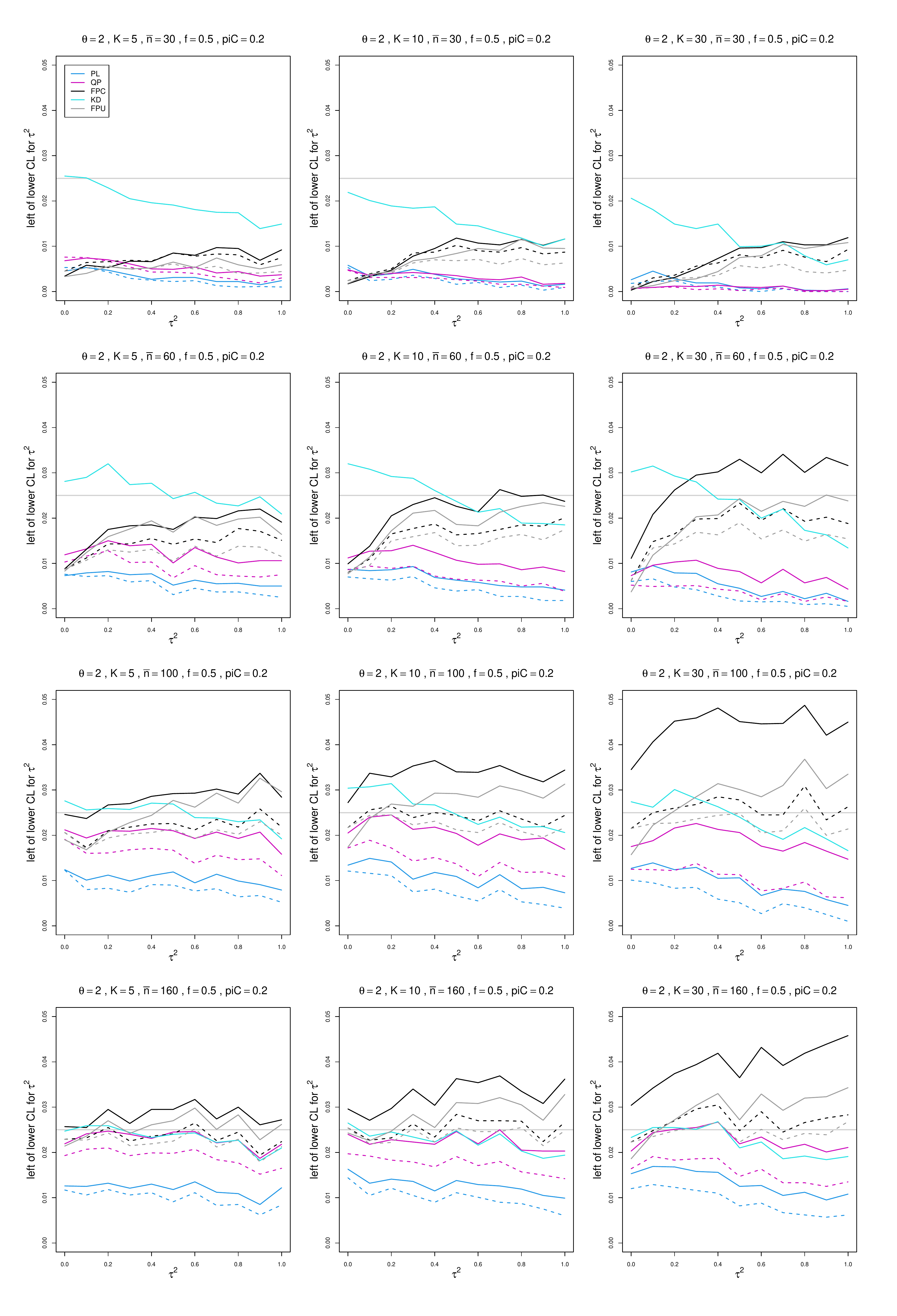}
	\caption{Miss-left probability  of  PL, QP, KD, FPC, and FPU 95\% confidence intervals for between-study variance of LOR vs $\tau^2$, for unequal sample sizes $\bar{n}=30,\;60,\;100$ and $160$, $p_{iC} = .2$, $\theta=2$ and  $f=0.5$.   Solid lines: PL, QP, and FPC \lq\lq only", FPU model-based, and KD. Dashed lines: PL, QP, and FPC \lq\lq always" and FPU na\"{i}ve.  }
	\label{PlotCovLeftOfTau2_piC_02theta=2_LOR_unequal_sample_sizes}
\end{figure}

\begin{figure}[ht]
	\centering
	\includegraphics[scale=0.33]{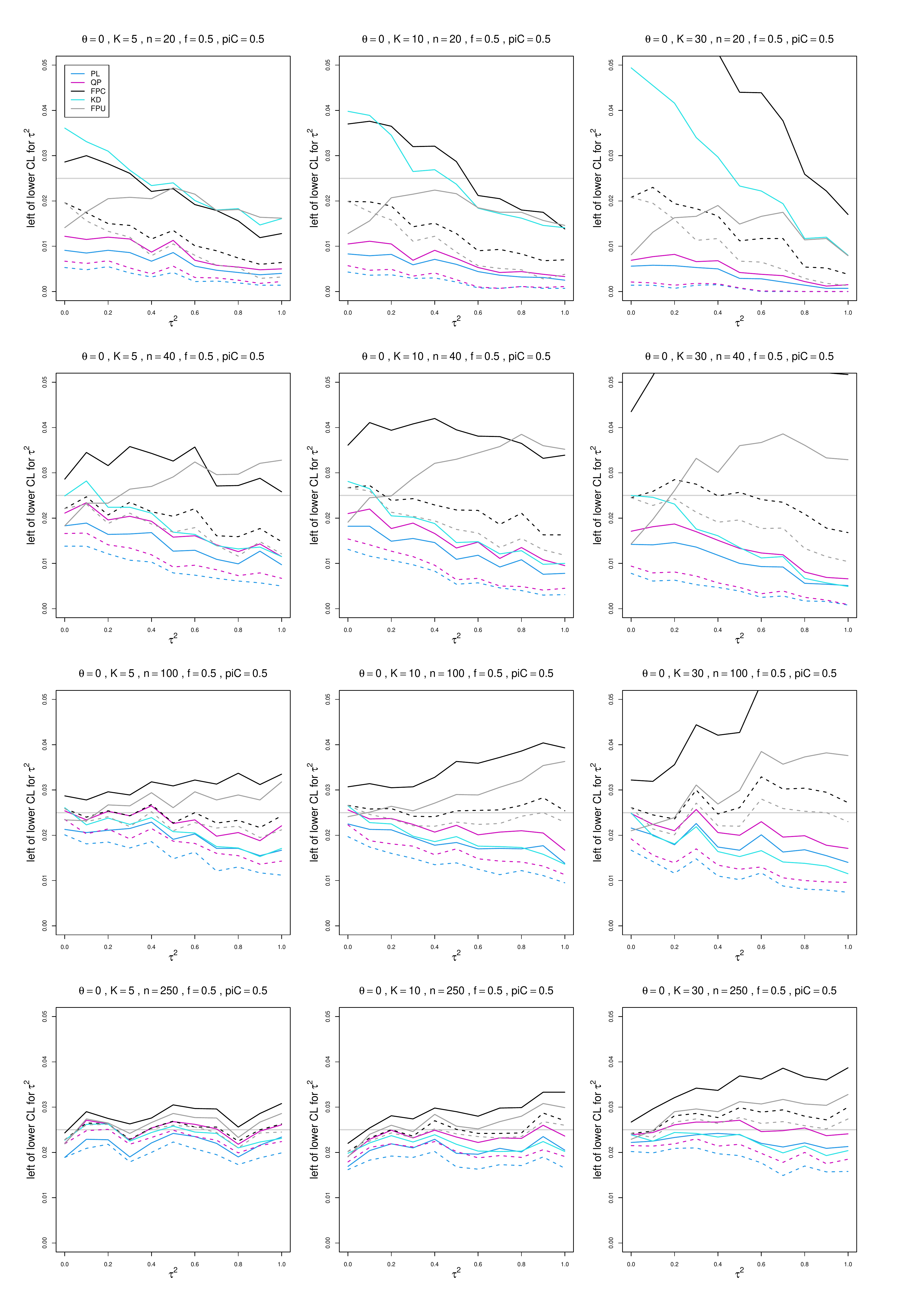}
	\caption{Miss-left probability  of  PL, QP, KD, FPC, and FPU 95\% confidence intervals for between-study variance of LOR vs $\tau^2$, for equal sample sizes $n=20,\;40,\;100$ and $250$, $p_{iC} = .5$, $\theta=0$ and  $f=0.5$.   Solid lines: PL, QP, and FPC \lq\lq only", FPU model-based, and KD. Dashed lines: PL, QP, and FPC \lq\lq always" and FPU na\"{i}ve.  }
	\label{PlotCovLeftOfTau2_piC_05theta=0_LOR_equal_sample_sizes}
\end{figure}

\begin{figure}[ht]
	\centering
	\includegraphics[scale=0.33]{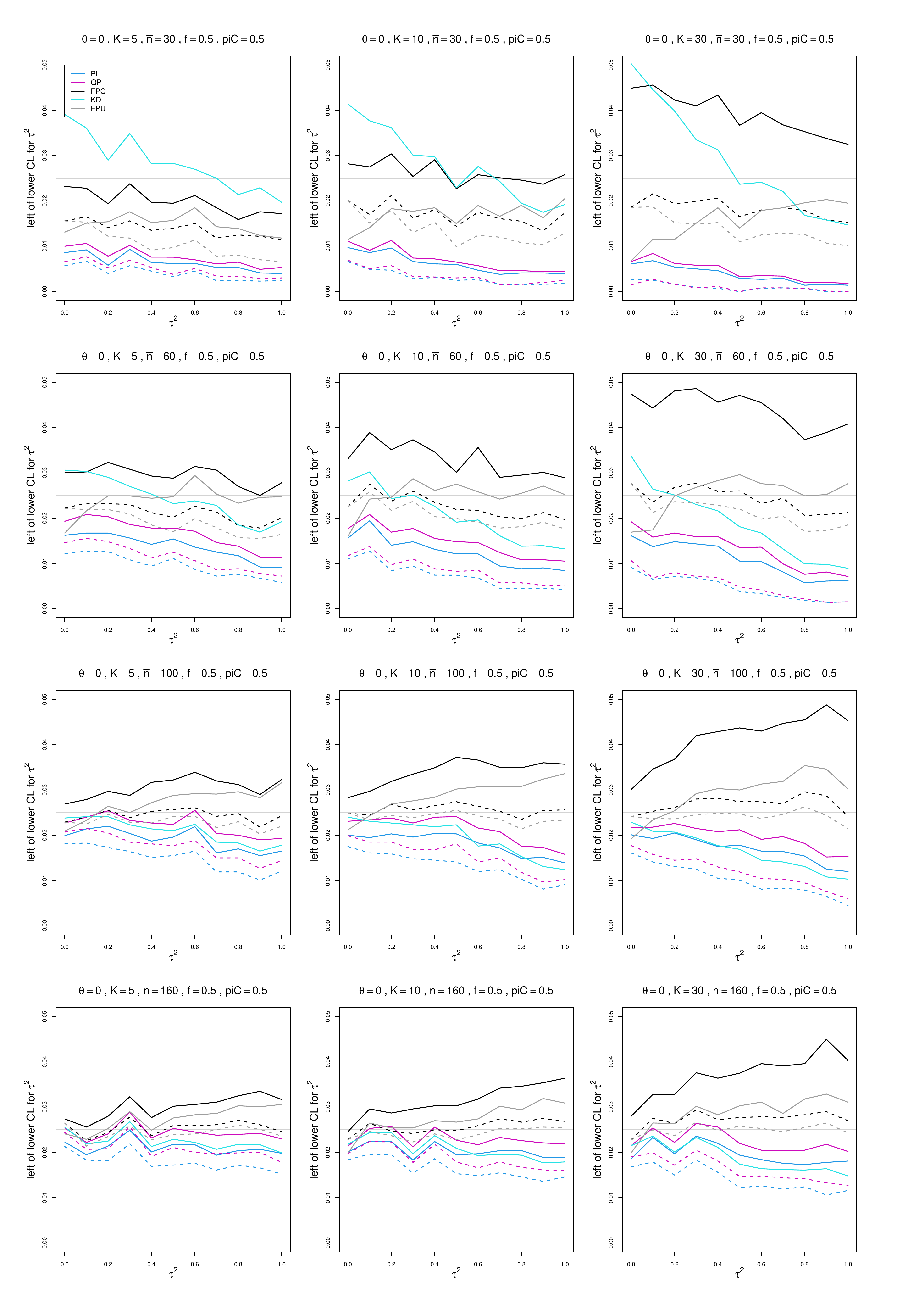}
	\caption{Miss-left probability  of  PL, QP, KD, FPC, and FPU 95\% confidence intervals for between-study variance of LOR vs $\tau^2$, for unequal sample sizes $\bar{n}=30,\;60,\;100$ and $160$, $p_{iC} = .5$, $\theta=0$ and  $f=0.5$.   Solid lines: PL, QP, and FPC \lq\lq only", FPU model-based, and KD. Dashed lines: PL, QP, and FPC \lq\lq always" and FPU na\"{i}ve.  }
	\label{PlotCovLeftOfTau2_piC_05theta=0_LOR_unequal_sample_sizes}
\end{figure}

\begin{figure}[ht]
	\centering
	\includegraphics[scale=0.33]{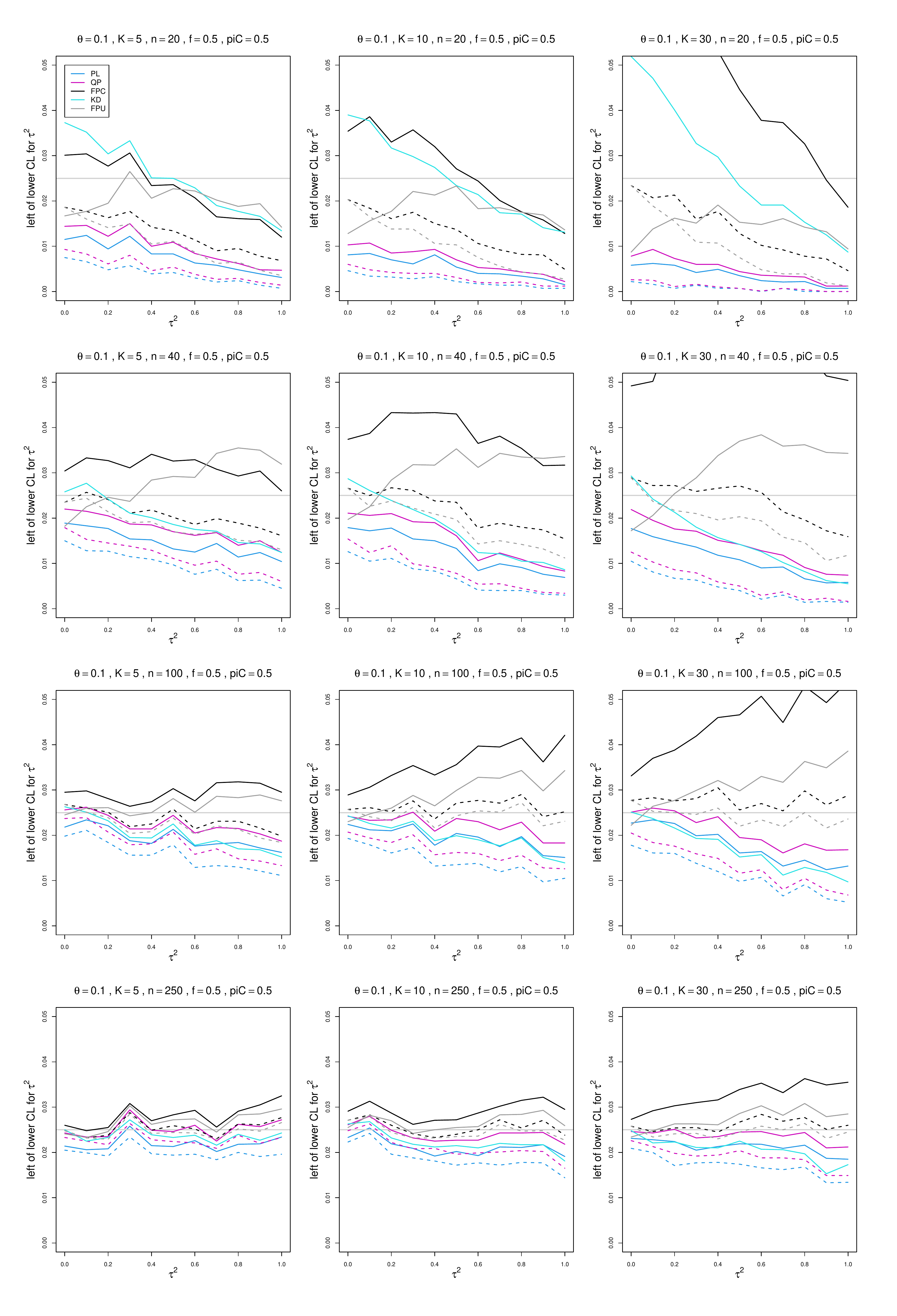}
	\caption{Miss-left probability  of  PL, QP, KD, FPC, and FPU 95\% confidence intervals for between-study variance of LOR vs $\tau^2$, for equal sample sizes $n=20,\;40,\;100$ and $250$, $p_{iC} = .5$, $\theta=0.1$ and  $f=0.5$.   Solid lines: PL, QP, and FPC \lq\lq only", FPU model-based, and KD. Dashed lines: PL, QP, and FPC \lq\lq always" and FPU na\"{i}ve.  }
	\label{PlotCovLeftOfTau2_piC_05theta=0.1_LOR_equal_sample_sizes}
\end{figure}

\begin{figure}[ht]
	\centering
	\includegraphics[scale=0.33]{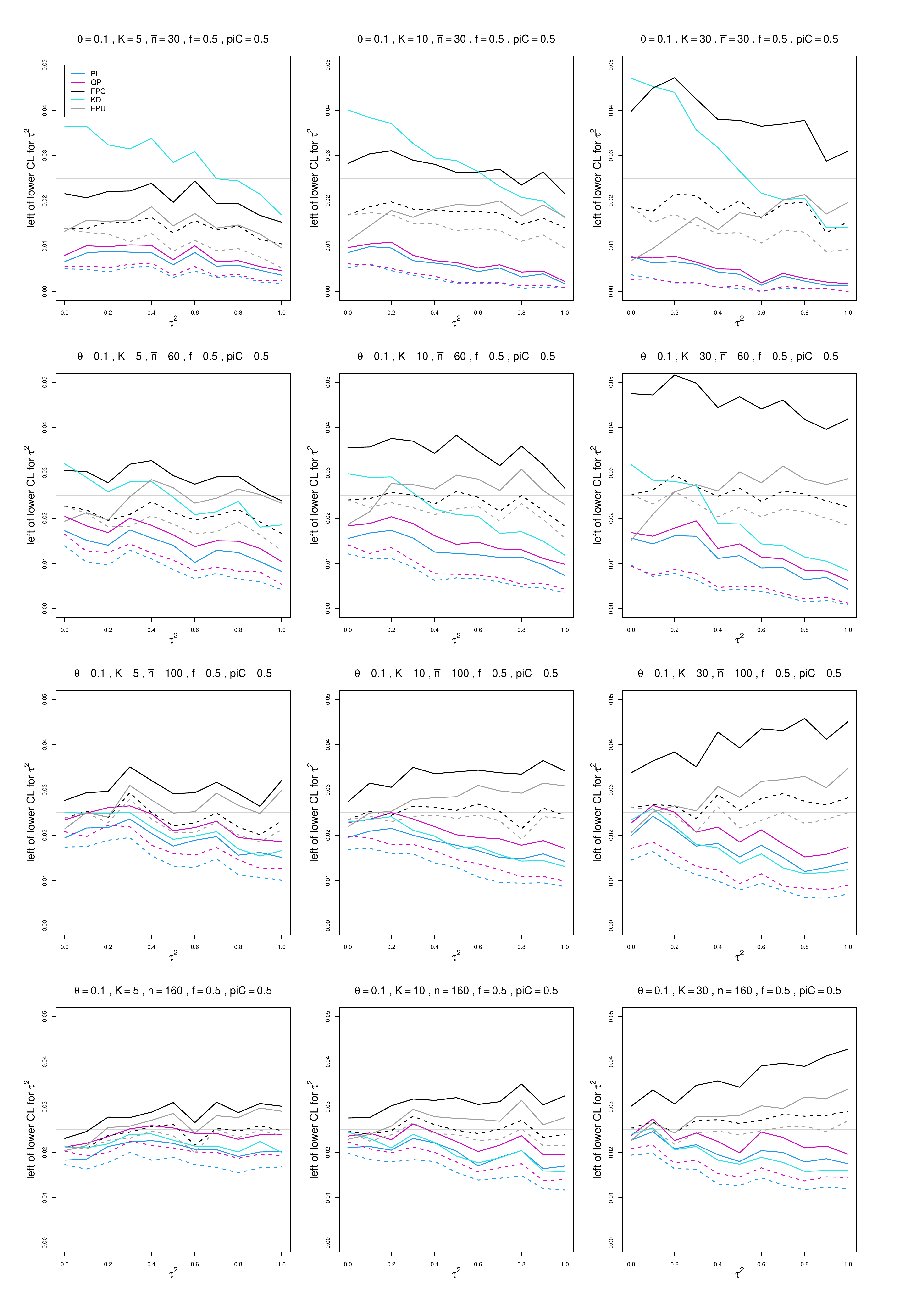}
	\caption{Miss-left probability  of  PL, QP, KD, FPC, and FPU 95\% confidence intervals for between-study variance of LOR vs $\tau^2$, for unequal sample sizes $\bar{n}=30,\;60,\;100$ and $160$, $p_{iC} = .5$, $\theta=0.1$ and  $f=0.5$.   Solid lines: PL, QP, and FPC \lq\lq only", FPU model-based, and KD. Dashed lines: PL, QP, and FPC \lq\lq always" and FPU na\"{i}ve.  }
	\label{PlotCovLeftOfTau2_piC_05theta=0.1_LOR_unequal_sample_sizes}
\end{figure}

\begin{figure}[ht]
	\centering
	\includegraphics[scale=0.33]{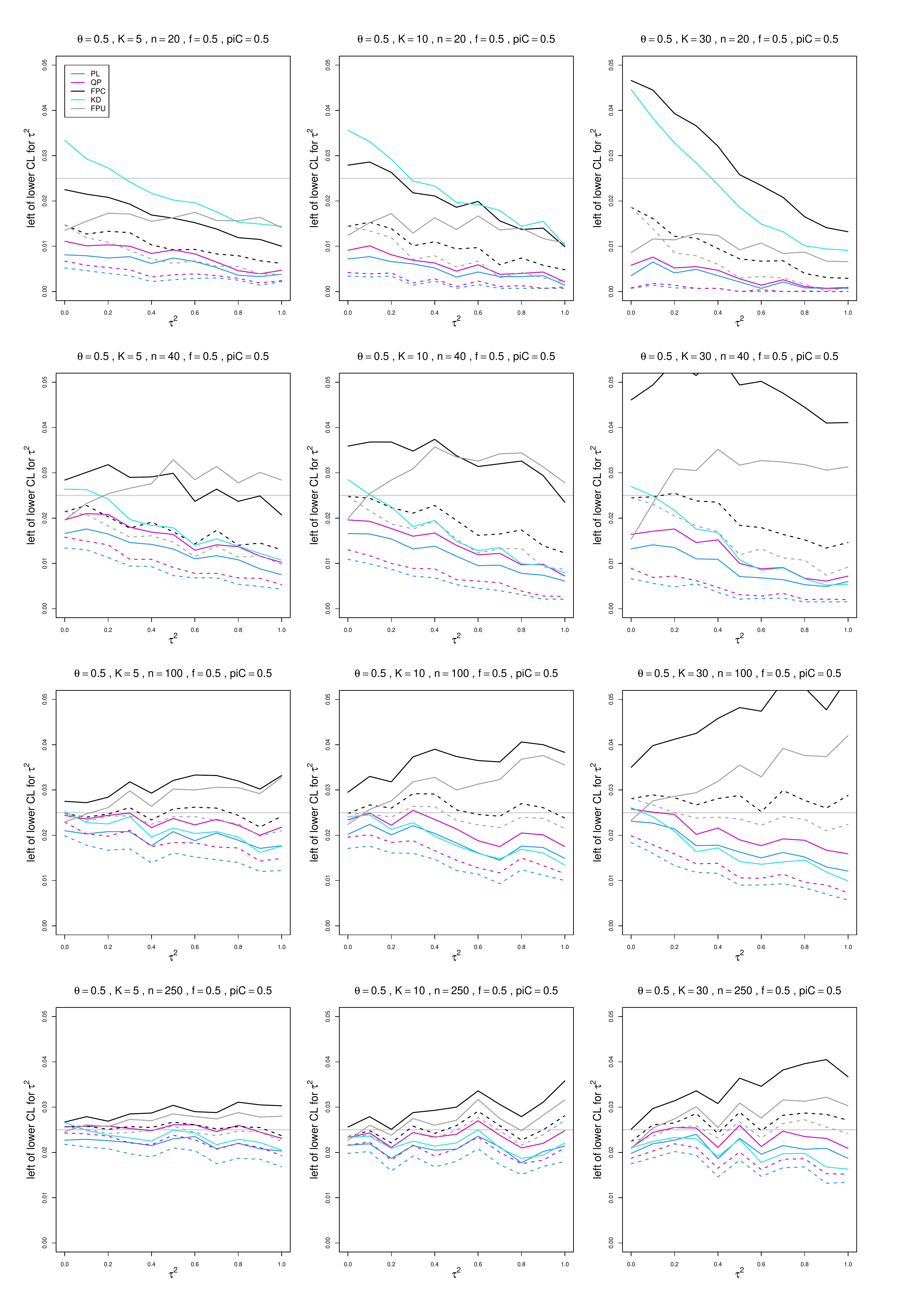}
	\caption{Miss-left probability  of  PL, QP, KD, FPC, and FPU 95\% confidence intervals for between-study variance of LOR vs $\tau^2$, for equal sample sizes $n=20,\;40,\;100$ and $250$, $p_{iC} = .5$, $\theta=0.5$ and  $f=0.5$.   Solid lines: PL, QP, and FPC \lq\lq only", FPU model-based, and KD. Dashed lines: PL, QP, and FPC \lq\lq always" and FPU na\"{i}ve.  }
	\label{PlotCovLeftOfTau2_piC_05theta=0.5_LOR_equal_sample_sizes}
\end{figure}

\begin{figure}[ht]
	\centering
	\includegraphics[scale=0.33]{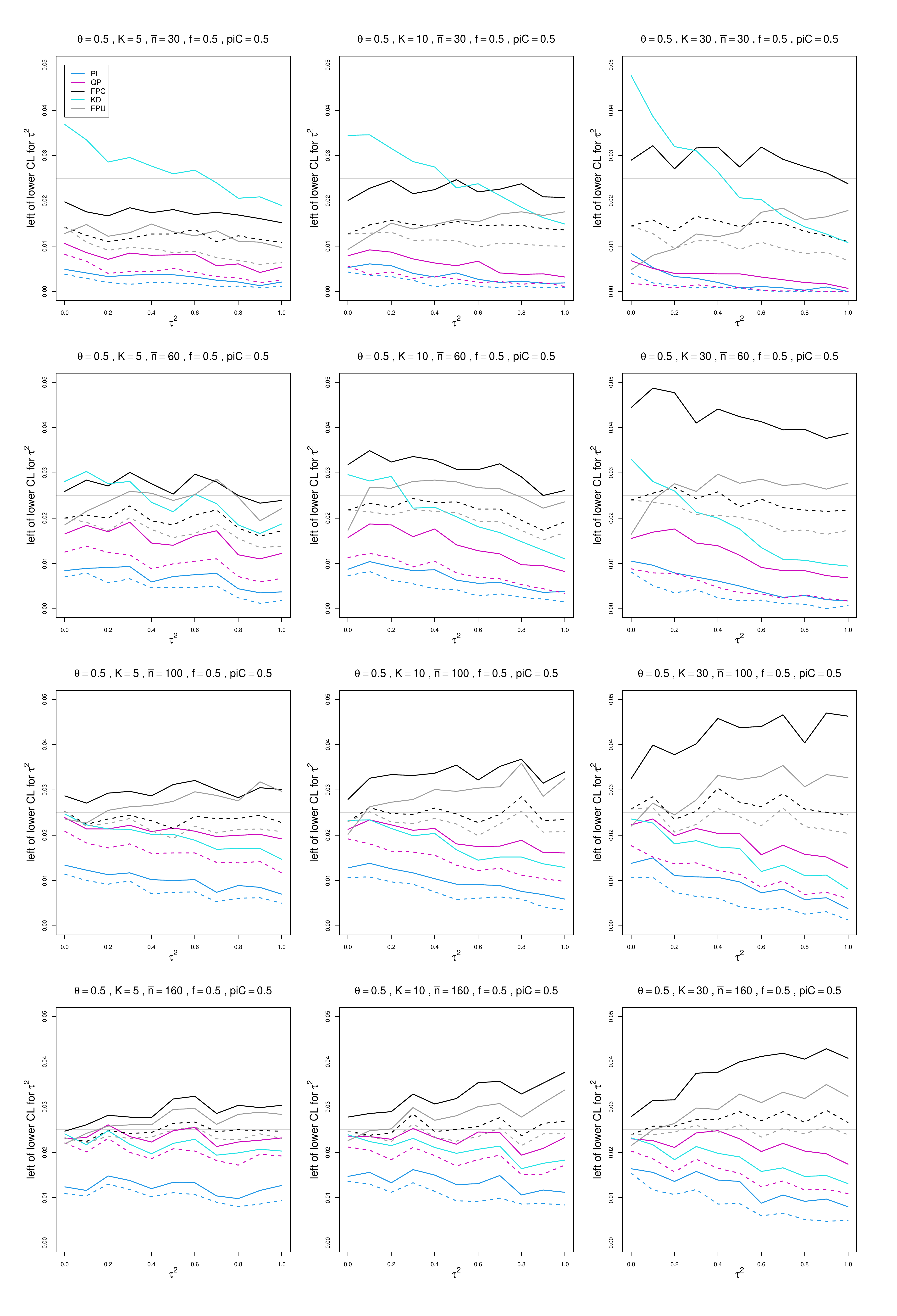}
	\caption{Miss-left probability  of  PL, QP, KD, FPC, and FPU 95\% confidence intervals for between-study variance of LOR vs $\tau^2$, for unequal sample sizes $\bar{n}=30,\;60,\;100$ and $160$, $p_{iC} = .5$, $\theta=0.5$ and  $f=0.5$.   Solid lines: PL, QP, and FPC \lq\lq only", FPU model-based, and KD. Dashed lines: PL, QP, and FPC \lq\lq always" and FPU na\"{i}ve.  }
	\label{PlotCovLeftOfTau2_piC_05theta=0.5_LOR_unequal_sample_sizes}
\end{figure}

\begin{figure}[ht]
	\centering
	\includegraphics[scale=0.33]{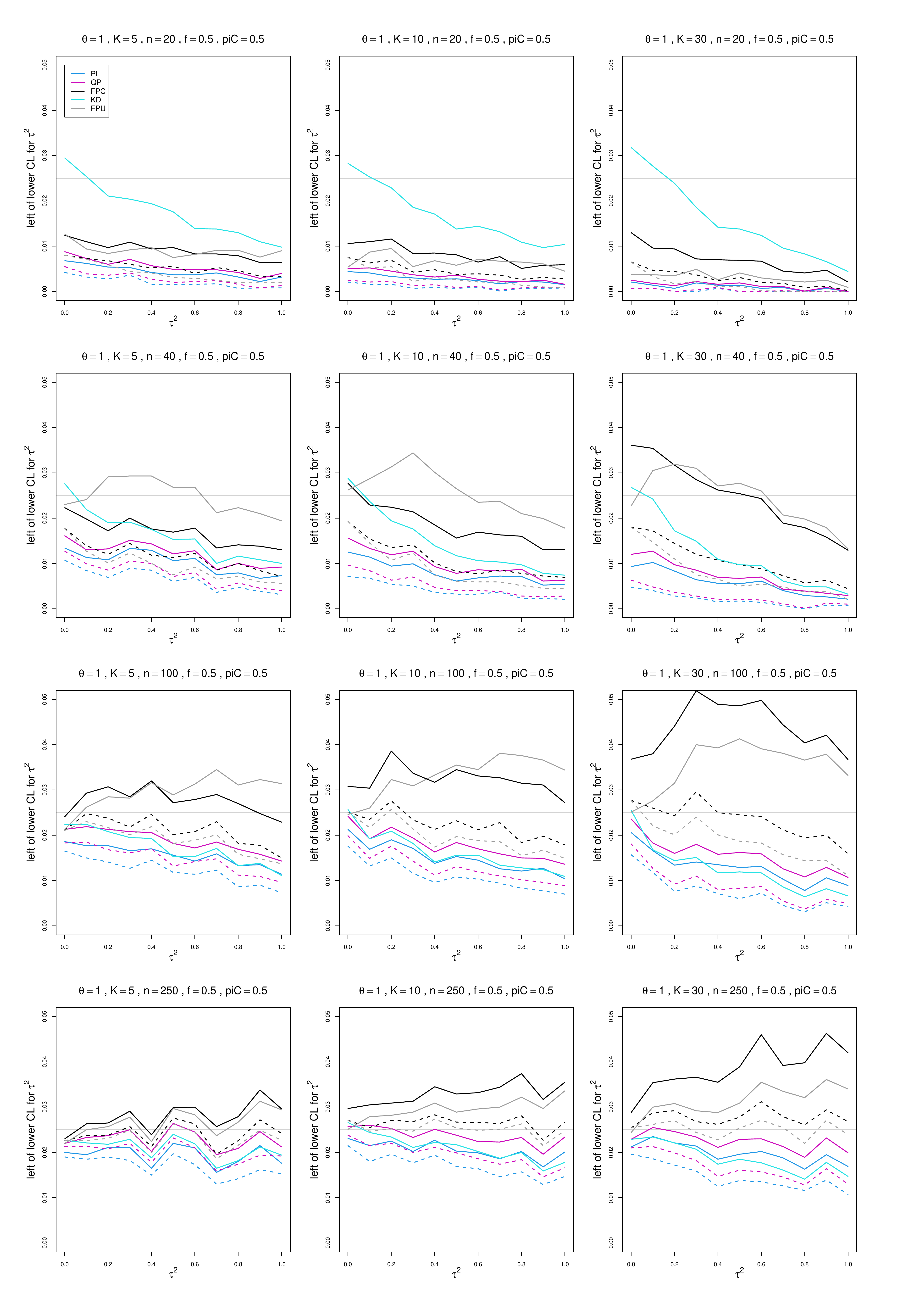}
	\caption{Miss-left probability  of  PL, QP, KD, FPC, and FPU 95\% confidence intervals for between-study variance of LOR vs $\tau^2$, for equal sample sizes $n=20,\;40,\;100$ and $250$, $p_{iC} = .5$, $\theta=1$ and  $f=0.5$.   Solid lines: PL, QP, and FPC \lq\lq only", FPU model-based, and KD. Dashed lines: PL, QP, and FPC \lq\lq always" and FPU na\"{i}ve.  }
	\label{PlotCovLeftOfTau2_piC_05theta=1_LOR_equal_sample_sizes}
\end{figure}

\begin{figure}[ht]
	\centering
	\includegraphics[scale=0.33]{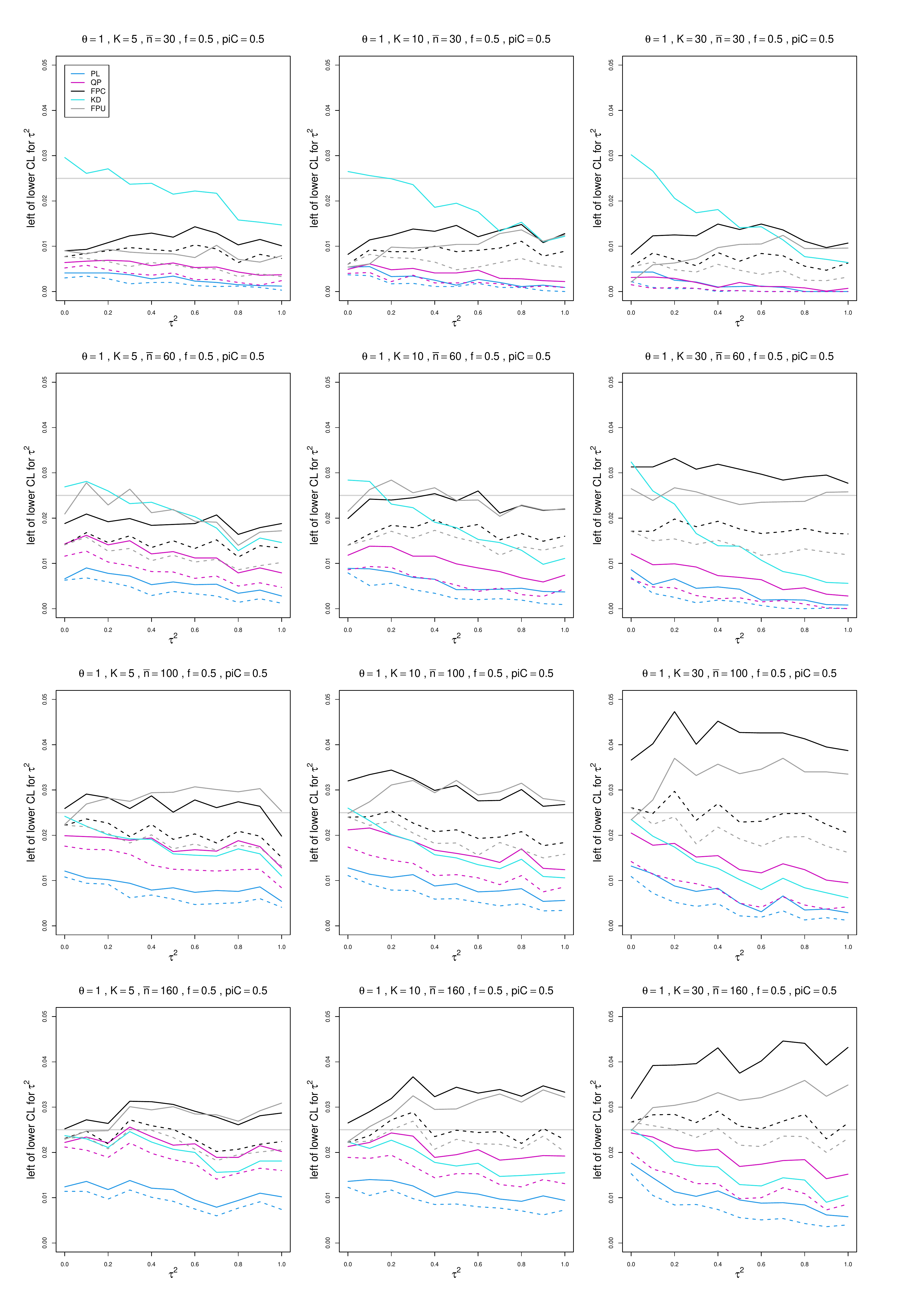}
	\caption{Miss-left probability  of  PL, QP, KD, FPC, and FPU 95\% confidence intervals for between-study variance of LOR vs $\tau^2$, for unequal sample sizes $\bar{n}=30,\;60,\;100$ and $160$, $p_{iC} = .5$, $\theta=1$ and  $f=0.5$.   Solid lines: PL, QP, and FPC \lq\lq only", FPU model-based, and KD. Dashed lines: PL, QP, and FPC \lq\lq always" and FPU na\"{i}ve.  }
	\label{PlotCovLeftOfTau2_piC_05theta=1_LOR_unequal_sample_sizes}
\end{figure}

\begin{figure}[ht]
	\centering
	\includegraphics[scale=0.33]{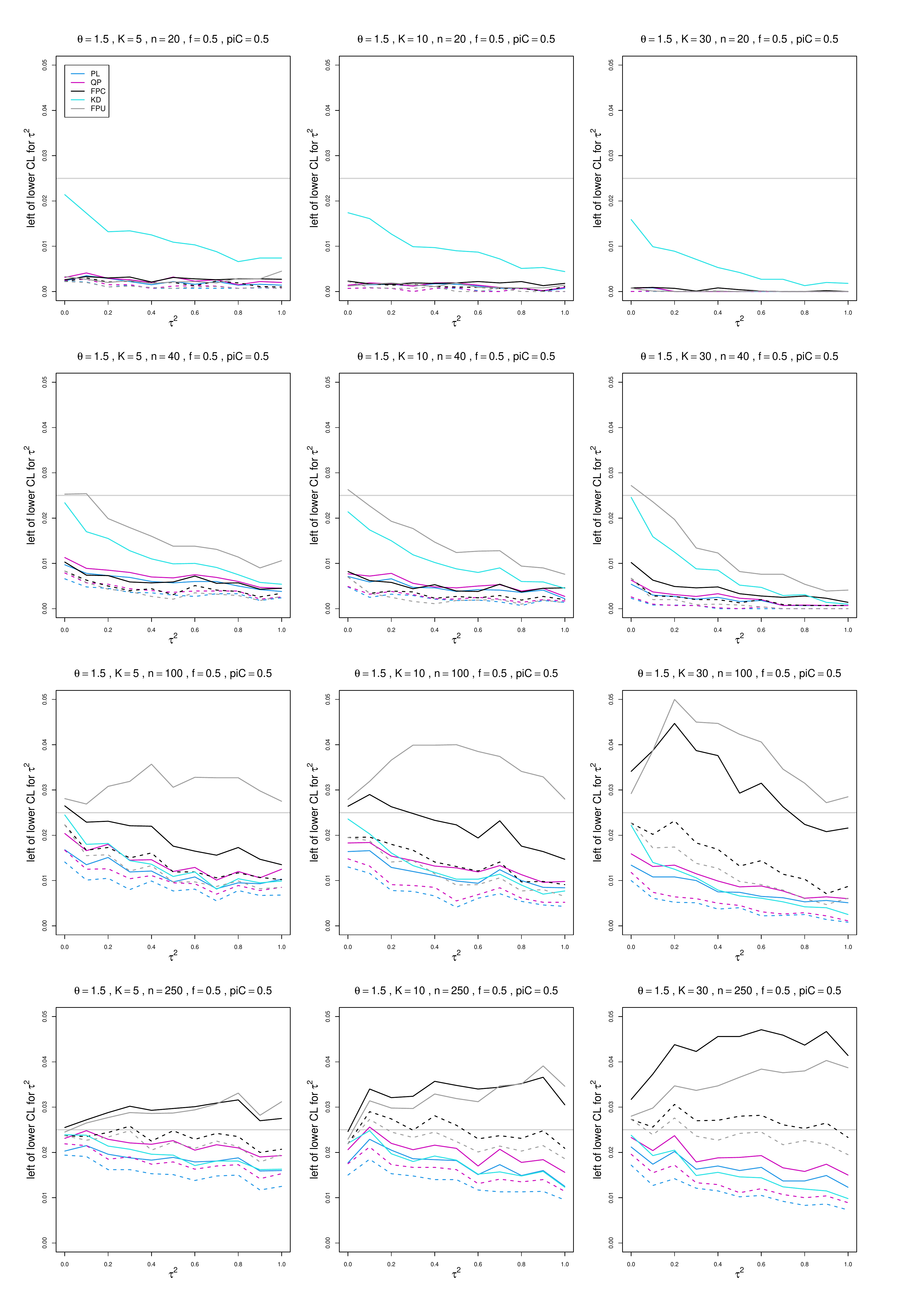}
	\caption{Miss-left probability  of  PL, QP, KD, FPC, and FPU 95\% confidence intervals for between-study variance of LOR vs $\tau^2$, for equal sample sizes $n=20,\;40,\;100$ and $250$, $p_{iC} = .5$, $\theta=1.5$ and  $f=0.5$.   Solid lines: PL, QP, and FPC \lq\lq only", FPU model-based, and KD. Dashed lines: PL, QP, and FPC \lq\lq always" and FPU na\"{i}ve.  }
	\label{PlotCovLeftOfTau2_piC_05theta=1.5_LOR_equal_sample_sizes}
\end{figure}

\begin{figure}[ht]
	\centering
	\includegraphics[scale=0.33]{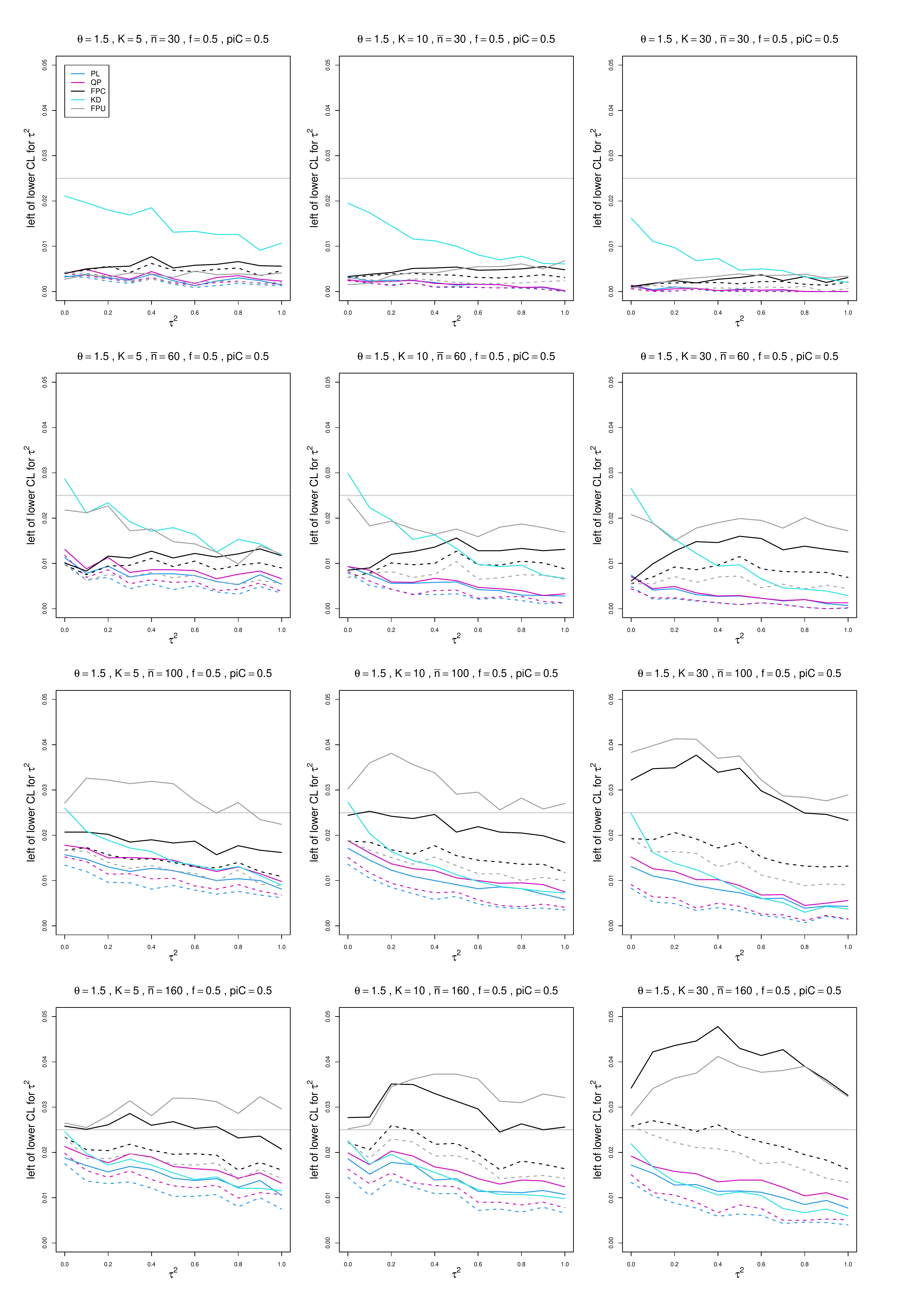}
	\caption{Miss-left probability  of  PL, QP, KD, FPC, and FPU 95\% confidence intervals for between-study variance of LOR vs $\tau^2$, for unequal sample sizes $\bar{n}=30,\;60,\;100$ and $160$, $p_{iC} = .5$, $\theta=1.5$ and  $f=0.5$.   Solid lines: PL, QP, and FPC \lq\lq only", FPU model-based, and KD. Dashed lines: PL, QP, and FPC \lq\lq always" and FPU na\"{i}ve.  }
	\label{PlotCovLeftOfTau2_piC_05theta=1.5_LOR_unequal_sample_sizes}
\end{figure}

\begin{figure}[ht]
	\centering
	\includegraphics[scale=0.33]{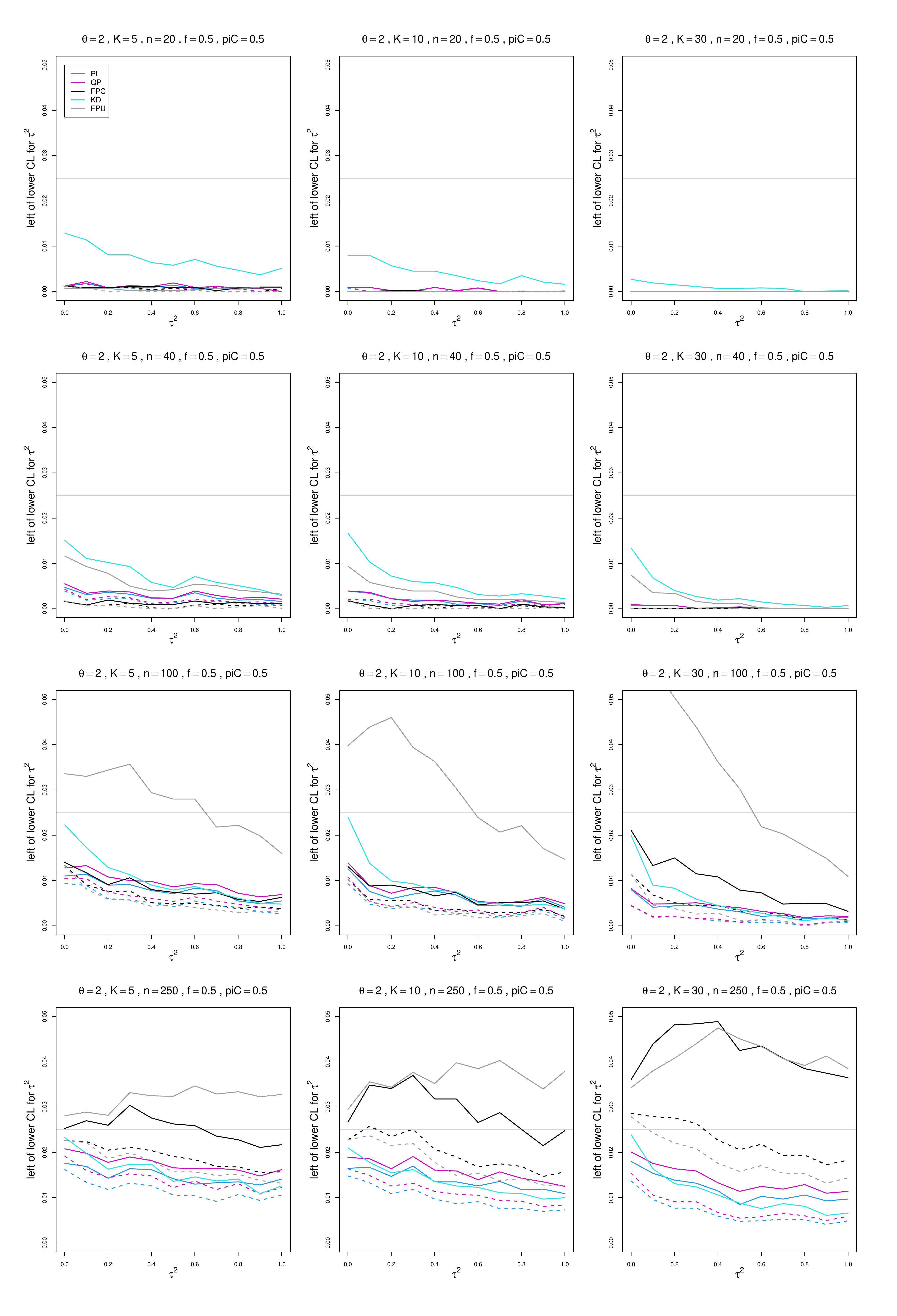}
	\caption{Miss-left probability  of  PL, QP, KD, FPC, and FPU 95\% confidence intervals for between-study variance of LOR vs $\tau^2$, for equal sample sizes $n=20,\;40,\;100$ and $250$, $p_{iC} = .5$, $\theta=2$ and  $f=0.5$.   Solid lines: PL, QP, and FPC \lq\lq only", FPU model-based, and KD. Dashed lines: PL, QP, and FPC \lq\lq always" and FPU na\"{i}ve.  }
	\label{PlotCovLeftOfTau2_piC_05theta=2_LOR_equal_sample_sizes}
\end{figure}

\begin{figure}[ht]
	\centering
	\includegraphics[scale=0.33]{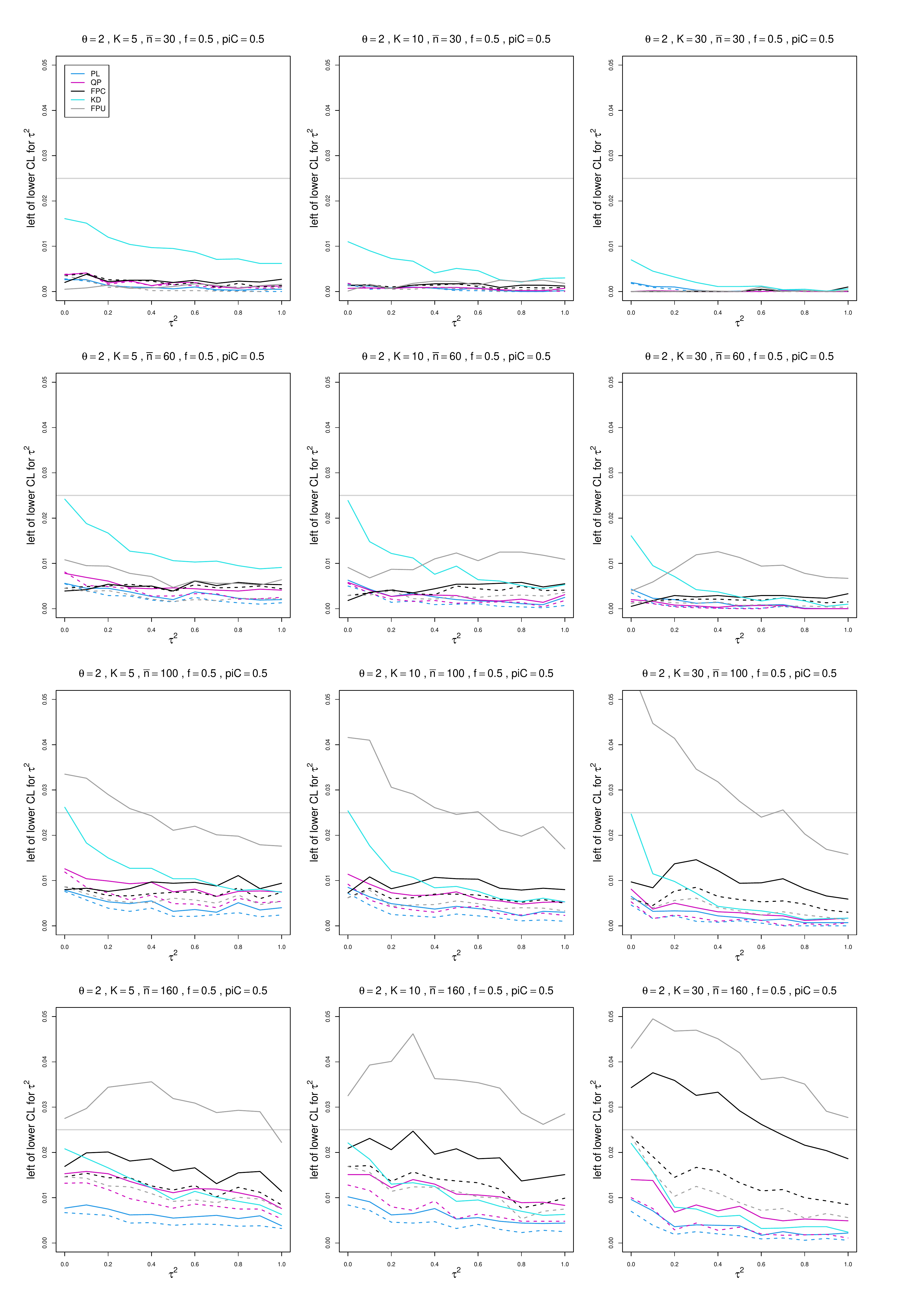}
	\caption{Miss-left probability  of  PL, QP, KD, FPC, and FPU 95\% confidence intervals for between-study variance of LOR vs $\tau^2$, for unequal sample sizes $\bar{n}=30,\;60,\;100$ and $160$, $p_{iC} = .5$, $\theta=2$ and  $f=0.5$.   Solid lines: PL, QP, and FPC \lq\lq only", FPU model-based, and KD. Dashed lines: PL, QP, and FPC \lq\lq always" and FPU na\"{i}ve.  }
	\label{PlotCovLeftOfTau2_piC_05theta=2_LOR_unequal_sample_sizes}
\end{figure}

\clearpage

\section*{Appendix E: Sample miss-right probability for 95\% confidence intervals for between-study variance}

Each figure corresponds to a value of the probability of an event in the Control arm $p_{iC}$  (= .1, .2, .5) . \\
The fraction of each study's sample size in the Control arm  $f$ is held constant at 0.5. For each combination of a value of $n$ (= 20, 40, 100, 250) or  $\bar{n}$=(= 30, 60, 100, 160) and a value of $K$ (= 5, 10, 30), a panel plots the probability that the parameter is to the right of the upper confidence limit of the confidence interval versus $\tau^2$ (= 0.0(0.1)1).\\
The confidence intervals are
\begin{itemize}
\item PL (Profile Likelihood), inverse-variance weights)
\item QP (Q profile, inverse-variance weights)
\item KD (based on  Kulinskaya-Dollinger (2015) approximation, inverse-variance weights)
\item FPC (based on Farebrother approximation, effective-sample-size weights, conditional variance of LOR)
\item FPU (based on Farebrother approximation, effective-sample-size weights, unconditional variance of LOR)
\end{itemize}
The plots include two versions of PL, QP, and FPC: adding $1/2$ to all four of $X_{iT},\;X_{iC},\; n_{iT}-X_{iT},\; n_{iC}-X_{iC}$ only when one of these is zero (solid lines) or always (dashed lines).\\
The plots also include two versions of FPU: model-based estimation of $p_{iT}$ (solid lines) or na\"{i}ve estimation (dashed lines).

\clearpage
\setcounter{figure}{0}
\renewcommand{\thefigure}{E.\arabic{figure}}

\begin{figure}[ht]
	\centering
	\includegraphics[scale=0.33]{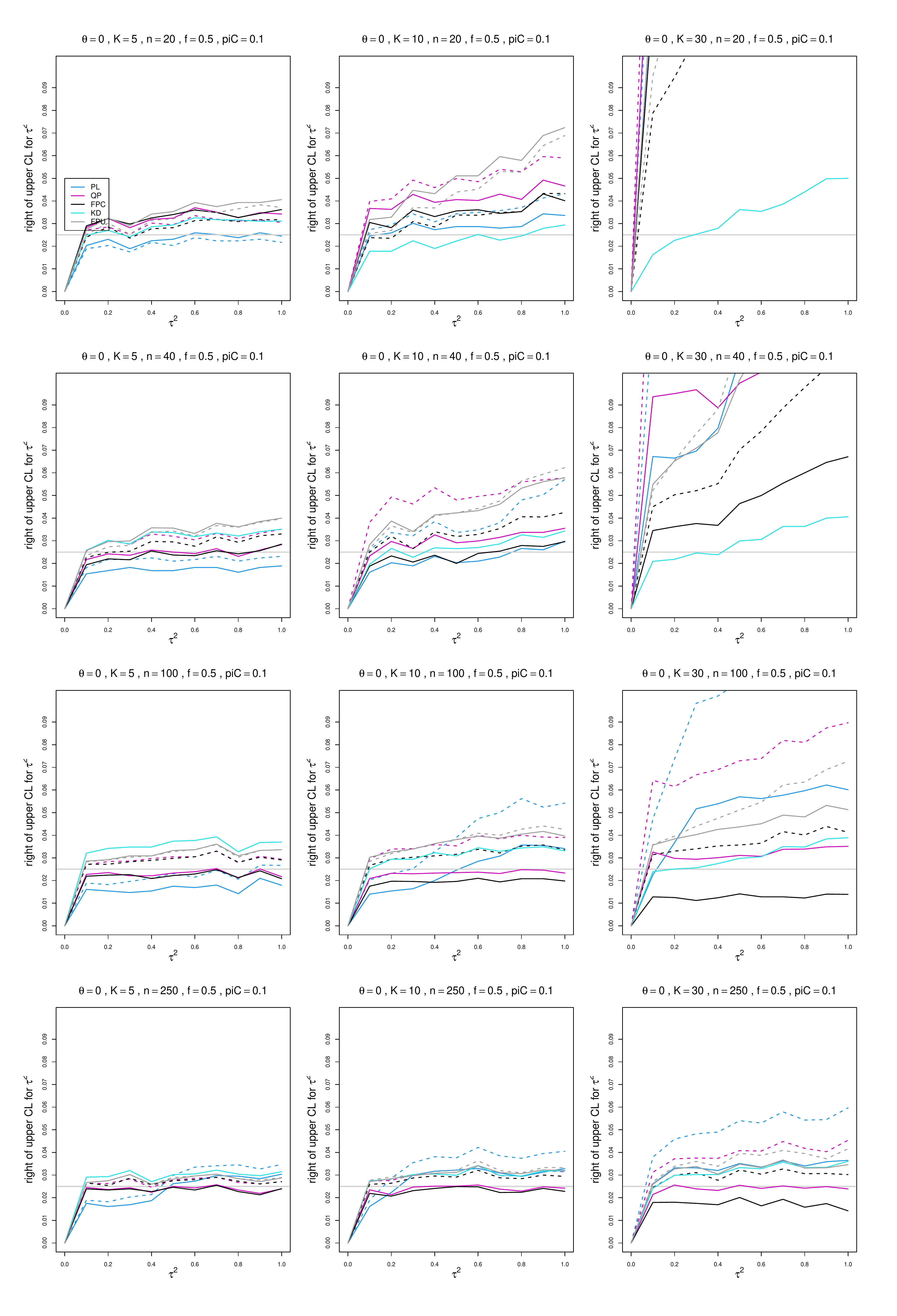}
	\caption{Miss-right probability  of  PL, QP, KD, FPC, and FPU 95\%  confidence intervals for between-study variance of LOR vs $\tau^2$, for equal sample sizes $n=20,\;40,\;100$ and $250$, $p_{iC} = .1$, $\theta=0$ and  $f=0.5$.   Solid lines: PL, QP, and FPC \lq\lq only", FPU model-based, and KD. Dashed lines:
PL, QP and FPC \lq\lq always" and FPU na\"{i}ve.   }
	\label{PlotCovRightOfTau2_piC_01theta=0_LOR_equal_sample_sizes}
\end{figure}

\begin{figure}[ht]
	\centering
	\includegraphics[scale=0.33]{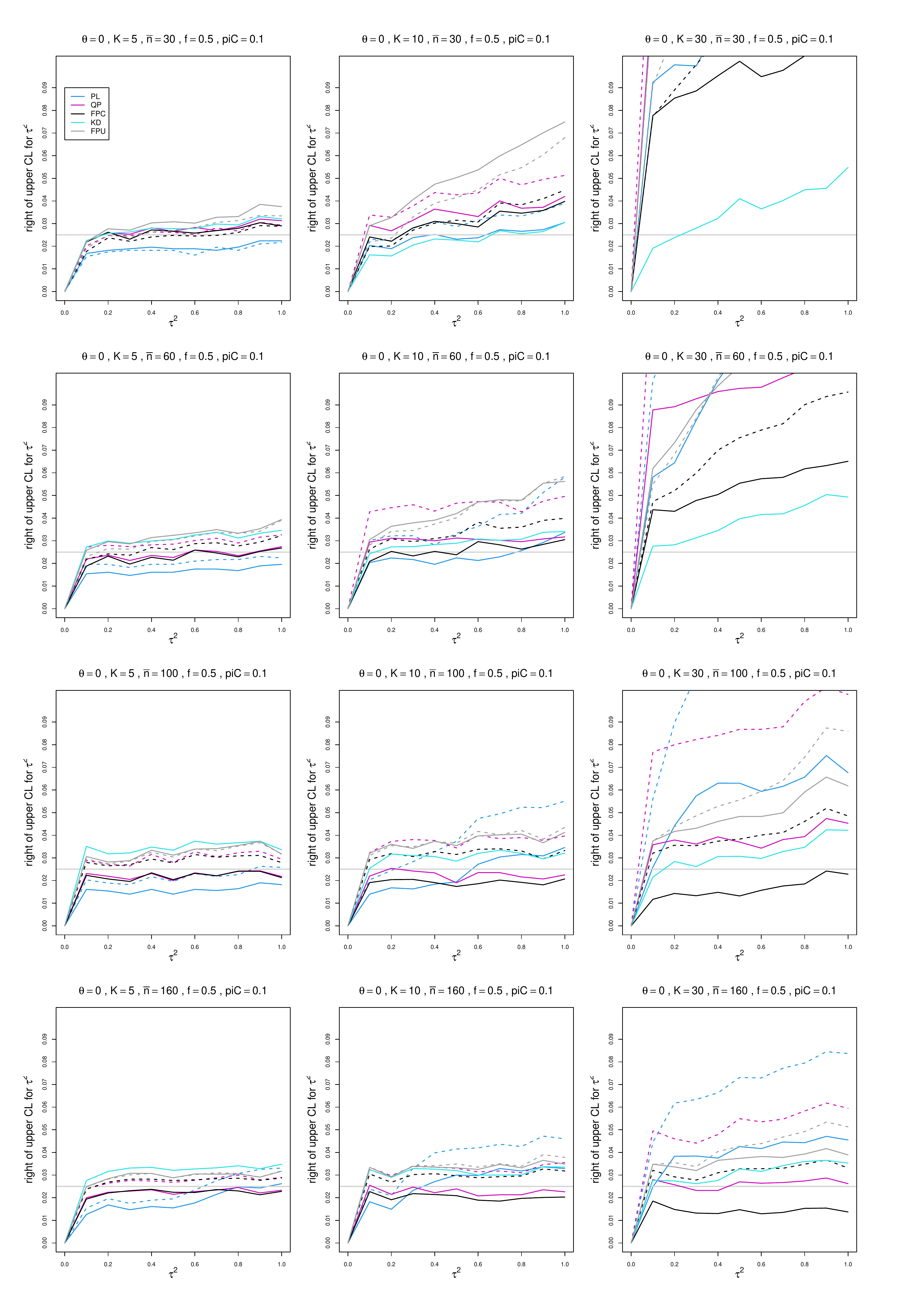}
	\caption{Miss-right probability  of  PL, QP, KD, FPC, and FPU 95\%  confidence intervals for between-study variance of LOR vs $\tau^2$, for unequal sample sizes $\bar{n}=30,\;60,\;100$ and $160$, $p_{iC} = .1$, $\theta=0$ and  $f=0.5$.   Solid lines: PL, QP, and FPC \lq\lq only", FPU model-based, and KD. Dashed lines: PL, QP, and FPC \lq\lq always" and FPU na\"{i}ve.   }
	\label{PlotCovRightOfTau2_piC_01theta=0_LOR_unequal_sample_sizes}
\end{figure}

\begin{figure}[ht]
	\centering
	\includegraphics[scale=0.33]{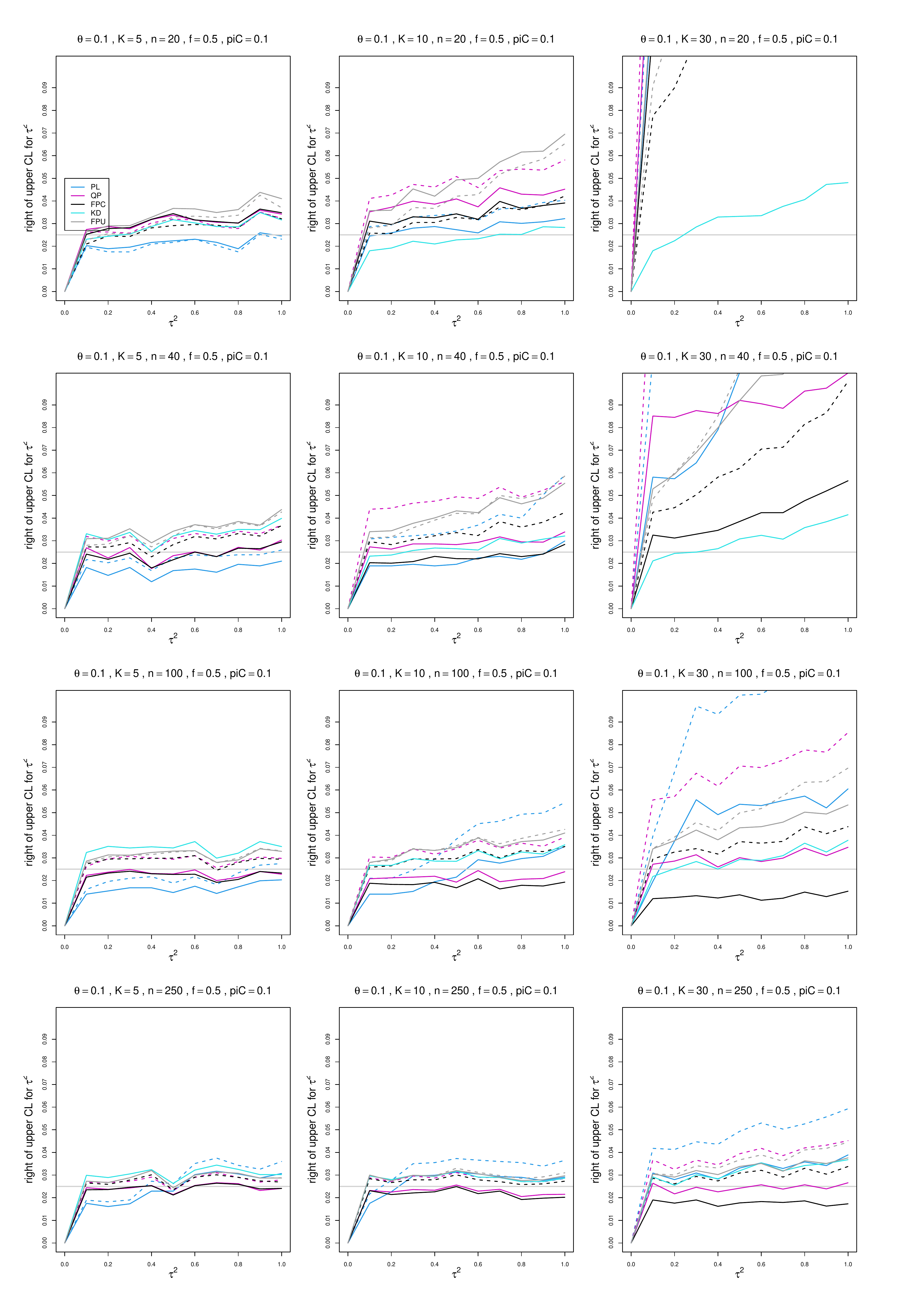}
	\caption{Miss-right probability  of  PL, QP, KD, FPC, and FPU 95\%  confidence intervals for between-study variance of LOR vs $\tau^2$, for equal sample sizes $n=20,\;40,\;100$ and $250$, $p_{iC} = .1$, $\theta=0.1$ and  $f=0.5$.   Solid lines: PL, QP, and FPC \lq\lq only", FPU model-based, and KD. Dashed lines:
PL, QP and FPC \lq\lq always" and FPU na\"{i}ve.  }
	\label{PlotCovRightOfTau2_piC_01theta=0.1_LOR_equal_sample_sizes}
\end{figure}

\begin{figure}[ht]
	\centering
	\includegraphics[scale=0.33]{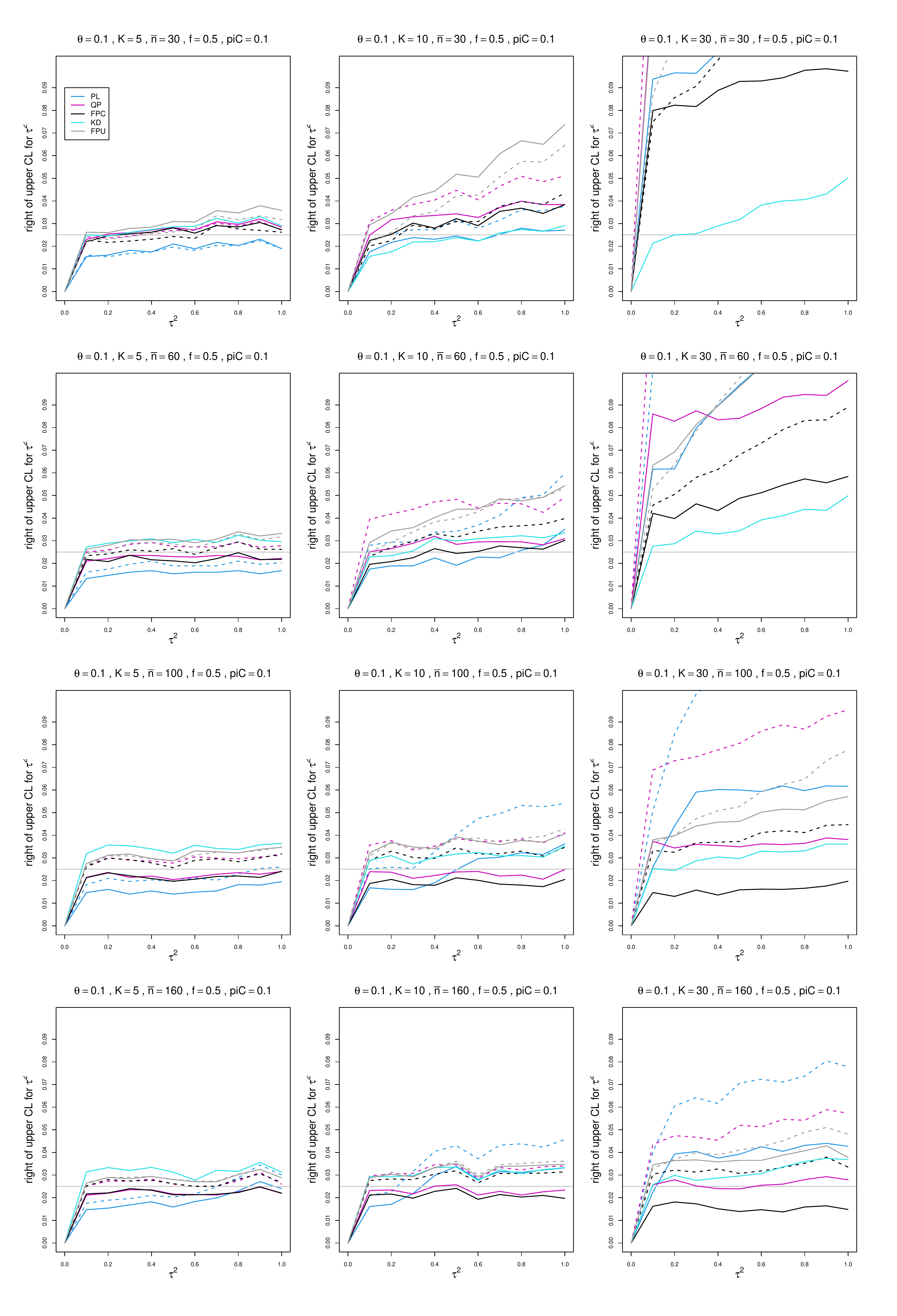}
	\caption{Miss-right probability  of  PL, QP, KD, FPC, and FPU 95\%  confidence intervals for between-study variance of LOR vs $\tau^2$, for unequal sample sizes $\bar{n}=30,\;60,\;100$ and $160$, $p_{iC} = .1$, $\theta=0.1$ and  $f=0.5$.   Solid lines: PL, QP, and FPC \lq\lq only", FPU model-based, and KD. Dashed lines: PL, QP, and FPC \lq\lq always" and FPU na\"{i}ve.   }
	\label{PlotCovRightOfTau2_piC_01theta=0.1_LOR_unequal_sample_sizes}
\end{figure}

\begin{figure}[ht]
	\centering
	\includegraphics[scale=0.33]{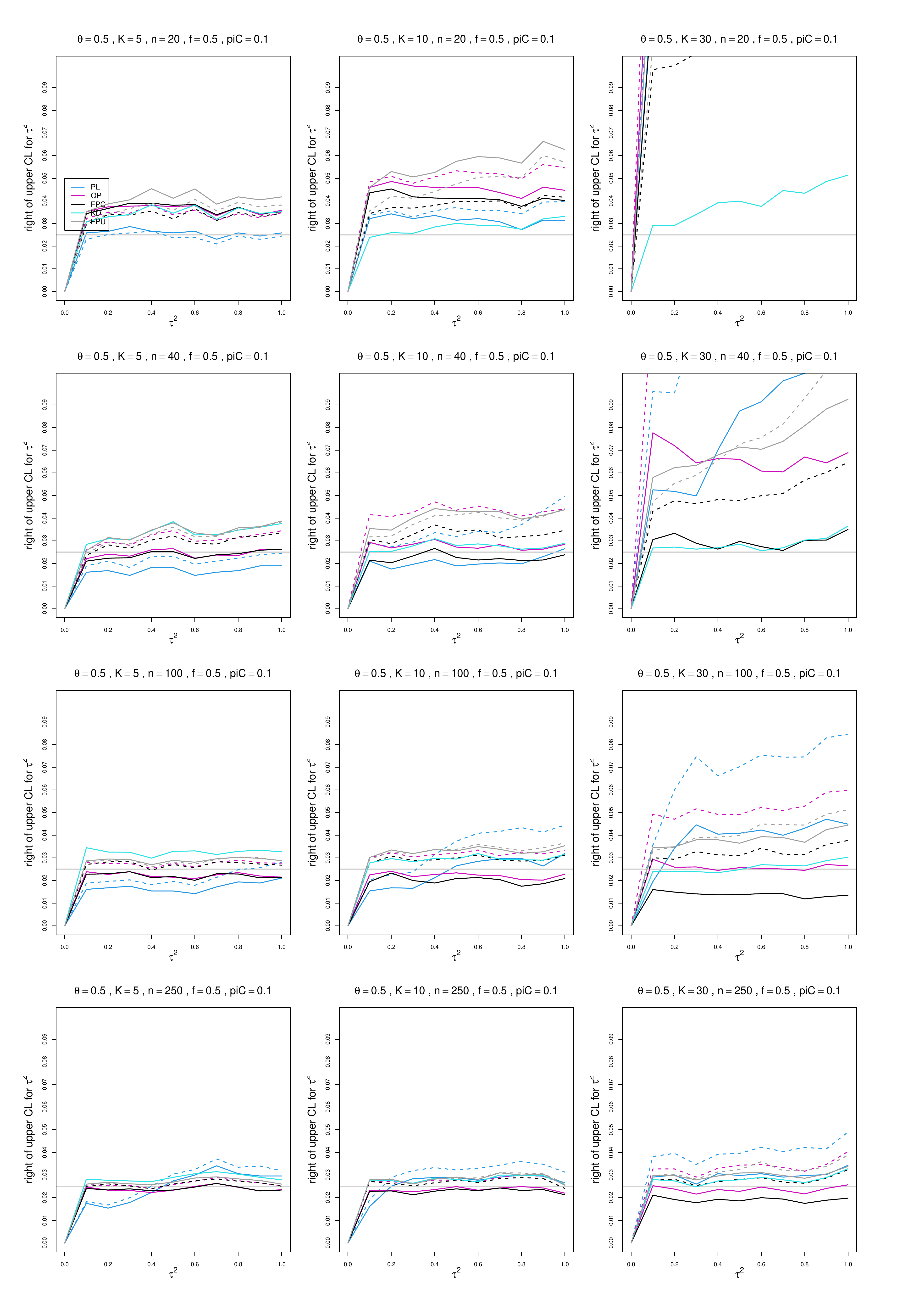}
	\caption{Miss-right probability  of  PL, QP, KD, FPC, and FPU 95\%  confidence intervals for between-study variance of LOR vs $\tau^2$, for equal sample sizes $n=20,\;40,\;100$ and $250$, $p_{iC} = .1$, $\theta=0.5$ and  $f=0.5$.   Solid lines: PL, QP, and FPC \lq\lq only", FPU model-based, and KD. Dashed lines:
PL, QP and FPC \lq\lq always" and FPU na\"{i}ve.  }
	\label{PlotRightOfTau2_piC_01theta=0.5_LOR_equal_sample_sizes}
\end{figure}

\begin{figure}[ht]
	\centering
	\includegraphics[scale=0.33]{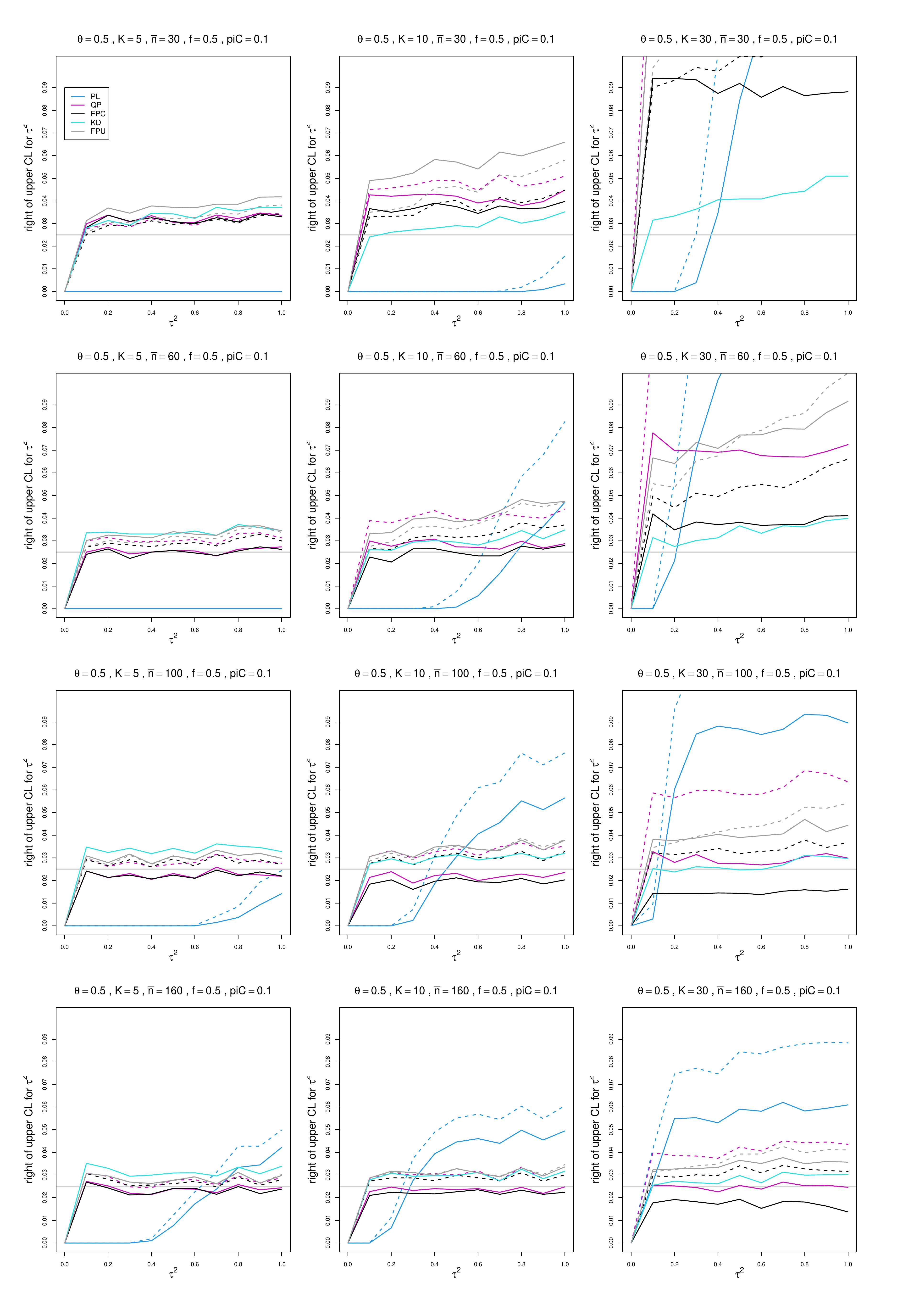}
	\caption{Miss-right probability  of  PL, QP, KD, FPC, and FPU 95\%  confidence intervals for between-study variance of LOR vs $\tau^2$, for unequal sample sizes $\bar{n}=30,\;60,\;100$ and $160$, $p_{iC} = .1$, $\theta=0.5$ and  $f=0.5$.   Solid lines: PL, QP, and FPC \lq\lq only", FPU model-based, and KD. Dashed lines: PL, QP, and FPC \lq\lq always" and FPU na\"{i}ve.   }
	\label{PlotCovRightOfTau2_piC_01theta=0.5_LOR_unequal_sample_sizes}
\end{figure}

\begin{figure}[ht]
	\centering
	\includegraphics[scale=0.33]{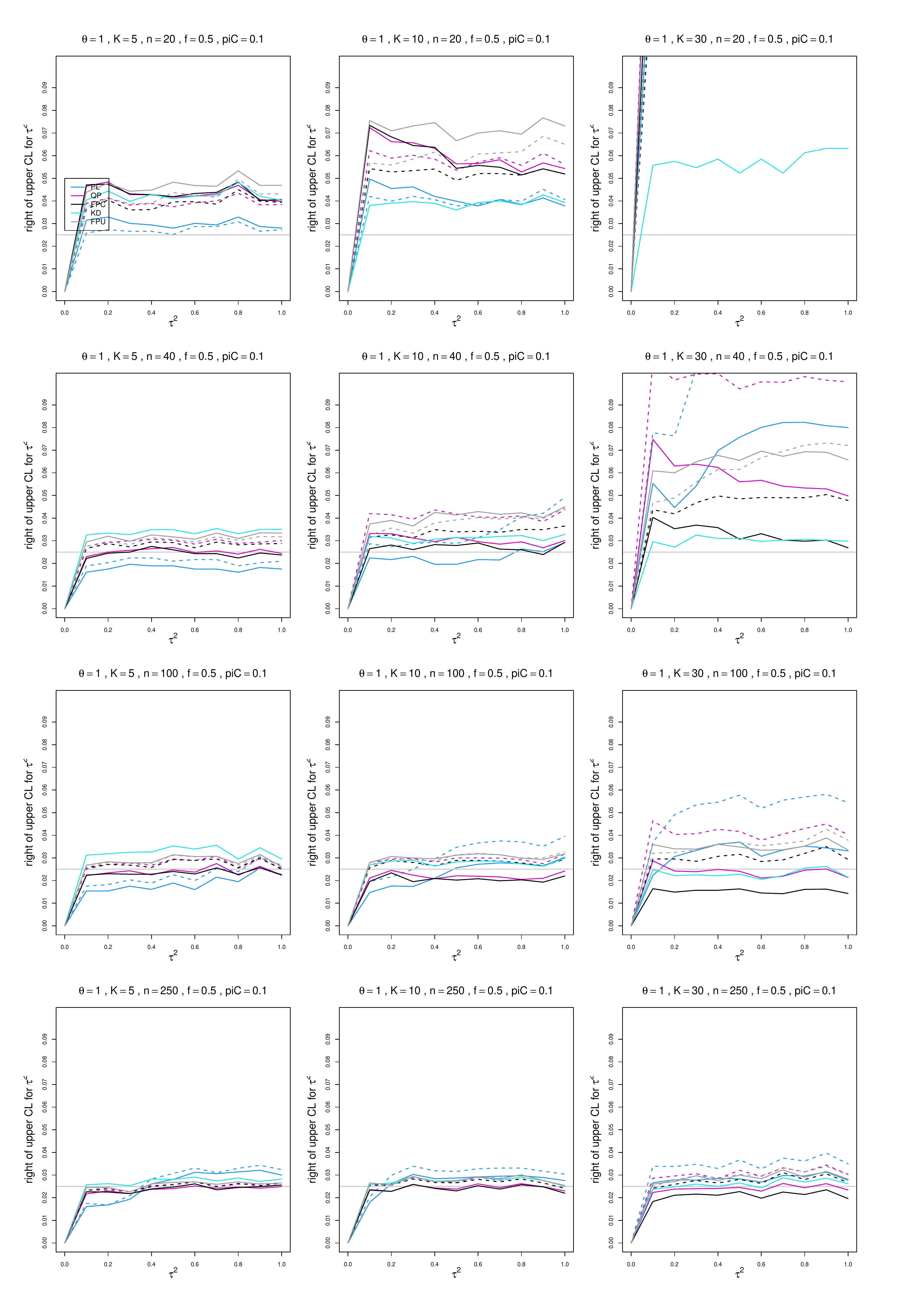}
	\caption{Miss-right probability  of  PL, QP, KD, FPC, and FPU 95\%  confidence intervals for between-study variance of LOR vs $\tau^2$, for equal sample sizes $n=20,\;40,\;100$ and $250$, $p_{iC} = .1$, $\theta=1$ and  $f=0.5$.   Solid lines: PL, QP, and FPC \lq\lq only", FPU model-based, and KD. Dashed lines:
PL, QP and FPC \lq\lq always" and FPU na\"{i}ve.  }
	\label{PlotCovRightOfTau2_piC_01theta=1_LOR_equal_sample_sizes}
\end{figure}

\begin{figure}[ht]
	\centering
	\includegraphics[scale=0.33]{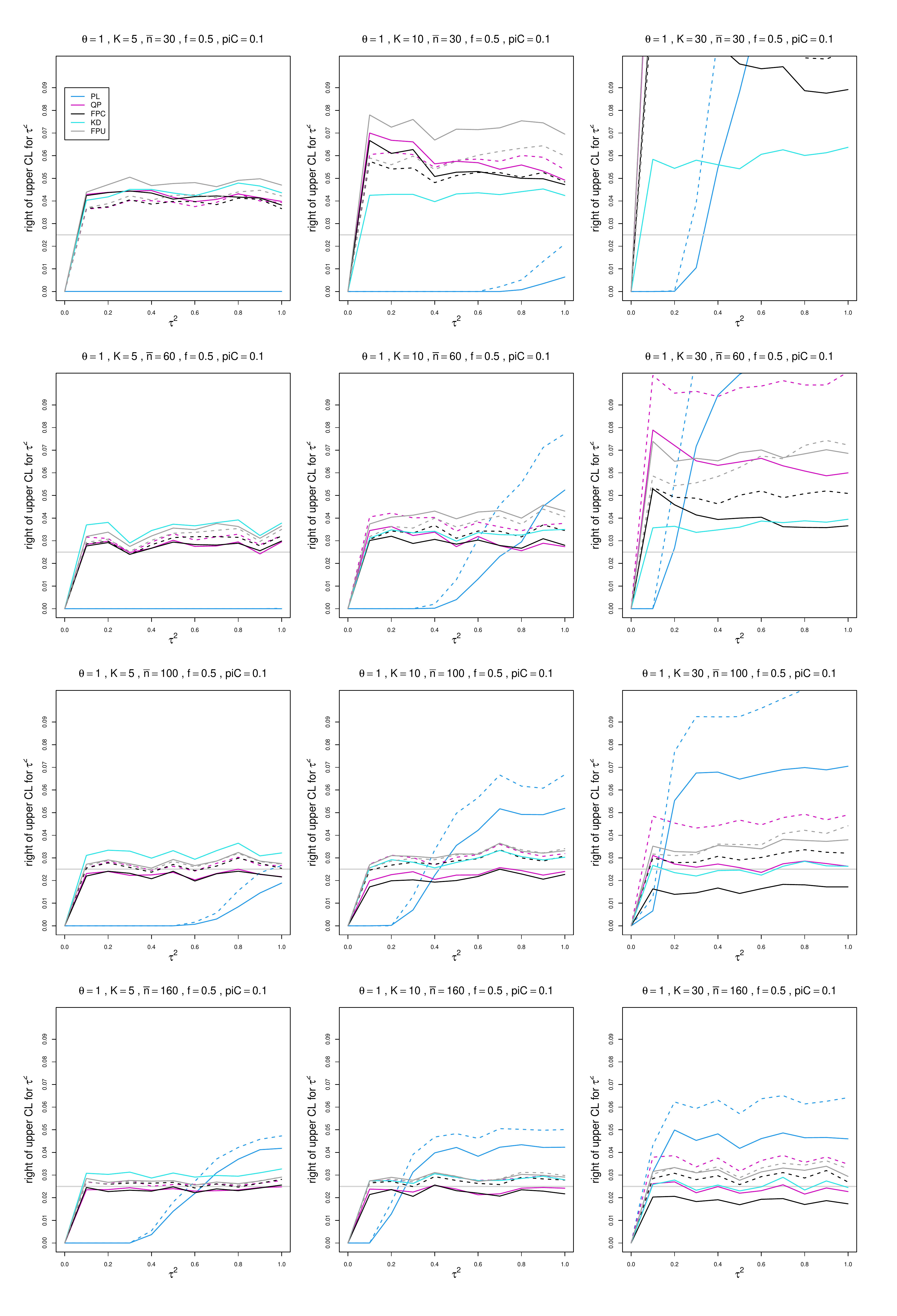}
	\caption{Miss-right probability  of  PL, QP, KD, FPC, and FPU 95\%  confidence intervals for between-study variance of LOR vs $\tau^2$, for unequal sample sizes $\bar{n}=30,\;60,\;100$ and $160$, $p_{iC} = .1$, $\theta=1$ and  $f=0.5$.   Solid lines: PL, QP, and FPC \lq\lq only", FPU model-based, and KD. Dashed lines: PL, QP, and FPC \lq\lq always" and FPU na\"{i}ve.   }
	\label{PlotCovRightOfTau2_piC_01theta=1_LOR_unequal_sample_sizes}
\end{figure}

\begin{figure}[ht]
	\centering
	\includegraphics[scale=0.33]{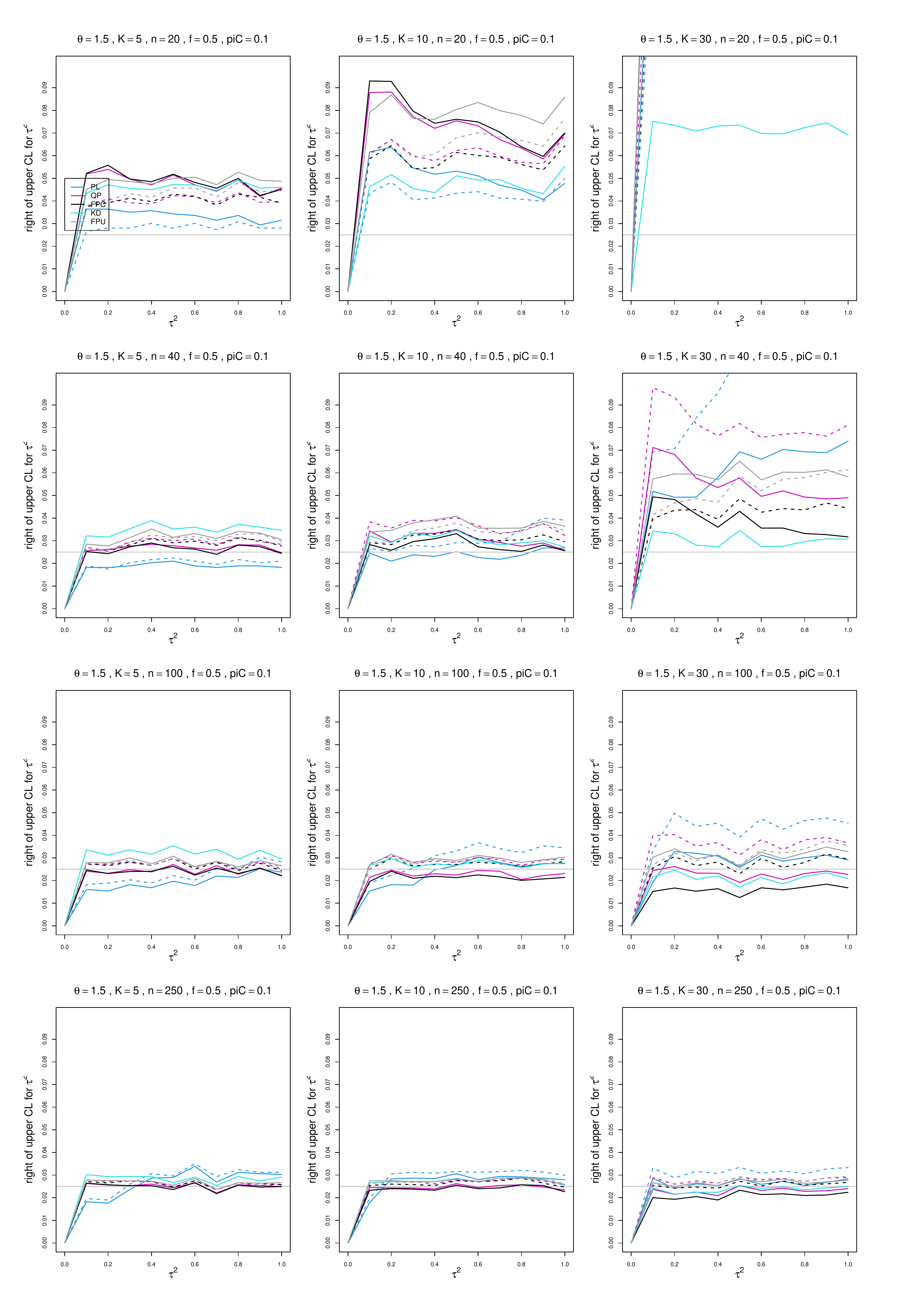}
	\caption{Miss-right probability  of  PL, QP, KD, FPC, and FPU 95\%  confidence intervals for between-study variance of LOR vs $\tau^2$, for equal sample sizes $n=20,\;40,\;100$ and $250$, $p_{iC} = .1$, $\theta=1.5$ and  $f=0.5$.   Solid lines: PL, QP, and FPC \lq\lq only", FPU model-based, and KD. Dashed lines:
PL, QP and FPC \lq\lq always" and FPU na\"{i}ve.   }
	\label{PlotCovRightOfTau2_piC_01theta=1.5_LOR_equal_sample_sizes}
\end{figure}

\begin{figure}[ht]
	\centering
	\includegraphics[scale=0.33]{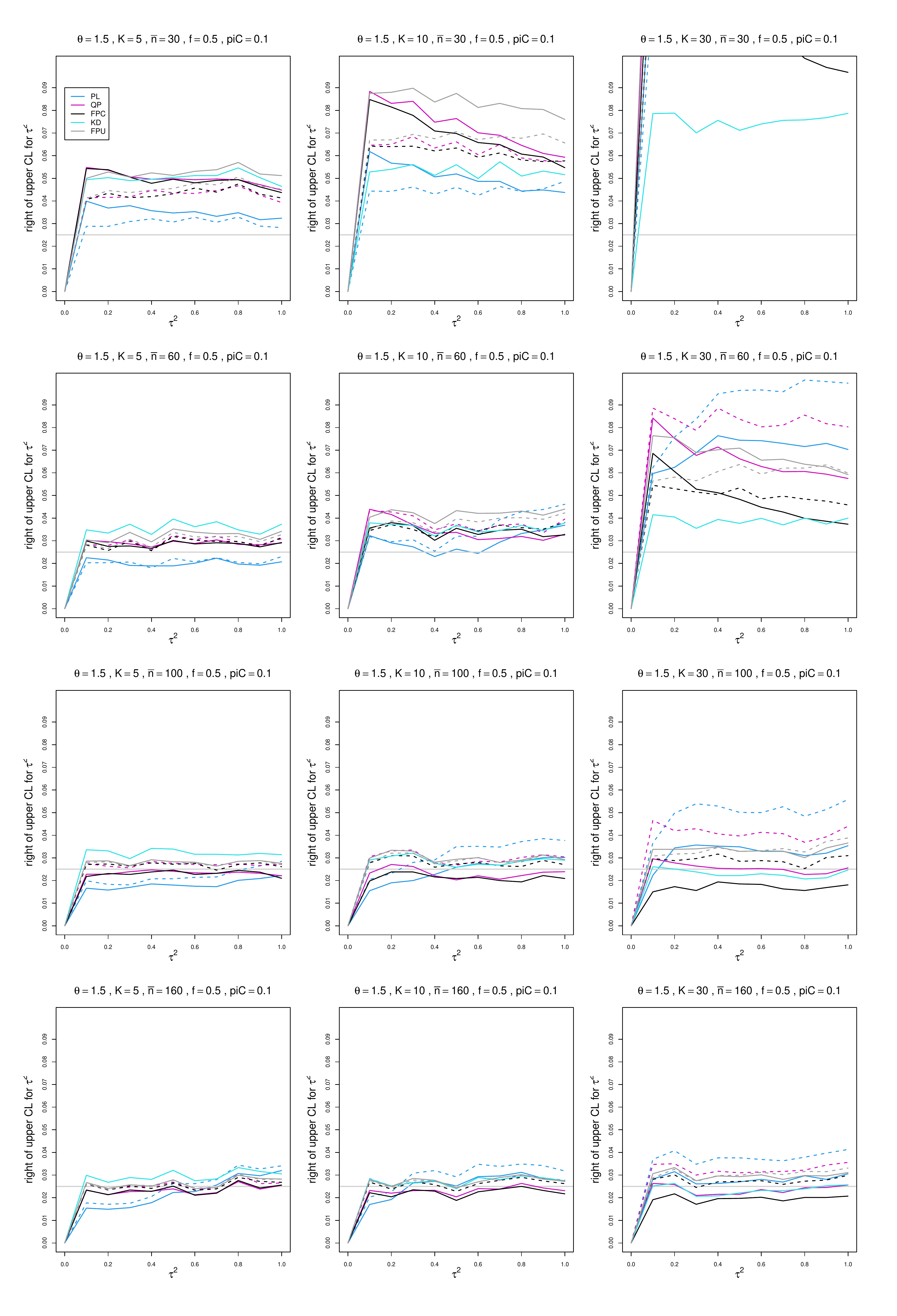}
	\caption{Miss-right probability  of  PL, QP, KD, FPC, and FPU 95\%  confidence intervals for between-study variance of LOR vs $\tau^2$, for unequal sample sizes $\bar{n}=30,\;60,\;100$ and $160$, $p_{iC} = .1$, $\theta=1.5$ and  $f=0.5$.   Solid lines: PL, QP, and FPC \lq\lq only", FPU model-based, and KD. Dashed lines: PL, QP, and FPC \lq\lq always" and FPU na\"{i}ve.  }
	\label{PlotCovRightOfTau2_piC_01theta=1.5_LOR_unequal_sample_sizes}
\end{figure}

\begin{figure}[ht]
	\centering
	\includegraphics[scale=0.33]{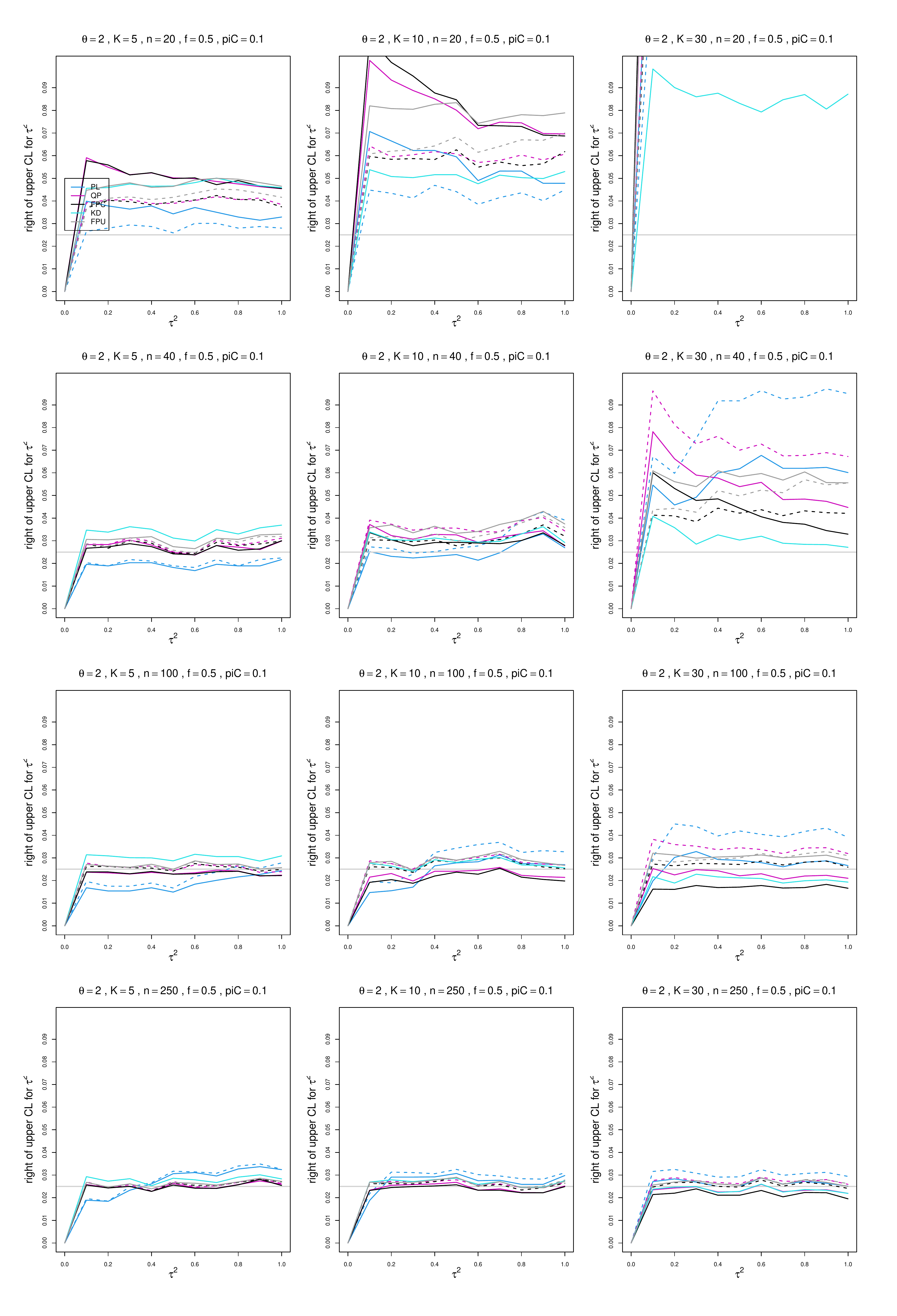}
	\caption{Miss-right probability  of  PL, QP, KD, FPC, and FPU 95\%  confidence intervals for between-study variance of LOR vs $\tau^2$, for equal sample sizes $n=20,\;40,\;100$ and $250$, $p_{iC} = .1$, $\theta=2$ and  $f=0.5$.   Solid lines: PL, QP, and FPC \lq\lq only", FPU model-based, and KD. Dashed lines:
PL, QP and FPC \lq\lq always" and FPU na\"{i}ve.  }
	\label{PlotCovRightOfTau2_piC_01theta=2_LOR_equal_sample_sizes}
\end{figure}

\begin{figure}[ht]
	\centering
	\includegraphics[scale=0.33]{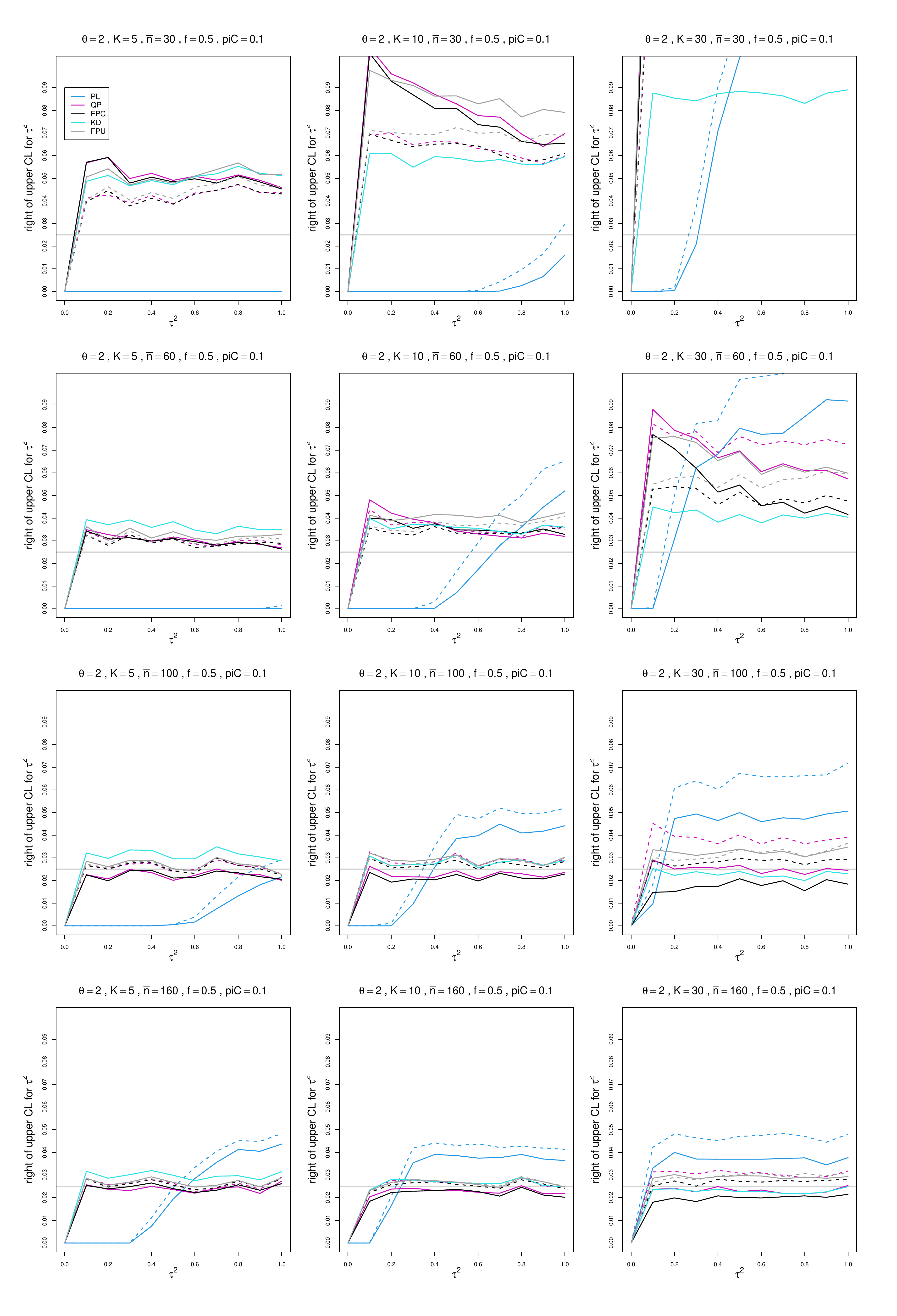}
	\caption{Miss-right probability  of  PL, QP, KD, FPC, and FPU 95\%  confidence intervals for between-study variance of LOR vs $\tau^2$, for unequal sample sizes $\bar{n}=30,\;60,\;100$ and $160$, $p_{iC} = .1$, $\theta=2$ and  $f=0.5$.   Solid lines: PL, QP, and FPC \lq\lq only", FPU model-based, and KD. Dashed lines: PL, QP, and FPC \lq\lq always" and FPU na\"{i}ve.   }
	\label{PlotCovRightOfTau2_piC_01theta=2_LOR_unequal_sample_sizes}
\end{figure}

\begin{figure}[ht]
	\centering
	\includegraphics[scale=0.33]{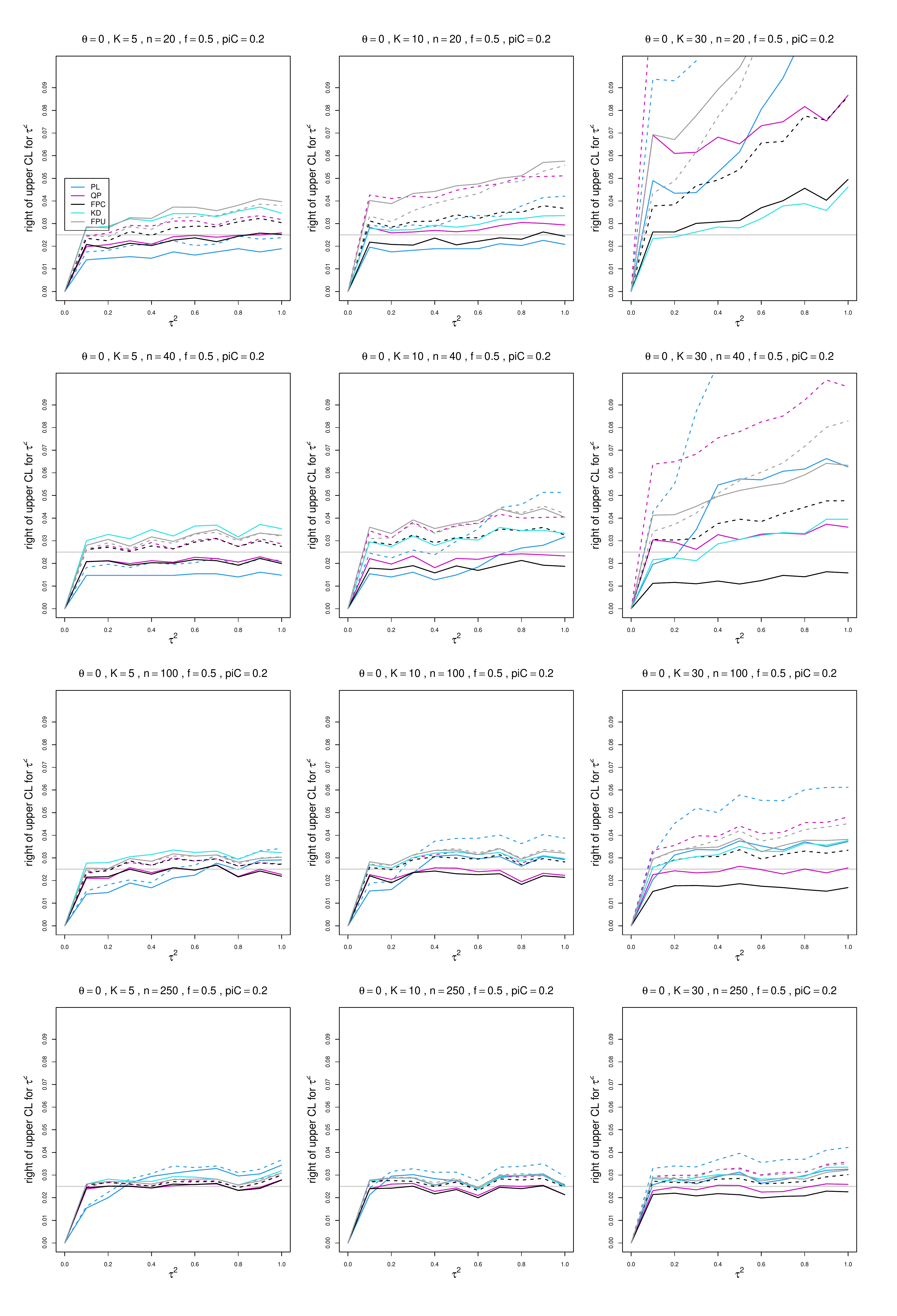}
	\caption{Miss-right probability  of  PL, QP, KD, FPC, and FPU 95\%  confidence intervals for between-study variance of LOR vs $\tau^2$, for equal sample sizes $n=20,\;40,\;100$ and $250$, $p_{iC} = .2$, $\theta=0$ and  $f=0.5$.   Solid lines: PL, QP, and FPC \lq\lq only", FPU model-based, and KD. Dashed lines:
PL, QP and FPC \lq\lq always" and FPU na\"{i}ve.   }
	\label{PlotCovRightOfTau2_piC_02theta=0_LOR_equal_sample_sizes}
\end{figure}

\begin{figure}[ht]
	\centering
	\includegraphics[scale=0.33]{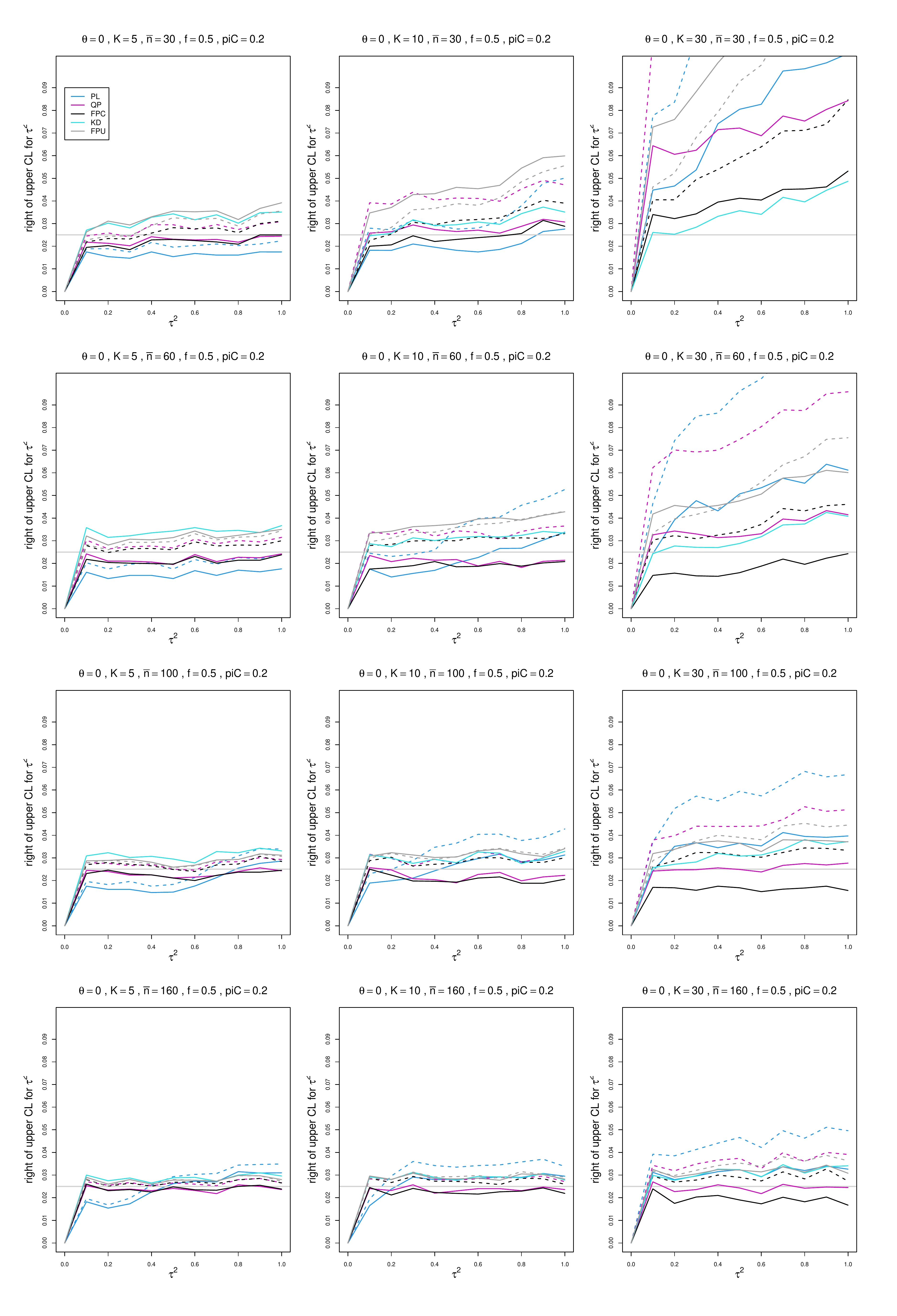}
	\caption{Miss-right probability  of  PL, QP, KD, FPC, and FPU 95\%  confidence intervals for between-study variance of LOR vs $\tau^2$, for unequal sample sizes $\bar{n}=30,\;60,\;100$ and $160$, $p_{iC} = .2$, $\theta=0$ and  $f=0.5$.   Solid lines: PL, QP, and FPC \lq\lq only", FPU model-based, and KD. Dashed lines: PL, QP, and FPC \lq\lq always" and FPU na\"{i}ve.   }
	\label{PlotCovRightOfTau2_piC_02theta=0_LOR_unequal_sample_sizes}
\end{figure}

\begin{figure}[ht]
	\centering
	\includegraphics[scale=0.33]{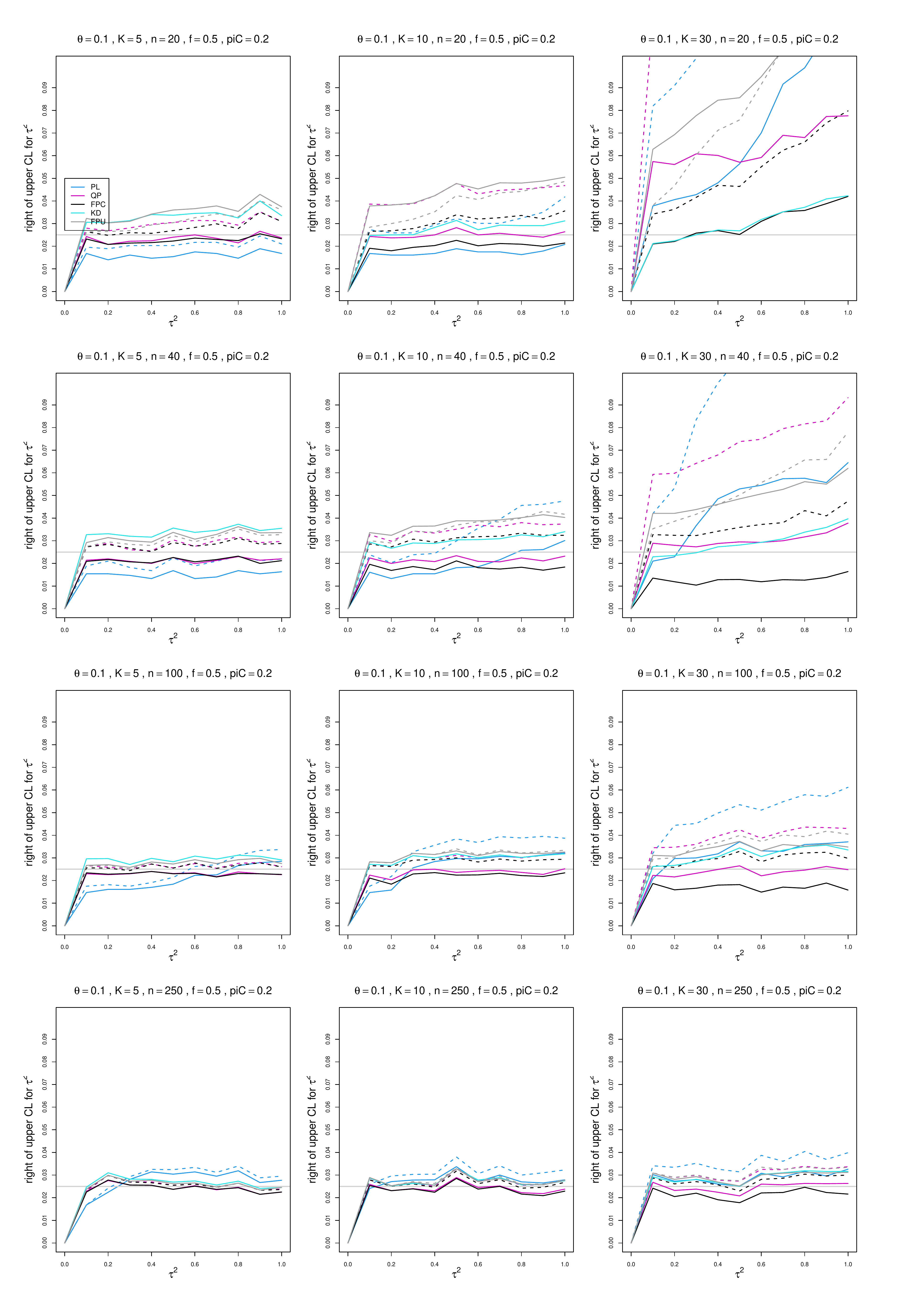}
	\caption{Miss-right probability  of  PL, QP, KD, FPC, and FPU 95\%  confidence intervals for between-study variance of LOR vs $\tau^2$, for equal sample sizes $n=20,\;40,\;100$ and $250$, $p_{iC} = .2$, $\theta=0.1$ and  $f=0.5$.   Solid lines: PL, QP, and FPC \lq\lq only", FPU model-based, and KD. Dashed lines:
PL, QP and FPC \lq\lq always" and FPU na\"{i}ve.   }
	\label{PlotCovRightOfTau2_piC_02theta=0.1_LOR_equal_sample_sizes}
\end{figure}

\begin{figure}[ht]
	\centering
	\includegraphics[scale=0.33]{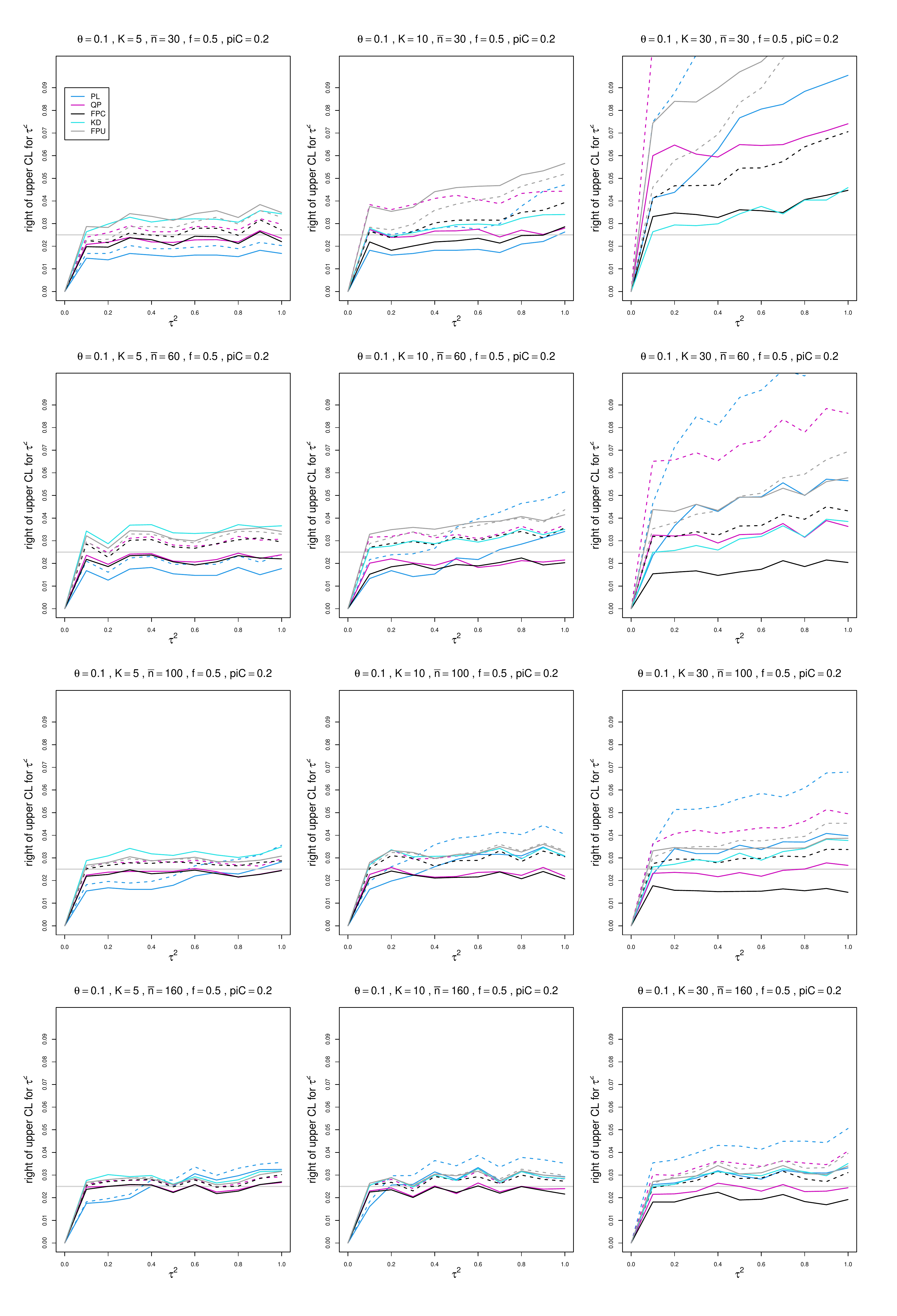}
	\caption{Miss-right probability  of  PL, QP, KD, FPC, and FPU 95\%  confidence intervals for between-study variance of LOR vs $\tau^2$, for unequal sample sizes $\bar{n}=30,\;60,\;100$ and $160$, $p_{iC} = .2$, $\theta=0.1$ and  $f=0.5$.   Solid lines: PL, QP, and FPC \lq\lq only", FPU model-based, and KD. Dashed lines: PL, QP, and FPC \lq\lq always" and FPU na\"{i}ve.   }
	\label{PlotCovRightOfTau2_piC_02theta=0.1_LOR_unequal_sample_sizes}
\end{figure}

\begin{figure}[ht]
	\centering
	\includegraphics[scale=0.33]{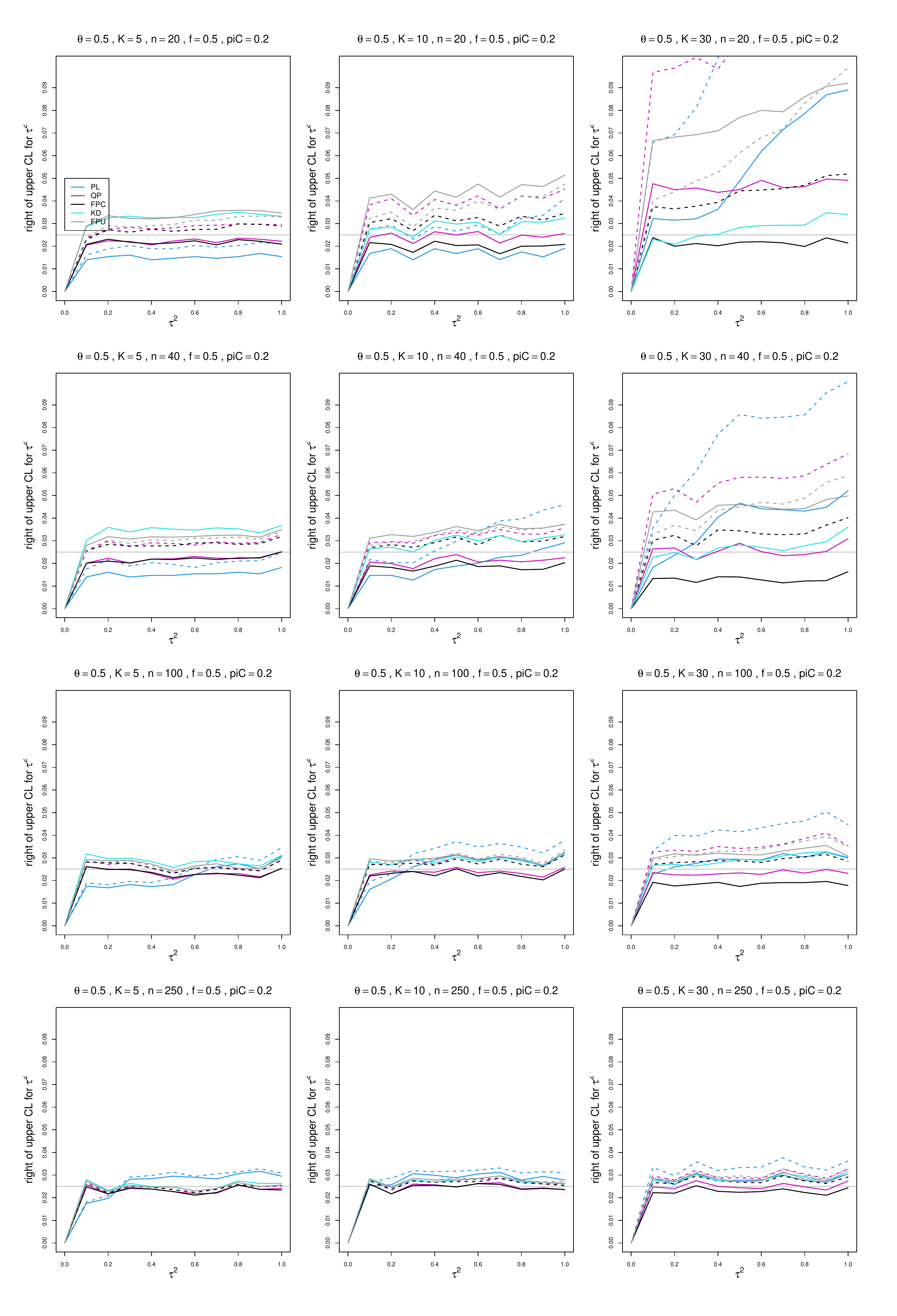}
	\caption{Miss-right probability  of  PL, QP, KD, FPC, and FPU 95\%  confidence intervals for between-study variance of LOR vs $\tau^2$, for equal sample sizes $n=20,\;40,\;100$ and $250$, $p_{iC} = .2$, $\theta=0.5$ and  $f=0.5$.   Solid lines: PL, QP, and FPC \lq\lq only", FPU model-based, and KD. Dashed lines:
PL, QP and FPC \lq\lq always" and FPU na\"{i}ve.   }
	\label{PlotCovRightOfTau2_piC_02theta=0.5_LOR_equal_sample_sizes}
\end{figure}

\begin{figure}[ht]
	\centering
	\includegraphics[scale=0.33]{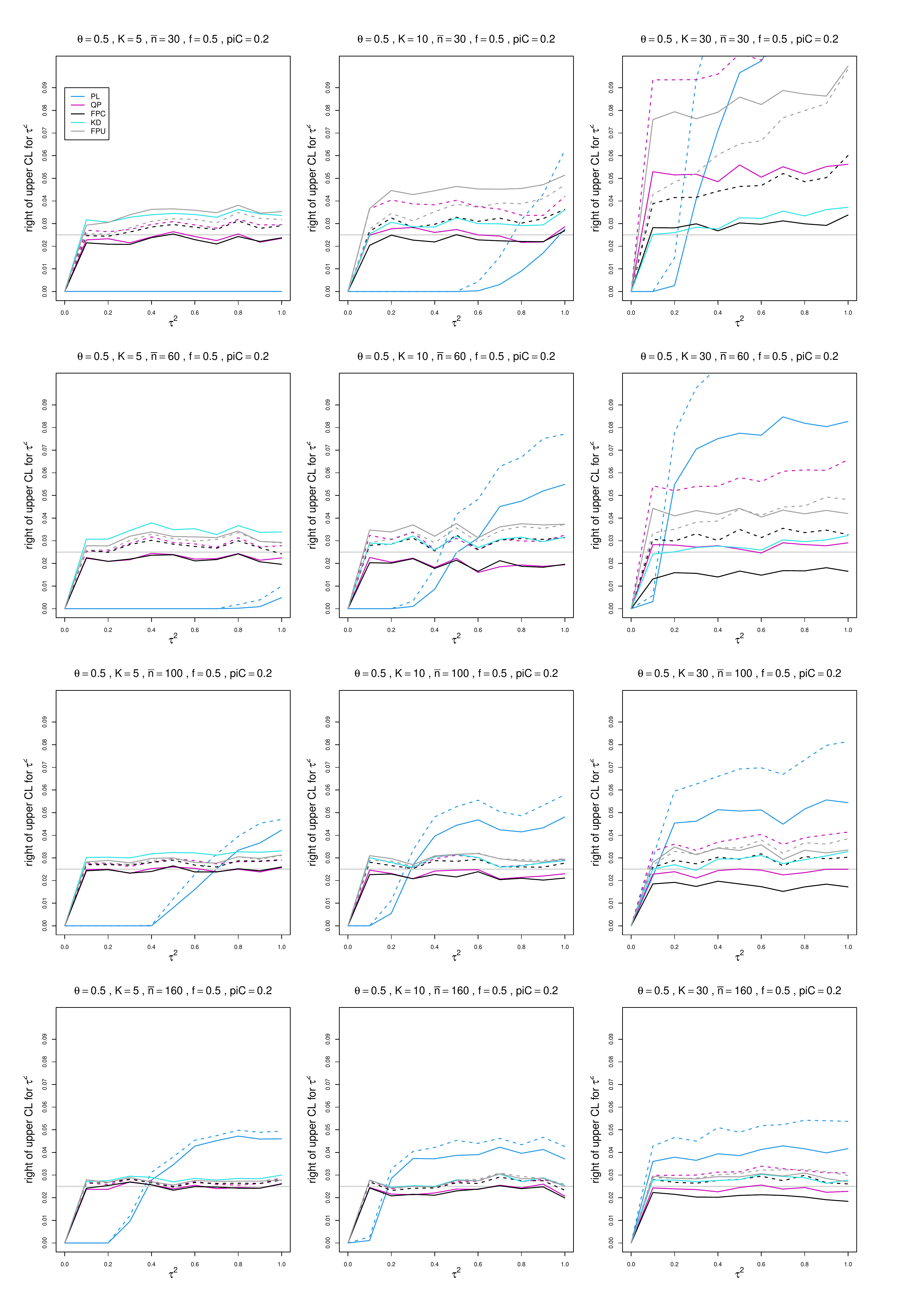}
	\caption{Miss-right probability  of  PL, QP, KD, FPC, and FPU 95\%  confidence intervals for between-study variance of LOR vs $\tau^2$, for unequal sample sizes $\bar{n}=30,\;60,\;100$ and $160$, $p_{iC} = .2$, $\theta=0.5$ and  $f=0.5$.   Solid lines: PL, QP, and FPC \lq\lq only", FPU model-based, and KD. Dashed lines: PL, QP, and FPC \lq\lq always" and FPU na\"{i}ve.    }
	\label{PlotCovRightOfTau2_piC_02theta=0.5_LOR_unequal_sample_sizes}
\end{figure}

\begin{figure}[ht]
	\centering
	\includegraphics[scale=0.33]{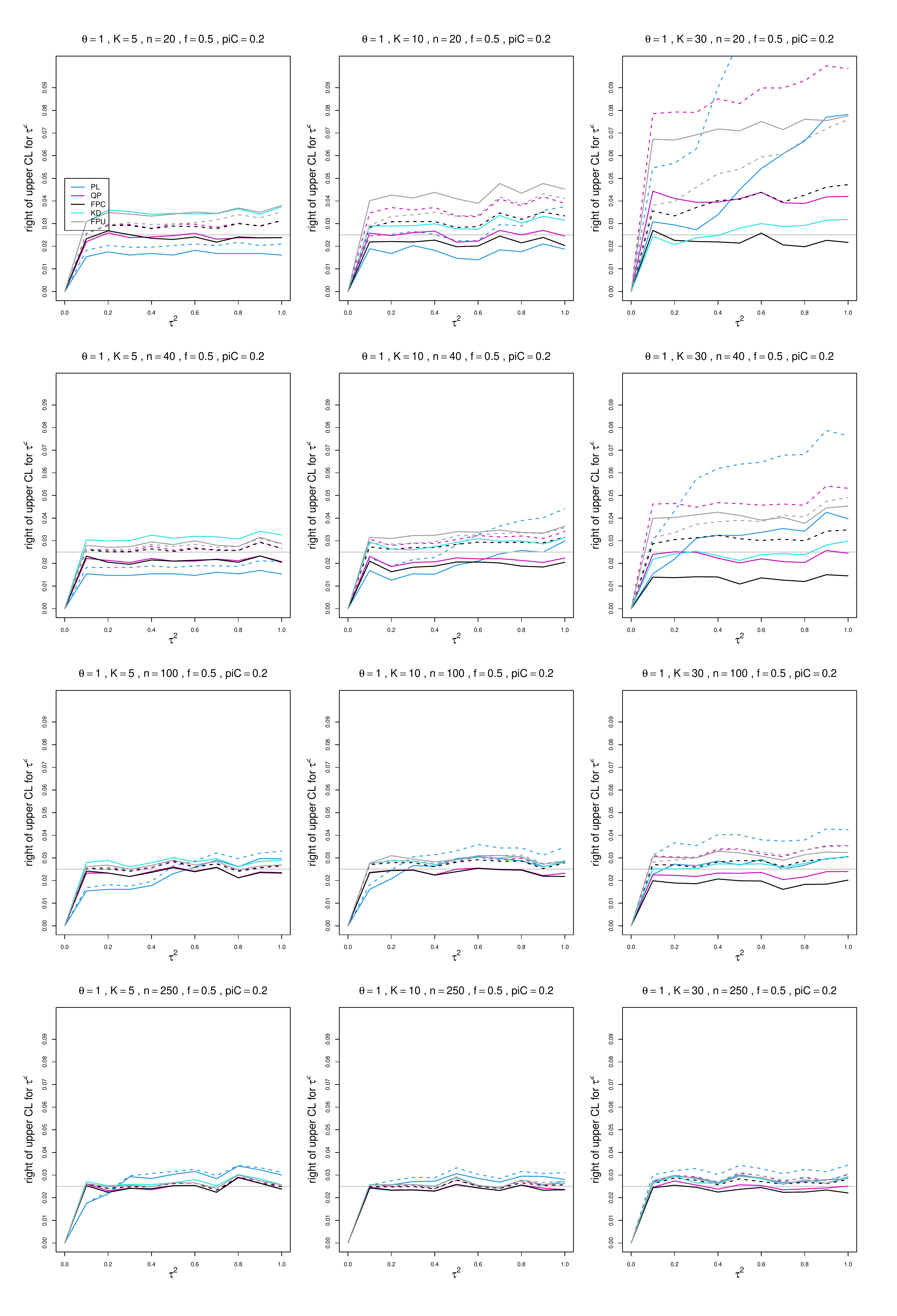}
	\caption{Miss-right probability  of  PL, QP, KD, FPC, and FPU 95\%  confidence intervals for between-study variance of LOR vs $\tau^2$, for equal sample sizes $n=20,\;40,\;100$ and $250$, $p_{iC} = .2$, $\theta=1$ and  $f=0.5$.   Solid lines: PL, QP, and FPC \lq\lq only", FPU model-based, and KD. Dashed lines:
PL, QP and FPC \lq\lq always" and FPU na\"{i}ve.    }
	\label{PlotCovRightOfTau2_piC_02theta=1_LOR_equal_sample_sizes}
\end{figure}

\begin{figure}[ht]
	\centering
	\includegraphics[scale=0.33]{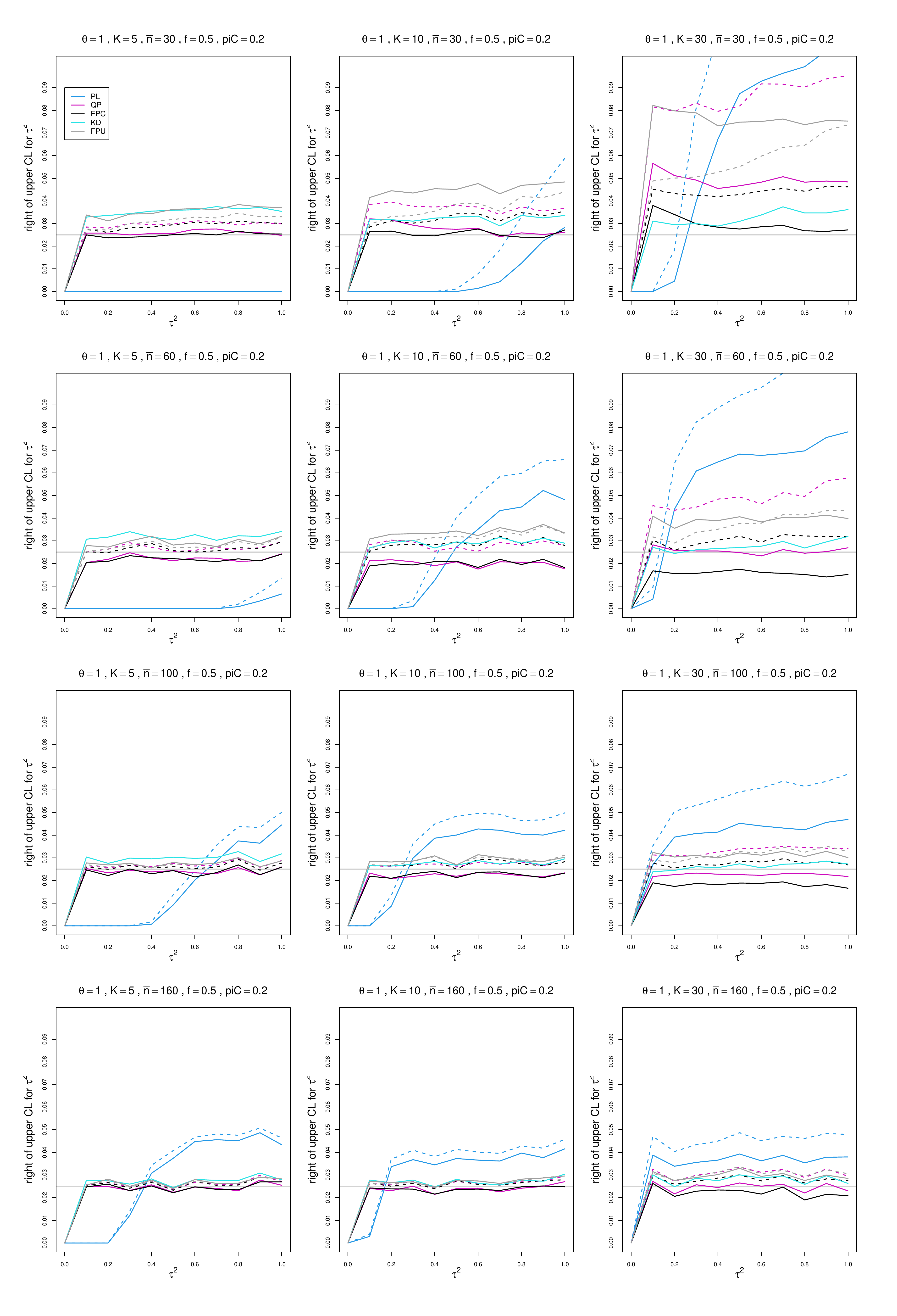}
	\caption{Miss-right probability  of  PL, QP, KD, FPC, and FPU 95\%  confidence intervals for between-study variance of LOR vs $\tau^2$, for unequal sample sizes $\bar{n}=30,\;60,\;100$ and $160$, $p_{iC} = .2$, $\theta=1$ and  $f=0.5$.   Solid lines: PL, QP, and FPC \lq\lq only", FPU model-based, and KD. Dashed lines: PL, QP, and FPC \lq\lq always" and FPU na\"{i}ve.    }
	\label{PlotCovRightOfTau2_piC_02theta=1_LOR_unequal_sample_sizes}
\end{figure}

\begin{figure}[ht]
	\centering
	\includegraphics[scale=0.33]{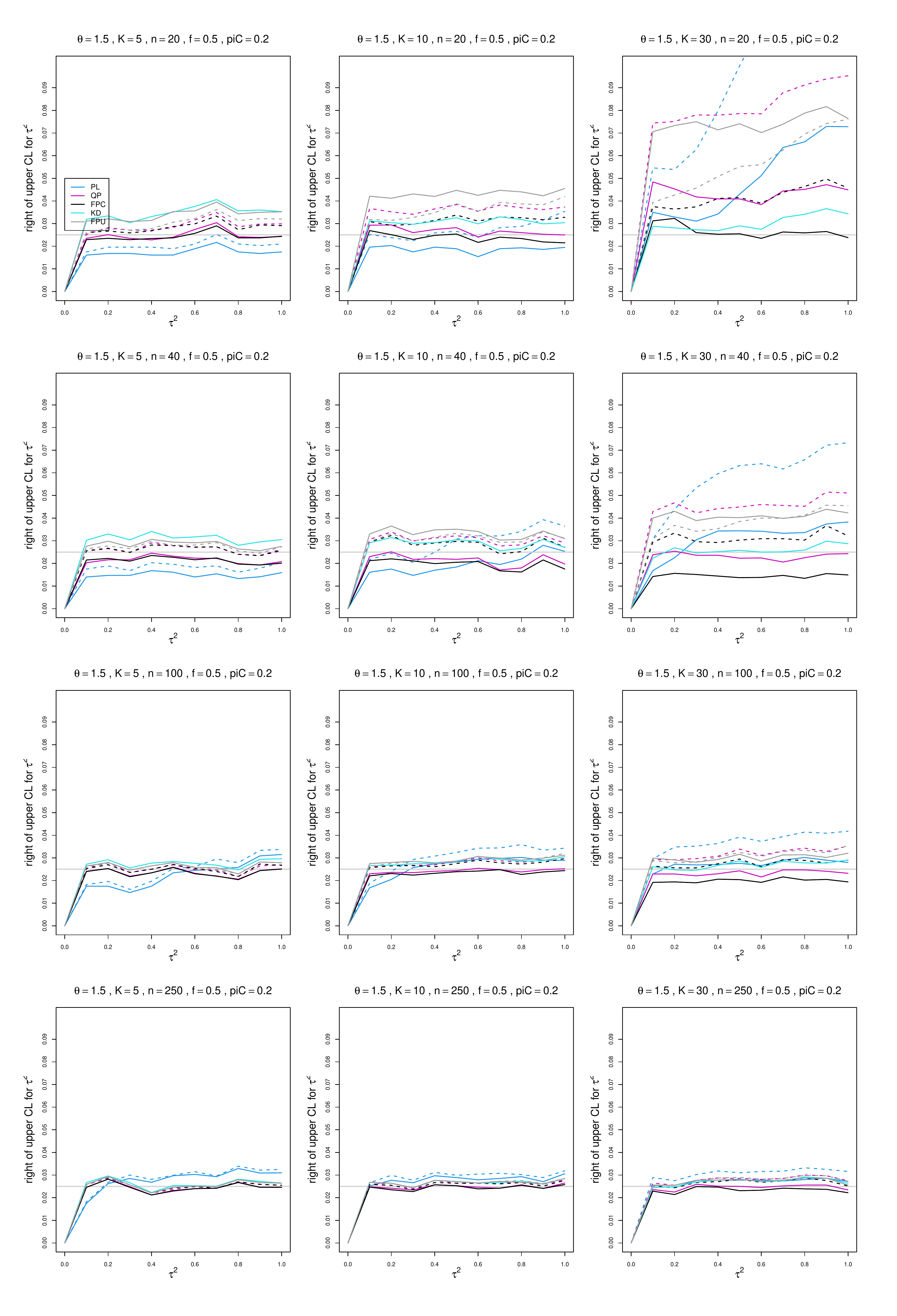}
	\caption{Miss-right probability  of  PL, QP, KD, FPC, and FPU 95\%  confidence intervals for between-study variance of LOR vs $\tau^2$, for equal sample sizes $n=20,\;40,\;100$ and $250$, $p_{iC} = .2$, $\theta=1.5$ and  $f=0.5$.   Solid lines: PL, QP, and FPC \lq\lq only", FPU model-based, and KD. Dashed lines:
PL, QP and FPC \lq\lq always" and FPU na\"{i}ve.    }
	\label{PlotCovRightOfTau2_piC_02theta=1.5_LOR_equal_sample_sizes}
\end{figure}

\begin{figure}[ht]
	\centering
	\includegraphics[scale=0.33]{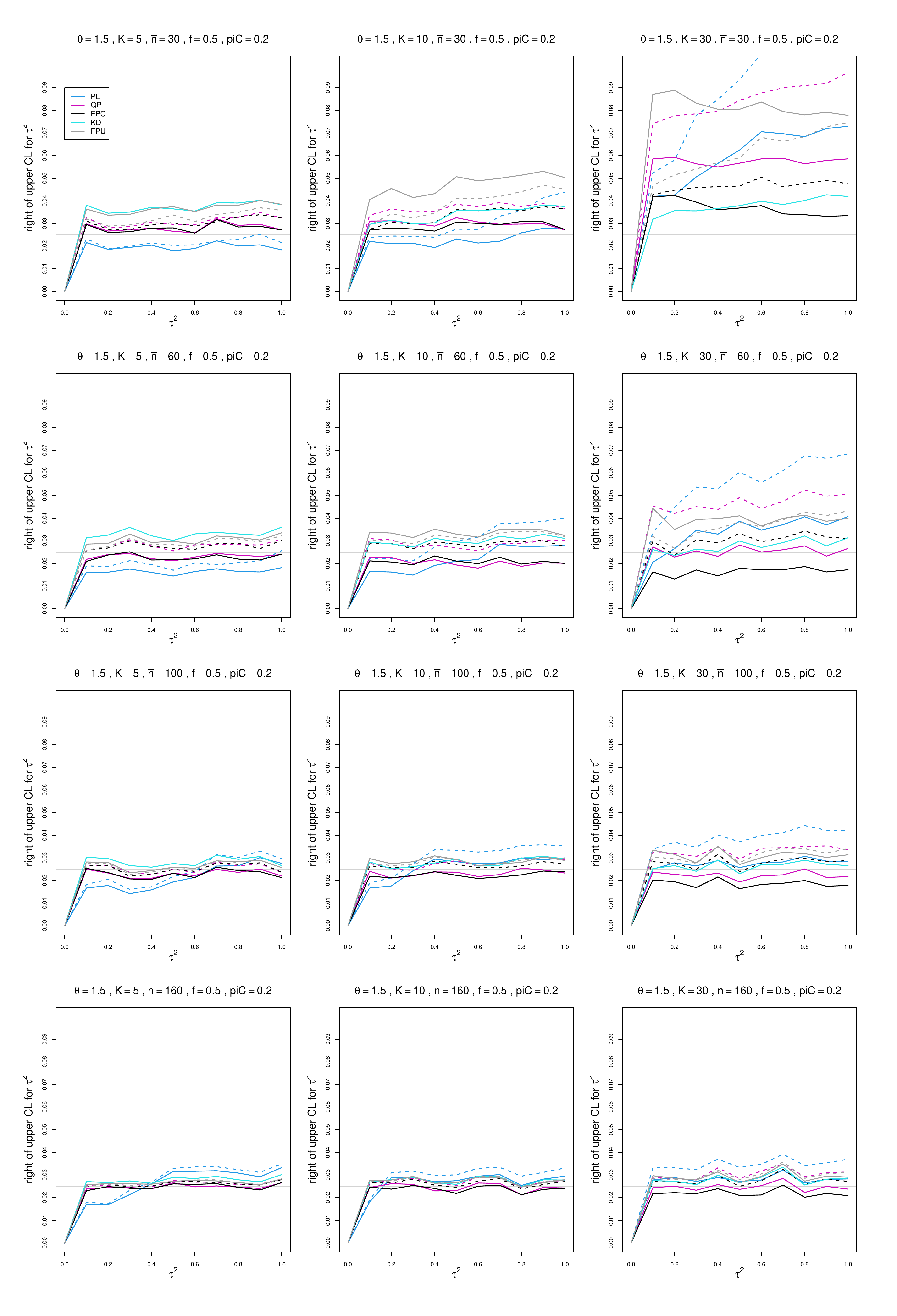}
	\caption{Miss-right probability  of  PL, QP, KD, FPC, and FPU 95\%  confidence intervals for between-study variance of LOR vs $\tau^2$, for unequal sample sizes $\bar{n}=30,\;60,\;100$ and $160$, $p_{iC} = .2$, $\theta=1.5$ and  $f=0.5$.   Solid lines: PL, QP, and FPC \lq\lq only", FPU model-based, and KD. Dashed lines: PL, QP, and FPC \lq\lq always" and FPU na\"{i}ve.   }
	\label{PlotCovRightOfTau2_piC_02theta=1.5_LOR_unequal_sample_sizes}
\end{figure}

\begin{figure}[ht]
	\centering
	\includegraphics[scale=0.33]{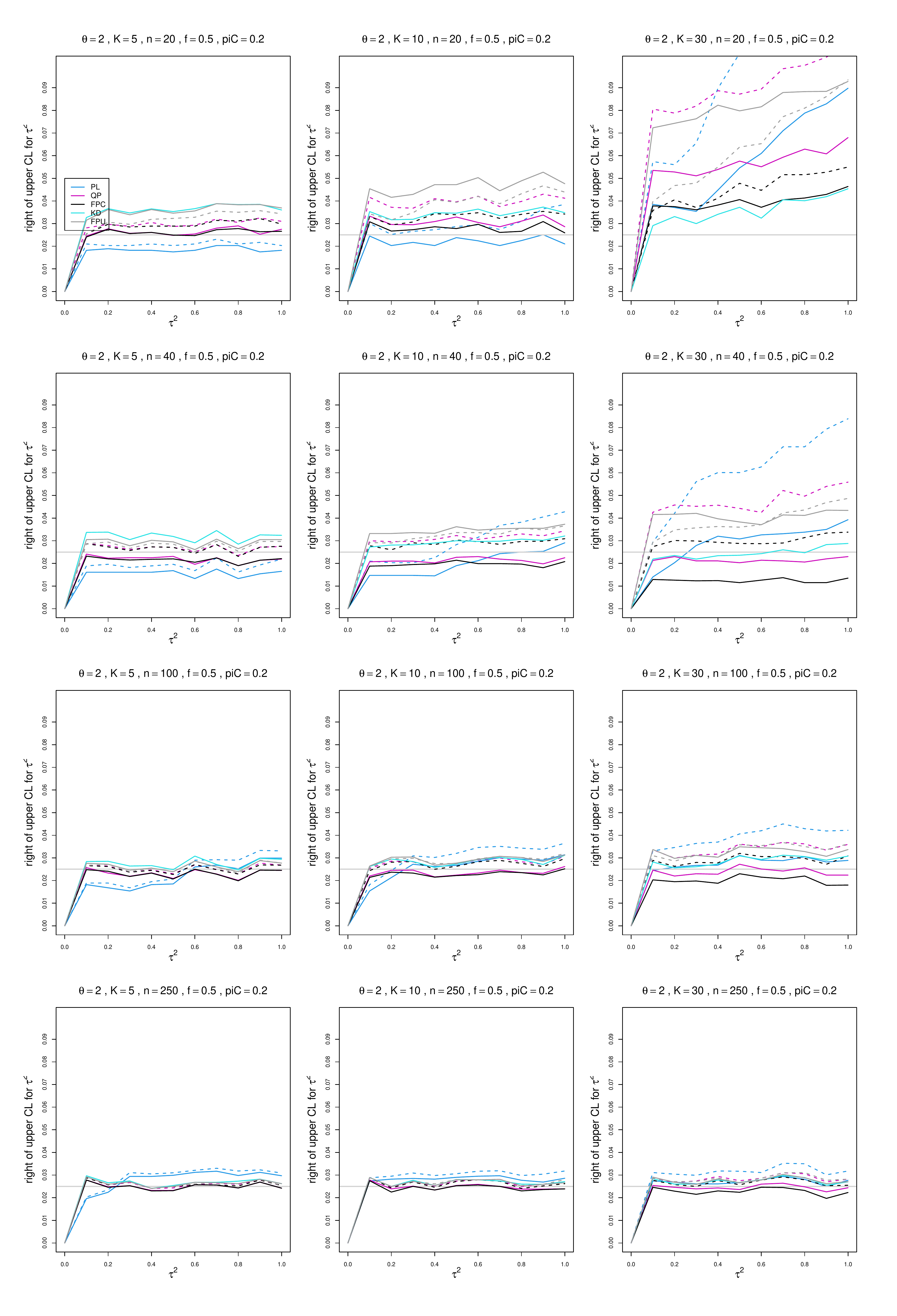}
	\caption{Miss-right probability  of  PL, QP, KD, FPC, and FPU 95\%  confidence intervals for between-study variance of LOR vs $\tau^2$, for equal sample sizes $n=20,\;40,\;100$ and $250$, $p_{iC} = .2$, $\theta=2$ and  $f=0.5$.   Solid lines: PL, QP, and FPC \lq\lq only", FPU model-based, and KD. Dashed lines:
PL, QP and FPC \lq\lq always" and FPU na\"{i}ve.   }
	\label{PlotCovRightOfTau2_piC_02theta=2_LOR_equal_sample_sizes}
\end{figure}

\begin{figure}[ht]
	\centering
	\includegraphics[scale=0.33]{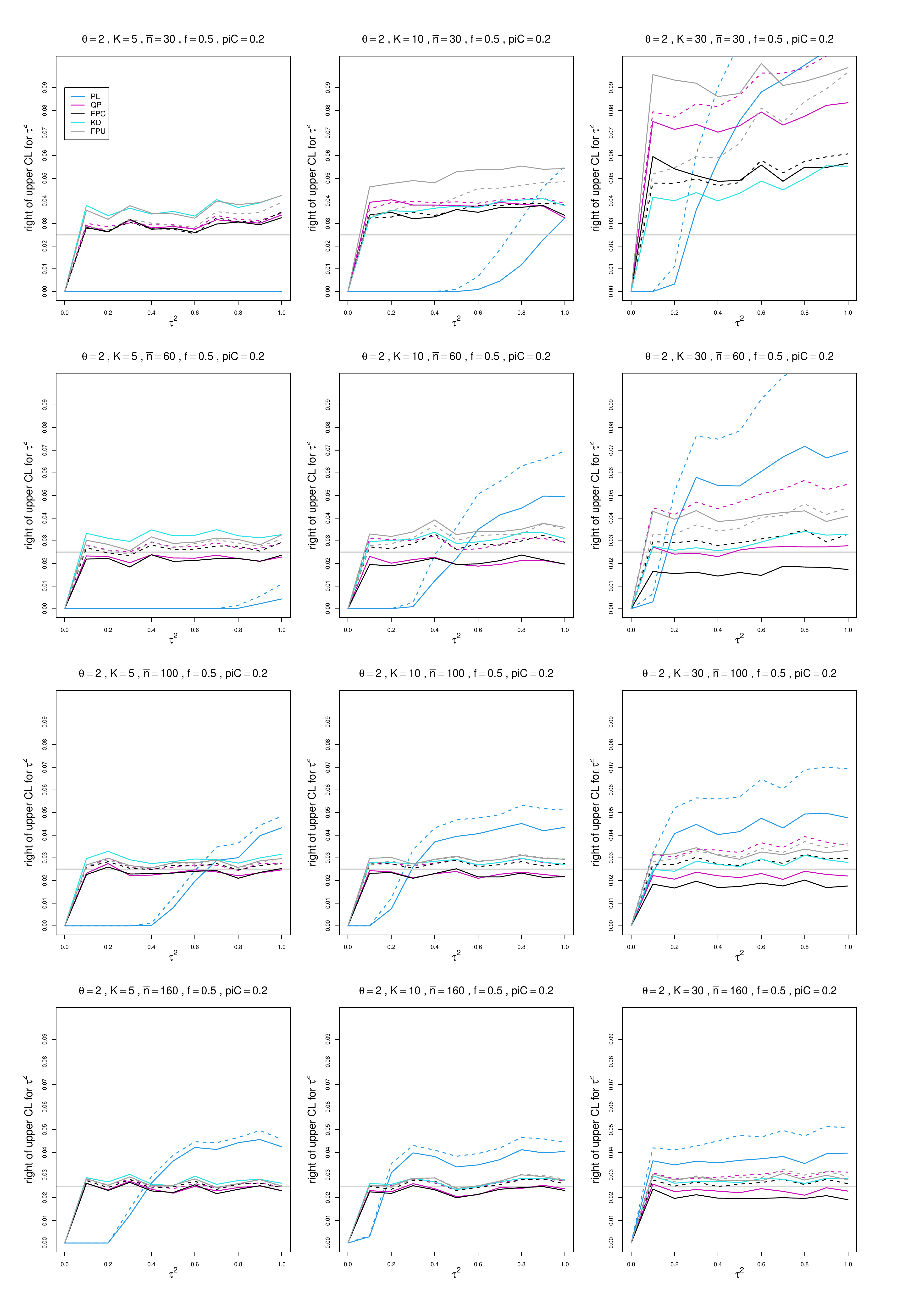}
	\caption{Miss-right probability  of  PL, QP, KD, FPC, and FPU 95\%  confidence intervals for between-study variance of LOR vs $\tau^2$, for unequal sample sizes $\bar{n}=30,\;60,\;100$ and $160$, $p_{iC} = .2$, $\theta=2$ and  $f=0.5$.   Solid lines: PL, QP, and FPC \lq\lq only", FPU model-based, and KD. Dashed lines: PL, QP, and FPC \lq\lq always" and FPU na\"{i}ve.   }
	\label{PlotCovRightOfTau2_piC_02theta=2_LOR_unequal_sample_sizes}
\end{figure}

\begin{figure}[ht]
	\centering
	\includegraphics[scale=0.33]{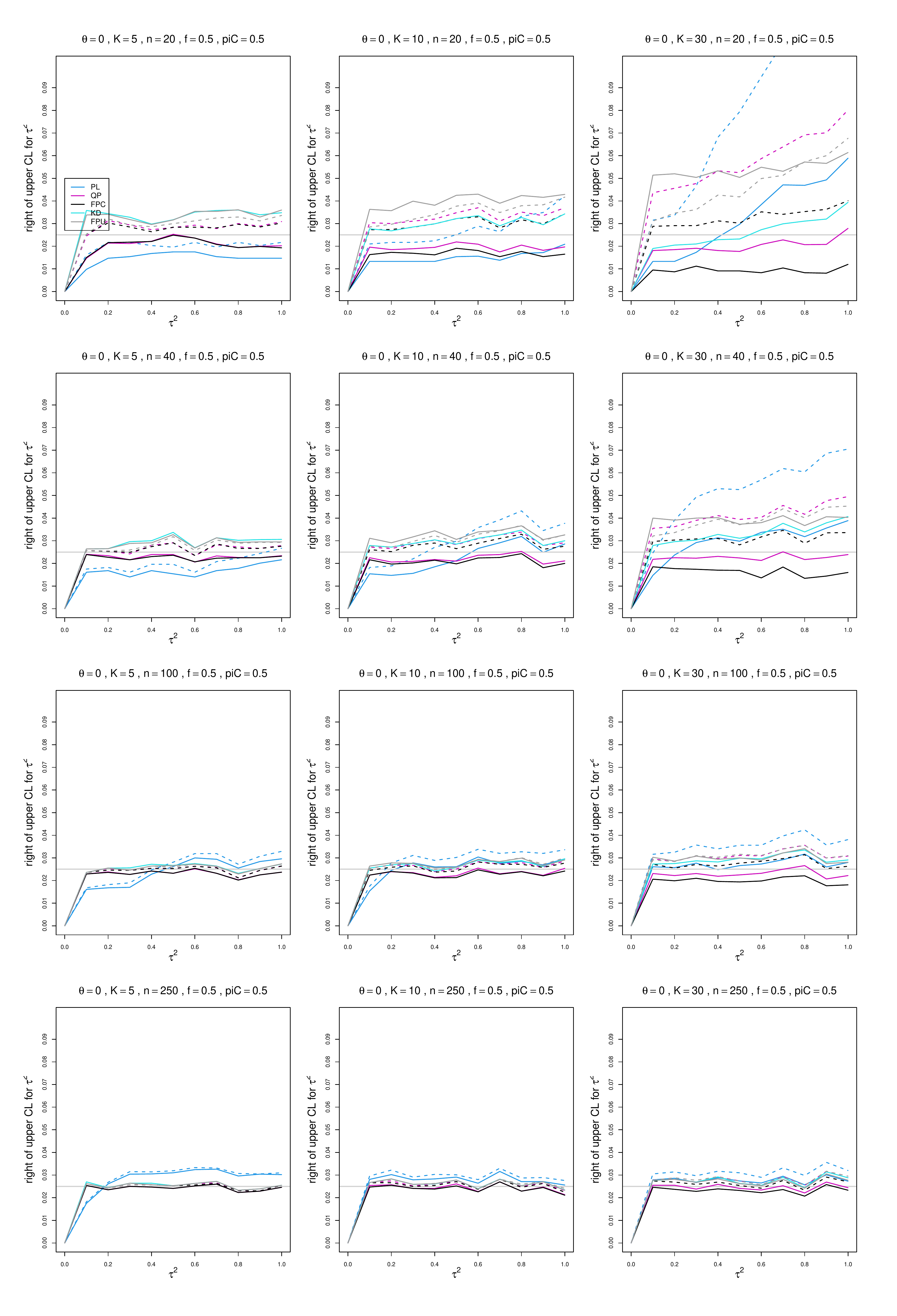}
	\caption{Miss-right probability  of  PL, QP, KD, FPC, and FPU 95\%  confidence intervals for between-study variance of LOR vs $\tau^2$, for equal sample sizes $n=20,\;40,\;100$ and $250$, $p_{iC} = .5$, $\theta=0$ and  $f=0.5$.   Solid lines: PL, QP, and FPC \lq\lq only", FPU model-based, and KD. Dashed lines:
PL, QP and FPC \lq\lq always" and FPU na\"{i}ve.   }
	\label{PlotCovRightOfTau2_piC_05theta=0_LOR_equal_sample_sizes}
\end{figure}

\begin{figure}[ht]
	\centering
	\includegraphics[scale=0.33]{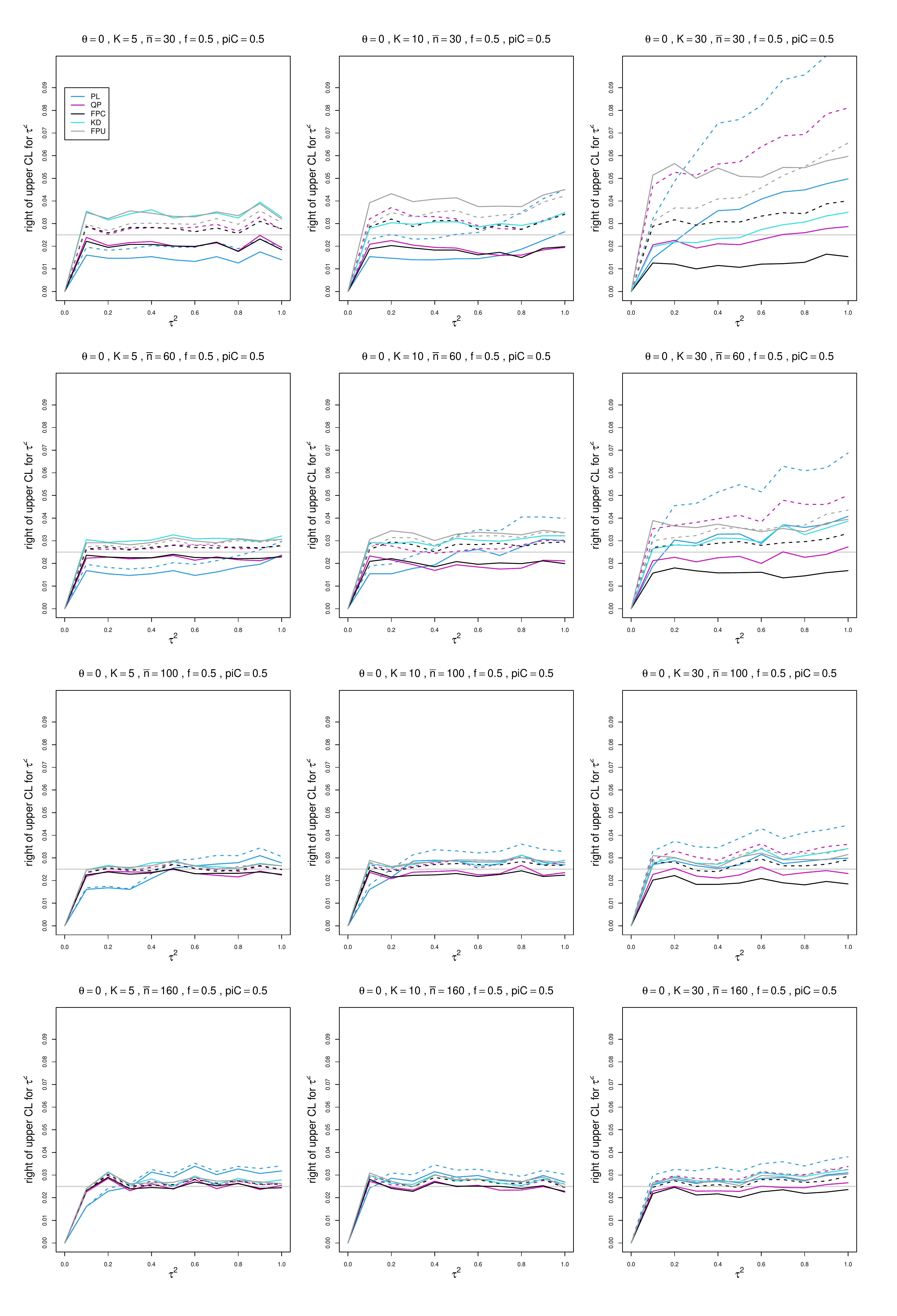}
	\caption{Miss-right probability  of  PL, QP, KD, FPC, and FPU 95\%  confidence intervals for between-study variance of LOR vs $\tau^2$, for unequal sample sizes $\bar{n}=30,\;60,\;100$ and $160$, $p_{iC} = .5$, $\theta=0$ and  $f=0.5$.   Solid lines: PL, QP, and FPC \lq\lq only", FPU model-based, and KD. Dashed lines: PL, QP, and FPC \lq\lq always" and FPU na\"{i}ve.    }
	\label{PlotCovRightOfTau2_piC_05theta=0_LOR_unequal_sample_sizes}
\end{figure}

\begin{figure}[ht]
	\centering
	\includegraphics[scale=0.33]{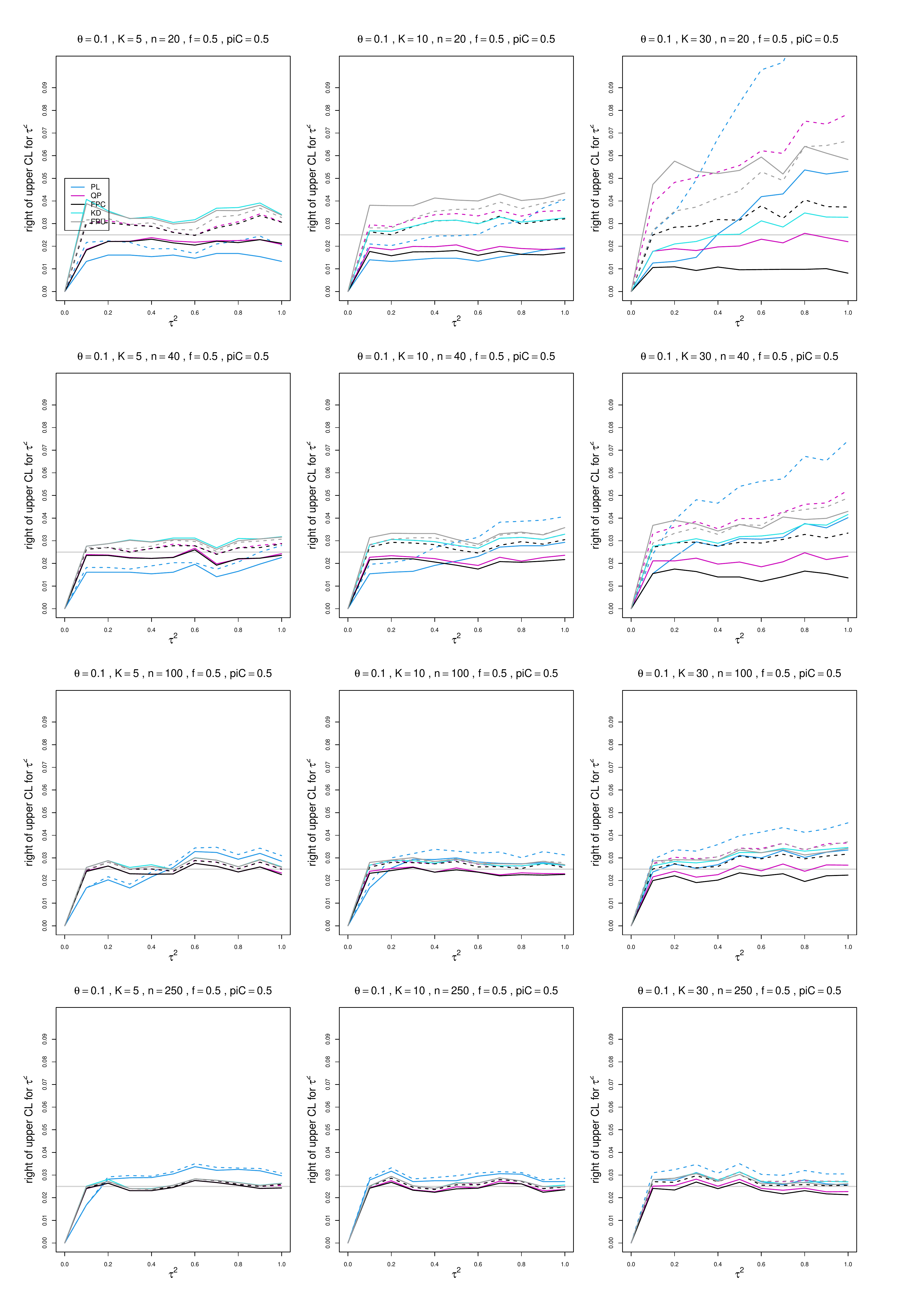}
	\caption{Miss-right probability  of  PL, QP, KD, FPC, and FPU 95\%  confidence intervals for between-study variance of LOR vs $\tau^2$, for equal sample sizes $n=20,\;40,\;100$ and $250$, $p_{iC} = .5$, $\theta=0.1$ and  $f=0.5$.   Solid lines: PL, QP, and FPC \lq\lq only", FPU model-based, and KD. Dashed lines:
PL, QP and FPC \lq\lq always" and FPU na\"{i}ve.    }
	\label{PlotCovRightOfTau2_piC_05theta=0.1_LOR_equal_sample_sizes}
\end{figure}

\begin{figure}[ht]
	\centering
	\includegraphics[scale=0.33]{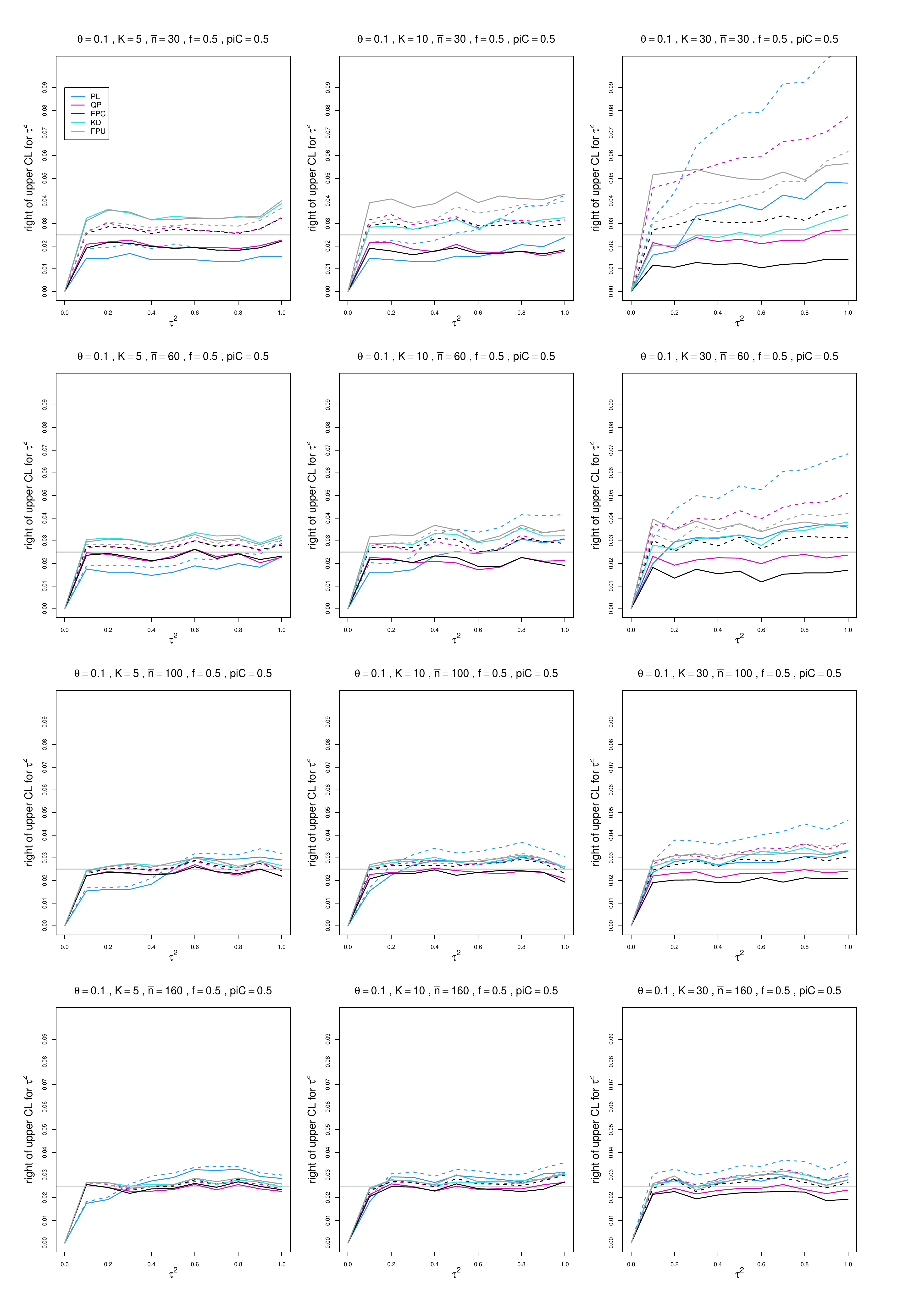}
	\caption{Miss-right probability  of  PL, QP, KD, FPC, and FPU 95\%  confidence intervals for between-study variance of LOR vs $\tau^2$, for unequal sample sizes $\bar{n}=30,\;60,\;100$ and $160$, $p_{iC} = .5$, $\theta=0.1$ and  $f=0.5$.   Solid lines: PL, QP, and FPC \lq\lq only", FPU model-based, and KD. Dashed lines: PL, QP, and FPC \lq\lq always" and FPU na\"{i}ve.   }
	\label{PlotCovRightOfTau2_piC_05theta=0.1_LOR_unequal_sample_sizes}
\end{figure}

\begin{figure}[ht]
	\centering
	\includegraphics[scale=0.33]{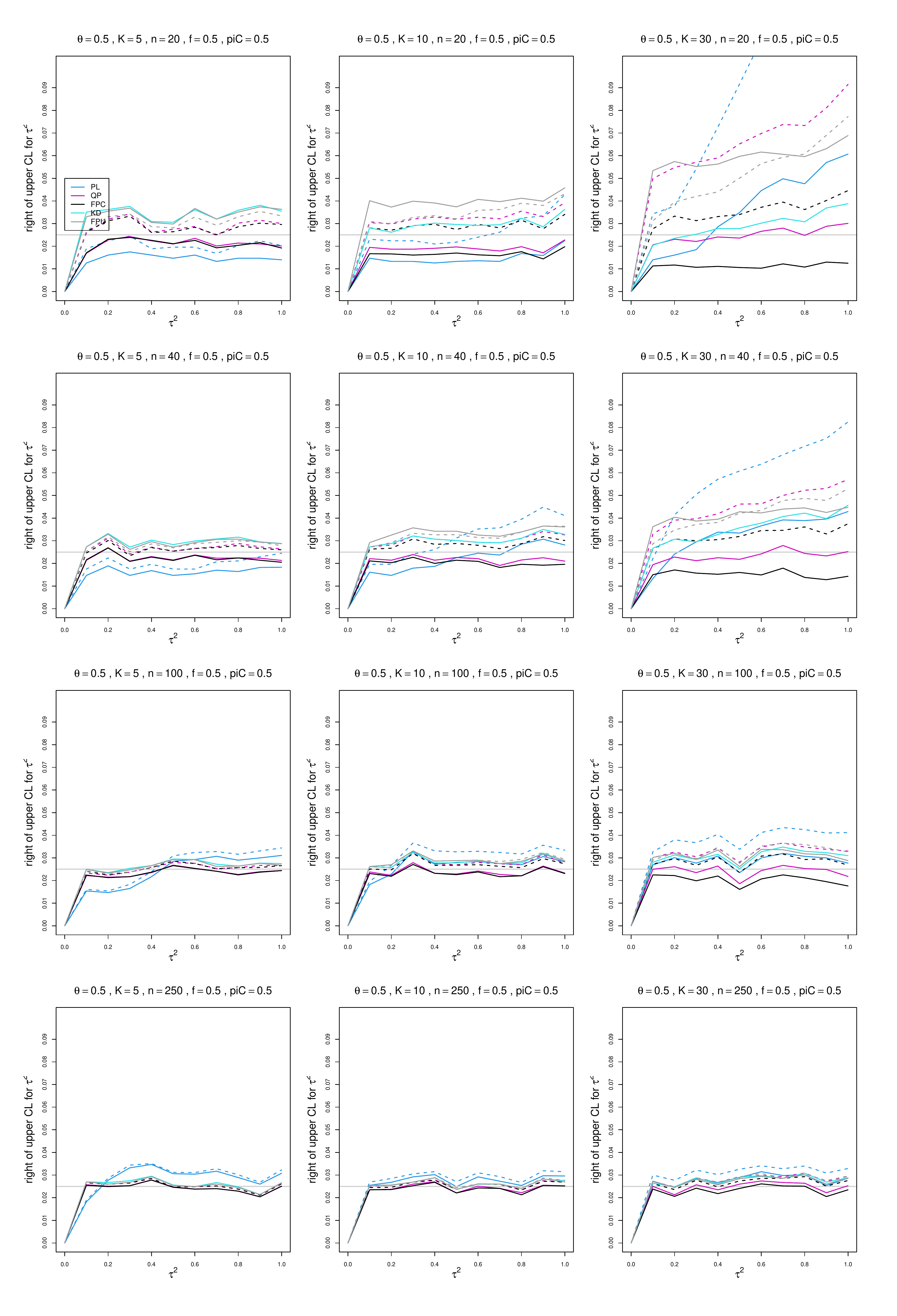}
	\caption{Miss-right probability  of  PL, QP, KD, FPC, and FPU 95\%  confidence intervals for between-study variance of LOR vs $\tau^2$, for equal sample sizes $n=20,\;40,\;100$ and $250$, $p_{iC} = .5$, $\theta=0.5$ and  $f=0.5$.   Solid lines: PL, QP, and FPC \lq\lq only", FPU model-based, and KD. Dashed lines:
PL, QP and FPC \lq\lq always" and FPU na\"{i}ve.   }
	\label{PlotCovRightOfTau2_piC_05theta=0.5_LOR_equal_sample_sizes}
\end{figure}
\begin{figure}[ht]
	\centering
	\includegraphics[scale=0.33]{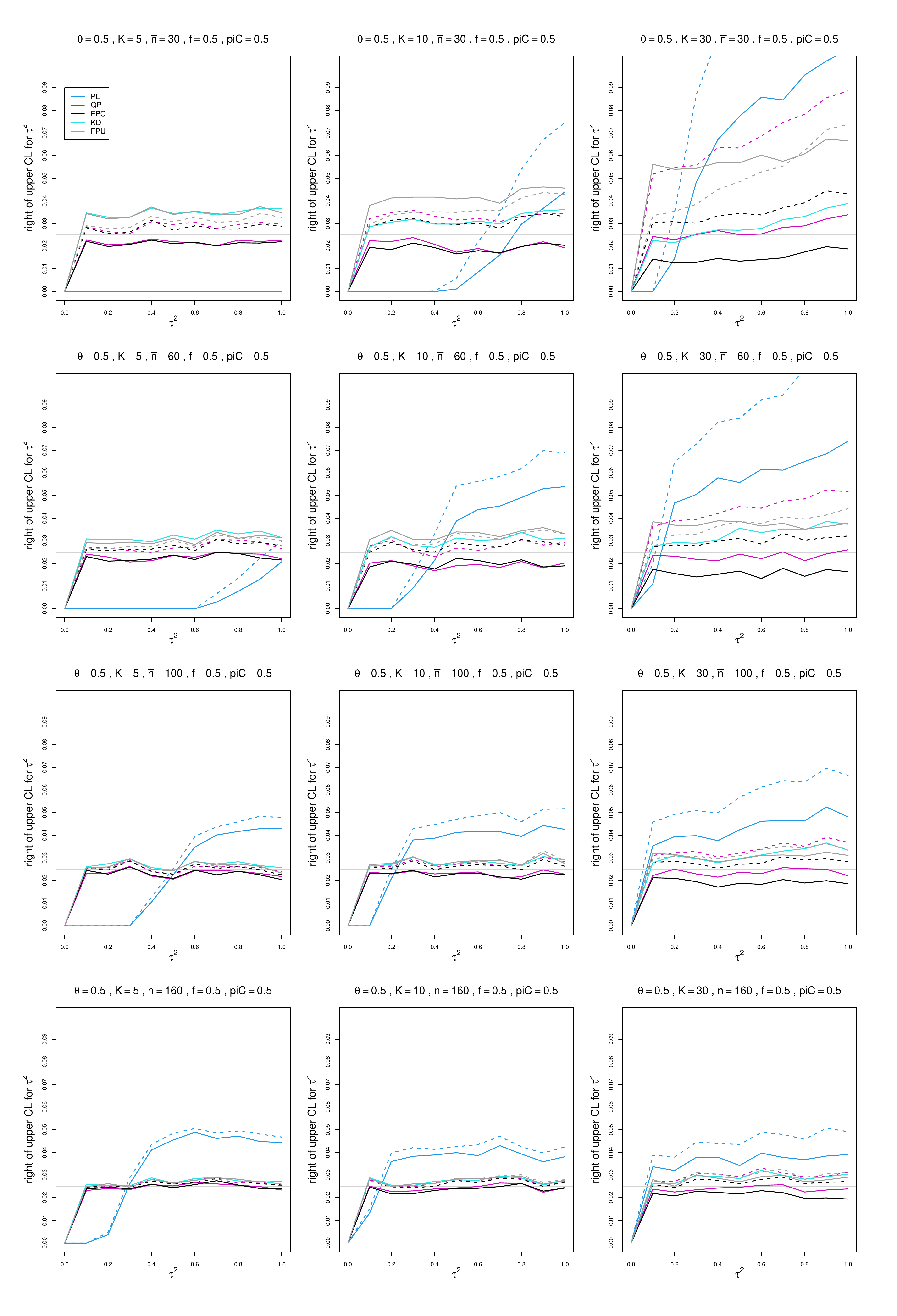}
	\caption{Miss-right probability  of  PL, QP, KD, FPC, and FPU 95\%  confidence intervals for between-study variance of LOR vs $\tau^2$, for unequal sample sizes $\bar{n}=30,\;60,\;100$ and $160$, $p_{iC} = .5$, $\theta=0.5$ and  $f=0.5$.   Solid lines: PL, QP, and FPC \lq\lq only", FPU model-based, and KD. Dashed lines: PL, QP, and FPC \lq\lq always" and FPU na\"{i}ve.   }
	\label{PlotCovRightOfTau2_piC_05theta=0.5_LOR_unequal_sample_sizes}
\end{figure}

\begin{figure}[ht]
	\centering
	\includegraphics[scale=0.33]{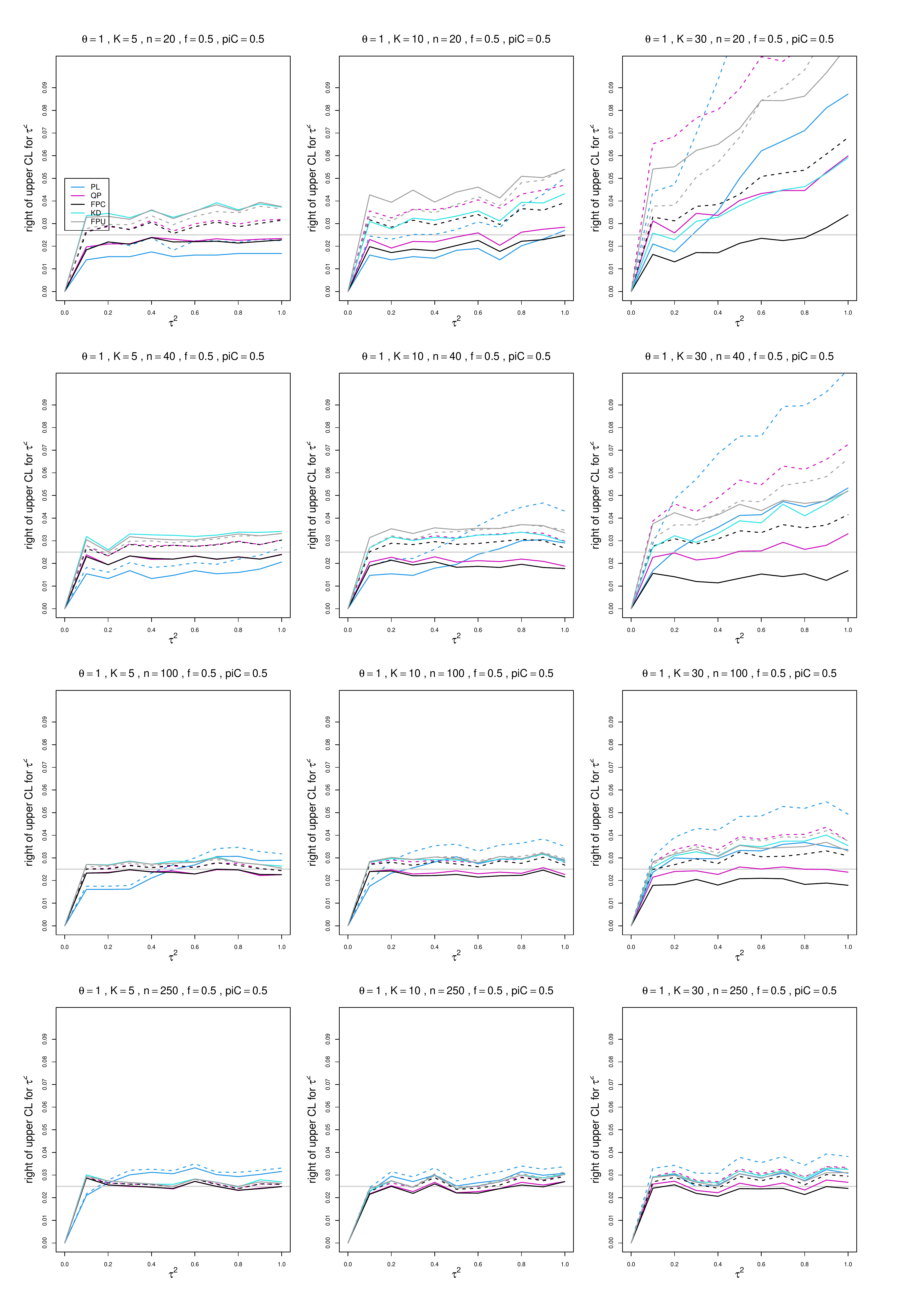}
	\caption{Miss-right probability  of  PL, QP, KD, FPC, and FPU 95\%  confidence intervals for between-study variance of LOR vs $\tau^2$, for equal sample sizes $n=20,\;40,\;100$ and $250$, $p_{iC} = .5$, $\theta=1$ and  $f=0.5$.   Solid lines: PL, QP, and FPC \lq\lq only", FPU model-based, and KD. Dashed lines:
PL, QP and FPC \lq\lq always" and FPU na\"{i}ve.   }
	\label{PlotCovRightOfTau2_piC_05theta=1_LOR_equal_sample_sizes}
\end{figure}

\begin{figure}[ht]
	\centering
	\includegraphics[scale=0.33]{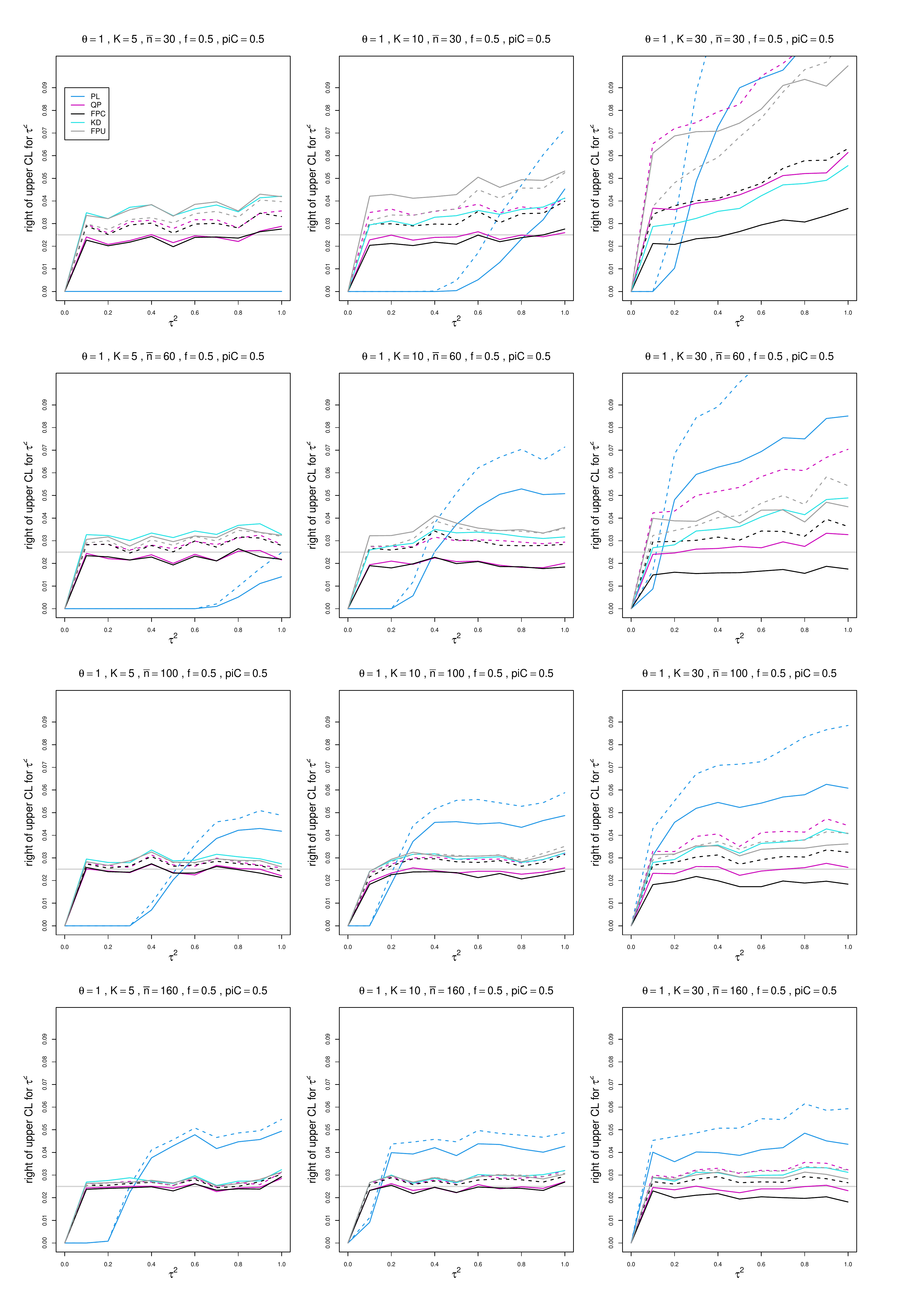}
	\caption{Miss-right probability  of  PL, QP, KD, FPC, and FPU 95\%  confidence intervals for between-study variance of LOR vs $\tau^2$, for unequal sample sizes $\bar{n}=30,\;60,\;100$ and $160$, $p_{iC} = .5$, $\theta=1$ and  $f=0.5$.   Solid lines: PL, QP, and FPC \lq\lq only", FPU model-based, and KD. Dashed lines: PL, QP, and FPC \lq\lq always" and FPU na\"{i}ve.   }
	\label{PlotCovRightOfTau2_piC_05theta=1_LOR_unequal_sample_sizes}
\end{figure}

\begin{figure}[ht]
	\centering
	\includegraphics[scale=0.33]{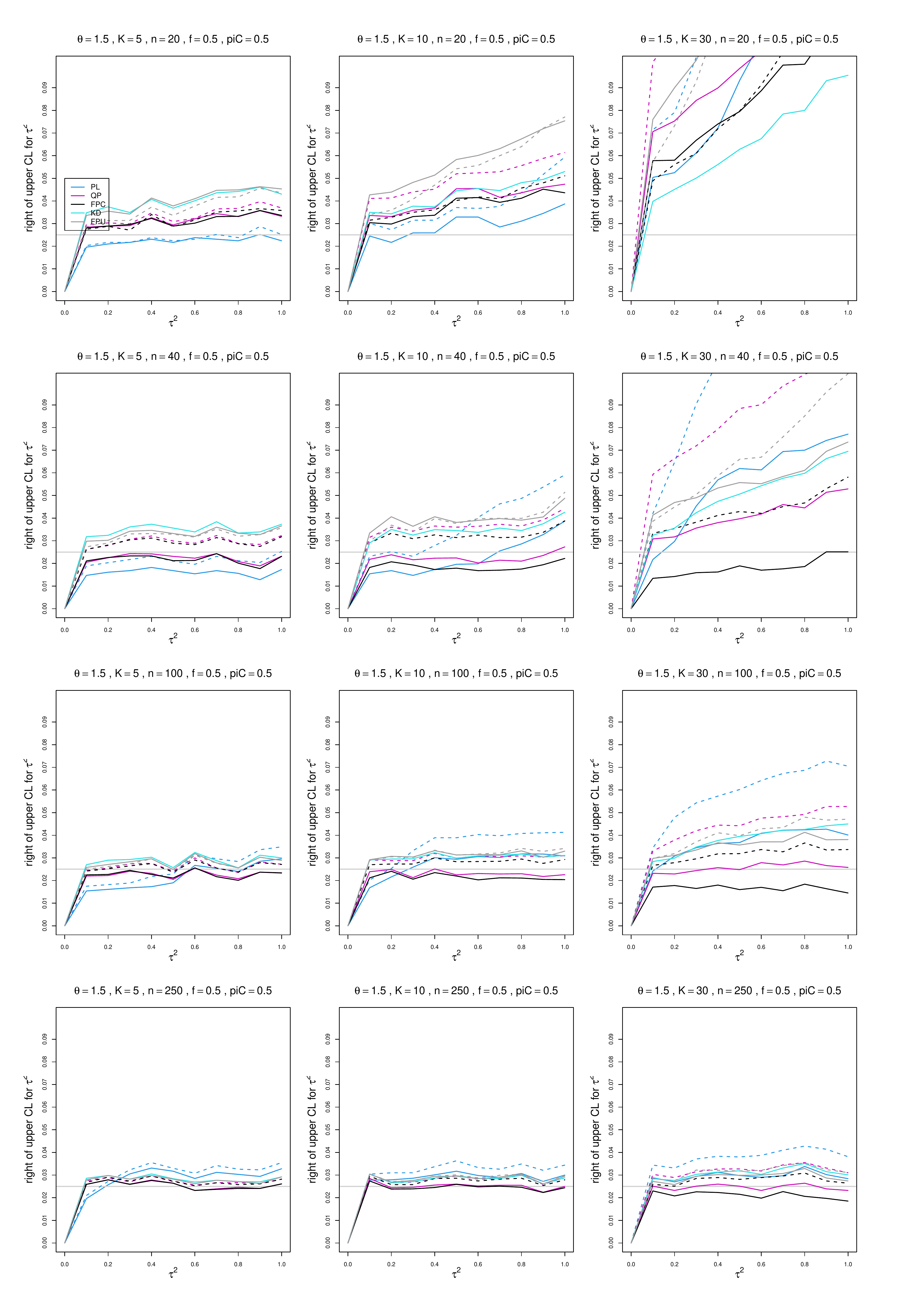}
	\caption{Miss-right probability  of  PL, QP, KD, FPC, and FPU 95\%  confidence intervals for between-study variance of LOR vs $\tau^2$, for equal sample sizes $n=20,\;40,\;100$ and $250$, $p_{iC} = .5$, $\theta=1.5$ and  $f=0.5$.   Solid lines: PL, QP, and FPC \lq\lq only", FPU model-based, and KD. Dashed lines:
PL, QP and FPC \lq\lq always" and FPU na\"{i}ve.    }
	\label{PlotCovRightOfTau2_piC_05theta=1.5_LOR_equal_sample_sizes}
\end{figure}

\begin{figure}[ht]
	\centering
	\includegraphics[scale=0.33]{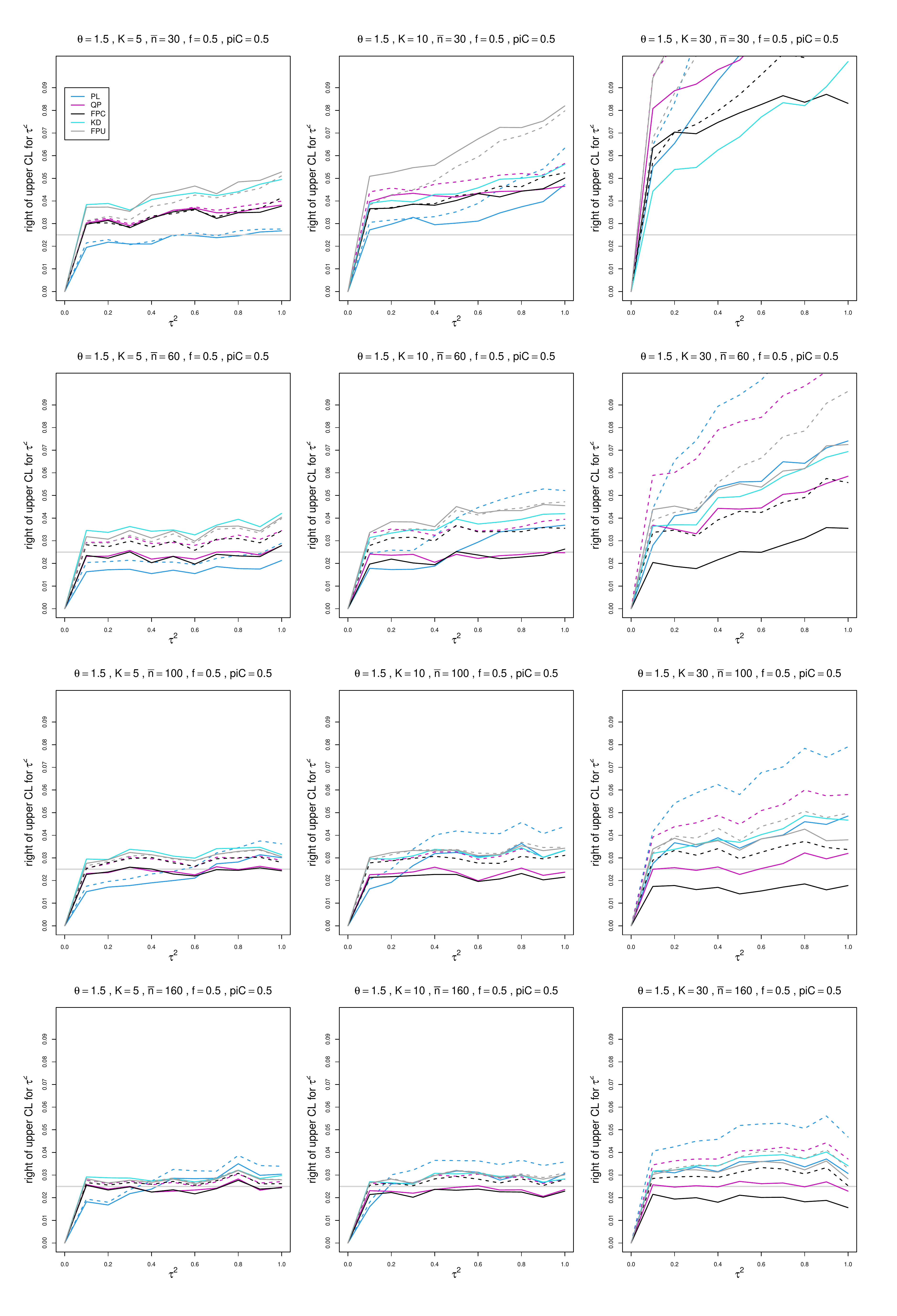}
	\caption{Miss-right probability  of  PL, QP, KD, FPC, and FPU 95\%  confidence intervals for between-study variance of LOR vs $\tau^2$, for unequal sample sizes $\bar{n}=30,\;60,\;100$ and $160$, $p_{iC} = .5$, $\theta=1.5$ and  $f=0.5$.   Solid lines: PL, QP, and FPC \lq\lq only", FPU model-based, and KD. Dashed lines: PL, QP, and FPC \lq\lq always" and FPU na\"{i}ve.   }
	\label{PlotCovRightOfTau2_piC_05theta=1.5_LOR_unequal_sample_sizes}
\end{figure}

\begin{figure}[ht]
	\centering
	\includegraphics[scale=0.33]{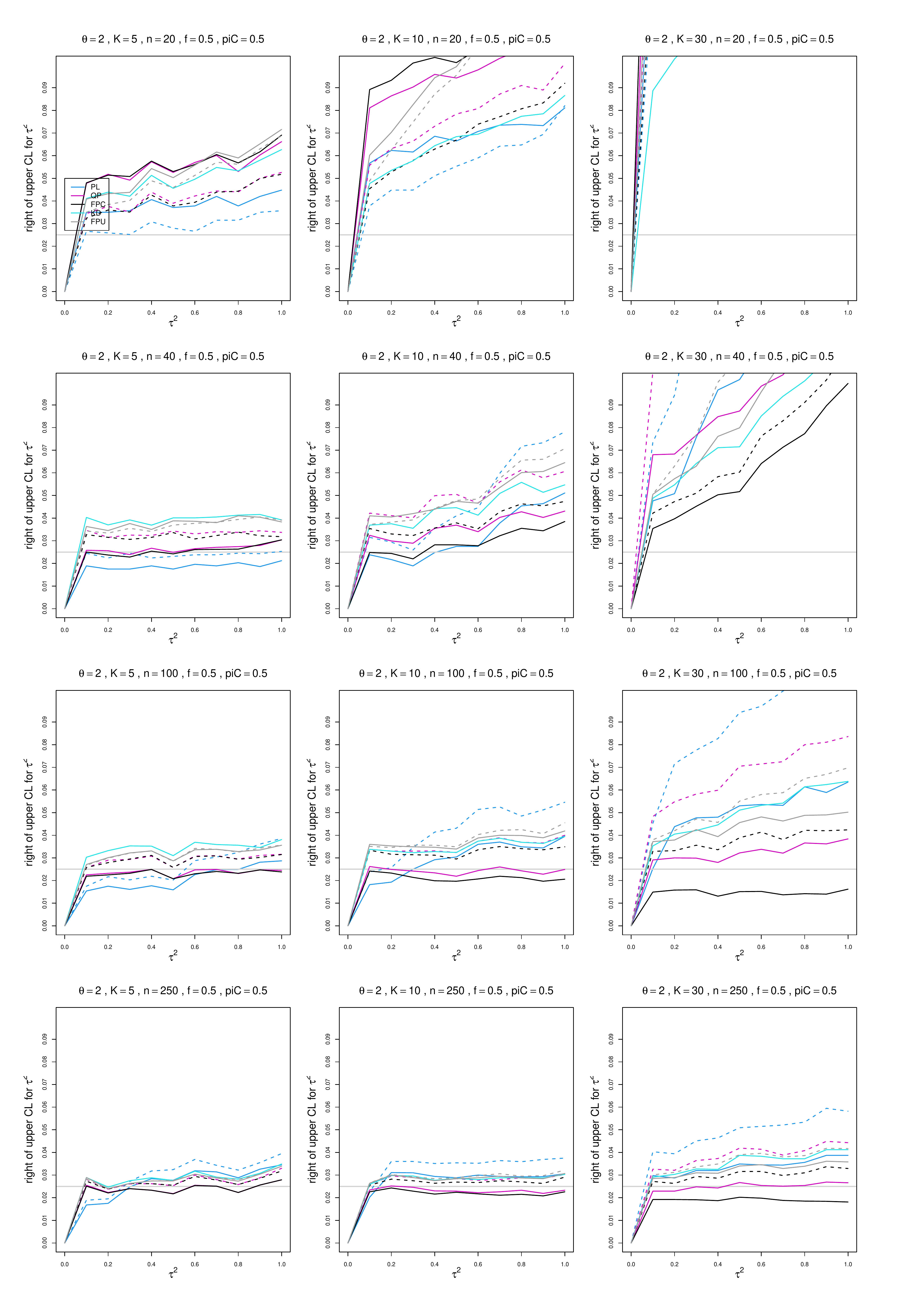}
	\caption{Miss-right probability  of  PL, QP, KD, FPC, and FPU 95\%  confidence intervals for between-study variance of LOR vs $\tau^2$, for equal sample sizes $n=20,\;40,\;100$ and $250$, $p_{iC} = .5$, $\theta=2$ and  $f=0.5$.   Solid lines: PL, QP, and FPC \lq\lq only", FPU model-based, and KD. Dashed lines:
PL, QP and FPC \lq\lq always" and FPU na\"{i}ve.   }
	\label{PlotCovRightOfTau2_piC_05theta=2_LOR_equal_sample_sizes}
\end{figure}

\begin{figure}[ht]
	\centering
	\includegraphics[scale=0.33]{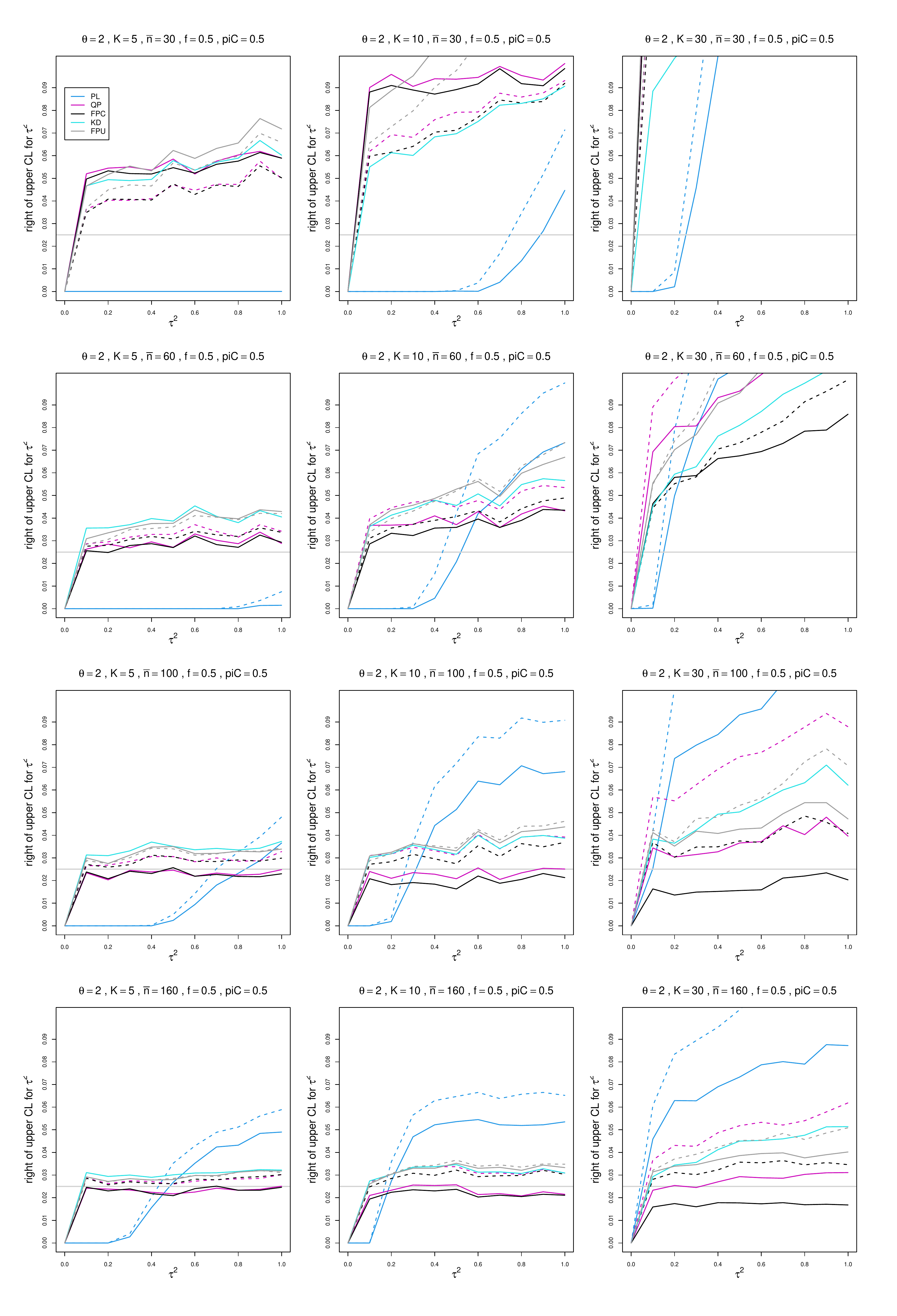}
	\caption{Miss-right probability  of  PL, QP, KD, FPC, and FPU 95\%  confidence intervals for between-study variance of LOR vs $\tau^2$, for unequal sample sizes $\bar{n}=30,\;60,\;100$ and $160$, $p_{iC} = .5$, $\theta=2$ and  $f=0.5$.   Solid lines: PL, QP, and FPC \lq\lq only", FPU model-based, and KD. Dashed lines: PL, QP, and FPC \lq\lq always" and FPU na\"{i}ve.   }
	\label{PlotCovRightOfTau2_piC_05theta=2_LOR_unequal_sample_sizes}
\end{figure}

\clearpage

\end{document}